\documentclass[a4paper,11pt]{article}

\usepackage{PRIMEarxiv}
\usepackage[utf8]{inputenc} % allow utf-8 input
\usepackage[T1]{fontenc}    % use 8-bit T1 fonts
\usepackage{hyperref}       % hyperlinks
\usepackage{url}            % simple URL typesetting
\usepackage{booktabs}       % professional-quality tables
\usepackage{amsfonts}       % blackboard math symbols
\usepackage{nicefrac} 
\usepackage{dsfont}         % compact symbols for 1/2, etc.
\usepackage{microtype}      % micro typography
\usepackage{lipsum}
\usepackage{bbold}
\usepackage{fancyhdr}       % header
\usepackage{graphicx}       % graphics
\graphicspath{{media/}}     % organize your images and other figures under media/ folder
\graphicspath{ {./images/} }
\usepackage{amsfonts}
\usepackage{amsmath}
\usepackage{amssymb}
\usepackage{bbm}
\usepackage{floatrow}
\usepackage{xcolor}
\usepackage{xcoffins}
\NewCoffin\tablecoffin
\usepackage{verbatim}
\usepackage{lettrine}
\usepackage{stmaryrd}
\usepackage{latexsym}
\usepackage{array}
\usepackage{manfnt}
\usepackage{tikz}
\usepackage[makeroom]{cancel}
\usepackage{color}
\usepackage{float}
\usepackage[ps,matrix,arrow,curve]{xy}

\def\ie{{\sl i.e.\/}}
\def\dd{{\cal D}}

\newcommand{\cD}{{\cal D}}
\newcommand{\cF}{{\cal F}}
\newcommand{\cG}{{\cal G}}
\newcommand{\cL}{{\cal L}}
\newcommand{\cM}{{\cal M}}
\newcommand{\cH}{{\cal H}}

\newcommand{\cZ}{{\cal Z}}
\newcommand{\cT}{{\cal T}}
\newcommand{\cC}{{\cal C}}

\usepackage{amsthm}
\theoremstyle{definition}
\newtheorem{theorem}{Theorem}[section]
\newtheorem{lemma}[theorem]{Lemma}

\newtheorem{definition}[theorem]{Definition}

\newtheorem{example}[theorem]{Example}

\def\emph#1{{\sl #1\/}}

\newcommand{\hpeprint}[2]{%
  \href{http://www.arxiv.org/abs/#1}{\texttt{arxiv:#1#2}}}%
\newcommand{\hpspires}[1]{%
  \href{http://www.slac.stanford.edu/spires/find/hep/www?#1}{SPIRES Link}}%
\newcommand{\hpmathsci}[1]{%
  \href{http://www.ams.org/mathscinet-getitem?mr=#1}{\texttt{MR #1}}}%
\newcommand{\hpdoi}[1]{%
  \href{http://dx.doi.org/#1}{\ Journal Link}}%

\newcommand{\kompleksni}{\ensuremath{\mathbb{C}}}

\newcommand{\ds}{\displaystyle}
\newcommand{\del}{\partial}

\newcommand{\bra}[1]{{\langle{#1}\vert}}
\newcommand{\ket}[1]{{\vert{#1}\rangle}}
\newcommand{\bracket}[2]{\langle #1 \vert #2 \rangle}
\newcommand{\im}{\mathop{\rm im}\nolimits}

\title{The 3BF theory as a TQFT}

\author{
  T. Radenkovi\' c,\\
  Institute of Physics Belgrade,\\
  University of Belgrade, \\
  Pregrevica 118, 11000 Belgrade, Serbia,\\
  \texttt{rtijana@ipb.ac.rs} \\
   \And
  M. Vojinovi\' c,\\
  Institute of Physics Belgrade,\\
  University of Belgrade, \\
  Pregrevica 118, 11000 Belgrade, Serbia,\\
  \texttt{vmarko@ipb.ac.rs} \\
}

\usepackage[normalem]{ulem}
\begin{document}
\maketitle

\begin{abstract}
%\lipsum[1]
We study the path integral quantization of the topological $3BF$ theory, whose gauge symmetry is described by a 3-group. This theory is relevant for the quantization of general relativity coupled to Standard Model of elementary particles. We explicitly construct a state sum corresponding to the discretized path integral of a $3BF$ action. Being a topological invariant of 4-dimensional manifolds with boundary, this state sum gives rise to a topological quantum field theory (TQFT), realized as a functor between the category of cobordisms and the category of Hilbert spaces. After an introduction to appropriate category theory concepts and the construction of the state sum, we provide an explicit proof that it satisfies all Atiyah's axioms, and thus represents a genuine TQFT. The formulation of this TQFT represents a major step in the spinfoam quantization programme for a realistic theory of quantum gravity with matter.
\end{abstract}

% keywords can be removed
\keywords{quantum gravity \and topological quantum field theory \and higher gauge theory \and $3$-groups \and state sum models}

\section{Introduction}

Quantization of the gravitational field represents one of the main open problems of modern theoretical physics. Over the years, many strategies to tackle this problem have been proposed, and some have grown to vast research areas, such as String Theory (ST) \cite{StringTheory, StringTheory1, StringTheory2} and Loop Quantum Gravity (LQG) \cite{Ashtekar,Thiemann,Rovelli2004,RovelliVidotto2014, zakopane}. In many of these approaches, the notion of a \emph{topological quantum field theory} (TQFT) became one of the prominent and very useful mathematical tools \cite{Y3, PonzanoRegge, Baez1996, GirelliPfeifferPopescu2008, Witten1, Witten2, CMR}. In addition, TQFT became an object of interest in several areas of mathematics, such as algebraic topology and category theory, one of the main applications being the construction of topological invariants for $D$-dimensional manifolds (for $D=3,4,5,\dots$) \cite{QuinnNotes, Sozer, CraneKauffmanYetter, BaezDolan}.

One of the first rigorous definitions of a TQFT was the collection of axioms put out by Michael Atiyah \cite{Atiyah1988}. With the advent of category theory, these axioms have been reorganized into a very succinct definition of a TQFT --- it is a \emph{functor} from the category of cobordisms to the category of vector spaces. In the context of gravity, we are mostly interested in the $4$-dimensional cobordisms between $3$-dimensional boundary manifolds, since they represent \emph{spacetime} and \emph{spatial hypersurfaces}, respectively. On the other hand, in the context of quantum theory, the \emph{quantum state} of a system is described by a vector from a Hilbert space, while its \emph{evolution} is described by an operator over a Hilbert space. Therefore, when discussing quantum gravity (QG), we are mainly interested in defining a TQFT as a functor between the category $3Cob$ and the category $Hilb$:
\smallskip
\begin{center}
 \begin{tikzpicture}[>=stealth, baseline=(current  bounding  box.center)]
  % Nodes
  \node (A1) at (0,0) {$3Cob$};
  \node (A2) at (3,0) {$Hilb\,.$};
  % Arrows
  \draw[->] (A1) -- node[above]{$\scriptstyle TQFT$} (A2);
\end{tikzpicture}
\end{center}
\smallskip
In the next Section we will introduce and unpack all details of this definition, in a hopefully pedagogical manner, for a reader who is familiar with quantum theory and smooth manifolds, but maybe not with category theory. Just for a qualitative overview, Table \ref{tab1} illustrates the correspondence between elements of differential topology and quantum theory in the context of TQFT \cite{Baez}.
\begin{table}[!ht]\label{tab1}
\begin{center}
\begin{tabular}{ |c|c| } 
 \hline
$\vphantom{\ds\int}$ Differential topology & Quantum theory \\ \hline \hline
% $\vphantom{\ds\int}$
$3D$ manifold (space) & Hilbert space (states) \\ \hline
% $\vphantom{\ds\int}$
cobordism between $3$-dimensional manifolds
(spacetime) & operator (process)  \\ \hline
% $\vphantom{\ds\int}$
composition of cobordisms & composition of operators \\ \hline
% $\vphantom{\ds\int}$
identity cobordism & identity operator \\ \hline
% $\vphantom{\ds\int}$
disjoint union of manifolds & tensor product \\ \hline
% $\vphantom{\ds\int}$
empty $3$-dimensional manifold & $1$-dimensional Hilbert space \\ \hline
% $\vphantom{\ds\int}$
orientation reversal of the $4$-dimensional manifold & Hermitian adjoint of an operator \\ \hline
% $\vphantom{\ds\int}$
orientation reversal of the $3$-dimensional manifold & dual vector space \\ \hline
\end{tabular}
\caption{Correspondence between differential topology and quantum theory defined by a TQFT functor.}
\end{center}
\end{table}

In a recent approach to quantum gravity based on the ideas of \emph{higher category theory} \cite{BaezHuerta}, the central starting point for the formulation of the theory is the so-called $3BF$ action,
$$
S_{3BF} = \int_{\cM_4} \langle B \wedge \cF \rangle_{\mathfrak{g}} + \langle C \wedge \cG \rangle_{\mathfrak{h}} + \langle D \wedge \cH \rangle_{\mathfrak{l}}\,,
$$
where $B$, $C$, $D$ are Lagrange multipliers, while $(\cF, \cG, \cH)$ is the so-called $3$-curvature corresponding to a $3$-connection of a $3$-group based on the three Lie groups $G$, $H$ and $L$ and their corresponding Lie algebras $\mathfrak{g}$, $\mathfrak{h}$ and $\mathfrak{l}$, see \cite{Radenkovic2019} for details. The notion of a $3$-group represents the generalized gauge symmetry of the theory, so that the $3BF$ action for a suitable choice of a $3$-group, with added simplicity constraint terms, can successfully describe the relevant classical theory for physics --- namely general relativity coupled to the Standard Model of elementary particle physics (for the explicit form of the action, see \cite{Radenkovic2019,StipsicVojinovic2024}, see also \cite{MikovicVojinovic2021} for the $4$-group approach). The quantization of this action can then be performed using the so-called \emph{spinfoam quantization programme}, which consists of three main steps:
\begin{enumerate}
\item the classical theory is written in the form of a topological $3BF$ action with appropriate simplicity constraints,
\item the $3$-group gauge structure of the $3BF$ action is leveraged to quantize the topological sector by defining an appropriate state sum (\ie, a discretization of a path integral) and its corresponding TQFT,
\item the resulting TQFT is then deformed into a non-topological theory by suitably implementing the simplicity constraints, promoting it into a path integral for a physically relevant theory.
\end{enumerate}
Step $1$ of the spinfoam quantization programme has been completed in \cite{Radenkovic2019}. A major part of Step $2$ has been established in~\cite{Radenkovic2022_2}, where a path integral for the $3BF$ action, formally written as
$$
\cZ_\varnothing = \int \dd \phi \; e^{iS_{3BF}[\phi]}\,,
$$
has been rigorously defined as a discrete state sum over simplicial complexes (see \cite{Radenkovic2022_2} and also Section \ref{SecIII}, specifically equation (\ref{eq_partition_3bf}) for details). Assuming that a given simplicial complex is homeomorphic to a compact $4$-dimensional manifold $\cM$ without boundary (hence the subscript $\varnothing$), the state sum $\cZ_\varnothing$ represents a topological invariant of $\cM$, for every explicit choice of a $3$-group.

The purpose of this paper is to complete Step 2, by extending the definition of the state sum $\cZ_\varnothing$ to a new state sum $\cZ_\partial$, corresponding to $4$-dimensional manifolds with boundary (hence the subscript $\partial$), and to demonstrate that this extended definition gives rise to a TQFT, by explicitly verifying the axioms. It inherently defines a functor between two dagger symmetric monoidal categories equipped with a dual, in the sense of the above definition of a TQFT. More explicitly, we provide a detailed construction of the functor
\smallskip
\begin{center}
 \begin{tikzpicture}[>=stealth, baseline=(current  bounding  box.center)]
  % Nodes
  \node (A1) at (0,0) {$3TCob$};
  \node (A2) at (3,0) {$Hilb\,,$};
  % Arrows
  \draw[->] (A1) -- node[above]{$\scriptstyle Z$} (A2);
\end{tikzpicture}
\end{center}
\smallskip
where $3TCob$ is the category of triangulations of cobordisms (introduced in detail in the next Section), while $Z$ is the state sum representing the functor between this category and the category of Hilbert spaces. By ``forgetting'' the explicit choice of a triangulation in $3TCob$, the functor $Z$ also induces a similar functor $\tilde{Z}$ between the usual category of cobordisms $3Cob$ and the category of Hilbert spaces,
\smallskip
\begin{center}
 \begin{tikzpicture}[>=stealth, baseline=(current  bounding  box.center)]
  % Nodes
  \node (A1) at (0,0) {$3Cob$};
  \node (A2) at (3,0) {$Hilb\,,$};
  % Arrows
  \draw[->] (A1) -- node[above]{$\scriptstyle \tilde{Z}$} (A2);
\end{tikzpicture}
\end{center}
\smallskip
corresponding to the usual definition of a TQFT.

Our results are important both for physics and mathematics. On the physics side, our construction of the state sum $\cZ_\partial$ gives a rigorous definition of a physically relevant path integral --- one corresponding to a class of theories which includes general relativity coupled to Standard Model \cite{Radenkovic2019}. Thus, it provides a firm mathematical ground for the construction of a physically relevant model of quantum gravity.

On the mathematics side, our construction of the state sum $\cZ_\partial$ and functors $Z$ and $\tilde{Z}$ is an explicit realization of Porter’s TQFT for $d = 4$ and $n = 3$, see \cite{Porter98, Porter96}. It represents a genuine topological invariant of $4$-dimensional manifolds with boundary, defined using the higher categorical structure of a semistrict $3$-group. It is the first explicit construction of this kind of invariant. One should emphasize, however, that in $4$ dimensions there are manifolds which do not admit a triangulation (i.e., there is no simplicial complex which is homeomorphic to such a manifold, the most famous example is the $E_8$ manifold), and vice versa, there are simplicial complexes which do not correspond to any piecewise linear (PL) manifold. Our construction of the state sum $\cZ_\partial$ explicitly depends on the simplicial complex structure, so it represents an invariant only for a class of $4$-dimensional manifolds that can be triangulated.

The layout of the paper is as follows. In Section \ref{SecII}, we provide a pedagogical overview of the foundational concepts of category theory required to define a TQFT within this specific framework. The overview is aimed at researchers with a background in physics, so we have accordingly adapted the notation, conventions, and terminology. We begin by introducing the notion of a dagger symmetric monoidal category with a dual, and construct three explicit example categories --- $3TCob$, $3Cob$ and $Hilb$. We then introduce the notion of a functor between two categories, and apply it to our examples in order to explicitly spell out all axioms that define a TQFT. In Section \ref{SecIII} we present the main results of the paper. Namely, we provide an explicit definition of the state sums $\cZ_\varnothing $ and $\cZ_\partial$, and use the latter to construct the functors $Z$ and $\tilde{Z}$. The functor $Z$ depends on the choice of a triangulation of the boundary manifolds, while $\tilde{Z}$ corresponds to an equivalence class of all possible choices of triangulations. Then, in Section \ref{SecIV}, we demonstrate that our definition of $Z$ satisfies all necessary axioms, hence verifying that it is indeed a TQFT. Our concluding remarks are presented in Section \ref{SecV}. Since certain parts of the proofs rely on lengthy calculations, they are presented in the Appendices.

Notations and conventions throughout the paper are as follows. If $G$ is a finite group, with cardinality denoted as $|G|$, the integral of a function over a group denotes the normalized sum over all group elements, namely:
\begin{equation} \label{eq:GroupIntegralDef}
 \int_G dg \, f(g) \equiv \frac{1}{|G|} \sum_{g\in G} f(g)\,.
\end{equation}
On the other hand, if $G$ is a Lie group, the integral is defined using the standard Haar measure. Next, the symbol $\delta_G$ denotes the corresponding $\delta$-distribution on $G$. If $G$ is a finite group, it is defined as:
\begin{equation} \label{eq:DiracDeltaDef}
\delta_G(g)  = \left\{
\begin{array}{cl}
  |G| & \text{ if }g=e\,, \\
  0 & \text{ if }g\neq e\,.
\end{array}
\right. 
\end{equation}
Here $e$ denotes the neutral element of $G$. On the other hand, if $G$ is a Lie group, $\delta_G$ is the usual Dirac delta distribution on $G$. As a consequence of (\ref{eq:GroupIntegralDef}) and (\ref{eq:DiracDeltaDef}), we can simply write
$$
\int_G dg \, \delta_G(g) = 1\,, \qquad
\int_G dg \, \delta_G(gg_0^{-1}) \, f(g) = f(g_0)\,,
$$
for both finite and Lie groups. Next, given a simplicial complex (representing the triangulation of some $d$-dimensional manifold), the set of all $k$-dimensional simplices (where $k\in\{0,\dots,d \}$) is denoted by $\Lambda_k$, while its cardinality is denoted as $|\Lambda_k|$. The set of vertices of a triangulation, denoted as $\Lambda_0$, is both finite and ordered. Each $k$-simplex is labeled by $(k+1)$-tuple of vertices, denoted as $(i_0 \ldots i_k)$, where $i_0, \ldots, i_k \in\Lambda_0$ and $i_0 < \cdots < i_k$. All other notation is explicitly introduced in the main text, at the place of first appearance.

\section{Axioms for a TQFT}\label{SecII}

Let us begin with a short review of some of the basic concepts of the category theory. Our main aim is to introduce the notion of a \emph{dagger symmetric monoidal category with a dual}, and then the notion of a \emph{functor} between two such categories, as well as three examples, namely the categories $3TCob$, $3Cob$ and $Hilb$. Having in mind a reader not too familiar with any of these notions, we proceed in several steps:
\begin{itemize}
\item we first introduce the notion of a \emph{category},
\item we then equip a category with additional structure, namely \emph{symmetric monoidal} structure,
\item we next equip a category with a \emph{dagger} structure,
\item and finally, we equip a category with a \emph{dual} structure.
\end{itemize}
The relevance of these additional structures will become obvious once we discuss the relevant example categories --- in the $Hilb$ category, the symmetric monoidal structure enables one to talk about tensor products of vector spaces, the dagger structure introduces the adjoint operator over a vector space, while the dual structure introduces the dual vector space. On the other hand, in $3TCob$ and $3Cob$ categories, the symmetric monoidal structure gives rise to a disjoint union of manifolds, while the dagger and the dual represent orientation reversal of cobordisms and their boundaries, respectively. Finally, once all of the above is introduced, we will define the notion of a functor between such categories. An explicit construction of a functor will then be taken up in Section \ref{SecIII}.

All these notions have a very rich structure and provide a language which is very expressive. In fact, one of the purposes of the whole framework of category theory is to leverage this expressiveness. In particular, it allows us to define a TQFT in a particularly simple fashion, by saying that it is a functor from $3TCob$ (or $3Cob$) to $Hilb$. This statement then encodes a wealth of properties and axioms that in fact define a TQFT.

It should also be emphasized that the definition of a TQFT based on Atiyah's axioms (which we will describe in this Section) is not the only possible one. An alternative definition of a TQFT has been given by Witten \cite{Witten1}, often called cohomological TQFT, and is generally considered to be inequivalent to Atiyah's approach. We will not be discussing Witten's definition of TQFT in this work.

\subsection{Definitions}

The first notion we need to introduce is that of a category. It is a fairly general structure, useful for studying the notion of compositions of abstract maps between abstract objects. There is a lot of literature on categories and their properties, and here we restrict ourselves only to the definition itself.

\begin{definition}A \emph{category} $\mathcal{C}$ consists of the following components: the \emph{objects} of the category $\mathcal{C}$, denoted by $A,B,...\in Obj_\mathcal{C}$, and the \emph{morphisms} between objects, denoted by $f,g,... \in Mor_\mathcal{C}$. For an ordered pair $\cC = (Obj_{\cC}, Mor_{\cC})$ to be called a category, the following properties must hold:
\begin{itemize}
\item There exist maps $s,t: Mor_\mathcal{C} \to Obj_\mathcal{C}$, called \emph{source} and \emph{target maps} that assign to each morphism its source and target object. For each morphism $f\in Mor_\mathcal{C}$ between objects $A$ and $B$, denoted $f: A \to B$, one has $s(f)=A$ and $t(f)=B$.
\item For every two morphisms $f$ and $g$, $f: s(f)\to t(f)$ and $g: s(g)\to t(g)$, such that the source object of $g$ is equal to the target object of $f$, $s(g)=t(f)$, the \emph{composition of morphisms} $f$ and $g$ gives a new morphism $g\circ f : s(f) \to t(g)$.
\item For each object 
$A\in Obj_\cC$ of the category, there exists an \emph{identity morphism}  $\text{id}_A:A \to A$. Then, for each morphism $f:s(f)\to t(f)$, one has 
$\text{id}_{t(f)}\circ f=f$ and $f \circ \text{id}_{s(f)}=f$.
\item The composition is \emph{associative}, meaning that 
$h\circ(g\circ f)=(h\circ g)\circ f$, for any three morphisms $f$, $g$, and $h$ which are composable, \ie\,, when $s(g)=t(f)$ and $s(h)=t(g)$.
\end{itemize}
\end{definition}

At the outset, from the above definition one can see that the main structures in a category are the composition and identities. While this is very general, most of the time we are interested in describing categories with additional structure, primarily the categorical analog of a tensor product. To describe those more special and more rich structures, we introduce the notion of a symmetric monoidal category.

\begin{definition}\label{Def:smcategory}A \emph{symmetric monoidal category} $(\cC, \boxtimes, \mathbb{1})$ is a category $\mathcal{C}$ with objects ${Obj}_{\mathcal{C}}$ and morphisms $Mor_{\mathcal{C}}$, equipped with the following structure:
\begin{itemize}
\item The \emph{tensor product} $\boxtimes : \mathcal{C} \times \mathcal{C} \to \mathcal{C}$ that assigns to each pair of objects $A, B \in Obj_\mathcal{C}$ an object $A \boxtimes B$, and to each pair of morphisms $f,g\in Mor_\cC$, such that $f: A_1 \to B_1$ and $g: A_2 \to B_2$, a morphism $f \boxtimes g: A_1 \boxtimes A_2 \to B_1 \boxtimes B_2$.
\item For all objects $A, B, C \in Obj_\mathcal{C}$, there exists a natural isomorphism, the \emph{associator map},
$$
\alpha_{A, B, C}: (A \boxtimes B) \boxtimes C {\to} A \boxtimes (B \boxtimes C)\,,
$$
such that the \emph{Pentagon identity} is satisfied, \ie\,, the following diagram commutes:
 \begin{equation}
 \text{\begin{tikzpicture}[>=stealth, baseline=(current  bounding  box.center)]
  % Nodes
  \node (A1) at (0,0) {$((A \boxtimes B) \boxtimes C) \boxtimes D$};
  \node (A2) at (5,0) {$(A \boxtimes B) \boxtimes (C \boxtimes D)$};
  \node (A3) at (10,0) {$A \boxtimes (B \boxtimes (C \boxtimes D))$};
  \node (A4) at (1.3,-2) {$(A \boxtimes (B \boxtimes C)) \boxtimes D$};
  \node (A5) at (8.2,-2) {$A \boxtimes ((B \boxtimes C) \boxtimes D)\,.$};
  % Arrows
  \draw[->] (A1) -- node[above]{$\alpha_{A \boxtimes B, C, D}$} (A2);
  \draw[->] (A2) -- node[above]{$\alpha_{A, B, C \boxtimes D}$} (A3);
  \draw[->] (A4) -- node[above]{$\alpha_{A, B \boxtimes C, D}$} (A5);
  \draw[->] (A1) -- node[left]{$\alpha_{A, B, C} \boxtimes \, \text{id}_D$} (A4);
  \draw[->] (A5) -- node[right]{$\text{id}_A \boxtimes \,\alpha_{B, C, D}$} (A3);
\end{tikzpicture}
}
\end{equation}
\item There exists an object $\mathbb{1} \in Obj_\mathcal{C}$ called the \emph{unit object}, and natural isomorphisms the \emph{left unitor map} $\lambda_A$ and the \emph{right unitor map} $\rho_A$,
$$
\lambda_A: \mathbb{1} \boxtimes A {\to} A \quad \text{and} \quad \rho_A: A \boxtimes \mathbb{1} {\to} A\,,
$$
for all objects $A \in Obj_\mathcal{C}$, such that the following \emph{triangle diagram} commutes:
\begin{equation}
\text{\begin{tikzpicture}[>=stealth, baseline=(current  bounding  box.center)]
  % Nodes
  \node (A1) at (0,0) {$(A \boxtimes \mathbb{1}) \boxtimes B$};
  \node (A2) at (4,0) {$A \boxtimes B$};
 \node (A3) at (2,-2) {$A \boxtimes (\mathbb{1} \boxtimes B)\,.$};
  % Arrows
  \draw[->] (A1) -- node[above]{$\rho_A \boxtimes \,\text{id}_{B}$} (A2);
  \draw[->] (A1) -- node[below left]{$\alpha_{A,\mathbb{1},B}$} (A3);
  \draw[->] (A3) -- node[below right]{$\text{id}_A\boxtimes \, \lambda_B$} (A2);
\end{tikzpicture}
}
\end{equation}
\item For all objects $A, B \in Obj_\mathcal{C}$, there exists a natural isomorphism called the \emph{swap map},
$$
\sigma_{A, B}: A \boxtimes B {\to} B \boxtimes A,
$$
such that for all morphisms $f: A \to C$ and $g: B \to D$, the following diagram commutes:
\begin{equation}
\text{\begin{tikzpicture}[>=stealth, baseline=(current  bounding  box.center)]
  % Nodes
  \node (A1) at (0,0) {$A \boxtimes B$};
  \node (A2) at (3,0) {$B \boxtimes A$};
  \node (A3) at (0,-2) {$C \boxtimes D$};
  \node (A4) at (3,-2) {$D \boxtimes C\,.$};
  % Arrows
  \draw[->] (A1) -- node[above]{$\sigma_{A, B}$} (A2);
  \draw[->] (A3) -- node[below]{$\sigma_{C, D}$} (A4);
  \draw[->] (A1) -- node[left]{$f \boxtimes g$} (A3);
  \draw[->] (A2) -- node[right]{$g \boxtimes f$} (A4);
\end{tikzpicture}
}
\end{equation}
\item The associator and the swap map must satisfy the \emph{hexagon identity}, \ie\,, the following diagram commutes,
\begin{equation}
\text{\begin{tikzpicture}[>=stealth, baseline=(current  bounding  box.center)]
  % Nodes
  \node (A1) at (0,0) {$(A \boxtimes B) \boxtimes C$};
  \node (A2) at (4,0) {$A \boxtimes (B \boxtimes C)$};
  \node (A3) at (8,0) {$(B\boxtimes C)\boxtimes A$};
  \node (A4) at (0,-2) {$(B\boxtimes A)\boxtimes C$};
  \node (A5) at (4,-2) {$B\boxtimes (A\boxtimes C)$};
  \node (A6) at (8,-2) {$B\boxtimes (C\boxtimes A)\,,$};
  % Arrows
  \draw[->] (A1) -- node[above]{$\alpha_{A,B,C}$} (A2);
  \draw[->] (A2) -- node[above]{$\sigma_{A,B\boxtimes C}$} (A3);
  \draw[->] (A4) -- node[below]{$\alpha_{B,A,C}$} (A5);
  \draw[->] (A5) -- node[below]{$\text{id}_B\boxtimes\sigma_{A,C}$} (A6);
  \draw[->] (A1) -- node[left]{$\sigma_{A,B}\boxtimes \text{id}_C$} (A4);
  \draw[->] (A3) -- node[right]{$\alpha_{B,C,A}$} (A6);
\end{tikzpicture}
}
\end{equation}
and, the swap map and left and right unitor maps must satisfy the following diagram:
\begin{equation}
\text{\begin{tikzpicture}[>=stealth, baseline=(current  bounding  box.center)]
  % Nodes
  \node (A1) at (0,0) {$A \boxtimes \mathbb{1}$};
  \node (A2) at (3,0) {$\mathbb{1} \boxtimes A$};
  \node (A3) at (1.5,-2) {$A\,.$};
  % Arrows
  \draw[->] (A1) -- node[above]{$\sigma_{A,\mathbb{1}}$} (A2);
  \draw[->] (A1) -- node[left]{$\rho_A$} (A3);
  \draw[->] (A2) -- node[right]{$\lambda_A$} (A3);
\end{tikzpicture}
}
\end{equation}
 \item Finally, the swap map satisfies the \emph{inverse law}, \ie\,, for all $A,B \in Obj_{\mathcal{C}}$:
\begin{equation}
\sigma_{B,A} \circ \sigma_{A,B}=\text{id}_{A\boxtimes B}\,.
\end{equation}
\end{itemize}
\end{definition}

From the above definition we can see that the new introduced notions are the tensor product $\boxtimes$ and the unit object $\mathbb{1}$. The remaining structures (namely the associator map $\alpha$, the unitor maps $\lambda$ and $\rho$, and the swap map $\sigma$) are there to ensure that $\boxtimes$ and $\mathbb{1}$ behave as we expect them to behave in a category, and to be compatible with the original structures inherited from the category --- the composition and the identities.

The notions such as a tensor product are however not enough, and we want to have yet more structure in the category. Hence, we define the notion of a dagger category.
 
\begin{definition}\label{def:dagger}
A \emph{dagger category} $(\cC, \dagger)$ is a category equipped with a dagger operation that reverses the directions of the morphisms, and does nothing to objects. The following properties must hold:
\begin{itemize}
\item For each morphism $f\in Mor_\mathcal{C}$, the target map of the dagger of a morphism is the source of the morphism, \ie\,, $t(f^\dagger)=s(f)$, and the source map of the dagger of a morphism is the target of that morphism $s(f^\dagger)=t(f)$. For a morphism $f:A\to B$, the dagger functor gives the morphism $f^\dagger:B \to A$.
  \item For every object $A \in Obj_\mathcal{C}$, the dagger of the identity morphism $\text{id}_A: A \to A$ is equal to the identity morphism itself, i.e., $(\text{id}_A)^\dagger = \text{id}_A$.
  
  \item The dagger of the composition of two morphisms $f: A \to B$ and $g: B \to C$ is the composition of their daggers in reverse order, \ie\,, $(g \circ f)^\dagger = f^\dagger \circ g^\dagger$.
  
  \item The dagger of the dagger of a morphism is the original morphism, \ie, $(f^\dagger)^\dagger = f$.
  \item The dagger of an object is trivial, \ie\,, it is the object itself, $A^\dagger=A$.
\end{itemize}
\end{definition}
Let us note that, formally speaking, a dagger can also be introduced as a functor from an opposite category $\mathcal{C}^{op}$ to the category $\mathcal{C}$ itself, \ie\,, $\dagger: \mathcal{C}^{op} \to \mathcal{C}$, where the opposite category $\cC^{op}$ is constructed from $\cC$ by reversing the sources and targets of all morphisms in $\cC$.

The only new structure in the dagger category is obviously the dagger itself, and the above rules define its behavior with respect to the original inherited structures of a category, the composition and the identities. In this sense, the above definition holds for any category. However, if we want our category to be simultaneously both a dagger category and a symmetric monoidal category, certain compatibility conditions must be required to hold between the dagger and the tensor product as well. Hence, a \emph{dagger symmetric monoidal category} $(\cC, \boxtimes, \mathbb{1}, \dagger)$ has to satisfy all requirements of definitions \ref{Def:smcategory} and \ref{def:dagger}, and in addition the following compatibility properties must also hold:
\begin{itemize}
  \item For each pair of morphisms $f,g\in Mor_\mathcal{C}$, the dagger of a tensor product of morphisms $f$ and $g$ is the tensor product of daggers, \ie\,, \[(f\boxtimes g)^\dagger= f^\dagger \boxtimes g^\dagger\,.\] 
  Since the dagger acts trivially on objects, the analogous identity obviously holds for the tensor product of objects. 
  \item For any objects $A, B, C \in Obj_\mathcal{C}$, the dagger of the associator map is equal to the inverse of the associator, \ie\,, 
  \[\alpha_{A, B, C}^\dagger=\alpha_{A, B, C}^{-1}:A \boxtimes (B \boxtimes C) \to  (A \boxtimes B) \boxtimes C\,.\]
  \item For each object $A \in Obj_\mathcal{C}$, the dagger of the left and right unitor maps are the inverses of those maps, \ie\,, 
  \[
\lambda_A^\dagger=\lambda_A^{-1}: A \to \mathbb{1} \boxtimes A  \quad \text{and} \quad \rho_A^\dagger=\rho_A^{-1}: A \to A \boxtimes \mathbb{1} \,.
    \]
  \item For any two objects $A, B\in Obj_\cC$, the dagger of the swap map is equal to the inverse of that map\,, \ie\,, 
  \[\sigma_{A, B}^\dagger=\sigma_{A, B}^{-1}: B \boxtimes A {\to} A \boxtimes B\,.\]
\end{itemize}

Finally, we introduce one more structure element, namely the dual. In order to keep the definition short, we introduce it on top of a dagger symmetric monoidal category.

\begin{definition}\label{def:dual} A dagger symmetric monoidal category with a \emph{dual} $(\cC, \boxtimes, \mathbb{1}, \dagger, * )$ is a category where every object has its corresponding dual object. The following properties must hold:
\begin{itemize}
\item For every object $A \in Obj_\mathcal{C}$ of a monoidal category, there is a \emph{dual object} $A^* \in  Obj_\cC$.

\item There exist the \emph{unit map} $\eta_{A}:\mathbb {1} \to A^*\boxtimes A$ and the \emph{counit map} $\epsilon_{A}:A\boxtimes A^*\to \mathbb {1}$, such that the following \emph{snake diagrams} are identity morphisms (compatibility of the dual with the symmetric monoidal structure):
\begin{equation}
\text{\begin{tikzpicture}[>=stealth, baseline=(current  bounding  box.center)]
  % Nodes
  \node (A1) at (0,0) {$A$};
  \node (A11) at (2,0) {$A \boxtimes \mathbb{1} $};
  \node (A2) at (6,0) {$A \boxtimes (A^* \boxtimes A) $};
  \node (A3) at (6,-2) {$(A \boxtimes A^*) \boxtimes A\,,$};
  \node (A5) at (0,-2) {$ \mathbb{1} \boxtimes A$};
  % Arrows
  \draw[->] (A1) -- node[above]{$\rho_A^{-1}$} (A11);
  \draw[->] (A11) -- node[above]{$ \text{id}_A \boxtimes \eta_A$} (A2);
  \draw[->] (A2) -- node[right]{$\alpha_{A,A^*, A}^{-1}$} (A3);
  \draw[->] (A3) -- node[below]{$\epsilon_A \boxtimes \text{id}_A$} (A5);
  \draw[->] (A5) -- node[left]{$\lambda_A$} (A1);
\end{tikzpicture}
}
\end{equation}
\begin{equation}
\text{\begin{tikzpicture}[>=stealth, baseline=(current  bounding  box.center)]
  % Nodes
  \node (A1) at (0,0) {$A^*$};
  \node (A11) at (2,0) {$ \mathbb{1} \boxtimes A^*$};
  \node (A2) at (6,0) {$(A^* \boxtimes A) \boxtimes A^*$};
  \node (A3) at (6,-2) {$A^* \boxtimes (A \boxtimes A^*)\,.$};
  \node (A5) at (0,-2) {$ A^* \boxtimes \mathbb{1}$};
  % Arrows
  \draw[->] (A1) -- node[above]{$\lambda_{A^*}^{-1}$} (A11);
  \draw[->] (A11) -- node[above]{$  \eta_{A} \boxtimes \text{id}_{A^*}$} (A2);
  \draw[->] (A2) -- node[right]{$\alpha_{A^*,A, A^*}$} (A3);
  \draw[->] (A3) -- node[below]{$ \text{id}_{A^*} \boxtimes \epsilon_{A}$} (A5);
  \draw[->] (A5) -- node[left]{$\rho_{A^*}$} (A1);
\end{tikzpicture}
}
\end{equation}

\item For all $A \in Obj_\mathcal{C}$, the following diagram commutes (compatibility of the dual with the dagger):
 \begin{equation}
\text{\begin{tikzpicture}[>=stealth, baseline=(current  bounding  box.center)]
  % Nodes
  \node (A1) at (0,0) {$\mathbb{1}$};
  \node (A2) at (3,0) {$A \boxtimes A^*$};
  \node (A3) at (1.5,-2) {$A^* \boxtimes A\,.$};
  % Arrows
  \draw[->] (A1) -- node[above]{$\epsilon_A^\dagger$} (A2);
  \draw[->] (A1) -- node[left]{$\eta_A$} (A3);
  \draw[->] (A2) -- node[right]{$\sigma_{A,A^*}$} (A3);
\end{tikzpicture} 
}
\end{equation}
\item Since the dagger acts trivially on objects, for all objects $A\in Obj_{\cC}$, one has the identity $(A^*){}^\dagger=(A^\dagger){}^*$.  
\end{itemize}
\end{definition}

This concludes the category theory definitions necessary for introducing a TQFT. In what follows, we will focus on three relevant categories that satisfy all of the above definitions.

\subsection{Examples}

We will discuss three examples of a dagger symmetric monoidal category with a dual. These are the category of finite-dimensional Hilbert spaces (denoted by $Hilb$), the category of oriented cobordisms between $3$-dimensional manifolds (denoted by $3Cob$), and the category of oriented cobordisms between triangulated $3$-dimensional manifolds (denoted by $3TCob$).  The standard notation for the general case and these three examples is shown in Table \ref{Tabela2}.
\begin{table}[!ht]
\begin{center}
{\small
\begin{tabular}{|c|c|c|c|c|c|c|c|c|}
\hline
$\vphantom{\ds\int}$ & objects & morphisms & composition & \begin{tabular}{@{}c@{}}
                   identity\\
                   morphism\\
                 \end{tabular}  & \begin{tabular}{@{}c@{}}
                   tensor \\
                product \\
                 \end{tabular}  & \begin{tabular}{@{}c@{}}
                   unit\\
                   object\\
                 \end{tabular}  & dagger & dual \\
\hline \hline 
$\vphantom{\ds \int}\mathcal{C}$ & $A,B,\ldots$ & $f,g,\ldots$ & $\circ$ & $\text{id}_A$ &$\boxtimes$ & $\mathbb{1}$ & $\dagger$ & $*$\\ \hline
 $\vphantom{\ds \int}Hilb$ & $\mathcal{H}_A,\mathcal{H}_B,\ldots $ & $\hat{M},\hat{N},\ldots$ & $\cdot$ & $\hat{I}_\cH$ & $\otimes$&  $\mathbb{C}$& $\dagger$  & $*$\\ \hline  
$\vphantom{\ds \int}3Cob$ & $\Sigma_1,\Sigma_2,\ldots$ & $\cM_A, \cM_B,\ldots$ & $\#$ & $\text{id}_\Sigma$& $\sqcup$& $\varnothing$& $\dagger$& $\bar{\Sigma}$\\ \hline
$\vphantom{\ds \int}3TCob$ & $\cT(\Sigma_1),\cT(\Sigma_2),\ldots$ & $\cM_A, \cM_B,\ldots$ & $\#$ & $\text{id}_\Sigma $& $\sqcup$& $\varnothing $& $\dagger$ & $\cT(\bar{\Sigma})$\\ \hline
\end{tabular}
}
\caption{\label{Tabela2} The general dagger symmetric monoidal category with a dual and three relevant examples: the categories $Hilb$, $3Cob$ and $3TCob$.}
\end{center}
\end{table}

We now turn to each of these three categories, apply the general categorical definitions, and discuss their individual specific properties. First, let us consider the category $Hilb$ of finite-dimensional Hilbert spaces. The Hilbert spaces are equipped with the additional structure of the \emph{Hermitian form} (also called a scalar product) $\bracket{\,\_\,}{\,\_\,} : \cH \times \cH \to \mathbb{C}$ that provides a way to define the \emph{dual space} of a Hilbert space $\mathcal{H}$, denoted by $\mathcal{H}^*$, which consists of all continuous linear functionals $\phi :\mathcal{H}\to\mathbb{C}$. The \emph{Riesz representation theorem} establishes a one-to-one correspondence between elements in a finite-dimensional Hilbert space and linear functionals in its dual space. It states that for every linear functional $\phi(\ket{\psi})$ in the dual space $\mathcal{H}^*$, there exists a unique vector $\ket{\phi}$ in the Hilbert space $\mathcal{H}$, such that $\phi(\ket{\psi})=\bracket{\phi}{\psi}$ for all $\ket{\psi}\in \mathcal{H}$. The linear functional $\phi$ is then commonly denoted as $\bra{\phi}\in \cH^*$.

\medskip

\begin{example}
The category $Hilb$ of finite-dimensional Hilbert vector spaces is a dagger symmetric monoidal category with a dual, given as follows. 
\begin{itemize}
\item The objects of the category $Hilb$ are finite-dimensional Hilbert spaces $\cH\in Obj_{Hilb}$, and the morphisms $\hat{M}\in Mor_{Hilb}$ are linear operators between Hilbert spaces $\hat{M}: \cH_A\to \cH_B$. 
\item The source and target maps of morphisms are the domain and the codomain of the operator, \ie\,, for each $\hat{M}:\cH_A \to \cH_B$, one has $s(\hat{M})=\cH_A$ and $t(\hat{M})=\cH_B$. 
\item The composition of morphisms $\hat{M},\hat{N}\in Mor_{Hilb}$ is the composition of operators $\hat{M} \cdot \hat{N}$ (sometimes called matrix multiplication). 
\item For each object in the category, $\cH\in Obj_{Hilb}$, the identity morphism is the identity operator $\hat{I}_{\cH}$ over that Hilbert space.
\item The tensor product $\boxtimes$ of the category is the ordinary tensor product $\otimes$ of linear algebra, so that for any two objects $\cH_A, \cH_B\in Obj_{Hilb}$ we have the tensor product of vector spaces $\cH_A \otimes \cH_B$, and for any two morphisms $\hat{M}, \hat{N}\in Mor_{Hilb}$ we have the tensor product of operators $\hat{M}\otimes \hat{N}$. 
\item The tensor product is associative, \ie\,, the associator is trivial, $(\cH_A \otimes \cH_B)\otimes \cH_C= \cH_A \otimes (\cH_B \otimes \cH_C)$. 
\item The unit object is the $1$-dimensional Hilbert space, namely the set of complex numbers, $\mathbb{C}$.
\item The left and right unitor maps are trivial, \ie\,, for each $\cH\in Obj_{Hilb}$ we have $\mathbb{C}\otimes \cH= \cH$ and $\cH \otimes \mathbb{C}=\cH$.
\item For any two objects $\cH_A,\cH_B\in Obj_{Hilb}$, the swap map is defined as $\sigma: \cH_A\otimes \cH_B \to \cH_B\otimes \cH_A$.
\item The dagger is the Hermitian adjoint of the operator $\hat{M}:\cH_A \to \cH_B$, denoted $\dagger: \hat{M} \to \hat{M}^\dagger$, where the Hermitian adjoint $\hat{M}^\dagger:\cH_B \to \cH_A$ is the unique linear map that satisfies the property $\bracket{\phi}{ \hat{M} \psi } = \bracket{\hat{M}^\dagger \phi}{\psi}$ for all vectors $\ket{\phi}\in\cH_B$ and $\ket{\psi}\in\cH_A$. The dagger acts trivially on objects, $\cH^\dagger = \cH$.
\item For every object $\cH \in Obj_{Hilb}$ of the category, the dual is just the dual Hilbert space $\cH^*$, defined via the Riesz representation theorem.
\item The unit and counit maps $\eta_\cH:\mathbb{C} \to \cH^*\otimes \cH$ and $\epsilon_\cH: \cH\otimes \cH^* \to \mathbb{C}$, are realized as
\begin{equation} \label{eq:HilbDefUnitCounit}
z \;\; \mathop{\longmapsto}\limits^{\eta_\cH} \;\; z \sum_i \bra{i} \otimes \ket{i}\,, \qquad
\ket{i}\otimes \bra{j} \;\; \mathop{\longmapsto}\limits^{\epsilon_\cH} \;\; \bracket{j}{i}\,,
\end{equation}
for any $z\in\kompleksni$ and any orthonormal basis $\{\,\ket{i} \,|\, i=1,\dots,\dim\cH\, \}$ of the Hilbert space $\cH$. Note that one can always define a function $\epsilon_\cH \cdot \sigma \cdot \eta_\cH : \kompleksni \to \kompleksni$, and then using (\ref{eq:HilbDefUnitCounit}) to evaluate it on an arbitrary number $z\in\kompleksni$ as follows:
$$
z \;\; \mathop{\longmapsto}\limits^{\eta_\cH} \;\; z \sum_i \bra{i} \otimes \ket{i}
\;\; \mathop{\longmapsto}\limits^{\sigma} \;\; z \sum_i \ket{i} \otimes \bra{i}
\;\; \mathop{\longmapsto}\limits^{\epsilon_\cH} \;\; z \sum_i \bracket{i}{i}
\;\; = \;\; z\,\dim\cH \,.
$$
\end{itemize}
Essentially, $\sigma\cdot\eta_\cH$ multiplies the number $z$ with an identity operator $\hat{I}_\cH$, and then $\epsilon_\cH$ takes its \emph{trace}.
\end{example}

Let us now introduce the category $nCob$ of oriented cobordisms between $n$-dimensional manifolds, specifically for the case $n=3$. This category is another example of a dagger symmetric monoidal category with a dual. The category of oriented cobordisms, discussed below, is defined for manifolds with an additional structure of \emph{orientation}. The orientation of the manifold is defined using the notion of an \emph{orientation group} structure. The case when the orientation group is the orthogonal group corresponds to unoriented cobordisms, while the case when the orientation group is the special orthogonal group gives rise to oriented cobordisms.

\medskip

\begin{example}
The category $3Cob$ of oriented cobordisms between $3$-dimensional manifolds is a dagger symmetric monoidal category with a dual, defined as follows. 
\begin{itemize}
\item The objects $\Sigma\in Obj_{3Cob}$ of the category $3Cob$ are closed oriented $3$-dimensional manifolds.

\item For any two objects $\Sigma_1, \Sigma_2 \in Obj_{3Cob}$,  the set of morphisms $Mor_{3Cob}(\Sigma_1, \Sigma_2)$ consists of compact oriented $4$-dimensional manifolds $\mathcal{M}$ (called \emph{cobordisms}) with the boundary $\partial\mathcal{M}=\bar{\Sigma}_1\sqcup{\Sigma}_2$, where $\sqcup$ denotes the disjoint union, and $\bar{\Sigma}_1$ denotes the manifold $\Sigma_1$ with the opposite orientation. The set $\cM\setminus \partial\cM$ is often called the \emph{bulk}, distinguishing the interior of $\cM$ from its boundary $\partial\cM$.
\item The orientation of the boundary provides a way to understand a $4$-dimensional manifold $\cM$ with the boundary $\partial\mathcal{M}=\bar{\Sigma}_1\sqcup{\Sigma}_2$ as a morphism from its incoming to its outgoing boundary component, \ie\,, from $\Sigma_1$ to $\Sigma_2$, as shown in Figure \ref{Figure1}, and to define the source and target maps of the morphism as $s(\cM)=\Sigma_1$ and $t(\cM)=\Sigma_2$.
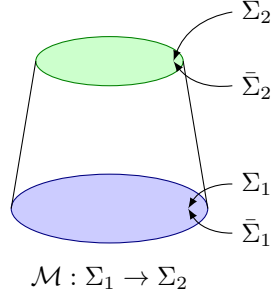
\begin{figure}[!ht]
\begin{center}
\begin{tikzpicture}[scale=0.65, >=latex] 
  % Bottom surface (colored blue)
  \filldraw[fill=blue!20, draw=blue!50!black]
    (0,-3) ellipse (2 and 0.7);
  
  % Top surface (colored green)
  \filldraw[fill=green!20, draw=green!50!black]
    (0,0) ellipse (1.5 and 0.5);
  \node at (3, -3.5) (label1) {$\bar{\Sigma}_1$};
  \node at (3, -0.5) (label2) {$\bar{\Sigma}_2$};
  \node at (3, 1) (label3) {${\Sigma}_2$};
  \node at (3, -2.5) (label3) {${\Sigma}_1$};
  % Cylinder outline
  \draw (-2,-3) -- (-1.5,0);
  \draw (2,-3) -- (1.5,0);
  \draw[->] (2.5,-3.5) to[bend left] (1.6,-3);
  \draw[->] (2.5,-0.5) to[bend left] (1.3,0);
  \draw[->] (2.5,1) to[bend right] (1.3,0);
  \draw[->] (2.5,-2.5) to[bend right] (1.6,-3);
  \node[below] at (0,-4) {$\mathcal{M}: \Sigma_1 \to \Sigma_2$};
\end{tikzpicture}
\caption{\label{Figure1} The oriented $4$-dimensional manifold $\mathcal{M}:\Sigma_1 \to \Sigma_2$ with the boundary $\partial\mathcal{M}=\bar{\Sigma}_1\sqcup{\Sigma}_2$.}
\end{center}
\end{figure}
\item The composition of morphisms is defined by the \emph{connected sum} --- \emph{gluing} of manifolds along a common boundary. For any two $4$-dimensional manifolds $\mathcal{M}_A \in Mor_{3Cob}(\Sigma_1, \Sigma)$ and $\mathcal{M}_B \in Mor_{3Cob}(\Sigma, \Sigma_2)$ that are composable, \ie\,, that satisfy $t(\cM_A)=s(\cM_B)$, the composition $\mathcal{M}_B \# \mathcal{M}_A \in Mor_{3Cob}(\Sigma_1, \Sigma_2)$ is the connected sum of $\mathcal{M}_A$ and $\mathcal{M}_B$ along the $3$-dimensional manifold $\Sigma$, as shown in Figure \ref{MM}.
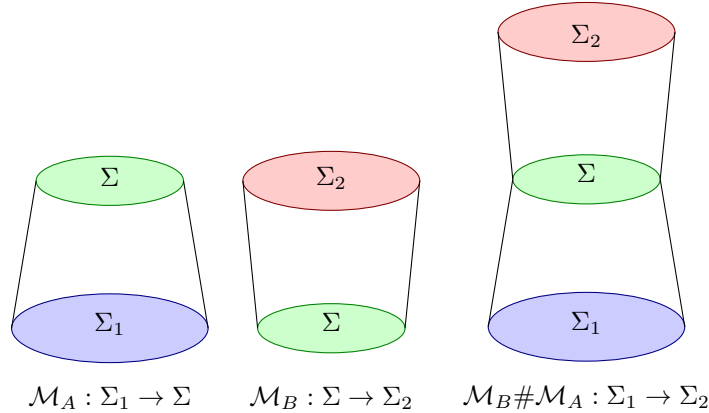
\begin{figure}[!ht]
\begin{center} 
\begin{tabular}{ccc}
% Drawing W1: M to N
\begin{tikzpicture}[scale=0.65]
  % Bottom surface labeled with M (colored blue)
  \filldraw[fill=blue!20, draw=blue!50!black]
    (0,-3) ellipse (2 and 0.7); % Adjusted ellipse size
  \node[below, black] at (0,-2.5) {$\Sigma_1$};
  
  % Top surface labeled with N (colored green)
  \filldraw[fill=green!20, draw=green!50!black]
    (0,0) ellipse (1.5 and 0.5);
  \node[above, black] at (0,-0.3) {$\Sigma$};
  
  % Cylinder outline
  \draw (-2,-3) -- (-1.5,0); % Adjusted left boundary
  \draw (2,-3) -- (1.5,0);   % Adjusted right boundary
  \node[below] at (0,-4) {$\mathcal{M}_A: \Sigma_1 \to \Sigma$};
\end{tikzpicture}
&
% Drawing W2: N to P
\begin{tikzpicture}[scale=0.65]
  % Bottom surface labeled with N (colored green)
  \filldraw[fill=green!20, draw=green!50!black]
    (0,-3) ellipse (1.5 and 0.5);
  \node[below, black] at (0,-2.5) {$\Sigma$};
  
  % Top surface labeled with P (colored red)
  \filldraw[fill=red!20, draw=red!50!black]
    (0,0) ellipse (1.8 and 0.6);
  \node[above, black] at (0,-0.35) {$\Sigma_2$};
  
  % Cylinder outline
  \draw (-1.5,-3) -- (-1.8,0);
  \draw (1.5,-3) -- (1.8,0);
  \node[below] at (0,-4) {$\mathcal{M}_B: \Sigma \to \Sigma_2$};
\end{tikzpicture}
&
% Drawing the composition W2 \circ W1
\begin{tikzpicture}[scale=0.65]
  % Bottom surface labeled with M (colored blue)
  \filldraw[fill=blue!20, draw=blue!50!black]
    (0,-6) ellipse (2 and 0.7);
  \node[below, black] at (0,-5.5) {$\Sigma_1$};
  
  % Middle surface labeled with N (colored green)
  \filldraw[fill=green!20, draw=green!50!black]
    (0,-3) ellipse (1.5 and 0.5);
  \node[below, black] at (0,-2.5) {$\Sigma$};
  
  % Top surface labeled with P (colored red)
  \filldraw[fill=red!20, draw=red!50!black]
    (0,0) ellipse (1.8 and 0.6);
  \node[above, black] at (0,-0.5) {$\Sigma_2$};
  
  % Cylinder outlines
  \draw (-2,-6) -- (-1.5,-3);
  \draw (2,-6) -- (1.5,-3);
  \draw (-1.5,-3) -- (-1.8,0);
  \draw (1.5,-3) -- (1.8,0);
  \node[below] at (0,-7) {$\mathcal{M}_B \# \mathcal{M}_A:\Sigma_1 \to \Sigma_2$};
\end{tikzpicture}

\end{tabular}
\caption{\label{MM}The composition of cobordisms $\mathcal{M}_B \# \mathcal{M}_A$.}
\end{center}
\end{figure}
\item For each object, a $3$-dimensional manifold $\Sigma$, there exists an identity morphism $\text{id}_\Sigma \in Mor_{3Cob}(\Sigma, \Sigma)$ given by a $4$-dimensional cylinder $[0,1] \times \Sigma$ with the boundary $\partial([0,1]\times \Sigma)=\bar{\Sigma}\sqcup {\Sigma}$, as shown in Figure \ref{figure:identity}. 
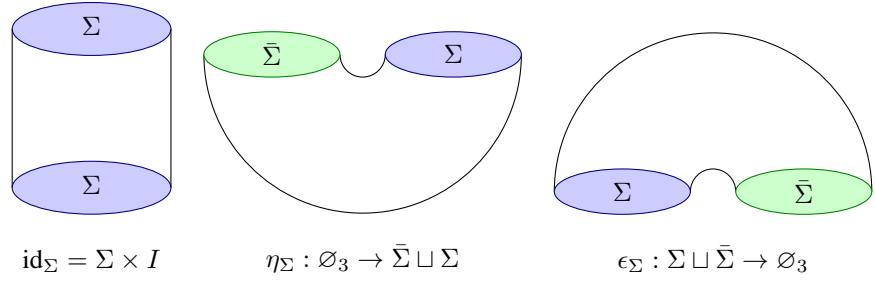
\begin{figure}[!ht]
\begin{center}
\begin{tabular}{ccc}
  \begin{tikzpicture}[scale=0.7]
  % Bottom surface labeled with N (colored green)
  \filldraw[fill=blue!20, draw=blue!50!black]
    (0,-3) ellipse (1.5 and 0.5);
  \node[below, black] at (0,-2.6) {$\Sigma$};
  
  % Top surface labeled with P (colored red)
  \filldraw[fill=blue!20, draw=blue!50!black]
    (0,0) ellipse (1.5 and 0.5);
  \node[above, black] at (0,-0.3) {$\Sigma$};
  
  % Cylinder outline
  \draw (-1.5,-3) -- (-1.5,0);
  \draw (1.5,-3) -- (1.5,0);
  \node[below] at (0,-4) {$\text{id}_{\Sigma}=\Sigma \times I$};
\end{tikzpicture}
&
\begin{tikzpicture}[scale=0.6]
  % Drawing two balls
  \draw[black, fill=white] (3.5,0) arc (0:-180:3.5);
  \draw[black, fill=white] (0.5,0) arc (0:-180:0.5);

  % Left surface labeled BarSigma
  \filldraw[fill=green!20, draw=green!50!black]
    (-2,0) ellipse (1.5 and 0.5);
  \node[black] at (-2,0) {$\bar{\Sigma}$};

  % Drawing the surface Sigma (ellipse)
  \filldraw[fill=blue!20, draw=blue!50!black]
    (2,0) ellipse (1.5 and 0.5);
  \node[black] at (2,0) {$\Sigma$};

  \node[below] at (0,-4) {$\eta_{\Sigma}:\varnothing_3 \to \bar{\Sigma} \sqcup \Sigma$};
\end{tikzpicture}
&
\begin{tikzpicture}[scale=0.6]
  % Drawing two balls
  \draw[black, fill=white] (3.5,0) arc (0:180:3.5);
  \draw[black, fill=white] (0.5,0) arc (0:180:0.5);

  % Left surface labeled BarSigma
  \filldraw[fill=blue!20, draw=blue!50!black]
    (-2,0) ellipse (1.5 and 0.5);
  \node[black] at (-2,0) {$\Sigma$};

  % Drawing the surface Sigma (ellipse)
  \filldraw[fill=green!20, draw=green!50!black]
    (2,0) ellipse (1.5 and 0.5);
  \node[black] at (2,0) {$\bar{\Sigma}$};

  \node[below] at (0,-1) {$\epsilon_{\Sigma}: \Sigma \sqcup \bar{\Sigma} \to \varnothing_3 $};
\end{tikzpicture}
\end{tabular}
\caption{The identity cobordism, the unit map and the counit map.\label{figure:identity}}
\end{center}
\end{figure}
\item For any two objects $\Sigma_1, \Sigma_2 \in Obj_{3Cob}$, the tensor product $\boxtimes$ is a disjoint union of manifolds $\Sigma_1\sqcup \Sigma_2$. For any two cobordisms $\cM_A:\Sigma_1\to\Sigma_3$ and $\cM_B:\Sigma_2 \to \Sigma_4$, the tensor product gives the disjoint union $\cM_A\sqcup \cM_B:\Sigma_1\sqcup \Sigma_2\to\Sigma_3\sqcup \Sigma_4$. 
\item The disjoint union of manifolds is associative, \ie\,, the associator is trivial, $(\Sigma_1 \sqcup \Sigma_2)\sqcup\Sigma_3= \Sigma_1 \sqcup (\Sigma_2 \sqcup \Sigma_3)$. 
\item The unit object is the empty manifold, denoted by $\varnothing_3$. Note that the empty manifold $\varnothing$ is a manifold of any dimension, but we have added the dimension in the subscript in order to notationally distinguish between $\varnothing_3$ which is an object in the category, and $\varnothing_4$ which is a morphism in the category. The latter is in fact the identity morphism for the unit object, $\text{id}_{\varnothing_3} = \varnothing_4$. Formally, like all other $4$-dimensional manifolds without boundary, $\varnothing_4$ can be regarded as a cobordism from $\varnothing_3$ to $\varnothing_3$.
\item The left and right unitor are trivial, \ie\,, for each $\Sigma\in Obj_{3Cob}$ one has $\varnothing_3\sqcup \Sigma= \Sigma$ and $\Sigma \sqcup \varnothing_3=\Sigma$.
\item The swap map is trivial, \ie\,, for any two objects $\Sigma_1,\Sigma_2\in Obj_{3Cob}$ one has $\Sigma_1\sqcup \Sigma_2 = \Sigma_2 \sqcup \Sigma_1$.
\item For every cobordism $\cM:\Sigma_1 \to \Sigma_2$, the dagger is a reversed cobordism $\cM^\dagger: \Sigma_2\to \Sigma_1$.
\item For every object $\Sigma \in Obj_{3Cob}$ of the category, the dual object is the $3$-dimensional manifold of the opposite orientation, denoted by $\bar{\Sigma}$.
\item For every object $\Sigma \in Obj_{3Cob}$ of the category, the unit map $\eta_\Sigma$ and the counit map $\epsilon_\Sigma$ are realized as $\eta_\Sigma: \varnothing_3 \to \bar{\Sigma} \sqcup \Sigma$ and $\epsilon_\Sigma:\Sigma \sqcup \bar{\Sigma}\to \varnothing_3$, as follows. One starts from the identity cobordism $\text{id}_\Sigma$ for the object $\Sigma$, and constructs the unit map $\eta_\Sigma$ by ``bending'' the identity cobordism into a $\cup$-shape, so that it does not map $\Sigma \to\Sigma$, but instead $\varnothing_3 \to \bar{\Sigma} \sqcup \Sigma$. Similarly, the identity cobordism can be bent into a $\cap$-shape, so that it maps $\Sigma \sqcup \bar{\Sigma} \to \varnothing_3$, thereby becoming the counit map $\epsilon_\Sigma$. The bending procedure reverses the orientation of the ``bottom'' boundary $\Sigma$ in the first case, while it reverses the orientation of the ``top'' boundary $\Sigma$ in the second case, respectively (see Figure \ref{figure:identity}).
\end{itemize}
\end{example}

Finally, we are particularly interested in the third example -- the category $3TCob$, which differs from the category $3Cob$ due to its additional structure of \emph{triangulation} of $3$-dimensional manifolds (a simplicial complex, technically speaking). Note that we introduce a triangulation on the cobordism boundaries (objects in the category), but not in the $4$-dimensional bulks (morphisms in the category). Instead, the equivalence class of all possible triangulations of the bulk of any given $4$-dimensional manifold (with fixed triangulation on its boundary) corresponds to a single morphism of the category $3TCob$. The reason for this definition is the following. If one tries to define a category of triangulated cobordisms, such that each triangulation of the bulk of $4$-dimensional manifold corresponds to a different morphism of the category, it can be seen that the identity morphism of any given object would not be unique, hence this structure would not form a category. The definition of the category $3TCob$ circumvents this issue by triangulating the boundary, but not the bulk.

\medskip

\begin{example}
The category $3TCob$ of cobordisms between triangulated $3$-dimensional manifolds is a dagger symmetric monoidal category with a dual, given as follows.
\begin{itemize}
\item The objects of the category $3TCob$ are closed oriented triangulated $3$-dimensional manifolds $\mathcal{T}(\Sigma)\in Obj_{3TCob}$.
\item For any two objects $\cT(\Sigma_1)$ and $\cT(\Sigma_2)$ in the category $3TCob$, the set of morphisms $Mor_{3TCob}(\cT(\Sigma_1), \cT(\Sigma_2))$ consists of compact oriented $4$-dimensional manifolds $\mathcal{M}$ with the boundary $\partial\mathcal{M}=\cT(\bar{\Sigma}_1)\sqcup\cT({\Sigma}_2)$, where $\sqcup$ as before denotes the disjoint union, and $\cT(\bar{\Sigma}_1)$ denotes the triangulated manifold $\cT(\Sigma_1)$ with the opposite orientation of the triangulation simplices. Note that a possible triangulation of the bulk of the cobordism is not fixed, that is, the equivalence class of all possible triangulations of the same $4$-dimensional manifold (that have the fixed triangulation on the boundary and that preserve the topology of the manifold) corresponds to a single morphism of the category.
\item The orientation of the boundary provides a way to understand a $4$-dimensional manifold $\mathcal{M}$ as a cobordism from its incoming to its outgoing boundary component, \ie\,, from $\cT(\Sigma_1)$ to $\cT(\Sigma_2)$. The source and target maps are defined as $s(\mathcal{M}) = \cT(\Sigma_1)$ and $t(\mathcal{M}) = \cT(\Sigma_2)$.
\item The composition of morphisms is given by the connected sum of manifolds, as in $3Cob$, but the difference is that it is defined only if the \emph{triangulation} of their common boundary matches, \ie\,, for $\mathcal{M}_A \in Mor_{3TCob}(\cT(\Sigma_1), \cT(\Sigma))$ and $\mathcal{M}_B \in Mor_{3TCob}(\cT'(\Sigma), \cT(\Sigma_2))$, the composition is defined only when $\cT(\Sigma)=\cT'(\Sigma)$, and results in the cobordism $\mathcal{M}_B \# \mathcal{M}_A \in Mor_{3TCob}(\cT(\Sigma_1), \cT(\Sigma_2))$. After gluing, the triangulation of $\Sigma$ becomes part of the bulk, and thus irrelevant.
\item For each object $\cT(\Sigma)\in Obj_{3TCob}$, there exists an identity morphism $\text{id}_{\cT(\Sigma)} \in {Mor}_{3TCob}(\cT(\Sigma), \cT(\Sigma))$ given by a cylinder, \ie, the manifold $[0,1] \times \cT(\Sigma)$. The boundary $\partial([0,1]\times \cT(\Sigma))=\cT(\bar{\Sigma})\sqcup \cT({\Sigma})$ has a fixed triangulation, while the bulk of the manifold does not.

As stated above, the motivation for introducing the category $3TCob$ in this manner lies in the uniqueness of the identity morphism for a given object. Namely, if one instead considers cobordisms that have a fixed triangulation both in bulk and on the boundaries, the identity morphism would not be unique and thus not well defined, and consequently such a structure would not form a category. Introducing the morphisms of the category as an equivalence class of all possible triangulations of the same $4$-dimensional manifold (while keeping the triangulation of its boundary and its topology fixed) circumvents this problem and renders the identity morphism unique (for a given object in the category). Thus in the category $3TCob$ the identity morphisms are well defined.
\item The tensor product $\boxtimes$ is the disjoint union of manifolds, like in $3Cob$. For any two objects $\cT(\Sigma_1), \cT(\Sigma_2) \in Obj_{3TCob}$, the product gives the disjoint union of triangulated $3$-dimensional manifolds $\cT(\Sigma_1)\sqcup \cT(\Sigma_2)$. Similarly for morphisms.
\item The unit object is the empty manifold $\varnothing_3$, like in $3Cob$. Its triangulation is trivial, and its corresponding identity morphism is $\varnothing_4$.
\item As in the case of the $3Cob$ category, the associator, the left and the right unitor, and the swap map are all trivial.
\item For every morphism $\cM: \cT(\Sigma_1) \to \cT(\Sigma_2)$, with a boundary $\partial \cM=\cT(\bar{\Sigma}_1) \sqcup \cT(\Sigma_2)$, the dagger gives the morphism $\cM^\dagger:\cT(\Sigma_2) \to \cT(\Sigma_1)$, with the boundary $\partial \cM^\dagger=\cT(\bar{\Sigma}_2) \sqcup \cT(\Sigma_1)$, namely the reverse cobordism. 
\item For every object $\cT(\Sigma) \in Obj_{3TCob}$ of the category, the dual operation gives the dual object, the triangulated $3$-dimensional manifold of opposite orientation $\cT(\bar{\Sigma})$, where the orientation reversal of a manifold is manifested in its triangulation by reversing the orientation of all links, triangles, and tetrahedrons.
\item For every object $\cT(\Sigma) \in Obj_{3TCob}$ of the category, the unit map $\eta_{\cT(\Sigma)}$ and the counit map $\epsilon_{\cT(\Sigma)}$ are realized as $\eta_{\cT(\Sigma)}: \varnothing \to  \cT(\bar{\Sigma}) \sqcup \cT(\Sigma)$ and $\epsilon_{\cT(\Sigma)}:\cT(\Sigma) \sqcup \cT(\bar{\Sigma})\to \varnothing$ in the same way as for the category $3Cob$.
\end{itemize}
\end{example}

This concludes the definition of the $3TCob$ category, as the final example of a dagger symmetric monoidal category with a dual. Our next point of interest is to establish relationships between these categories.

\subsection{Functors}

In category theory, a \emph{functor} is a map between two categories that maps objects and morphisms from one category to objects and morphisms in another category, in a way that preserves all of the structure that is defined within the categories. The formal definition of a functor is as follows.

\medskip

\begin{definition}
Let $F: \mathcal{C}\to\mathcal{D}$ be a \emph{functor} between two dagger symmetric monoidal categories with a dual. The following must hold:
\begin{enumerate}
\item The functor $F$ maps the objects of one category to the objects of another, such that for each object $A\in Obj_\mathcal{C}$, there is an object $F(A)\in Obj_\mathcal{D}$, and the morphisms from one category to the morphisms of another, such that for each morphism $f: A \to B$ in $Mor_\mathcal{C}$, there is a morphism $F(f): F(A) \to F(B)$ in $Mor_\mathcal{D}$. 
\item For every morphism $f\in Mor_\cC$, the image of the source of $f$ under $F$ is equal to the source of the image of $f$ under $F$, and the image of the target of $f$ under $F$ is equal to the target of the image of $f$ under $F$, \ie\,,
$$
F(s(f))=s(F(f))\,, \qquad F(t(f))=t(F(f))\,.
$$
\item The functor $F$ must preserve the composition of morphisms, \ie\,, for every pair of morphisms $f: A \to B$ and $g: B \to C$ in $\mathcal{C}$ the following condition is satisfied:
$$
F(g \circ f) = F(g) \circ F(f) \,.
$$
Note that we denote the composition in $\cC$ and the composition in $\cD$ with the same symbol, $\circ$, for simplicity.
\item The identity morphism in the source category must map to the identity morphism of the corresponding object in the target category, \ie, for every object $A\in Obj_\mathcal{C}$ and its identity morphism $\text{id}_A$, the functor $F$ satisfies the condition
$$
F(\text{id}_A) = \text{id}_{F(A)}\,.
$$
\item The functor $F$ must preserve the tensor product structure. For any two objects $A, B\in Obj_\cC$, the image of the tensor product of $A$ and $B$ under $F$ is equal to the tensor product of the images of $A$ and $B$ under $F$, the objects $F(A), F(B)\in Obj_\cD$, \ie\,,
$$
F(A \boxtimes_\mathcal{C} B)=F(A)\boxtimes_\mathcal{D} F(B)\,.
$$
Likewise, for any two morphisms $f,g\in Mor_\cC$, the image of the tensor product of $f$ and $g$ under $F$ is equal to the tensor product of the images of $f$ and $g$ under $F$, the morphisms $F(f), F(g)\in Mor_\cD$, \ie\,,
$$
F(f \boxtimes_\mathcal{C} g)=F(f)\boxtimes_\mathcal{D} F(g)\,.
$$
\item The functor $F$ maps the unit object in category $\mathcal{C}$ to the unit object in category $\mathcal{D}$ , \ie\,,
$$
F(\mathbb{1}_\cC)=\mathbb{1}_\mathcal{D}\,,
$$
and as a consequence of property 4 above, it also maps the identity morphism of the unit object in $\mathcal{C}$ to the identity morphism of the unit object in $\mathcal{D}$,
$$
F(\text{id}_{\mathbb{1}_\cC}) = \text{id}_{F(\mathbb{1}_\cC)} = \text{id}_{\mathbb{1}_\mathcal{D}}\,.
$$
\item The functor $F$ maps the associator map $\alpha_{A,B,C}$, the swap map $\sigma_{A,B}$, and the left and right unitor maps $\lambda_A$ and $\rho_A$ from the source category, to their corresponding maps in the target category, \ie\,,
$$
F(\alpha_{A,B,C})=\alpha_{F(A),F(B),F(C)}\,, \qquad F(\sigma_{A,B})=\sigma_{F(A),F(B)}\,,
$$
$$
F(\lambda_A)=\lambda_{F(A)}\,, \qquad F(\rho_A)=\rho_{F(A)}\,.
$$
\item The functor $F$ commutes with the dagger, \ie\,, for any morphism $f\in Mor_\cC$ the functor satisfies the condition
$$
F(f^\dagger)=F(f)^\dagger\,.
$$
We keep in mind that the dagger acts trivially on objects.
\item Similarly, the functor $F$ commutes with the dual, \ie\,, for any object $A\in Obj_\cC$ the functor satisfies the condition
$$
F(A^*)=F(A)^*\,.
$$
We keep in mind that the dual acts trivially on morphisms.
\item For every object $A \in Obj_\mathcal{C}$ in the source category, the functor $F$ maps the unit map
  $\eta_{A}:\mathbb {1}_\cC \to A^*\boxtimes A$ and the counit map $\epsilon_{A}:A\boxtimes A^*\to \mathbb {1}_\cC$, to the unit and the counit map $\eta_{F(A)}:\mathbb {1}_\mathcal{D} \to F(A)^*\boxtimes F(A)$ and $\epsilon_{F(A)}:F(A)\boxtimes F(A)^*\to \mathbb {1}_\mathcal{D}$ of the corresponding object $F(A)\in Obj_\mathcal{D}$ in the target category, respectively, \ie\,,
$$
F(\eta_{A})=\eta_{F(A)}\,, \qquad F(\epsilon_{A})=\epsilon_{F(A)}\,.
$$
\end{enumerate}
\end{definition}

At this point we are finally ready to give a definition of a TQFT --- it is a functor, traditionally denoted $Z$, from the category of cobordisms to the category of vector spaces. In our case, there are two flavors, since we have introduced two different categories of cobordisms:
\smallskip
\begin{center}
 \begin{tikzpicture}[>=stealth, baseline=(current  bounding  box.center)]
  % Nodes
  \node (A1) at (0,0) {$3TCob$};
  \node (A2) at (3,0) {$Hilb\,,$};
  % Arrows
  \draw[->] (A1) -- node[above]{$\scriptstyle Z$} (A2);
\end{tikzpicture}
\end{center}
\smallskip
and
\smallskip
\begin{center}
 \begin{tikzpicture}[>=stealth, baseline=(current  bounding  box.center)]
  % Nodes
  \node (A1) at (0,0) {$3Cob$};
  \node (A2) at (3,0) {$Hilb\,.$};
  % Arrows
  \draw[->] (A1) -- node[above]{$\scriptstyle \tilde{Z}$} (A2);
\end{tikzpicture}
\end{center}
\smallskip
One can in principle construct many functors of either flavor, but they all need to satisfy the rules expected from a functor. Let us therefore spell out the axioms for $Z$. The axioms for $\tilde{Z}$ will be analogous.

\medskip

\begin{definition}\label{TQFTfunctor}
If $Z: 3TCob \to Hilb $ is a functor between two dagger symmetric monoidal categories with a dual $3TCob$ and $Hilb$, the following axioms must hold:
\begin{enumerate}
\item[1.] \emph{Objects.} The functor $Z$ maps objects to objects. Specifically, each triangulated $3$-dimensional manifold $\cT(\Sigma) \in Obj_{3TCob}$ is mapped to some Hilbert space $\mathcal{H}\in Obj_{Hilb}$:
\begin{equation} \label{eq:functorAxiomsBeginning}
Z(\cT(\Sigma)) = \cH\,.
\end{equation}
\item[2.] \emph{Morphisms.}  The functor $Z$ maps morphisms to morphisms. Specifically, each $4$-dimensional manifold $\cM$ (that maps the triangulated manifold $\cT(\Sigma_1)$ to the triangulated manifold $\cT(\Sigma_2)$) is mapped to an operator $\hat{M}: \cH_1 \to \cH_2$,
\begin{equation} \label{eq:def13}
Z(\cM) = \hat{M}\,,
\end{equation}
provided that $Z(\cT(\Sigma_1))=\cH_1$ and $Z(\cT(\Sigma_2))=\cH_2$.
\item[3.] \emph{Composition of morphisms --- functoriality.} The functor of a composition is a composition of functors. Specifically, given two composable manifolds $\cM_A: \cT(\Sigma_1) \to \cT(\Sigma)$ and $\cM_B: \cT(\Sigma) \to \cT(\Sigma_2)$, their connected sum $\cM_B \# \cM_A $ along the common boundary $\cT(\Sigma)$ is mapped into the composition of operators corresponding to functors of individual manifolds
$\hat{M}_A = Z({\cM_A}) : Z(\cT({\Sigma_1})) \to Z(\cT({\Sigma}))$ and $\hat{M}_B = Z({\cM_2}) : Z(\cT({\Sigma}))\to Z(\cT({\Sigma_2}))$:
\begin{equation}\label{eq:functoriality}
Z({\mathcal{M}_{B}\# \mathcal{M}_{A}})=\hat{M}_B \cdot \hat{M}_A\,.
\end{equation}
Note that the operators $\hat{M}_A$ and $\hat{M}_B$ are often represented as rectangular matrices, and their composition $\hat{M}_B \cdot \hat{M}_A$ is understood to be ordinary matrix multiplication.
\item[4.] \emph{Identity morphism}. For each object, the functor $Z$ maps the identity morphism into the identity morphism. Specifically, given a triangulated manifold $\cT(\Sigma)$ and its corresponding Hilbert space $\cH$, their identity morphisms are related via the functor $Z$ as:
\begin{equation}\label{eq:def2}
Z(\text{id}_{\cT(\Sigma)})=\hat{I}_{\cH}\,.
\end{equation}
\item[5.] \emph{Tensor product --- multiplicativity.} The functor of a tensor product is a tensor product of functors, both for objects and morphisms. Specifically, given two triangulated manifolds $\cT(\Sigma_1)$ and $\cT(\Sigma_2)$, which are mapped by the functor $Z$ to Hilbert spaces $\cH_1$ and $\cH_2$ respectively, we have:
\begin{equation}\label{eq:def3}
Z(\cT(\Sigma_1)\sqcup \cT(\Sigma_2))=\cH_1 \otimes \cH_2\,.
\end{equation}
Also, given two $4$-dimensional manifolds $\cM_A: \cT(\Sigma_1)\to \cT(\Sigma_1')$ and $\cM_B: \cT(\Sigma_2)\to \cT(\Sigma_2')$, we have 
\begin{equation}\label{eq:def4}
Z(\cM_A \sqcup \cM_B)=\hat{M}_A\otimes \hat{M}_B\,,
\end{equation}
where $\hat{M}_A = Z(\cM_A)$ and $\hat{M}_B = Z(\cM_B)$.
\item[6.] \emph{Unit object --- normalization.} The functor maps the unit object into the unit object. Specifically, the empty manifold $\varnothing_3$ is mapped to the the one-dimensional Hilbert space $\mathbb{C}$:
\begin{equation}\label{eq:def5}
Z({\varnothing_3})=\mathbb{C}\,.
\end{equation}
Given this, one can see that also the \emph{identity morphism of the unit object}, $\varnothing_4 = \text{id}_{\varnothing_3} : \varnothing_3 \to \varnothing_3$, maps to the identity morphism $\hat{I}_\mathbb{C}: \mathbb{C} \to \mathbb{C}$:
\begin{equation}\label{eq:def6}
Z(\text{id}_{\varnothing_3})=\hat{I}_\mathbb{C}\,.
\end{equation}
The identity operator over a one-dimensional Hilbert space $\mathbb{C}$ is of course simply the function $\hat{I}(z) = z$, for every $z\in\mathbb{C}$. Therefore, the above equation is sometimes written in the form of a \emph{normalization condition},
\begin{equation} \label{eq:def6alt}
Z(\varnothing_4)=1\,,
\end{equation}
where the number $1$ on the right-hand side is understood both as the identity operator over the Hilbert space $\mathbb{C}$ and as the neutral element for multiplication in the field of complex numbers $\mathbb{C}$.
\item[7.] \emph{Symmetric monoidal maps.} The associator map,  the swap map, and the left and right unitor maps are all trivial in both categories, so the functor $Z$ maps them in a trivial manner as well.
\item[8.] \emph{Dagger --- hermitivity.} The functor of a dagger is the dagger of a functor. Specifically, given a $4$-dimensional manifold $\cM: \cT(\Sigma_1) \to \cT(\Sigma_2)$ and its corresponding operator $\hat{M} = Z(\cM)$, we have that the reverse manifold $\cM^\dagger$ maps into a Hermitian adjoint operator $\hat{M}^\dagger$:
\begin{equation} \label{eq:def11}
Z({\cM^\dagger})=\hat{M}^\dagger\,.
\end{equation}
\item[9.] \emph{Dual --- involution.} The functor of a dual is the dual of a functor. Specifically, given a triangulated manifold $\cT(\Sigma)$ and its corresponding Hilbert space $\cH$, we have that the oppositely oriented manifold $\cT(\bar{\Sigma})$ maps into a dual Hilbert space $\cH^*$:
\begin{equation}\label{eq:def8}
Z(\cT(\bar{\Sigma}))=\cH^*\,.
\end{equation}
Note that reversing the orientation of a triangulated manifold twice gives us back the original triangulated manifold, $\cT(\bar{\bar{\Sigma}}) = \cT(\Sigma)$. The action of the functor then implies that
\begin{equation} \label{eq:def10}
\cH^{**} = \cH\,,
\end{equation}
which is satisfied only if the Hilbert space is finite-dimensional, according to the Riesz representation theorem.

\item[10.] \emph{Unit and counit maps.} For every triangulated manifold $\cT(\Sigma)$ and its corresponding Hilbert space $\cH$, the unit map $\eta_{\cT(\Sigma)}:\varnothing_3 \to \cT(\bar{\Sigma})\sqcup \cT(\Sigma)$ and the counit map $\epsilon_{\cT(\Sigma)}:\cT(\Sigma)\sqcup \cT(\bar{\Sigma})\to \varnothing_3$ are mapped by the functor $Z$ to their counterparts $\eta_\cH:\mathbb{C} \to \cH^*\otimes \cH$ and $\epsilon_\cH:\cH\otimes \cH^*\to \mathbb{C}$, respectively:
\begin{equation} \label{eq:functorAxiomsEnd}
Z(\eta_{\cT(\Sigma)})=\eta_\cH\,, \qquad Z(\epsilon_{\cT(\Sigma)})=\epsilon_\cH\,.
\end{equation}
\end{enumerate}
\end{definition}

From Definition \ref{TQFTfunctor} one can see that the functorial conditions stated above essentially reflect Atiyah's axioms for topological quantum field theories \cite{Atiyah1988}. That said, one can also appreciate the elegance and expressiveness of the category theory language, naturally encoding all of the above properties into a compact statement that a TQFT is a functor between the category of cobordisms and the category of Hilbert spaces. Let us also note that the entire set of statements in Definition \ref{TQFTfunctor} can be repeated for the category $3Cob$, simply by not mentioning that the $3$-dimensional manifolds are equipped with triangulations. In Section \ref{SecIII} we will explore the differences between the functor $Z$ from $3TCob$ to $Hilb$, and the functor $\tilde{Z}$ from $3Cob$ to $Hilb$.

Obviously, many different functors $Z$ satisfying requirements (\ref{eq:functorAxiomsBeginning})--(\ref{eq:functorAxiomsEnd}) can in principle exist, corresponding to different TQFTs. In what follows, we shall explicitly formulate one such functor, discuss its properties, and later demonstrate that it indeed satisfies the requirements (\ref{eq:functorAxiomsBeginning})--(\ref{eq:functorAxiomsEnd}).

\section{TQFT functor based on a 3-group}\label{SecIII} 

\subsection{Preliminaries}

In our previous work \cite{Radenkovic2022_2}, we have constructed a topological invariant for compact closed $4$-dimensional manifolds, based on the algebraic structure of a $3$-group, within the framework of higher category theory \cite{BaezHuerta}. A semistrict $3$-group is most commonly represented as a $2$-crossed module $(L\stackrel{\delta}{\to} H \stackrel{\partial}{\to}G,\,\rhd,\,\{\_,\,\_\}_\mathrm{p})$, which consists of three groups $G$, $H$ and $L$, and four maps between them --- the action $\rhd$ of group $G$ on the other three groups,
$$
\rhd : G \times X \to X\,, \quad (X = G,H,L)
$$
the homomorphisms $\partial$ and $\delta$ between the three groups to make a chain complex,
$$
L\stackrel{\delta}{\to} H \stackrel{\partial}{\to}G\,,
$$
and the bracket map
$$
\{\_,\,\_\}_\mathrm{p} : H \times H \to L\,,
$$
called the \emph{Peiffer lifting}. In order to be a valid $2$-crossed module, this structure needs to satisfy a certain number of axioms, detailed in Appendix \ref{AppI}.

Given a $2$-crossed module and a compact closed $4$-dimensional manifold $\cM$ that admits a triangulation, one introduces a triangulation $\cT(\cM)$ which consists of simplices with dimensions from $0$ up to $4$, respectively:
\begin{itemize}
\item \emph{vertices}, labeled with positive integers $j = 1,2,\dots$,
\item \emph{edges}, labeled with pairs of vertices $(jk)$,
\item \emph{triangles}, labeled with three vertices $(jk\ell)$,
\item \emph{tetrahedra}, labeled with four vertices $(jk\ell m)$, and
\item \emph{$4$-simplices}, labeled with five vertices $(jk\ell mn)$.  
\end{itemize}
Then, to each edge one assigns a group element $g_{jk} \in G$, to each triangle one assigns $h_{jk\ell} \in H$, and to each tetrahedron one assigns $l_{jk\ell m} \in L$. Each set of assignments is called a \emph{state} of the triangulation $\cT(\cM)$, sometimes also called a \emph{configuration}, or a \emph{coloring}.

One then defines a \emph{state sum} (also called a \emph{configuration integral} or \emph{path integral}) as an integral over all possible configurations of the Dirac delta functions of certain group elements, as follows:
\begin{equation} \label{eq_partition_3bf}
\begin{array}{lcl}
  \cZ_\varnothing &=& {|G|}^{-|\Lambda_0|+|\Lambda_1|-|\Lambda_2|}\,
  {|H|}^{|\Lambda_0|-|\Lambda_1|+|\Lambda_2|-|\Lambda_3|}\,
  {|L|}^{-|\Lambda_0|+|\Lambda_1|-|\Lambda_2|+|\Lambda_3|-|\Lambda_4|} \vphantom{\ds\int} \\
  & & \ds \times\biggl[\prod_{(jk)\in\Lambda_1}\int_G dg_{jk}\biggr] 
  \biggl[\prod_{(jk\ell)\in\Lambda_2}\int_H dh_{jk\ell}\biggr] 
  \biggl[\prod_{(jk\ell m)\in\Lambda_3}\int_L dl_{jk\ell m}\biggr] \vphantom{\ds\int} \\
  & & \ds \times \biggl[\prod_{(jk\ell)\in\Lambda_2}\delta_G\bigl( g_{jk\ell} \bigr)\biggr] 
  \biggl[\prod_{(jk\ell m)\in\Lambda_3}\delta_H\bigl( h_{jk\ell m} \bigr)\biggr] 
  \biggl[\prod_{(jk\ell mn)\in\Lambda_4}\delta_L\bigl( l_{jk\ell mn} \bigr)\biggr]\,,\vphantom{\ds\int} \\
\end{array}
\end{equation}
Here $\Lambda_1$ is the set of edges, $\Lambda_2$ is the set of triangles, $\Lambda_3$ is the set of tetrahedrons, and $\Lambda_4$ is the set of $4$-simplices of the triangulation $\cT(\cM)$, and the arguments of the Dirac delta functions are given as:
\begin{equation} \label{eq:DiracDeltaArguments}
\begin{array}{lcl}
g_{jk\ell} \!\!\!\! & \equiv & \!\!\! \partial(h_{jk\ell})\,g_{k\ell}\,g_{jk}\,g_{j\ell}^{-1} \,, \vphantom{\ds\int} \\
h_{jk\ell m} \!\!\!\! & \equiv & \!\!\! \delta(l_{jk\ell m})h_{j\ell m}\,(g_{\ell m }\rhd h_{jk\ell})\,h_{k\ell m}^{-1}\,h_{jkm}^{-1} \,,\vphantom{\ds\int} \\
l_{jk\ell mn} \!\!\!\! & \equiv & \!\!\! l_{j\ell m n}^{-1}\,h_{j\ell n}\rhd'\{h_{\ell m n}, (g_{m n}g_{\ell m})\rhd h_{jk\ell }\}_{\mathrm{p}}\,l_{jk\ell n}^{-1}(h_{jkn}\rhd'l_{k\ell m n}) l_{jkm n} h_{jm n}\rhd'(g_{m n}\rhd l_{jk\ell m }) \,,  \vphantom{\ds\int} \\
\end{array}
\end{equation}
where the map $\rhd':H\times L \to L$ is defined as
\begin{equation}
h \rhd' l = l \, \{\delta(l){}^{-1},\,h\}_\mathrm{p}\,,
\end{equation}
see (\ref{identitet2}) and Appendix \ref{AppI} for details.

The expressions (\ref{eq:DiracDeltaArguments}) have been specially constructed so that the state sum \eqref{eq_partition_3bf} remains invariant under Pachner moves \cite{Pachner}, and thus independent of the particular choice of the triangulation $\cT(\cM)$. Namely, while $\cZ_\varnothing$ is defined and can be evaluated for some particular triangulation $\cT(\cM)$, its value does not depend on this choice, but depends only on the topology of the underlying manifold $\cM$ itself, and thus represents a topological invariant of $\cM$. See \cite{Radenkovic2022_2} for details and proof of invariance.

We now want to extend the definition of the above state sum to include manifolds with boundaries. Given a $4$-dimensional manifold $\cM$ with a possibly nontrivial boundary $\Sigma =\partial \cM$, we introduce the state sum $\cZ_\partial$ as follows:
\begin{equation}\label{ZSigma1Sigma2}
\begin{array}{lcl}
  \cZ_\partial &=& {|G|}^{-(|\Lambda_{0}|-\frac{|\Lambda_{0,\Sigma}|}{2})+|\Lambda_{1}|-|\Lambda_{2}|}\,{|H|}^{(|\Lambda_{0}|-\frac{|\Lambda_{0,\Sigma}|}{2})-(|\Lambda_{1}|-\frac{|\Lambda_{1,\Sigma}|}{2})+|\Lambda_{2}|-|\Lambda_{3}|} \vphantom{\ds\int} \\
  & & {|L|}^{-(|\Lambda_{0}|-\frac{|\Lambda_{0,\Sigma}|}{2})+(|\Lambda_{1}|-\frac{|\Lambda_{1,\Sigma}|}{2})-(|\Lambda_{2}|-\frac{|\Lambda_{2,\Sigma}|}{2})+|\Lambda_{3}|-|\Lambda_{4}|} \vphantom{\ds\int} \\
  & & \ds \biggl[\prod_{(jk)\in\Lambda_{1}^*}\int_G dg_{jk}\biggr] \biggl[\prod_{(jk\ell)\in\Lambda_{2}^*}\int_H dh_{jk\ell}\biggr]  \biggl[\prod_{(jk\ell m)\in\Lambda_{3}^*}\int_L dl_{jk\ell m}\biggr] \vphantom{\ds\int} \\
  & & \ds \biggl[\prod_{(jk\ell)\in\Lambda_{2}^*}\delta_G\bigl(g_{jk\ell}\bigr)\biggr]  \biggl[\prod_{(jk\ell m)\in\Lambda_{3}^*}\delta_H\bigl(h_{jk\ell m}\bigr)\biggr]  \biggl[\prod_{(jk\ell mn)\in\Lambda_{4}}\delta_L\bigl(l_{jk\ell mn}\bigr)\biggr] \vphantom{\ds\int} \\
  & & \ds \biggl[\prod_{(jk\ell)\in\Lambda_{2,\Sigma}}\delta_G\bigl(g_{jk\ell}\bigr)\biggr]  \biggl[\prod_{(jk\ell m)\in\Lambda_{3,\Sigma}}\delta_H\bigl(h_{jk\ell m}\bigr)\biggr]\,, \vphantom{\ds\int} \\
\end{array}
\end{equation}
where $\Lambda_{i,\Sigma}$ are the sets of simplices of dimension $i$ on the boundary $\Sigma$ (here $i\in \{0,1,2,3\}$). The sets $\Lambda_i^*$ contain simplices in the bulk only, namely $\Lambda_{1}^*=\Lambda_{1}\setminus\Lambda_{1,\Sigma}$ is the set of edges of $\cT(\cM)$ that are not the edges of $\cT(\Sigma)$, and similarly for $\Lambda_{2}^*$ and $\Lambda_{3}^*$.

Compared to \eqref{eq_partition_3bf}, one can observe three new features of the state sum \eqref{ZSigma1Sigma2}. First, the exponents in the top two rows have been modified by the presence of the boundary. Second, the integration (third row) is performed only over the group elements living on the simplices in the bulk, but not over the ones living on the simplices on the boundary. And third, the bottom row contains the Dirac delta terms corresponding to the boundary. Since the integration is performed only over the bulk, the state sum $\cZ_\partial$ is not a constant like $\cZ_\varnothing$, but instead a function of the group elements living on the boundary.

When the boundary is empty, one can see that the state sum \eqref{ZSigma1Sigma2} reduces to \eqref{eq_partition_3bf},
\begin{equation} \label{eq:ZemptyBoundaryCorr}
\cZ_{\partial} \Big|_{\partial\cM = \varnothing} =\cZ_\varnothing\,.
\end{equation}
By looking at the structure of the state sum (\ref{ZSigma1Sigma2}), one can observe that it is identical to the structure of (\ref{eq_partition_3bf}), up to additional factors related to the boundary. Therefore, it is straightforward to show that $\cZ_\partial $ is also invariant with respect to Pachner moves, provided that they do not modify the triangulation of the boundary, but act only on the bulk. In this sense, the state sum (\ref{ZSigma1Sigma2}) depends on the choice of the triangulation of the boundary, but not on the choice of the triangulation in the bulk.

\subsection{Construction of the \texorpdfstring{$Z$}{Z} functor}

Having the definitions of $\cZ_\varnothing$ and $\cZ_\partial$ in hand, we can proceed to the explicit construction of the functor $Z$ from the category $3TCob$ to the category $Hilb$.

According to axiom 1 of Definition \ref{TQFTfunctor}, to each triangulated $3$-dimensional manifold $\cT(\Sigma)\in Obj_{3TCob}$ the functor $Z$ assigns a Hilbert space $\cH_\Sigma\in Obj_{Hilb}$. Denoting the numbers of edges, triangles and tetrahedra of the triangulation $\cT(\Sigma)$ as $|\Lambda_{1,\Sigma}|$, $|\Lambda_{2,\Sigma}|$, and $|\Lambda_{3,\Sigma}|$ respectively, we choose the following Hilbert space of square-integrable functions over the three groups:
\begin{equation}\label{eq:HilbertSpace}
Z(\cT(\Sigma))=\cH=\mathcal{L}^2(G{}^{|\Lambda_{1,\Sigma}|}\times H{}^{|\Lambda_{2,\Sigma}|}\times L{}^{|\Lambda_{3,\Sigma}| })\,.
\end{equation}
If $G$, $H$ and $L$ are finite groups, the dimensions of their corresponding spaces of square-integrable functions are finite. Namely, in $\cL^2(G)$ one can introduce a natural choice of basis vectors $\{ \,\ket{g}\, | \,g\in G\, \}$, whose components are indicator-functions $\chi_g:G\to\kompleksni$,
\begin{equation} \label{eq:IndicatorFunction}
\bracket{g'}{g} \equiv \chi_g(g') = \left\{ \begin{array}{cc}
  1 & \text{ if } g'=g\,, \\
  0 & \text{ if } g'\neq g\,, \\
\end{array}
\right. \qquad \forall g' \in G\,,
\end{equation}
and similarly for $\cL^2(H)$ and $\cL^2(L)$, so we have
$$
\dim \cL^2(G) = |G|\,, \qquad \dim \cL^2(H) = |H|\,, \qquad \dim \cL^2(L) = |L|\,.
$$
Then, a basis element of the whole space $\cH$ can be written as a tensor product of appropriate number of copies of the basis vectors from these three spaces,
$$
\ket{g,h,l} \equiv \bigotimes_{i=1}^{|\Lambda_{1,\Sigma}|} \ket{g_i}
\bigotimes_{j=1}^{|\Lambda_{2,\Sigma}|} \ket{h_j}
\bigotimes_{k=1}^{|\Lambda_{3,\Sigma}|} \ket{l_k} \in \cH\,,
$$
and the total dimension of $\cH$ is given as:
$$
\dim \cH=|G|^{|\Lambda_{1,\Sigma}|}|H|^{|\Lambda_{2,\Sigma}|}|L|^{|\Lambda_{3,\Sigma}|} \,.
$$
The dual space $\cH^*$ can be defined in the usual way, as a space of linear functionals over $\cH$, with a typical basis element denoted as $\bra{g,h,l} \in \cH^*$.

According to axiom 2 of Definition \ref{TQFTfunctor}, to each $4$-dimensional manifold $\cM$ with the boundary $\partial \cM=\cT(\bar{\Sigma}_1)\sqcup \cT(\Sigma_2)$, the functor $Z$ assigns the operator $Z(\cM)=\hat{M} : \cH_1 \to \cH_2$, where $\cH_1 = Z(\cT(\Sigma_1))$ and $\cH_2 = Z(\cT(\Sigma_2))$ as above. Then, the operator $\hat{M}$ is introduced as follows:
\begin{equation}\label{McZ}
\begin{array}{lcl}
\hat{M}&=& \ds \biggl[\prod_{(jk)\in\Lambda_{1,\Sigma_1}}\int_G dg'_{jk}\biggr]
\biggl[\prod_{(jk\ell)\in\Lambda_{2,\Sigma_1}}\int_H dh'_{jk\ell}\biggr] \biggl[\prod_{(jk\ell m)\in\Lambda_{3,\Sigma_1}}\int_L dl'_{jk\ell m}\biggr] \vphantom{\ds\int} \\
 & & \ds \biggl[\prod_{(jk)\in\Lambda_{1,\Sigma_2}}\int_G dg_{jk}\biggr]
\biggl[\prod_{(jk\ell)\in\Lambda_{2,\Sigma_2}}\int_H dh_{jk\ell}\biggr] \biggl[\prod_{(jk\ell m)\in\Lambda_{3,\Sigma_2}}\int_L dl_{jk\ell m}\biggr]\;\cZ_\partial\;\ket{g,h,l}\otimes \bra{g',h',l'} \,. \vphantom{\int\ds} \\
\end{array}
\end{equation}
Several comments should be made regarding this definition. First, note that $\cZ_\partial= \bra{g,h,l} Z(\cM) \ket{g',h',l'}$ is the kernel (\ie, the matrix element) of the operator $\hat{M}$, while keeping in mind that by its definition (\ref{ZSigma1Sigma2}), $\cZ_\partial$ is a function of the group elements living on the boundary. More explicit notation would be $\cZ_\partial (g,h,l,g',h',l')$, but we denote it just as $\cZ_\partial$ for brevity.

Second, the definition (\ref{ZSigma1Sigma2}) of the kernel $\cZ_\partial$ assumes that the whole $4$-dimensional manifold $\cM$ has been assigned some triangulation, $\cT(\cM)$, while according to axiom 2, only the boundary $\partial\cM$ is assigned a triangulation. Nevertheless, given the fact that $\cZ_\partial$ is in fact invariant with respect to Pachner moves (as long as they do not modify the boundary), it has the same value for any choice of the triangulation in the bulk of $\cM$. Therefore, the operator $\hat{M}$ does not in fact depend on the triangulation in the bulk, and for its explicit evaluation any choice $\cT(\cM)$ will do, as long as it is compatible with the fixed triangulation of the boundary, \ie, $\partial \cT(\cM) = \cT(\bar{\Sigma}_1)\sqcup \cT(\Sigma_2)$.

Finally, note that the boundary $\cT(\bar{\Sigma}_1)\sqcup \cT(\Sigma_2)$ of the manifold $\cM$ has been chosen as a disjoint union because this provides a convenient set of Hilbert spaces $\cH_1$ and $\cH_2$ over which the operator $\hat{M}$ is defined. However, no generality is lost in this way, since it is perfectly valid for either disjoint component of the boundary to be empty, or to be a disjoint union of multiple other components. In other words, any boundary $\partial\cM$ of any manifold $\cM$ can be represented as a disjoint union of two (possibly empty) boundaries. We shall discuss some interesting examples below.

Returning to the axioms of Definition \ref{TQFTfunctor}, once we have satisfied axioms 1 and 2 by the functor $Z$ via (\ref{eq:HilbertSpace}) and (\ref{McZ}), we need to make sure that the remaining axioms 3--10 are also satisfied by this choice. This task is taken up in Section \ref{SecIV}, and the result is that all axioms are indeed satisfied, leading to our main result:

\bigskip

\begin{theorem}
The map $Z$, specified by (\ref{ZSigma1Sigma2}), (\ref{eq:HilbertSpace}) and (\ref{McZ}), is a functor between two dagger symmetric monoidal categories with a dual,
\begin{center}
 \begin{tikzpicture}[>=stealth, baseline=(current  bounding  box.center)]
  % Nodes
  \node (A1) at (0,0) {$3TCob$};
  \node (A2) at (3,0) {$Hilb\,,$};
  % Arrows
  \draw[->] (A1) -- node[above]{$\scriptstyle Z$} (A2);
\end{tikzpicture}
\end{center}
and thus represents a TQFT.
\end{theorem}

\subsection{\label{SubSecExamples}Examples}

Let us now look at two specific examples. The first example is the functor $Z$ corresponding to a manifold $\cM$ with no boundary, $\partial \cM = \varnothing$. This can be understood as a special case of a manifold with the general boundary $\cT(\bar{\Sigma}_1)\sqcup \cT(\Sigma_2)$, with the choices $\Sigma_1 = \Sigma_2 = \varnothing_3$. The triangulations of both are then trivial, so there are no vertices, edges or tetrahedra living on the boundary. Then, according to (\ref{eq:HilbertSpace}), we have
\begin{equation} \label{eq:exampleOneH}
Z(\cT(\Sigma_1))=\cH_1=\mathcal{L}^2(G{}^0\times H{}^0\times L{}^0) = \cL^2(\{e\}) = \kompleksni\,.
\end{equation}
Namely, the resulting Hilbert space is the space of square-integrable functions over a trivial group, and is therefore just the space of complex numbers, with $\dim\cH_1 = 1$. The same holds for the $\cT(\Sigma_2)$ boundary --- we have $\cH_2 = \kompleksni$ and $\dim\cH_2 = 1$. Noting that the sole basis vector of $\kompleksni$ is the number $1$, according to (\ref{McZ}) the action of the functor on the whole manifold $\cM$ gives the following operator:
\begin{equation} \label{eq:exampleOneZ}
Z(\cM) = \hat{M}= \cZ_\varnothing\,1\otimes 1 \equiv \cZ_\varnothing \,,
\end{equation}
taking into account (\ref{eq:ZemptyBoundaryCorr}). Since the manifold $\cM$ is closed, one assigns to it an arbitrary triangulation $\cT(\cM)$ and applies (\ref{eq_partition_3bf}) to evaluate the state sum $\cZ_\varnothing$. The value of $\cZ_\varnothing$ depends on the topology of $\cM$, but not on the particular choice of the triangulation $\cT(\cM)$.

The second example corresponds to the choice $\Sigma_1 = \varnothing_3$, with $\Sigma_2 \equiv \Sigma$ remaining nontrivial and having some particular triangulation $\cT(\Sigma)$ (see Figure \ref{FigureState}).
\begin{figure}[!ht]
\begin{center} 
  \begin{tikzpicture}
  % Drawing the ball
  \draw[black, fill=white] (0,0) -- (2,0) arc (0:-180:2) -- cycle;
  
  % Drawing the surface Sigma (ellipse)
  \filldraw[fill=blue!20, draw=blue!50!black]
    (0,0) ellipse (2 and 0.7);
  \node[black] at (0,0) {$\cT(\Sigma)$};
  \node[black] at (0,-1.3) {$\cM$};
  
\end{tikzpicture}
\caption{\label{FigureState} The cobordism $\cM:\varnothing\to \cT(\Sigma)$.}
\end{center}
\end{figure}
Then we have the Hilbert space $\cH_1 = \kompleksni$, while $\cH_2$ is given by (\ref{eq:HilbertSpace}). According to (\ref{McZ}), the action of the functor on the manifold $\cM$ gives the ``operator'' $\ket{\Psi}:\kompleksni \to \cH_2$ as follows:
$$
Z(\cM) = \ket{\Psi} = \biggl[\prod_{(jk)\in\Lambda_{1,\Sigma_2}}\int_G dg_{jk}\biggr]
\biggl[\prod_{(jk\ell)\in\Lambda_{2,\Sigma}}\int_H dh_{jk\ell}\biggr] \biggl[\prod_{(jk\ell m)\in\Lambda_{3,\Sigma}}\int_L dl_{jk\ell m}\biggr]\;\cZ_\partial\;\ket{g,h,l} \,,
$$
where the kernel (\ie, the \emph{wavefunction}) is:
$$
\Psi(g,h,l) = \bracket{g,h,l}{\Psi} = \cZ_\partial\,.
$$
Here $\cZ_\partial$ can be obtained from (\ref{ZSigma1Sigma2}) by choosing a particular triangulation $\cT(\cM)$ such that $\partial\cT(\cM) = \cT(\Sigma)$. This wavefunction is traditionally called the \emph{Hartle-Hawking} wavefunction for the TQFT defined by the functor $Z$. There is one such wavefunction for each choice of the manifold $\cM$, its boundary $\partial\cM = \Sigma$ and for each particular triangulation $\cT(\Sigma)$ of this boundary, but is independent of the choice of the triangulation of the bulk, $\cT(\cM)$.

\subsection{Construction of the \texorpdfstring{$\tilde{Z}$}{\~Z} functor}

We now wish to remove the dependence of the $3$-dimensional manifold on the triangulation, that is, to construct a functor $\tilde{Z}:3Cob\to Hilb$, from the category $3Cob$ of oriented cobordisms between $3$-dimensional manifolds to the category $Hilb$ of finite-dimensional Hilbert spaces. To this end, the functor $Z:3TCob\to Hilb$ introduced above serves as an intermediate step. The construction follows the key ideas by Yetter. However, for the purpose of this work, it is enough to just outline the basic ideas of the construction, while a more rigorous treatment can be found in \cite{Y3, Y1, Y2, PorterNotes}.

The first thing to note is that different triangulations of the same $3$-dimensional manifold form an equivalence class. Namely, given two different triangulations $\cT(\Sigma)$ and $\cT'(\Sigma)$ of the same manifold $\Sigma$, one can always transform between them using $3$-dimensional Pachner moves (these are to be distinguished from the $4$-dimensional Pachner moves acting on the bulk triangulations $\cT(\cM)$). Conversely, for any two triangulations $\cT(\Sigma)$ and $\cT'(\Sigma')$, if $\Sigma'$ has different topology from $\Sigma$, their triangulations cannot be connected using Pachner moves, since the latter conserve the topology of a manifold. Let us therefore denote the set of all possible triangulations of a given manifold $\Sigma$ as:
$$
Trng(\Sigma) = \{ \;\cT(\Sigma)\; | \; \Sigma\in Obj_{3Cob} \; \}\,.
$$
This set is infinite, since one can always refine any given triangulation without bound. But it can be subdivided into subsets of triangulations which consist of the same numbers of vertices, edges, triangles and tetrahedra. Specifically, each triangulation $\cT(\Sigma)$ can be assigned a $4$-tuple of positive integers,
$$
\lambda(\cT(\Sigma)) = ( |\Lambda_{0,\Sigma}|, |\Lambda_{1,\Sigma}| , |\Lambda_{2,\Sigma}| , |\Lambda_{3,\Sigma}| ) \,.
$$
Finally, one can introduce an order relation over all such $4$-tuples (for example a lexicographic order), and since all of its integers are positive, there will exist a \emph{minimal element}. There are possibly several triangulations in $Trng(\Sigma)$ that correspond to this minimal $4$-tuple --- pick any of those and denote it as $\cT_m(\Sigma)$. Such a triangulation can be qualitatively described as the ``minimal'' triangulation of the manifold $\Sigma$, since it contains the smallest number of vertices, edges, triangles and tetrahedra (in that order, based on the choice of lexicographic ordering of $4$-tuples).

The purpose of the above construction is to demonstrate that each set $Trng(\Sigma)$ contains at least one such distinguished ``simplest'' triangulation $\cT_m(\Sigma) \in Trng(\Sigma)$. This now enables us to construct the functor $\tilde{Z}:3Cob\to Hilb$ as follows. Given any object $\Sigma\in Obj_{3Cob}$, find its minimal triangulation $\cT_m(\Sigma)$, and then define the action of the functor $\tilde{Z}$ on $\Sigma$ to be equal to the action of the functor $Z$ on $\cT_m(\Sigma) \in Obj_{3TCob}$:
\begin{equation} \label{eq:defTildeZobj}
\tilde{Z}(\Sigma) = Z(\cT_m(\Sigma))\,.
\end{equation}
Note that, while there may be several triangulations that can be labeled as $\cT_m(\Sigma)$, they are all mapped by $\tilde{Z}$ to the same Hilbert space (see (\ref{eq:HilbertSpace})), since they have the same $4$-tuple $\lambda(\cT_m(\Sigma))$, \ie, they have the same numbers of vertices, edges, triangles and tetrahedra. Next, given any cobordism $\tilde{\cM}: \Sigma_1 \to \Sigma_2$ which is a morphism in category $3Cob$, one can find minimal triangulations of its source and target $\cT_m(\Sigma_1)$ and $\cT_m(\Sigma_2)$ respectively, and then define the action of the functor $\tilde{Z}$ on $\tilde{\cM}$ to be equal to the action of the functor $Z$ on a corresponding morphism $\cM:\cT_m(\Sigma_1) \to \cT_m(\Sigma_2)$ from the category $3TCob$:
\begin{equation} \label{eq:defTildeZmor}
\tilde{Z}(\tilde{\cM}) = Z(\cM)\,,
\end{equation}
where $Z(\cM)$ is given by (\ref{McZ}).

The equations (\ref{eq:defTildeZobj}) and (\ref{eq:defTildeZmor}) thus define a map
\smallskip
\begin{center}
 \begin{tikzpicture}[>=stealth, baseline=(current  bounding  box.center)]
  % Nodes
  \node (A1) at (0,0) {$3Cob$};
  \node (A2) at (3,0) {$Hilb\,,$};
  % Arrows
  \draw[->] (A1) -- node[above]{$\scriptstyle \tilde{Z}$} (A2);
\end{tikzpicture}
\end{center}
\smallskip
which is obviously a functor, since $Z$ is a functor.

Regarding the above construction, one should note that the functor $\tilde{Z}$ depends on the choice of the ordering imposed on the $4$-tuples $\lambda(\cT(\Sigma))$ within the set $Trng(\Sigma)$. Such a choice is not unique. Namely, we have adopted the lexicographic ordering of $4$-tuples, which essentially means that a triangulation $\cT(\Sigma)$ is called \emph{minimal}, and denoted $\cT_m(\Sigma)$ if:
\begin{itemize}
\item it has the smallest number of vertices among all triangulations in the set $Trng(\Sigma)$, and given that,
\item it has the smallest number of edges in $Trng(\Sigma)$, and given that,
\item it has the smallest number of triangles in $Trng(\Sigma)$, and given that,
\item it has the smallest number of tetrahedra in $Trng(\Sigma)$.
\end{itemize}
Alternatively, one could have instead adopted, for example, the reverse lexicographic ordering --- a triangulation $\cT(\Sigma)$ is called \emph{minimal}, and denoted $\cT_m(\Sigma)$ if:
\begin{itemize}
\item it has the smallest number of tetrahedra among all triangulations in the set $Trng(\Sigma)$, and given that,
\item it has the smallest number of triangles in $Trng(\Sigma)$, and given that,
\item it has the smallest number of edges in $Trng(\Sigma)$, and given that,
\item it has the smallest number of vertices in $Trng(\Sigma)$.
\end{itemize}
In general, depending on the topology of $\Sigma$, the set of minimal triangulations in the latter case need not coincide with the set of minimal triangulations in the former case. Moreover, they may even be completely disjoint, so that not a single triangulation could be considered minimal with respect to both ordering schemes. In this sense, each choice of the ordering scheme gives rise to a functor $\tilde{Z}$, and different choices may correspond to different functors.

One should also note that instead of a minimal triangulation $\cT_m(\Sigma)$, we could have chosen a completely random triangulation from the set $Trng(\Sigma)$, denoted $\cT_r(\Sigma)$, and used that instead of $\cT_m(\Sigma)$ in the definitions (\ref{eq:defTildeZobj}) and (\ref{eq:defTildeZmor}). Each such random choice would give rise to one well-defined functor $\tilde{Z}$. However, this would be way too arbitrary. The intuition behind the requirement of choosing a minimal triangulation lies in the idea that every $3$-manifold $\Sigma$ should be assigned a \emph{smallest possible} Hilbert space (\ie, with the smallest possible dimension), such that a TQFT remains well defined. Namely, enlarging the dimension of a Hilbert space can always accommodate the topological properties of a manifold, but reducing the dimension is not always possible in this sense, so
 minimizing the sizes of Hilbert spaces can be considered worthwhile.

 \section{\label{SecIV}Proof of the TQFT axioms}
 
 In this section, we will show that the functor $Z$ defined in Section \ref{SecIII} satisfies the axioms 3--10 given in Definition \ref{TQFTfunctor}. The axioms will be discussed one by one, in order. Throughout the text we assume that every $4$-dimensional manifold being discussed belongs to the class of manifolds that admit a triangulation, so that the state sums $\cZ_\varnothing $ and $\cZ_\partial $ can be defined.

\subsection{Functoriality axiom}
Let us start with the functoriality axiom, equation \eqref{eq:functoriality}. The main strategy of the proof is as follows. First, we will study the form of the operators on the right-hand side of (\ref{eq:functoriality}), and compare their product to the corresponding operator on the left-hand side of (\ref{eq:functoriality}). Using the definition (\ref{McZ}), this comparison will give us a reformulation of the functoriality axiom in terms of the three state sums $\cZ_\partial$ which are the kernels of the corresponding operators. Second, we will apply the definition (\ref{ZSigma1Sigma2}) to explicitly verify the obtained reformulation of the functoriality axiom in terms of the kernels. This will demonstrate that (\ref{eq:functoriality}) is indeed satisfied. We will use this general procedure for other axioms as well.

Consider two $4$-manifolds $\mathcal{M}_A:\cT(\Sigma_A) \to \cT(\Sigma)$ and $\mathcal{M}_B: \cT(\Sigma) \to \cT(\Sigma_B)$ with a common boundary $\cT(\Sigma)$ as shown in Figure \ref{MM}. According to (\ref{McZ}), the corresponding operators are:
\begin{equation}
\begin{array}{lcl}
\hat{M}_A&=& \ds \biggl[\prod_{(jk)\in\Lambda_{1,\Sigma_A}}\int_G dg^A_{jk}\biggr]
\biggl[\prod_{(jk\ell)\in\Lambda_{2,\Sigma_A}}\int_H dh^A_{jk\ell}\biggr] \biggl[\prod_{(jk\ell m)\in\Lambda_{3,\Sigma_A}}\int_L dl^A_{jk\ell m}\biggr] \vphantom{\ds\int} \\
 & & \ds \biggl[\prod_{(jk)\in\Lambda_{1,\Sigma}}\int_G dg_{jk}\biggr]
\biggl[\prod_{(jk\ell)\in\Lambda_{2,\Sigma}}\int_H dh_{jk\ell}\biggr] \biggl[\prod_{(jk\ell m)\in\Lambda_{3,\Sigma}}\int_L dl_{jk\ell m}\biggr]\;\cZ_\partial^A\;\ket{g,h,l}\otimes \bra{g^A,h^A,l^A} \,, \vphantom{\int\ds} \\
\end{array}
\end{equation}
and
\begin{equation}
\begin{array}{lcl}
\hat{M}_B&=& \ds \biggl[\prod_{(jk)\in\Lambda_{1,\Sigma}}\int_G dg'_{jk}\biggr]
\biggl[\prod_{(jk\ell)\in\Lambda_{2,\Sigma}}\int_H dh'_{jk\ell}\biggr] \biggl[\prod_{(jk\ell m)\in\Lambda_{3,\Sigma}}\int_L dl'_{jk\ell m}\biggr] \vphantom{\ds\int} \\
 & & \ds \biggl[\prod_{(jk)\in\Lambda_{1,\Sigma_B}}\int_G dg^B_{jk}\biggr]
\biggl[\prod_{(jk\ell)\in\Lambda_{2,\Sigma_B}}\int_H dh^B_{jk\ell}\biggr] \biggl[\prod_{(jk\ell m)\in\Lambda_{3,\Sigma_B}}\int_L dl^B_{jk\ell m}\biggr]\;\cZ_\partial^B\;\ket{g^B,h^B,l^B}\otimes \bra{g',h',l'} \,. \vphantom{\int\ds} \\
\end{array}
\end{equation}
The composition $\hat{M}_B \cdot \hat{M}_A$ features two integrations over the simplices of $\cT(\Sigma)$, namely over nonprimed and primed group elements, and a scalar product $\bracket{g',h',l'}{g,h,l}$. Let us look at these integrations for a single given edge in $\Lambda_{1,\Sigma}$, and perform the integration over the primed variable as follows:
$$
\begin{array}{lcl}
  \ds \int_G dg \int_G dg'\, \cZ_\partial^A(g,\dots) \cZ_\partial^B(g',\dots)\, \bracket{g'}{g} & = & \ds
  \int_G dg \;\frac{1}{|G|}\sum_{g' \in G} \, \cZ_\partial^A(g,\dots) \cZ_\partial^B(g',\dots)\, \chi_{g}(g') = \\
 & = & \ds \frac{1}{|G|} \int_G dg \, \cZ_\partial^A(g,\dots) \cZ_\partial^B(g,\dots)\,. \\
\end{array}
$$
In the above we have assumed that $G$ is a finite group, and we have used (\ref{eq:GroupIntegralDef}) and (\ref{eq:IndicatorFunction}). In this way all integrals over primed group elements can be performed, leaving us with the resulting expression for the r.h.s.\ of (\ref{eq:functoriality}):
\begin{equation}
\begin{array}{lcl}
  \hat{M}_B\cdot\hat{M}_A&=& \ds \biggl[\prod_{(jk)\in\Lambda_{1,\Sigma_A}}\int_G dg^A_{jk}\biggr]
\biggl[\prod_{(jk\ell)\in\Lambda_{2,\Sigma_A}}\int_H dh^A_{jk\ell}\biggr] \biggl[\prod_{(jk\ell m)\in\Lambda_{3,\Sigma_A}}\int_L dl^A_{jk\ell m}\biggr] \vphantom{\ds\int} \\
 & & \ds \biggl[\prod_{(jk)\in\Lambda_{1,\Sigma_B}}\int_G dg^B_{jk}\biggr]
\biggl[\prod_{(jk\ell)\in\Lambda_{2,\Sigma_B}}\int_H dh^B_{jk\ell}\biggr] \biggl[\prod_{(jk\ell m)\in\Lambda_{3,\Sigma_B}}\int_L dl^B_{jk\ell m}\biggr] \vphantom{\int\ds} \\
 & & \ds \frac{1}{\dim\cH} \biggl[\prod_{(jk)\in\Lambda_{1,\Sigma}}\int_G dg_{jk}\biggr]
\biggl[\prod_{(jk\ell)\in\Lambda_{2,\Sigma}}\int_H dh_{jk\ell}\biggr] \biggl[\prod_{(jk\ell m)\in\Lambda_{3,\Sigma}}\int_L dl_{jk\ell m}\biggr] \vphantom{\int\ds} \\
 & & \ds \cZ_\partial^B(g^B,h^B,l^B;g,h,l) \cZ_\partial^A(g,h,l;g^A,h^A,l^A)\;\ket{g^B,h^B,l^B}\otimes \bra{g^A,h^A,l^A} \,, \vphantom{\int\ds} \\
\end{array}
\end{equation}
where $\dim\cH=|G|^{|\Lambda_{1,\Sigma}|}|H|^{|\Lambda_{2,\Sigma}|}|L|^{|\Lambda_{3,\Sigma}|}$ is the dimension of the Hilbert space $\cH$ of the joint boundary $\cT(\Sigma)$. On the other hand, according to (\ref{McZ}), the l.h.s.\ of (\ref{eq:functoriality}) gives us the following operator:
\begin{equation}
\begin{array}{lcl}
\hat{M}_{B\#A}&=& \ds \biggl[\prod_{(jk)\in\Lambda_{1,\Sigma_A}}\int_G dg^A_{jk}\biggr]
\biggl[\prod_{(jk\ell)\in\Lambda_{2,\Sigma_A}}\int_H dh^A_{jk\ell}\biggr] \biggl[\prod_{(jk\ell m)\in\Lambda_{3,\Sigma_A}}\int_L dl^A_{jk\ell m}\biggr] \vphantom{\ds\int} \\
 & & \ds \biggl[\prod_{(jk)\in\Lambda_{1,\Sigma_B}}\int_G dg^B_{jk}\biggr]
\biggl[\prod_{(jk\ell)\in\Lambda_{2,\Sigma_B}}\int_H dh^B_{jk\ell}\biggr] \biggl[\prod_{(jk\ell m)\in\Lambda_{3,\Sigma_B}}\int_L dl^B_{jk\ell m}\biggr] \vphantom{\int\ds} \\
 & & \ds \cZ_\partial^{B\#A}\;\ket{g^B,h^B,l^B}\otimes \bra{g^A,h^A,l^A} \,. \vphantom{\int\ds} \\
\end{array}
\end{equation}
Comparing the two expressions, we see that (\ref{eq:functoriality}) will be satisfied iff the respective kernels satisfy the following equation:
\begin{equation}\label{eq:functorcondition}
\cZ_\partial^{B\#A}  = \frac{1}{\dim\cH} \vphantom{\ds\int} \biggl[\prod_{(jk)\in\Lambda_{1,\Sigma}}\int_G dg_{jk}\biggr] \biggl[\prod_{(jk\ell)\in\Lambda_{2,\Sigma}}\int_H dh_{jk\ell}\biggr] \biggl[\prod_{(jk\ell m)\in\Lambda_{3,\Sigma}}\int_L dl_{jk\ell m}\biggr] \; \cZ_\partial^B\cZ_\partial^A\,.
\end{equation}

This equation represents the reformulation of (\ref{eq:functoriality}) in terms of state sums $\cZ_\partial^A$, $\cZ_\partial^B$ and $\cZ_\partial^{B\#A}$. The proof of the functoriality axiom is now reduced to verifying that (\ref{eq:functorcondition}) holds. In order to do so, we proceed as follows. Using the equation \eqref{ZSigma1Sigma2}, the state sum $\cZ_\partial^A$ is given as,
\begin{equation}
\begin{array}{lcl}
\label{eq_partition_3bf*}
 \cZ_\partial^A &=& \ds \mathcal{N}_A \vphantom{\ds\int} \biggl[\prod_{(jk)\in\Lambda_{1,A}^*}\int_G dg_{jk}\biggr] \biggl[\prod_{(jk\ell)\in\Lambda_{2,A}^*}\int_H dh_{jk\ell}\biggr] \biggl[\prod_{(jk\ell m)\in\Lambda_{3,A}^*}\int_L dl_{jk\ell m}\biggr] 
 \vphantom{\ds\int} \\
 && \ds \biggl[\prod_{(jk\ell)\in\Lambda_{2,A}}\delta_G\bigl(g_{jk\ell}\bigr)\biggr] \biggl[\prod_{(jk\ell m)\in\Lambda_{3,A}}\delta_H\bigl(h_{jk\ell m}\bigr)\biggr]\vphantom{\ds\int}\biggl[\prod_{(jk\ell mn)\in\Lambda_{4,A}}\delta_L\bigl(l_{jk\ell mn}\bigr)\biggr] ,\vphantom{\ds\int}\end{array}
\end{equation}
where $\Lambda_{k,A}^*=\Lambda_{k,A}\setminus\Lambda_{k,\Sigma_A\sqcup\Sigma}$ is the set of bulk $k$-simplices ($k\in\{0,1,2,3\}$), \ie, those that are not on the boundary $\cT(\Sigma_A)\sqcup \cT(\Sigma)$. The factor $\mathcal{N}_A$ is defined as:
\begin{equation}\label{eq:N1}
\begin{array}{lcl}
  \mathcal{N}_A&=& \ds {|G|}^{-(|\Lambda_{0,A}|-\frac{|\Lambda_{0,\Sigma_A\sqcup\Sigma}|}{2})+|\Lambda_{1,A}|-|\Lambda_{2,A}|} \\
  & & \ds {|H|}^{(|\Lambda_{0,A}|-\frac{|\Lambda_{0,\Sigma_A\sqcup\Sigma}|}{2})-(|\Lambda_{1,A}|-\frac{|\Lambda_{1,\Sigma_A\sqcup\Sigma}|}{2})+|\Lambda_{2,A}|-|\Lambda_{3,A}|}\,\\
  & & \ds {|L|}^{-(|\Lambda_{0,A}|-\frac{|\Lambda_{0,\Sigma_A\sqcup\Sigma}|}{2})+(|\Lambda_{1,A}|-\frac{|\Lambda_{1,\Sigma_A\sqcup\Sigma}|}{2})-(|\Lambda_{2,A}|-\frac{|\Lambda_{2,\Sigma_A\sqcup\Sigma}|}{2})+|\Lambda_{3,A}|-|\Lambda_{4,A}|}\,. \\
\end{array}
\end{equation}
The state sum $\cZ_\partial^B$ is defined similarly. Also, $\cZ_\partial^{B\#A}$ is given by the connected sum of manifolds ${\mathcal{M}_A}$ and $\mathcal{M}_B$ along their common boundary $\cT(\Sigma)$. 
Now, let us substitute in the equation \eqref{eq:functorcondition} the state sums $\cZ_\partial^A$ and $\cZ_\partial^B$ and simplify the resulting expression. We have that the r.h.s. of \eqref{eq:functorcondition} is equal to:
\begin{equation}\label{eq:dokaz1}
\begin{array}{lcl}
  r.h.s. &=& \ds \frac{\mathcal{N}_A\mathcal{N}_B}{\dim\cH} \biggl[\prod_{(jk)\in\Lambda_{1,\Sigma}}\int_G dg_{jk}\biggr] \biggl[\prod_{(jk\ell)\in\Lambda_{2,\Sigma}}\int_H dh_{jk\ell}\biggr] \biggl[\prod_{(jk\ell m)\in\Lambda_{3,\Sigma}}\int_L dl_{jk\ell m}\biggr] \vphantom{\ds\int} \\
  && \ds \biggl[\prod_{(jk)\in\Lambda_{1,A}^*}\int_G dg_{jk}\biggr] \biggl[\prod_{(jk)\in\Lambda_{1,B}^*}\int_G dg_{jk}\biggr]\biggl[\prod_{(jk\ell)\in\Lambda_{2,A}^*}\int_H dh_{jk\ell}\biggr] \vphantom{\ds\int}\\
  &&\ds \biggl[\prod_{(jk\ell)\in\Lambda_{2,B}^*}\int_H dh_{jk\ell}\biggr] \,\biggl[\prod_{(jk\ell m)\in\Lambda_{3,A}^*}\int_L dl_{jk\ell m}\biggr] \biggl[\prod_{(jk\ell m)\in\Lambda_{3,B}^*}\int_L dl_{jk\ell m}\biggr] \vphantom{\ds\int}\\
  && \ds  \biggl[\prod_{(jk\ell)\in\Lambda_{2,A}}\delta_G\bigl(g_{jk\ell}\bigr)\biggr] \biggl[\prod_{(jk\ell m)\in\Lambda_{3,A}}\delta_H\bigl(h_{jk\ell m}\bigr)\biggr]\biggl[\prod_{(jk\ell mn)\in\Lambda_{4,A}}\delta_L\bigl(l_{jk\ell mn}\bigr)\biggr] \vphantom{\ds\int} \\
 && \ds \biggl[\prod_{(jk\ell)\in\Lambda_{2,B}}\delta_G\bigl(g_{jk\ell}\bigr)\biggr] \,\biggl[\prod_{(jk\ell m)\in\Lambda_{3,B}}\delta_H\bigl(h_{jk\ell m}\bigr)\biggr]\vphantom{\ds\int}\biggl[\prod_{(jk\ell mn)\in\Lambda_{4,B}}\delta_L\bigl(l_{jk\ell mn}\bigr)\biggr] \,.\vphantom{\ds\int} \\
\end{array}
\end{equation}
The integrations over the sets $\Lambda_{k,A}^*$, $\Lambda_{k,B}^*$, and $\Lambda_{k,\Sigma}$ join into the integrations over $\Lambda_k^*$ of the manifold $\cM_A\#\cM_B$, for $k\in\{0,1,2,3\}$. Also, the products over sets $\Lambda_{k,A}$ and $\Lambda_{k,B}$ join into the products over $\Lambda_k$ and $\Lambda_{k,\Sigma}$, for $k\in\{2,3\}$, since the simplices of $\cT(\Sigma)$ are counted two times. Finally, the products over sets $\Lambda_{4,A}$ and $\Lambda_{4,B}$ join into the products over the set $\Lambda_4$. The equation \eqref{eq:dokaz1} thus simplifies to:
  \begin{equation}
\begin{array}{lcl}
r.h.s.
&=& \ds \frac{\mathcal{N}_A\mathcal{N}_B}{\dim\cH} \biggl[\prod_{(jk)\in\Lambda_{1}^*}\int_G dg_{jk}\biggr] \biggl[\prod_{(jk\ell)\in\Lambda_{2}^*}\int_H dh_{jk\ell}\biggr]\biggl[\prod_{(jk\ell m)\in\Lambda_{3}^*}\int_L dl_{jk\ell m}\biggr] \vphantom{\ds\int}\\
&& \ds \biggl[\prod_{(jk\ell)\in\Lambda_{2}}\delta_G\bigl(g_{jk\ell}\bigr)\biggr]  \biggl[\prod_{(jk\ell m)\in\Lambda_{3}}\delta_H\bigl(h_{jk\ell m}\bigr)\biggr] \biggl[\prod_{(jk\ell mn)\in\Lambda_{4}}\delta_L\bigl(l_{jk\ell mn}\bigr)\biggr]  \vphantom{\ds\int}\\
 && \ds \biggl[\prod_{(jk\ell)\in\Lambda_{2,\Sigma}}\delta_G\bigl(g_{jk\ell}\bigr)\biggr] \biggl[\prod_{(jk\ell m)\in\Lambda_{3,\Sigma}}\delta_H\bigl(h_{jk\ell m}\bigr)\biggr] .\vphantom{\ds\int} \\
  \end{array}
  \end{equation}
Note that the squares of delta functions appear in the above expression, since the boundary $\cT(\Sigma)$ from manifold $\cM_A$ was glued to the identical boundary $\cT(\Sigma)$ from the manifold $\cM_B$, and each carried a set of identical delta functions. Nevertheless, for finite groups $G$ and $H$ we have
$$
\delta(g)^2 = \delta(e_G)\delta(g) = |G|\,\delta(g)\,, \qquad
\delta(h)^2=\delta(e_H)\delta(h)=|H|\,\delta(h)\,.
$$
Using the equation \eqref{eq:N1} for the factors $\mathcal{N}_A$ and $\mathcal{N}_B$, and substituting the dimension of the Hilbert space $\dim\cH$, we obtain:
  \begin{equation}\label{eq:dokaz2}
\begin{array}{lcl}
  r.h.s.&=& \ds \frac{1}{|G|^{|\Lambda_{1,\Sigma}|}|H|^{|\Lambda_{2,\Sigma}|}|L|^{|\Lambda_{3,\Sigma}|} } \,{|G|}^{-(|\Lambda_{0,A}|-\frac{|\Lambda_{0,\Sigma_A\sqcup\Sigma}|}{2})+|\Lambda_{1,A}|-|\Lambda_{2,A}|}\,{|G|}^{-(|\Lambda_{0,B}|-\frac{|\Lambda_{0,\Sigma_B\sqcup\Sigma}|}{2})+|\Lambda_{1,B}|-|\Lambda_{2,B}|} \vphantom{\ds\int}\\
  && \ds {|H|}^{(|\Lambda_{0,A}|-\frac{|\Lambda_{0,\Sigma_A\sqcup\Sigma}|}{2})-(|\Lambda_{1,A}|-\frac{|\Lambda_{1,\Sigma_A\sqcup\Sigma}|}{2})+|\Lambda_{2,A}|-|\Lambda_{3,A}|} \vphantom{\ds\int}\\
  && \ds {|H|}^{(|\Lambda_{0,B}|-\frac{|\Lambda_{0,\Sigma_B\sqcup\Sigma}|}{2})-(|\Lambda_{1,B}|-\frac{|\Lambda_{1,\Sigma_B\sqcup\Sigma}|}{2})+|\Lambda_{2,B}|-|\Lambda_{3,B}|}\vphantom{\ds\int}\\
  && \ds {|L|}^{-(|\Lambda_{0,A}|-\frac{|\Lambda_{0,\Sigma_A\sqcup\Sigma}|}{2})+(|\Lambda_{1,A}|-\frac{|\Lambda_{1,\Sigma_A\sqcup\Sigma}|}{2})-(|\Lambda_{2,A}|-\frac{|\Lambda_{2,\Sigma_A\sqcup\Sigma}|}{2})+|\Lambda_{3,A}|-|\Lambda_{4,A}|}\vphantom{\ds\int}\\
 && \ds {|L|}^{-(|\Lambda_{0,B}|-\frac{|\Lambda_{0,\Sigma_B\sqcup\Sigma}|}{2})+(|\Lambda_{1,B}|-\frac{|\Lambda_{1,\Sigma_B\sqcup\Sigma}|}{2})-(|\Lambda_{2,B}|-\frac{|\Lambda_{2,\Sigma_B\sqcup\Sigma}|}{2})+|\Lambda_{3,B}|-|\Lambda_{4,B}|} \vphantom{\ds\int}\\
 && \ds \biggl[\prod_{(jk)\in\Lambda_{1}^*}\int_Gdg_{jk}\biggr] \biggl[\prod_{(jk\ell)\in\Lambda_{2}^*}\int_H dh_{jk\ell}\biggr] \biggl[\prod_{(jk\ell m)\in\Lambda_{3}^*}\int_L dl_{jk\ell m}\biggr] \vphantom{\ds\int}\\
  && \ds \biggl[\prod_{(jk\ell)\in\Lambda_{2}}\delta_G\bigl(g_{jk\ell}\bigr)\biggr] \biggl[\prod_{(jk\ell m)\in\Lambda_{3}}\delta_H\bigl(h_{jk\ell m}\bigr)\biggr]\biggl[\prod_{(jk\ell mn)\in\Lambda_{4}}\delta_L\bigl(l_{jk\ell mn}\bigr)\biggr]|G|^{|\Lambda_{2,\Sigma}|}|H|^{|\Lambda_{3,\Sigma}|}\,.\vphantom{\ds\int} \\
\end{array}
\end{equation}
Finally, after straightforward calculation of the exponents, the equation \eqref{eq:dokaz2} reduces to
\begin{equation}
\begin{array}{lcl}
  r.h.s.&=& \ds \,{|G|}^{-(|\Lambda_0|-\frac{|\Lambda_{0,\Sigma_A\sqcup\Sigma_B}|}{2})+|\Lambda_{1}|-|\Lambda_{2}|}{|H|}^{(|\Lambda_0|-\frac{|\Lambda_{0,\Sigma_A\sqcup\Sigma_B}|}{2})-(|\Lambda_1|-\frac{|\Lambda_{1,\Sigma_A\sqcup\Sigma_B}|}{2})+|\Lambda_{2}|-|\Lambda_{3}|} \vphantom{\ds\int}\\
  &&\ds {|L|}^{-(|\Lambda_0|-\frac{|\Lambda_{0,\Sigma_A\sqcup\Sigma_B}|}{2})+(|\Lambda_1|-\frac{|\Lambda_{1,\Sigma_A\sqcup\Sigma_B}|}{2})-(|\Lambda_2|-\frac{|\Lambda_{2,\Sigma_A\sqcup\Sigma_B}|}{2})+|\Lambda_{3}|-|\Lambda_{4}|} \vphantom{\ds\int}\\
  && \ds\biggl[\prod_{(jk)\in\Lambda_{1}^*}\int_G dg_{jk}\biggr] \biggl[\prod_{(jk\ell)\in\Lambda_{2}^*}\int_H dh_{jk\ell}\biggr] \biggl[\prod_{(jk\ell m)\in\Lambda_{3}^*}\int_L dl_{jk\ell m}\biggr] \vphantom{\ds\int}\\
  &&\ds \biggl[\prod_{(jk\ell)\in\Lambda_{2}}\delta_G\bigl(g_{jk\ell}\bigr)\biggr] \biggl[\prod_{(jk\ell m)\in\Lambda_{3}}\delta_H\bigl(h_{jk\ell m}\bigr)\biggr]\biggl[\prod_{(jk\ell mn)\in\Lambda_{4}}\delta_L\bigl(l_{jk\ell mn}\bigr)\biggr] \,,  \vphantom{\ds\int}\\
\end{array}
\end{equation}
which is equal to the state sum $\cZ_\partial^{B\# A}$ for the manifold $\cM_B\#\cM_A$ with the boundary $\cT(\Sigma_A)\sqcup \cT(\Sigma_B)$, according to the definition (\ref{McZ}). In other words, this is precisely the l.h.s. of the equation \eqref{eq:functorcondition}.

This demonstrates that the functor $Z$ satisfies the functoriality axiom \eqref{eq:functoriality} in the definition \ref{TQFTfunctor}.

\subsection{\label{SubSec:Identity}Identity morphism axiom} 

Next we turn to the proof of the identity morphism axiom, namely the equation \eqref{eq:def2} in the definition \ref{TQFTfunctor}. We begin by recalling that, given any triangulated manifold $\cT(\Sigma) \in Obj_{3TCob}$, the cylinder cobordism $\text{id}_{\cT(\Sigma)} = [0,1] \times \cT(\Sigma)$ indeed acts as its identity morphism in the category $3TCob$ (see Figure \ref{figure:identity}). Namely, given any cobordisms $\cM: \cT(\Sigma) \to \cT(\Sigma')$ and $\cM': \cT(\Sigma') \to \cT(\Sigma)$ we have:
\begin{equation} \label{eq:IdentityCobordismDefEquations}
\text{id}_{\cT(\Sigma)} \,\#\,\cM' = \cM'\,, \qquad \cM \,\#\, \text{id}_{\cT(\Sigma)} = \cM\,.
\end{equation}
Assuming that the functor $Z$ assigns operators $\hat{M} = Z(\cM)$ and $\hat{M'} = Z(\cM')$ to the cobordisms $\cM$ and $\cM'$, and the operator $\hat{N} = Z(\text{id}_{\cT(\Sigma)})$ to the cylinder cobordism, one can apply the functor to (\ref{eq:IdentityCobordismDefEquations}), and employ the functoriality axiom (proved in previous subsection), to obtain
$$
\hat{N}\cdot \hat{M}' = \hat{M}'\,, \qquad \hat{M}\cdot \hat{N} = \hat{M}\,,
$$
for all possible cobordisms $\cM$ and $\cM'$. Specifically, choosing $\cM$ to be the cylinder cobordism itself, we obtain
$$
\hat{N} \cdot \hat{N} = \hat{N}\,,
$$
which means that $\hat{N}$ is a projector. Then, assuming that the kernel of $\hat{N}$ is trivial in the space $\cH = Z(\cT(\Sigma))$, elementary linear algebra gives us
$$
\hat{N} = \hat{I}_\cH \,,
$$
demonstrating that equation (\ref{eq:def2}) is satisfied.

Regarding the assumption of the triviality of the kernel, one should note that for any vector $\ket{\psi} \in \ker(\hat{N})$ we have
$$
\hat{M} \ket{\psi} = \hat{M}\cdot \hat{N} \ket{\psi} = 0\,,
$$
which means that $\ket{\psi} \in \ker(\hat{\cM})$. Therefore, the kernel of $\hat{N}$ is simultaneously the kernel of all operators from the set
$$
\{ \, \hat{M} = Z(\cM)\; | \; \forall \cM: \cT(\Sigma) \to \cT(\Sigma')\,, \, \forall \cT(\Sigma') \in Obj_{3TCob}\, \}\,.
$$
Given that there are infinitely many $4$-dimensional and $3$-dimensional manifolds, this set should be rather large, and despite the specific form of the functor $Z$, the common kernel (\ie, the intersection of all kernels) for all operators $\hat{M}$ from this set should be trivial. In the unlikely case that it is not trivial, we have demonstrated that it is equal to $\ker(\hat{N})$, so one could in principle redefine the functor $Z$ so that the Hilbert space $\cH=Z(\cT(\Sigma))$ is not given by (\ref{eq:HilbertSpace}), but is instead equal to the image of $\hat{N}$,
$$
Z(\cT(\Sigma)) = \im(\hat{N})\,.
$$
With this redefinition, the kernel of $\hat{N}$ becomes trivial, giving us again the result $\hat{N} = \hat{I}_\cH$ and demonstrating that equation (\ref{eq:def2}) is satisfied.

Let us also note that the identity operator $\hat{I}_\cH$ can be written in the standard form (\ref{McZ}) as
\begin{equation} \label{eq:StandardFormIdentity}
\begin{array}{lcl}
\hat{I}_\cH &=& \ds \biggl[\prod_{(jk)\in\Lambda_{1,\Sigma}}\int_G dg'_{jk}\biggr]
\biggl[\prod_{(jk\ell)\in\Lambda_{2,\Sigma}}\int_H dh'_{jk\ell}\biggr] \biggl[\prod_{(jk\ell m)\in\Lambda_{3,\Sigma}}\int_L dl'_{jk\ell m}\biggr] \vphantom{\ds\int} \\
 & & \ds \biggl[\prod_{(jk)\in\Lambda_{1,\Sigma}}\int_G dg_{jk}\biggr]
\biggl[\prod_{(jk\ell)\in\Lambda_{2,\Sigma}}\int_H dh_{jk\ell}\biggr] \biggl[\prod_{(jk\ell m)\in\Lambda_{3,\Sigma}}\int_L dl_{jk\ell m}\biggr]\;\cZ_\partial^{\text{id}}\;\ket{g,h,l}\otimes \bra{g',h',l'} \,, \vphantom{\int\ds} \\
\end{array}
\end{equation}
where the kernel $\cZ_\partial^{\text{id}}$ is given as follows:
\begin{equation} \label{eq:KernelForIdentity}
\cZ_\partial^{\text{id}} = \left(\dim\cH \right)^2 \prod_{(jk)\in\Lambda_{1,\Sigma}} \chi_{g_{jk}}(g'_{jk}) \prod_{(jk\ell)\in\Lambda_{2,\Sigma}} \chi_{h_{jk\ell}}(h'_{jk\ell}) \prod_{(jk\ell m)\in\Lambda_{3,\Sigma}} \chi_{l_{jk\ell m}}(l'_{jk\ell m})\,.
\end{equation}
Namely, for finite groups $G$, $H$ and $L$, upon substituting (\ref{eq:KernelForIdentity}) into (\ref{eq:StandardFormIdentity}), the normalization terms in (\ref{eq:GroupIntegralDef}) will cancel the factor $\left(\dim\cH \right)^2$ and the indicator functions will eliminate sums over primed variables, leaving us with the following expression:
\begin{equation} \label{eq:TraditionalFormIdentity}
\begin{array}{lcl}
\hat{I}_\cH &=& \ds \biggl[\prod_{(jk)\in\Lambda_{1,\Sigma}}\sum_{g_{jk} \in G}\biggr]
\biggl[\prod_{(jk\ell)\in\Lambda_{2,\Sigma}}\sum_{h_{jk\ell} \in H} \biggr] \biggl[\prod_{(jk\ell m)\in\Lambda_{3,\Sigma}}\sum_{l_{jk\ell m} \in L} \biggr]\;\ket{g,h,l}\otimes \bra{g,h,l} \,. \vphantom{\int\ds} \\
\end{array}
\end{equation}
This has precisely the form $\hat{I} = \sum_i \ket{i} \otimes \bra{i}$ of the identity operator in an orthonormal basis $\{\,\ket{i} \,|\, i=1,\dots,\dim\cH\, \}$ of the Hilbert space $\cH$.

\subsection{Multiplicativity axiom}

Next we turn to the proof of the multiplicativity axiom, namely the equations \eqref{eq:def3} and \eqref{eq:def4} in the definition \ref{TQFTfunctor}. Suppose that we are given an object in the category $3TCob$ that is a disjoint union of two triangulated $3$-dimensional manifolds $\cT(\Sigma_1) \sqcup \cT(\Sigma_2)$. Then, we want to show that its Hilbert space decomposes into a tensor product of subspaces corresponding to $\cT(\Sigma_1)$ and $\cT(\Sigma_2)$, \ie\,, that it can be written in terms of the configurations of the edges, triangles, and tetrahedra of the triangulations $\cT(\Sigma_1)$ and $\cT(\Sigma_2)$.

The proof proceeds by recalling that the triangulations of $\Sigma_1$ and $\Sigma_2$ can be combined into a common refinement, namely that
$$
\cT(\Sigma_1)\sqcup \cT(\Sigma_2) = \cT(\Sigma_1\sqcup \Sigma_2) \,.
$$
Denoting $\Sigma = \Sigma_1 \sqcup \Sigma_2$, we can write the sets of simplices of the common refinement as a disjoint union of the sets of simplices of $\cT(\Sigma_1)$ and $\cT(\Sigma_2)$, namely $\Lambda_{i,\Sigma}=\Lambda_{i,\Sigma_1}\sqcup\Lambda_{i,\Sigma_2}$, for $i=0,1,2,3$. As a consequence, we have that
$$
G{}^{|\Lambda_{1,\Sigma}|} = G{}^{|\Lambda_{1,\Sigma_1}|} \times G{}^{|\Lambda_{1,\Sigma_2}|}\,, \qquad
H{}^{|\Lambda_{2,\Sigma}|} = H{}^{|\Lambda_{2,\Sigma_1}|} \times H{}^{|\Lambda_{2,\Sigma_2}|}\,, \qquad
L{}^{|\Lambda_{3,\Sigma}|} = L{}^{|\Lambda_{3,\Sigma_1}|} \times L{}^{|\Lambda_{3,\Sigma_2}|}\,.
$$
Then, according to (\ref{eq:HilbertSpace}), the space $\cH = Z(\cT(\Sigma))$ is given as
$$
\cH = \cL^2( G{}^{|\Lambda_{1,\Sigma_1}|} \times G{}^{|\Lambda_{1,\Sigma_2}|} \times H{}^{|\Lambda_{2,\Sigma_1}|} \times H{}^{|\Lambda_{2,\Sigma_2}|} \times L{}^{|\Lambda_{3,\Sigma_1}|} \times L{}^{|\Lambda_{3,\Sigma_2}|})\,.
$$
Rearranging the terms, we have
$$
\cH = \cL^2( G{}^{|\Lambda_{1,\Sigma_1}|} \times H{}^{|\Lambda_{2,\Sigma_1}|} \times  L{}^{|\Lambda_{3,\Sigma_1}|} \times G{}^{|\Lambda_{1,\Sigma_2}|} \times H{}^{|\Lambda_{2,\Sigma_2}|} \times L{}^{|\Lambda_{3,\Sigma_2}|})\,,
$$
which can then be split into the tensor product as:
$$
\cH = \cL^2( G{}^{|\Lambda_{1,\Sigma_1}|} \times H{}^{|\Lambda_{2,\Sigma_1}|} \times  L{}^{|\Lambda_{3,\Sigma_1}|}) \otimes \cL^2( G{}^{|\Lambda_{1,\Sigma_2}|} \times H{}^{|\Lambda_{2,\Sigma_2}|} \times L{}^{|\Lambda_{3,\Sigma_2}|})\,.
$$
Again according to (\ref{eq:HilbertSpace}), the r.h.s.\ is precisely the tensor product of spaces $\cH_1 = Z(\cT_1(\Sigma_1))$ and $\cH_2 = Z(\cT_2(\Sigma_2))$. Therefore, equation \eqref{eq:def3} is satisfied.

Now we turn to equation (\ref{eq:def4}). Given the cobordisms
$$
\cM_A: \cT(\Sigma_1) \to \cT(\Sigma_1')\,, \qquad
\cM_B: \cT(\Sigma_2) \to \cT(\Sigma_2')\,, \qquad
\cM_A\sqcup \cM_B : \cT(\Sigma_1)\sqcup \cT(\Sigma_2) \to \cT(\Sigma_1') \sqcup \cT(\Sigma_2')\,,
$$
and their corresponding operators
$$
\hat{M}_A = Z(\cM_A) \,, \qquad
\hat{M}_B = Z(\cM_B) \,, \qquad
\hat{M}_{A\sqcup B} = Z(\cM_A \sqcup \cM_B) \,,
$$
we can use the definition \eqref{McZ} and the factorization of the respective Hilbert spaces, to rewrite the equation \eqref{eq:def4} into the following requirement for the kernels of the three operators:
\begin{equation}\label{multiplicativity:l-r}
\cZ_\partial^{A\sqcup B}=\cZ_\partial^A\cZ_\partial^B\,.
\end{equation}
The state sums $\cZ_\partial^A$ and $\cZ_\partial^B$ are defined using the equation \eqref{ZSigma1Sigma2}, where the boundaries are denoted by $\partial \cM_A=\cT(\bar{\Sigma}_1)\sqcup \cT({\Sigma}_1')$ and $\partial \cM_B=\cT(\bar{\Sigma}_2)\sqcup \cT(\Sigma_2')$. If $\Lambda_{k,\cM}$ denote the set of $k$-simplices of the triangulated manifold $\cM=\cM_A\sqcup \cM_B$, and $\Lambda_{k,\cM_A}$ and $\Lambda_{k,\cM_B}$ the sets of $k$-simplices of the triangulated manifolds $\cM_A$ and $\cM_B$, respectively, one can decompose the products of delta functions that appear on the left-hand side of the equation \eqref{multiplicativity:l-r} to the products of delta functions corresponding to the simplices of $\cM_A$ and $\cM_B$, 
\begin{equation}\label{eq:delta_G}
\prod_{(jk\ell)\in\Lambda_{2,\cM}} \delta_G(g_{jk\ell}) =  \prod_{(jk\ell)\in\Lambda_{2,\cM_A}}\delta_{G}(g_{jk\ell}) \prod_{(jk\ell)\in\Lambda_{2,\cM_B}}\delta_{G}(g_{jk\ell})\,,
\end{equation}
\begin{equation}\label{eq:delta_H:1}
\prod_{(jk\ell m)\in\Lambda_{3,\cM}}\delta_H\bigl(h_{jk\ell m}\bigr)=\prod_{(jk\ell m)\in\Lambda_{3,\cM_A}}\delta_H\bigl(h_{jk\ell m}\bigr)\prod_{(jk\ell m)\in\Lambda_{3,\cM_B}}\delta_H\bigl(h_{jk\ell m}\bigr)\,,
\end{equation}
\begin{equation}\label{eq:delta_H:2}
\prod_{(jk\ell mn)\in\Lambda_{4,\cM}}\delta_L\bigl(l_{jk\ell mn}\bigr)=\prod_{(jk\ell mn)\in\Lambda_{4,\cM_A}}\delta_L\bigl(l_{jk\ell mn}\bigr)\prod_{(jk\ell mn)\in\Lambda_{4,\cM_B}}\delta_L\bigl(l_{jk\ell mn}\bigr)\,.
\end{equation}
Similarly, the integration over the set $\Lambda_{k,\cM}^*$ decomposes into the integration over $\Lambda_{k,\cM_A}^*$, and $\Lambda_{k,\cM_B}^*$, so we have:
\begin{equation}
\prod_{(jk)\in\Lambda_{1,\cM}^*}\int dg_{jk}=\prod_{(jk)\in\Lambda_{1,\cM_A}^*}\int dg_{jk}\prod_{(jk)\in\Lambda_{1,\cM_B}^*}\int dg_{jk}\,,
\end{equation}
\begin{equation}
\prod_{(jk\ell)\in\Lambda_{2,\cM}^*}\int dh_{jk\ell}=\prod_{(jk\ell)\in\Lambda_{2,\cM_A}^*}\int dh_{jk\ell}\prod_{(jk\ell)\in\Lambda_{2,\cM_B}^*}\int dh_{jk\ell}\,,
\end{equation}
\begin{equation}
\prod_{(jk\ell m)\in\Lambda_{3,\cM}^*}\int dl_{jk\ell m}=\prod_{(jk\ell m)\in\Lambda_{3,\cM_A}^*}\int dl_{jk\ell m}\prod_{(jk\ell m)\in\Lambda_{3,\cM_B}^*}\int dl_{jk\ell m}\,.
\end{equation}
Finally, it is straightforward to verify that the product of constant factors $\mathcal{N}_A$ and $\mathcal{N}_B$, defined analogously to (\ref{eq:N1}), gives the constant factor $\mathcal{N}_\cM$. Having these identities in hand, one obtains that the equation \eqref{multiplicativity:l-r} is satisfied.

Thus the functor $Z$ satisfies the multiplicativity axiom given by equations \eqref{eq:def3} and \eqref{eq:def4} in the definition \ref{TQFTfunctor}.

\subsection{Normalization axiom}

Next we need to prove the equations \eqref{eq:def5} and \eqref{eq:def6} and equivalently (\ref{eq:def6alt}), in the definition \ref{TQFTfunctor}. First, according to (\ref{eq:HilbertSpace}), one can see that the Hilbert space corresponding to an empty manifold $\varnothing_3$ is:
\begin{equation}\label{eq_partition_empty}
\cH= Z(\varnothing_3)=\mathcal{L}^2(G{}^0\times H{}^0\times L{}^0) \equiv \kompleksni\,.
\end{equation}
In other words, when the boundary is empty, $\Sigma=\varnothing_3$, the Hilbert space $\cH$ is just the set of complex numbers $\kompleksni$. This has also already been evaluated in the first example in Subsection \ref{SubSecExamples}, see (\ref{eq:exampleOneH}). Thus, (\ref{eq:def5}) is satisfied.

Next, using the fact that $\text{id}_{\varnothing_3} = \varnothing_4$, according to (\ref{McZ}) or equivalently (\ref{eq:exampleOneZ}) the l.h.s. of the equation \eqref{eq:def6} is equal to $Z(\varnothing_4)=\cZ_\varnothing^{\varnothing_4} $ where $\cZ_{\varnothing}^{\varnothing_4}$ is the evaluation of \eqref{eq_partition_3bf} over an empty manifold $\varnothing_4$. The r.h.s. becomes $\hat{I}_\mathbb{C}=1 $, \ie\,, the equation \eqref{eq:def6} reduces to: 
\begin{equation}
\cZ_\varnothing=1\,,
\end{equation}
which is yet another way to rewrite (\ref{eq:def6}) that resembles (\ref{eq:def6alt}). Explicitly evaluating the state sum \eqref{eq_partition_3bf} over an empty $4$-manifold $\varnothing_4$, one indeed obtains
$$
\cZ_\varnothing^{\varnothing_4}={|G|}^{0}{|H|}^0\,{|L|}^{0}=1\,,
$$
thus verifying that
$$
Z(\varnothing_4)=1 \,.
$$
That is, equation \eqref{eq:def6} is satisfied. 

\subsection{Symmetric monoidal maps axiom}

Regarding this axiom, we merely note that since all relevant maps (associator, left and right unitors, and swap) are essentially trivial in both categories, the statement that the functor $Z$ maps them into each other is also trivially satisfied.

\subsection{Involutory axiom}

Next we consider the involutory axiom, given by the equation \eqref{eq:def8} in the definition \ref{TQFTfunctor}. Let $\mathcal{H}_{\Sigma}$ denote the Hilbert space associated with the triangulated $3$-dimensional manifold $\cT(\Sigma)$, namely $\mathcal{H}_{\Sigma}=Z(\cT(\Sigma))$, and let $\mathcal{H}_{\bar{\Sigma}}$ denote the Hilbert space associated with the triangulated manifold $\cT(\bar{\Sigma})$, namely $\mathcal{H}_{\bar{\Sigma}}=Z(\cT(\bar{\Sigma}))$, where $\cT(\bar{\Sigma})$ is the triangulation obtained from the triangulation $\cT(\Sigma)$ by reversing the orientation of all its links, triangles and tetrahedra. Then, the involutory axiom states that $\mathcal{H}_{\bar{\Sigma}}=\mathcal{H}_{\Sigma}^*$, where $\mathcal{H}_{\Sigma}^*$ denotes the dual Hilbert space of $\mathcal{H}_{\Sigma}$.

To demonstrate this, note that the triangulation $\cT(\Sigma)$ and the triangulation $\cT(\bar{\Sigma})$ have equal numbers of simplices, so their corresponding Hilbert spaces $\cH_\Sigma$ and $\cH_{\bar{\Sigma}}$ are isomorphic, and also both are isomorphic to $\mathcal{H}_{\Sigma}^*$,
$$
\cH_{\bar{\Sigma}} \cong \cH_\Sigma \cong \cH_{\Sigma}^*\,.
$$
Moreover, both the orientation reversal operation and the operation of taking the dual of a Hilbert space are involutions,
$$
\cT(\bar{\bar{\Sigma}}) = \cT(\Sigma)\,, \qquad \cH_{\Sigma}^{**} = \cH_\Sigma\,.
$$
Therefore, nothing prevents us from making an identification between orientation reversal and taking the dual, so we can basically define that
$$
Z(\cT(\bar{\Sigma})) = \cH_\Sigma^*\,.
$$
Thus the involutory axiom (\ref{eq:def8}) is satisfied.

\subsection{Hermitian axiom}

Next we consider the Hermitian axiom, given by the equation \eqref{eq:def11} in the definition \ref{TQFTfunctor}.  The dagger in the category $3TCob$ maps each $4$-dimensional manifold $\cM: \cT(\Sigma_1) \to \cT(\Sigma_2)$ to its reverse manifold $\cM^\dagger:\cT(\Sigma_2) \to \cT(\Sigma_1)$, while the dagger in the category $Hilb$ maps each operator $\hat{M}$ to its Hermitian adjoint operator $\hat{M}^\dagger$. Then, the Hermitian axiom requires that the functor $Z$ commutes with the dagger, \ie\,, $Z({\cM^\dagger})=\hat{M}^\dagger$, where $\hat{M}={Z}(\cM)$. Using the equation \eqref{McZ}, one can rewrite this condition in terms of the state sums as:
\begin{equation}
\cZ_\partial=\cZ_\partial^\dagger\,.
\end{equation}
Here, according to (\ref{ZSigma1Sigma2}), the state sum $\cZ_\partial$ corresponding to the manifold $\cM$ is
\begin{equation} \label{eq_partition_function}
\begin{array}{lcl}
 \cZ_\partial &=& \ds \mathcal{N}_\cM \biggl[\prod_{(jk)\in\Lambda_1^*}\int_G dg_{jk}\biggr] \biggl[\prod_{(jk\ell)\in\Lambda_2^*}\int_H dh_{jk\ell}\biggr] \biggl[\prod_{(jk\ell m)\in\Lambda_3^*}\int_L dl_{jk\ell m}\biggr] \vphantom{\ds\int} \\
 && \ds\biggl[\prod_{(jk\ell)\in\Lambda_{2}}\delta_G\bigl(g_{jk\ell}\bigr)\biggr] \biggl[\prod_{(jk\ell m)\in\Lambda_{3}}\delta_H\bigl(h_{jk\ell m}\bigr)\biggr]\biggl[\prod_{(jk\ell mn)\in\Lambda_{4}}\delta_L\bigl(l_{jk\ell m n}\bigr)\biggr]\vphantom{\ds\int}\,, \\
\end{array}
\end{equation}
while the state sum $\cZ_\partial^\dagger$ corresponding to the reversed manifold $\cM^\dagger$ is
\begin{equation} \label{eq_partition_function*}
\begin{array}{lcl}
  \cZ_\partial^\dagger &=& \ds \mathcal{N}_{\cM^\dagger} \biggl[\prod_{(jk)\in\Lambda_1^*}\int_G dg_{jk}\biggr] \biggl[\prod_{(jk\ell)\in\Lambda_2^*}\int_H dh_{jk\ell}\biggr] \biggl[\prod_{(jk\ell m)\in\Lambda_3^*}\int_L dl_{jk\ell m}\biggr] \vphantom{\ds\int} \\
  && \ds \biggl[\prod_{(jk\ell)\in\Lambda_{2}}\delta_G\bigl(\overline{g_{jk\ell}}\bigr)\biggr] \biggl[\prod_{(jk\ell m)\in\Lambda_{3}}\delta_H\bigl(\overline{h_{jk\ell m}}\bigr)\biggr]\biggl[\prod_{(jk\ell mn)\in\Lambda_{4}}\delta_L\bigl(\overline{l_{jk\ell m n}}\bigr)\biggr]\vphantom{\ds\int}\,. \vphantom{\ds\int} \\
\end{array}
\end{equation}
Note that in the reversed manifold all simplices in the triangulation are reversed (both in the bulk and on the boundary), which implies that the group elements associated to those simplices are replaced with their corresponding group inverses. This is denoted by bars over the group elements inside the delta functions.

The orientation reversal does not change the number of simplices of a triangulation, so we have $\mathcal{N}_{\cM}=\mathcal{N}_{\cM^\dagger}$. However, the consequences of orientation reversal on the group elements are not trivial, so in Lemmas \ref{Th01}, \ref{Th02} and \ref{Th:03}, we prove that
$$
\delta_G\bigl(g_{jk\ell}\bigr)=\delta_G\bigl(\overline{g_{jk\ell}}\bigr)\,, \qquad
\delta_H\bigl(h_{jk\ell m}\bigr)=\delta_H\bigl(\overline{h_{jk\ell m}}\bigr)\,, \qquad
\delta_L\bigl(l_{jk\ell mn}\bigr)=\delta_L\bigl(\overline{l_{jk\ell mn}}\bigr)\,,
$$
see Appendices \ref{appLemmas} and \ref{app}. Therefore, we have that (\ref{eq_partition_function}) and (\ref{eq_partition_function*}) are indeed equal, so the Hermitian axiom \eqref{eq:def11} is satisfied.

\subsection{Unit and counit axiom}

Finally, we consider the unit and counit axiom, given by the equations \eqref{eq:functorAxiomsEnd} in the definition \ref{TQFTfunctor}.

Looking at Figure \ref{figure:identity}, we can first observe that cobordisms $\text{id}_\Sigma$, $\eta_\Sigma$ and $\epsilon_\Sigma$ are formally different since they have different sources and targets, namely
$$
\text{id}_\Sigma : \cT(\Sigma) \to \cT(\Sigma)\,, \qquad
\eta_\Sigma: \varnothing_3 \to \cT(\bar{\Sigma}) \sqcup \cT(\Sigma)\,, \qquad
\epsilon_\Sigma: \cT(\Sigma) \sqcup \cT(\bar{\Sigma}) \to \varnothing_3\,.
$$
Nevertheless, they are all described by the same manifold $\cM$ (the cylinder) that has the same boundary $\partial\cM = \cT(\bar{\Sigma}) \sqcup \cT(\Sigma)$. Therefore, we have
\begin{equation} \label{eq:EqualitiesOfIdentityKernels}
\cZ_\partial^\text{id} = \cZ_\partial^\eta = \cZ_\partial^\epsilon \,,
\end{equation}
which can be trivially verified by explicitly writing down all three state sums according to the definition (\ref{ZSigma1Sigma2}). This means that the kernels of the operators $Z(\text{id}_\Sigma)$, $Z(\eta_\Sigma)$ and $Z(\epsilon_\Sigma)$ are identical, so the operators themselves differ only in the structure of their domain and codomain. Denoting the Hilbert space of $\cT(\Sigma)$ as $\cH \equiv Z(\cT(\Sigma))$, the Hilbert space of $\cT(\bar{\Sigma})$ is equal to $\cH^*$ according to the involutory axiom. Then, since $\partial\cM = \bar{\Sigma} \sqcup \Sigma$, using the multiplicativity axiom we can see that the domains and codomains of the three operators are given as:
$$
Z(\text{id}_\Sigma): \cH \to \cH\,, \qquad
Z(\eta_\Sigma): \kompleksni \to \cH^* \otimes \cH\,, \qquad
Z(\epsilon_\Sigma): \cH \otimes \cH^* \to \kompleksni\,.
$$

In Subsection \ref{SubSec:Identity} we have already seen that the identity operator has the form (\ref{eq:StandardFormIdentity}) with the kernel (\ref{eq:KernelForIdentity}). Regarding the unit operator, given its source and target, we can write it as:
\begin{equation}
\begin{array}{lcl}
Z(\eta_\Sigma)\!\!\!\! &=& \!\!\!\! \ds \biggl[\prod_{(jk)\in\Lambda_{1,\Sigma}}\int_G dg'_{jk}\biggr]
\biggl[\prod_{(jk\ell)\in\Lambda_{2,\Sigma}}\int_H dh'_{jk\ell}\biggr] \biggl[\prod_{(jk\ell m)\in\Lambda_{3,\Sigma}}\int_L dl'_{jk\ell m}\biggr] \vphantom{\ds\int} \\
 & & \!\!\!\! \ds \biggl[\prod_{(jk)\in\Lambda_{1,\Sigma}}\int_G dg_{jk}\biggr]
\biggl[\prod_{(jk\ell)\in\Lambda_{2,\Sigma}}\int_H dh_{jk\ell}\biggr] \biggl[\prod_{(jk\ell m)\in\Lambda_{3,\Sigma}}\int_L dl_{jk\ell m}\biggr]\;\cZ_\partial^{\eta}\;\bra{g,h,l}\otimes \ket{g',h',l'} \otimes 1 , \vphantom{\int\ds} \\
\end{array}
\end{equation}
where the tensor multiplication with the number $1$ indicates that the domain of the operator is the Hilbert space $\kompleksni$. To see that it has precisely the form as in the definition (\ref{eq:HilbDefUnitCounit}), we can act with it on an arbitrary number $z\in\kompleksni$, and remembering (\ref{eq:EqualitiesOfIdentityKernels}), perform the following calculation:
$$
\begin{array}{lcl}
Z(\eta_\Sigma) z\!\!\!\! &=& \!\!\!\! \ds z \biggl[\prod_{(jk)\in\Lambda_{1,\Sigma}}\int_G dg'_{jk}\biggr]
\biggl[\prod_{(jk\ell)\in\Lambda_{2,\Sigma}}\int_H dh'_{jk\ell}\biggr] \biggl[\prod_{(jk\ell m)\in\Lambda_{3,\Sigma}}\int_L dl'_{jk\ell m}\biggr] \vphantom{\ds\int} \\
 & & \!\!\!\! \ds \biggl[\prod_{(jk)\in\Lambda_{1,\Sigma}}\int_G dg_{jk}\biggr]
\biggl[\prod_{(jk\ell)\in\Lambda_{2,\Sigma}}\int_H dh_{jk\ell}\biggr] \biggl[\prod_{(jk\ell m)\in\Lambda_{3,\Sigma}}\int_L dl_{jk\ell m}\biggr]\;\cZ_\partial^{\text{id}}\;\bra{g,h,l}\otimes \ket{g',h',l'}\, . \vphantom{\int\ds} \\
\end{array}
$$
For the choice of finite groups $G$, $H$ and $L$ the integrals become normalized sums according to (\ref{eq:GroupIntegralDef}), and substituting (\ref{eq:KernelForIdentity}) normalizations from the sums cancel the term $(\dim\cH)^2$ while the indicator functions eliminate the sums over primed variables. The r.h.s.\ then reduces to the form
$$
\begin{array}{lcl}
Z(\eta_\Sigma) z\!\!\!\! &=& \!\!\!\! \ds z \biggl[\prod_{(jk)\in\Lambda_{1,\Sigma}}\sum_{g_{jk} \in G} \biggr]
\biggl[\prod_{(jk\ell)\in\Lambda_{2,\Sigma}}\sum_{h_{jk\ell} \in H} \biggr] \biggl[\prod_{(jk\ell m)\in\Lambda_{3,\Sigma}}\sum_{l_{jk\ell m} \in L} \biggr] \bra{g,h,l}\otimes \ket{g,h,l}\,, \vphantom{\int\ds} \\
\end{array}
$$
which has precisely the same form $\eta_\cH (z) = z\sum_i \bra{i} \otimes \ket{i}$ as the unit map in the definition (\ref{eq:HilbDefUnitCounit}). This concludes the proof of the first equation in (\ref{eq:functorAxiomsEnd}) for the unit and counit axiom of Definition \ref{TQFTfunctor}.

Regarding the counit operator, given its source and target, we can write it as:
\begin{equation}
\begin{array}{lcl}
Z(\epsilon_\Sigma)\!\!\!\! &=& \!\!\!\! \ds \biggl[\prod_{(jk)\in\Lambda_{1,\Sigma}}\int_G dg'_{jk}\biggr]
\biggl[\prod_{(jk\ell)\in\Lambda_{2,\Sigma}}\int_H dh'_{jk\ell}\biggr] \biggl[\prod_{(jk\ell m)\in\Lambda_{3,\Sigma}}\int_L dl'_{jk\ell m}\biggr] \vphantom{\ds\int} \\
 & & \!\!\!\! \ds \biggl[\prod_{(jk)\in\Lambda_{1,\Sigma}}\int_G dg_{jk}\biggr]
\biggl[\prod_{(jk\ell)\in\Lambda_{2,\Sigma}}\int_H dh_{jk\ell}\biggr] \biggl[\prod_{(jk\ell m)\in\Lambda_{3,\Sigma}}\int_L dl_{jk\ell m}\biggr]\;\cZ_\partial^{\epsilon}\; 1 \otimes \ket{g,h,l}\otimes \bra{g',h',l'} , \vphantom{\int\ds} \\
\end{array}
\end{equation}
where the tensor multiplication with the number $1$ indicates that the codomain of the operator is the Hilbert space $\kompleksni$. To see that it has precisely the form as in the definition (\ref{eq:HilbDefUnitCounit}), we can act with it on an arbitrary basis operator
$$
\ket{g^A,h^A,l^A} \otimes \bra{g^B,h^B,l^B}\,.
$$
Acting with $\ket{g,h,l}\otimes \bra{g',h',l'}$ on this operator is defined as a double scalar product of corresponding terms, namely $\bracket{g^B,h^B,l^B}{g,h,l}$ times $\bracket{g',h',l'}{g^A,h^A,l^A}$.

The action of the whole operator $Z(\epsilon_\Sigma)$ is then given as follows:
\begin{equation}
\begin{array}{lcl}
Z(\epsilon_\Sigma) \ket{g^A,h^A,l^A} \otimes \bra{g^B,h^B,l^B}\!\!\!\! &=& \!\!\!\! \ds \biggl[\prod_{(jk)\in\Lambda_{1,\Sigma}}\int_G dg'_{jk}\biggr]
\biggl[\prod_{(jk\ell)\in\Lambda_{2,\Sigma}}\int_H dh'_{jk\ell}\biggr] \biggl[\prod_{(jk\ell m)\in\Lambda_{3,\Sigma}}\int_L dl'_{jk\ell m}\biggr] \vphantom{\ds\int} \\
 & & \!\!\!\! \ds \biggl[\prod_{(jk)\in\Lambda_{1,\Sigma}}\int_G dg_{jk}\biggr]
\biggl[\prod_{(jk\ell)\in\Lambda_{2,\Sigma}}\int_H dh_{jk\ell}\biggr] \biggl[\prod_{(jk\ell m)\in\Lambda_{3,\Sigma}}\int_L dl_{jk\ell m}\biggr] \vphantom{\int\ds} \\
 & & \!\!\!\! \ds \cZ_\partial^{\epsilon}\; \bracket{g^B,h^B,l^B}{g,h,l} \; \bracket{g',h',l'}{g^A,h^A,l^A} \,. \vphantom{\int\ds} \\
\end{array}
\end{equation}
Now, remembering (\ref{eq:EqualitiesOfIdentityKernels}), substituting (\ref{eq:KernelForIdentity}) and employing (\ref{eq:GroupIntegralDef}) for the case of finite groups $G$, $H$ and $L$, we perform the calculation similar to the one for $Z(\eta_\Sigma)$ above, to obtain that the r.h.s.\ is equal to:
$$
r.h.s. = \biggl[\prod_{(jk)\in\Lambda_{1,\Sigma}}\sum_{g_{jk}\in G} \biggr]
\biggl[\prod_{(jk\ell)\in\Lambda_{2,\Sigma}}\sum_{h_{jk\ell}\in H} \biggr] \biggl[\prod_{(jk\ell m)\in\Lambda_{3,\Sigma}}\sum_{l_{jk\ell m}\in L}\biggr] 
\bracket{g^B,h^B,l^B}{g,h,l} \; \bracket{g,h,l}{g^A,h^A,l^A} \,.
$$
Looking at the form (\ref{eq:TraditionalFormIdentity}) of the unit operator $\hat{I}_\cH$, one can observe that by moving the vector $\bra{g^B,h^B,l^B}$ to the far left in front of all the sums, the r.h.s.\ in fact becomes
$$
r.h.s. = \bra{g^B,h^B,l^B} \; \hat{I}_\cH \; \ket{g^A,h^A,l^A}\,,
$$
which finally leads us to conclude that
$$
Z(\epsilon_\Sigma) \; \ket{g^A,h^A,l^A} \otimes \bra{g^B,h^B,l^B} = \bracket{g^B,h^B,l^B}{g^A,h^A,l^A} \,.
$$
This has precisely the same form $\epsilon_\cH \ket{i}\otimes\bra{j} = \bracket{j}{i}$ as the counit map in the definition (\ref{eq:HilbDefUnitCounit}). Thus, we have proved the second equation in (\ref{eq:functorAxiomsEnd}) of the unit and counit axiom of Definition \ref{TQFTfunctor}.

At this point, the proof that $Z$ satisfies all axioms of Definition \ref{TQFTfunctor} is complete, and we can say that $Z$ represents a genuine functor between categories $3TCob$ and $Hilb$.

\section{Conclusion}\label{SecV}

%\subsection{Summary}

Let us summarize the results of the paper. In Section \ref{SecII}, we provided an overview of the foundational concepts of category theory required to define Topological Quantum Field Theory within this specific framework.
Specifically, we gave a detailed description of the categories $Hilb$, $3Cob$, and $3TCob$ as three relevant examples of a dagger symmetric monoidal category with a dual. Given these, we have seen that Atiyah's axioms, initially given as abstract requirements for a TQFT, seamlessly translate into the criteria for a functor between the category of cobordisms and the category of vector spaces. In Section \ref{SecIII}, we presented our main results --- we introduced functors $Z$ and $\tilde{Z}$, using an extension of the $3BF$ state sum defined in \cite{Radenkovic2022_2}, discussed two illustrative examples, and formulated a theorem that $Z$ is a genuine functor from $3TCob$ to $Hilb$, and consequently that $\tilde{Z}$ is a genuine functor from $3Cob$ to $Hilb$. The resulting TQFT should coincide with Porter’s TQFT \cite{Porter98, Porter96} for $d = 4$ and $n = 3$. In Section \ref{SecIV}, we gave a detailed proof of our theorem, demonstrating that the functor $Z$ satisfies all axioms. Since certain parts of the proofs rely on lengthy calculations, they are presented in the Appendices.

%\subsection{Discussion}

The results presented in this paper complete the Step 2 of the spinfoam quantization programme and provide the necessary mathematical foundations underlying the physically relevant quantum gravity model, namely the one based on the constrained $3BF$ action for gravity coupled to matter fields, as introduced in \cite{Radenkovic2019}. We should mention that this is part of a wider research program of building a physically realistic quantum gravity model using the framework of higher gauge theory, and several other interesting results have been obtained so far \cite{MikovicVojinovic2012,MikovicVojinovic2011,MikovicVojinovicBook,Radenkovic2019,Radenkovic2020, Radenkovic2022_1, Radenkovic2022_2, StipsicVojinovic2024,MikovicVojinovic2021}. Investigation is underway regarding the Step 3 of the spinfoam quantization programme, as well as the study of the various properties of the resulting class of quantum gravity models.

The functor $Z$ is based on the state sum $\cZ_\partial$, defined in (\ref{ZSigma1Sigma2}), which represents a generalization of the state sum $\cZ_\varnothing $ (see (\ref{eq_partition_3bf})) introduced in our previous work \cite{Radenkovic2022_2}. The two state sums are topological invariants of a class of $4$-dimensional manifolds that admit a triangulation. Namely, the state sums are defined over a simplicial complex, and manifolds that are not homeomorphic to any simplicial complex are out of the scope of the definitions (\ref{eq_partition_3bf}) and (\ref{ZSigma1Sigma2}). While for the purpose of applications to physics and quantum gravity it can be satisfactory to restrict oneself to the class of manifolds that admit a triangulation, from the point of view of mathematics there is a natural question that can be asked --- is there an alternative definition of the topological invariant corresponding to $\cZ_\partial $ or $\cZ_\varnothing $ that does not rely on the structure of a simplicial complex, and can be applied to all $4$-dimensional manifolds? This is an open question, that deserves further research.

A related question one can also ask is the following --- are $\cZ_\partial $ and $\cZ_\varnothing $ novel topological invariants, or can they be related to the ones we already know? For example, for a simply-connected closed $4$-manifold $\cM$, one should be able to relate $\cZ_\varnothing $ to the intersection form of $\cM$, as per Freedman classification. This question also deserves further research.

As a final note, we should emphasize that the state sum $\cZ_\partial$ and the functor $Z$, as introduced in this work, have been rigorously defined (and the TQFT axioms have been rigorously verified) only for the choice of finite groups $G$, $H$ and $L$ that make up a $3$-group. Extending the TQFT to the choice of Lie groups, while formally possible, is more problematic due to the following issues:
\begin{itemize}
\item In the case of Lie groups, the cardinalities $|G|$, $|H|$, and $|L|$ of the three groups become infinite. This implies that the state sum $\cZ_\partial $ becomes divergent, a well known property of many other such manifold invariants (such as the Ponzano-Regge invariant \cite{PonzanoRegge}, and so on).
\item The Hilbert spaces, which are objects in $Hilb$, become infinite-dimensional. While this is not a problem in itself, it does break the involution axiom of a TQFT, since the dual $\cH^*$ of an infinite-dimensional Hilbert space $\cH$ is not isomorphic to $\cH$ anymore, and $\cH^{**} \neq \cH$, as per the Riesz representation theorem.
\item One of the steps in the verification of the functoriality axiom involved squares of Dirac delta functions over a group. While for finite groups such squares are well-defined and can be evaluated, for Lie groups these are in general not well defined, or at least are more delicate to handle and would require special attention.
\end{itemize}
One possible strategy to deal with the above issues is to consider a Lie group as some form of a limit of a finite group. For example, the group $SO(2)$ of arbitrary rotations in a plane could be regarded as the finite cyclic group $C_n$ of rotations by integer multiples of the angle $\theta= 2\pi / n$, in the limit $n\to\infty$. If such limits can be made well defined and rigorous, one could expect that the state sum $\cZ_\partial$ and the functor $Z$ could also be understood as a corresponding limit. However, these ideas require further research, if they are to be made more rigorous.

Another possible strategy to deal with the above issues would be to consider quantum groups instead of Lie groups. Namely, there is a well known example of generalizing the divergent Ponzano-Regge state sum for $3$-dimensional manifolds \cite{PonzanoRegge}, by passing from the Lie group $SU(2)$ to the quantum group $SU(2)_q$. When $q$ is chosen to be a root of unity, one obtains a convergent state sum, known as the Turaev-Viro invariant \cite{TuraevViro}. Although the generalization of the notion of a $3$-group to the corresponding ``quantum'' $3$-group has not been done yet (to the best of our knowledge), it is a priori possible that such a generalization could in principle also regularize the behavior of the state sum (\ref{ZSigma1Sigma2}). This would lead to a well-defined TQFT functor without assuming the finiteness of groups $G$, $H$ and $L$. Implementation of these ideas also requires further research.

\section*{Acknowledgments}

The authors would like to thank Tim Porter, Jo\~ao Faria Martins and Aleksandar Mikovi\'c for fruitful discussions.

This work was supported by the Ministry of Science, Technological development and Innovations of the Republic of Serbia. In addition, the authors were supported by the Science Fund of the Republic of Serbia, grant 7745968, ``Quantum Gravity from Higher Gauge Theory 2021'' --- QGHG-2021. The contents of this publication are the sole responsibility of the authors and can in no way be taken to reflect the views of the Science Fund of the Republic of Serbia.

\appendix
\section{\label{AppI}\texorpdfstring{$3$}{3}-groups}
In the framework of the category theory, a $2$-group is defined as a $2$-category consisting of only one object, where all the morphisms and $2$-morphisms are invertible. It is demonstrated that every strict $2$-group can be conveniently encoded by a crossed module structure, denoted by $(H \stackrel{\del}{\to}G \,, \rhd)$.

A \emph{pre-crossed module} $(H \stackrel{\del}{\to}G \,, \rhd)$ of groups $G$ and $H$, is given by a group map $\partial : H \to G$,
together with a left action $\rhd$ of $G$ on both groups, by automorphisms, such that the group $G$ acts on itself via conjugation, \ie\,, for each $g_1, g_2 \in G$,
$$g_1\rhd g_2 = g_1 g_2 g_1^{-1}\,, $$
and for each $h_1\,,h_2 \in H$ and $g \in G$ the following identity holds:
$$g \partial h g^{-1} = \partial (g \rhd h)\,.$$
In a pre-crossed module the \emph{Peiffer commutator} is defined as:
\begin{equation}\label{Peiffer_comm}
    \langle h_1\,,h_2 \rangle{}_{\mathrm{p}}=h_1h_2h_1^{-1} \partial(h_1) \rhd h_2^{-1}\,.
\end{equation}
A pre-crossed module is a \emph{crossed module} if all of its Peiffer commutators are trivial, \ie\,, the \emph{Peiffer identity} is satisfied:
\begin{equation}
  (\partial h_1) \rhd h_2 = h_1 h_2 h_1^{-1}\,.  
\end{equation}

Within the category theory formalism, one can continue the generalization one step further, and generalize the notion of a $2$-group to the notion of a $3$-group. The study of $3$-groups within the category theory provides a versatile framework to explore and generalize notions of symmetry. Similar to the definitions of a group and a $2$-group, a $3$-group is defined as a $3$-category with only one object, where all morphisms, $2$-morphisms, and $3$-morphisms are invertible. Drawing parallels to the equivalence between a strict $2$-group and a crossed-module, one can show that a $2$-crossed module encodes a semistrict $3$-group --- Gray group \cite{Con, martins2011}. 

A \emph{$2$-crossed module} $(L\stackrel{\delta}{\to} H \stackrel{\partial}{\to}G,\,\rhd,\,\{\_,\,\_\}_\mathrm{p})$ is a chain complex of three groups $G$, $H$, and $L$, 
\begin{equation*}
    L\stackrel{\delta}{\to} H \stackrel{\partial}{\to}G\,,
\end{equation*}
together with maps $\partial$ and $\delta$ (such that $\partial\delta=1_G$), an action $\rhd$ of the group $G$ on all three groups, and a map $\{ \_\,,\_ \}_\mathrm{p}$ called the Peiffer lifting:
\begin{equation*}
\{ \_\,,\_ \}_\mathrm{p} : H\times H \to L\,.
\end{equation*}
The maps $\partial$ and $\delta$, and the Peiffer lifting are $G$-equivariant, \ie\,, for each $ g \in G$ and $h \in H $
\begin{equation*}\label{eq:ekvivarijantnost_partial}
    g\rhd \partial (h)=\partial(g\rhd h)\,, \quad \quad \quad g\rhd \delta (l) = \delta(g \rhd l)\,,
\end{equation*}
and for each $h_1,\, h_2 \in H$ and $g \in G$:
$$ g\rhd\{h_1\,,h_2\}_\mathrm{p}=\{ g\rhd h_1, \,g \rhd h_2\}_\mathrm{p}\,.$$
The action of the group $G$ on the groups $H$ and $L$ is a smooth left action by automorphisms, \ie\,, for each $g,g_1,g_2 \in G$, $\;h_1,h_2 \in H$, $\;l_1,l_2 \in L$ and $k \in H, L$,
\begin{equation*}\label{eq:ekvivarijantnost_delta}
    g_1 \rhd (g_2 \rhd k) =(g_1 g_2) \rhd k\,, \quad \quad g \rhd (h_1 h_2) =(g \rhd h_1)(g \rhd h_2)\,, \quad \quad g \rhd (l_1 l_2) =(g \rhd l_1)(g \rhd l_2)\,. 
\end{equation*}
The action of the group $G$ on itself is again via conjugation. Further, the following identities are satisfied:
\begin{subequations}
\begin{align}
 \;\; & \delta(\{h_1,h_2 \}_\mathrm{p})= \langle h_1\,,h_2\rangle{}_{\mathrm{p}}\,, \qquad\qquad \forall h_1,h_2 \in H\,;\;\label{prop1}\\
 \;\; & [l_1,l_2]=\{\delta(l_1)\,,\delta(l_2) \}_\mathrm{p}\,, \qquad \qquad\forall l_1\,, l_2 \in L\,, \quad\text{where } [l,k]=lkl^{-1}k^{-1};\label{prop2}\\
  \;\; & \{h_1h_2,h_3 \}_\mathrm{p}=\{h_1,h_2h_3h_2^{-1} \}_\mathrm{p}\partial(h_1)\rhd \{h_2,h_3 \}_\mathrm{p}\,, \qquad  \forall h_1,h_2,h_3 \in H\,;\label{prop3}\\
 \;\;& \{h_1,h_2h_3 \}_\mathrm{p}= \{h_1,h_2\}_\mathrm{p} \{h_1,h_3\}_\mathrm{p}\{ \langle h_1,h_3 \rangle{}_{\mathrm{p}}^{-1}\,, \partial(h_1) \rhd h_2\}_\mathrm{p}\,, \quad{\forall h_1,h_2,h_3 \in H}\,;\label{prop4}\\
 \;\;&\{\delta(l),h\}_\mathrm{p}\{h, \delta(l) \}_\mathrm{p}=l(\partial(h) \rhd l^{-1})\,,\qquad\qquad \forall h\in H\,, \quad \forall l \in L\,.\label{prop5}
 \end{align}
\end{subequations}
In a $2$-crossed module the structure $(L\stackrel{\delta}{\to}H,\,\rhd')$ is a crossed module, where the action $\rhd'$ of the group $H$ on the group $L$ is defined for each $h \in H$ and $l \in L$ as:
\begin{equation}\label{identitet2}
h \rhd' l = l \, \{\delta(l){}^{-1},\,h\}_\mathrm{p}\,.
\end{equation}
As $(L\stackrel{\delta}{\to}H,\,\rhd')$ is a crossed module, it follows that the Peiffer identity is satisfied for each $l_1,l_2 \in L$:
\begin{equation}\label{PeifferidL}
\delta(l_1)\rhd' \,l_2= l_1 \,l_2\, l_1^{-1}\,. 
\end{equation}
However, the structure $(H \stackrel{\del}{\to}G \,, \rhd)$ in the general case does not form a crossed module, but a pre-crossed module, and for each $h,h'\in H$ the Peiffer commutator does not necessarily vanish.

The following identities hold, for each $h_1,h_2,h_3 \in H$ \cite{martins2011}:
\begin{equation}\label{eq:id01}
    \{h_1h_2,h_3\}_{\mathrm{p}}= (h_1 \rhd' \{h_2,h_3\}_{\mathrm{p}})\{h_1,\partial(h_2)\rhd h_3\}_{\mathrm{p}}\,,
\end{equation}
\begin{equation}\label{eq:id02}
    \begin{aligned}
    \{h_1,h_2h_3\}_{\mathrm{p}}=\{h_1,h_2\}_{\mathrm{p}}(\partial(h_1)\rhd h_2)\rhd'\{h_1,h_3\}_{\mathrm{p}}\,.
    \end{aligned}
\end{equation}
Using the condition \eqref{prop5} in the definition of a 2-crossed module, it follows that for each $h\in H$ and $l \in L$ one has the identity:
\begin{equation}\label{identitet}
    \{h, \delta(l)^{-1}\}_{\mathrm{p}}=(h \rhd' l^{-1}) (\partial(h)\rhd l)\,.
\end{equation}
Moreover, using the previous identities, one can derive the following. For each $h_1,h_2 \in H$,
\begin{equation}\label{eq:id03}
    \{h_1,h_2\}_{\mathrm{p}}^{-1}=h_1 \rhd'\{h_1^{-1},\partial(h_1)\rhd h_2\}_{\mathrm{p}}\,,
\end{equation}
\begin{equation}\label{eq:id04}
  \{h_1,h_2\}_{\mathrm{p}}^{-1}=\partial(h_1)\rhd\{h_1^{-1},h_1h_2h_1^{-1}\}_{\mathrm{p}}\,,
\end{equation}
\begin{equation}\label{eq:id05}
    \{h_1,h_2\}^{-1}_{\mathrm{p}}= (h_1h_2h_1^{-1})\rhd'\{h_1,h_2^{-1}\}_{\mathrm{p}}\,,
\end{equation}
\begin{equation}\label{eq:id06}
   \{h_1,h_2\}_{\mathrm{p}}^{-1}= (\partial(h_1)\rhd h_2)\rhd'\{h_1,h_2^{-1}\}_{\mathrm{p}}\,. 
\end{equation}
A reader interested in more details about $3$-groups is referred to \cite{Wang2014}.

\section{Three Lemmas}\label{appLemmas}

\begin{lemma}[Orientation reversal for a triangle]\label{Th01}
Given the first equation in (\ref{eq:DiracDeltaArguments}), we have
$$
\delta_G\bigl(g_{jk\ell}\bigr)=\delta_G\bigl(\overline{g_{jk\ell}}\bigr)\,,
$$
where $\overline{g_{jk\ell}}$ is the orientation reversal of $g_{jk\ell}$, namely:
$$
\overline{g_{jk\ell}} = g_{jl} g_{jk}^{-1}g_{kl}^{-1}\partial({h}_{jkl})^{-1}\,.
$$

To prove this, let us consider a triangle, $(jk\ell)$. The edges $(jk)\,, j < k$, are labeled by group elements $g_{jk}\in G$ and the triangle $(jk\ell)\,, j < k < \ell$, by element $h_{jk\ell}\in H$. Consider the diagram (\ref{diag1}). 

\begin{equation}\label{diag1}
  \xymatrix{l
\bullet & k\bullet
\ar@/_2ex/[l]_{g_{kl}}="g1" & \bullet j
\ar@/_2ex/[l]_{g_{jk}}="g2"
\ar@/^6ex/[ll]^{g_{jl}}="g3"
\ar@{=>}^{h_{jkl}} "g1"+<4.7ex,-4.8ex>;"g3"+<0ex,2.5ex>
}\quad = \quad \xymatrix{l \bullet && l \bullet 
\ar@/_2ex/[ll]_{1_\bullet}="g1"
\ar@/^2ex/[ll]^{\partial(h_{jkl})}="g2"
\ar@{=>}^{h_{jkl}} "g1"+<0ex,-2ex>;"g2"+<0ex,2ex>
& k\bullet
\ar@/_2ex/[l]_{g_{kl}}="g4" & \bullet j
\ar@/_2ex/[l]_{g_{jk}}="g5"
\ar@/^7ex/[ll]^{g_{kl}g_{jk}}="g3"
\ar@{=>}^(0.25){1_{g_{kl}g_{jk}}} "g4"+<4.7ex,-4.8ex>;"g3"+<0ex,2.5ex>
}\quad = \quad \xymatrix{l
\bullet & k\bullet
\ar@/_2ex/[l]_{g_{kl}}="g1" & \bullet j
\ar@/_2ex/[l]_{g_{jk}}="g2"
\ar@/^6ex/[ll]^{\partial(h_{jkl}) \,g_{kl}g_{jk}}="g3"
\ar@{=>}^{h_{jkl}} "g1"+<4.7ex,-4.8ex>;"g3"+<0ex,2.5ex>
}\,.
\end{equation}
The curve $\Sigma_1=g_{ k\ell}g_{j k}$ is the source and the curve $\Sigma_2=g_{j \ell}$ is the target of the surface morphism $\Sigma:\Sigma_1\to\Sigma_2$, labeled by the group element $h_{j k \ell}$, \ie\,,
\begin{equation}
g_{j\ell}=\partial(h_{jk\ell})g_{k\ell}g_{jk}\,,
\end{equation}
which gives rise to the $\delta$-function corresponding to the triangle 
\begin{equation}\label{eq:deltaG}
    \delta_G(g_{j\ell}^{-1}\partial(h_{jk\ell})g_{k\ell}g_{jk})\,.
\end{equation}
Let us reverse the orientation of the curves, edges of the triangle, and surface of the triangle. The reversal is shown on the diagram \eqref{diag11},
\begin{equation}\label{diag11}
  \xymatrix{ \ar@/^2ex/[r]^{g_{kl}^{-1}}="g1" \ar@/_6ex/[rr]_{g_{jl}^{-1}}="g3" l
\bullet & \ar@/^2ex/[r]^{g_{jk}^{-1}}="g2" k\bullet
 & \bullet j
\ar@{<=}^{\tilde{h}_{jkl}} "g1"+<4.7ex,-4.8ex>;"g3"+<0ex,2.5ex>
}\quad = \quad 
\xymatrix{ \ar@/^2ex/[rr]^{1_\bullet}="g1" \ar@/_2ex/[rr]_{\partial(h_{jkl})^{-1}}="g2" l \bullet && l \bullet 
\ar@{<=}^{h_{jkl}} "g1"+<0ex,-2ex>;"g2"+<0ex,2ex>
\ar@/^2ex/[r]^{g_{kl}^{-1}}="g4" \ar@/_7ex/[rr]_{g_{jk}^{-1}g_{kl}^{-1}}="g3" & k\bullet
\ar@/^2ex/[r]^{g_{jk}^{-1}}="g5" & \bullet j 
\ar@{<=}^(0.25){1_{g_{kl}g_{jk}}} "g4"+<4.7ex,-4.8ex>;"g3"+<0ex,2.5ex>
}\quad = \quad   \xymatrix{ \ar@/^2ex/[r]^{g_{kl}^{-1}}="g1" \ar@/_6ex/[rr]_{\partial(\tilde{h}_{jkl})^{-1}g_{jk}^{-1}g_{kl}^{-1}}="g3" l
\bullet & \ar@/^2ex/[r]^{g_{jk}^{-1}}="g2" k\bullet
 & \bullet j
\ar@{<=}^{\tilde{h}_{jkl}} "g1"+<4.7ex,-4.8ex>;"g3"+<0ex,2.5ex>
}\,,
\end{equation}
where $\tilde{h}_{jkl}=(g_{kl}g_{jk})^{-1}\rhd h_{jkl}$. Similarly as in the previous diagram, one can write that
\begin{equation}
g_{j\ell}^{-1}=\partial(\tilde{h}_{jk\ell})^{-1}g_{jk}^{-1}g_{k\ell}^{-1}\,.
\end{equation}
In other words, the $\delta$-function for the reversed orientation triangle is given as
\begin{equation}
    \begin{aligned}
    \delta_G(g_{jl}\partial(\tilde{h}_{jkl})^{-1}g_{jk}^{-1}g_{kl}^{-1})= \delta_G(g_{jl}g_{jk}^{-1}g_{kl}^{-1}\partial({h}_{jkl})^{-1}g_{kl}g_{jk}g_{jk}^{-1}g_{kl}^{-1})=\delta_G(g_{jl}g_{jk}^{-1}g_{kl}^{-1}\partial({h}_{jkl})^{-1}).
    \end{aligned}
\end{equation}
This expression is equivalent to the $\delta$-function given by the equation \eqref{eq:deltaG}.

Let us note also that due to the $G$-invariance of the Dirac delta function, $\delta_G(g'\rhd g) = \delta_G(g)$, the statement of the Lemma does not depend on the choice of the labeling $jkl$ of the vertices.

\end{lemma}
\begin{lemma}[Orientation reversal for a tetrahedron]\label{Th02}
Given the second equation in (\ref{eq:DiracDeltaArguments}), we have
$$
\delta_H\bigl(h_{jk\ell m}\bigr)=\delta_H\bigl(\overline{h_{jk\ell m}}\bigr)\,,
$$
where $\overline{h_{jk\ell m}}$ is the orientation reversal of $h_{jk\ell m}$.

To prove this, let us consider a tetrahedron, $(jk\ell m)$. The edges $(jk)\,, j < k$, are labeled by group elements $g_{jk}\in G$ and the triangles $(jk\ell)\,, j < k < \ell$, by elements $h_{jk\ell}\in H$, and the tetrahedron $(j k \ell m)\,, j<k<\ell <m$ by the group element $l_{j k \ell m}\in L$.  The triangles $(jk\ell)$ are oriented so that the source curve is $g_{k\ell}g_{jk}$ and the target curve is $g_{j\ell}$, \ie\ $g_{j\ell}=\partial(h_{jk\ell})g_{k\ell}g_{jk}$\,.

We cut the tetrahedron surface along the edge $(jm)$, which determines the ordering of the vertical composition of the constituent surfaces. 

Consider the diagram (\ref{diag4}). We first move the curve from $g_{k\ell} g_{jk}$ to the curve $g_{j\ell}$. To compose the result with the triangle $(j\ell m)$, one first has to whisker it from the left by $g_{\ell m}$.Then, the two morphisms are vertically composable, and the resulting $2$-morphism moves the curve to $g_{j m}$. The following $2$-morphism is obtained,
\begin{equation}\label{diag4}
\xymatrix{m\bullet & \bullet \ell
\ar[l]_{g_{\ell m}}="g1"
& \bullet k
\ar@/_2ex/[l]_{g_{k\ell}}="g2" 
& \bullet j
\ar@/_2ex/[l]_{g_{jk}}="g3"
\ar@/^6ex/[ll]^{g_{j\ell}}="g4"
\ar@{=>}^{h_{jk\ell}} "g2"+<4.7ex,-5ex>;"g4"+<0ex,2.5ex>
\ar@/^9ex/[lll]^{g_{jm}}="g5"
\ar@{=>}^{h_{j\ell m}} "g1"+<4.1ex,-3.7ex>;"g5"+<-6ex,5.3ex>
} \;\; = (g_{\ell m} g_{j\ell}, h_{j\ell m})\#_2\big(g_{\ell m}\#_1 (g_{k \ell}g_{jk}, h_{jk\ell})\big)=\big(g_{\ell m} g_{k\ell}g_{jk}, h_{j\ell m} (g_{\ell m}\rhd h_{jk\ell})\big)\,,
\end{equation}
or equivalently, 
\begin{equation*}\label{diag41}
\xymatrix{m \bullet & \ar@/_2ex/[l]_{1_\bullet}="g8" \ar@/^2ex/[l]^{\partial(h_{j\ell m})}="g10" \ar@{=>}^{h_{j\ell m}} "g8"+<-1ex,-2ex>;"g10"+<-1ex,2ex> m\bullet & \bullet \ell \ar[l]_{g_{\ell m}}="g0" & \bullet \ell
\ar@/_2ex/[l]_{1_\bullet}="g1"
\ar@/^2ex/[l]^{\partial(h_{j k \ell})}="g7"
\ar@{=>}^{h_{jk\ell}} "g1"+<-1ex,-2ex>;"g7"+<-1ex,2ex>
& \bullet k
\ar@/_2ex/[l]_{g_{k\ell}}="g2" 
& \bullet j
\ar@/_2ex/[l]_{g_{jk}}="g3"
\ar@/^6ex/[ll]^{{g_{k\ell} g_{jk}}}="g4"
\ar@/^12ex/[llll]^{g_{\ell m}\partial(h_{jk\ell})g_{k \ell} g_{j k}}="g5"
\ar@{=>}^{1} "g2"+<4.7ex,-5ex>;"g4"+<0ex,2.5ex>
\ar@{=>}^{1} "g1"+<4.7ex,-6ex>;"g5"+<0ex,3ex>
} \,,
\end{equation*}
which gives the interchanging $2$-arrow identity $3$-morphism:
\begin{equation}
    (1_G, h_{j\ell m})\#_1\big(g_{\ell m}\#_1 (g_{k \ell}g_{jk}, h_{jk\ell})\big)=\big(g_{\ell m} g_{k\ell}g_{jk}, h_{j\ell m} (g_{\ell m}\rhd h_{jk\ell}), 1_{L}\big)\,.
\end{equation}
Let us then consider the diagram (\ref{diag3:3}). We first move the curve from $g_{\ell m} g_{k\ell}$ to the curve $g_{k m}$. To compose the result with the triangle $(jkm)$, one first has to whisker it from the right by $g_{jk}$. Now the two morphisms are vertically composable, and the resulting $2$-morphism moves the curve to $g_{j m}$. The following $2$-morphism is obtained,
\begin{equation}\label{diag3:3}
\xymatrix{m\bullet & \bullet \ell
\ar@/_2ex/[l]_{g_{\ell m}}="g4" & \bullet k
\ar@/_2ex/[l]_{g_{k\ell}}="g3"
\ar@/^6ex/[ll]^{g_{km}}="g2"
\ar@{=>}^{h_{k\ell m}} "g4"+<4.7ex,-5ex>;"g2"+<0ex,2.5ex>
& \bullet j
\ar[l]_{g_{jk}}="g1"
\ar@/^9ex/[lll]^{g_{jm}}="g5"
\ar@{=>}^{h_{jkm}} "g3"+<4.6ex,-5.6ex>;"g5"+<6ex,5ex>
} \quad = (g_{km} g_{jk},h_{jkm})\#_2\big((g_{\ell m} g_{k \ell},h_{k \ell m})\#_1 g_{j k}\big)=(g_{\ell m} g_{k\ell} g_{jk},h_{jkm}h_{k\ell m}) \,,
\end{equation}
or, equivalently,
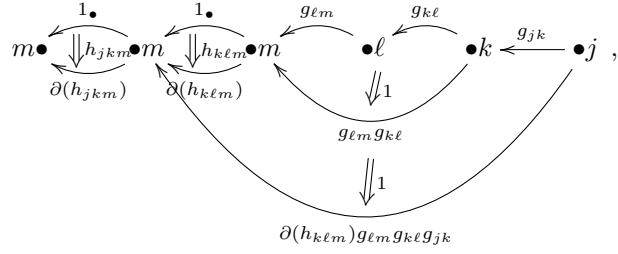
\begin{equation*}\label{diag31}
\xymatrix{m \bullet & \bullet m \ar@/_2ex/[l]_{1_\bullet}="g8" \ar@/^2ex/[l]^{\partial(h_{jk m})}="g10" \ar@{=>}^{h_{j k m}} "g8"+<-1ex,-2ex>;"g10"+<-1ex,2ex> & \bullet m  \ar@/_2ex/[l]_{1_\bullet}="g5" \ar@/^2ex/[l]^{\partial(h_{k\ell m})}="g6"
\ar@{=>}^{h_{k \ell m}} "g5"+<-1ex,-2ex>;"g6"+<-1ex,2ex> & \bullet \ell \ar@/_2ex/[l]_{g_{\ell m}}="g3"
%\ar@/^15ex/[llll]^{g_{\ell m}\partial(h_{jk\ell})g_{k \ell} g_{j k}}="g5"&  \ar@/_2ex/[l]_{g_{\ell m}}="g2"  \bullet \ell 
& \bullet k
\ar@/_2ex/[l]_{g_{k\ell}}="g44"
\ar@/^6ex/[ll]^{g_{\ell m}g_{k\ell}}="g33"
\ar@{=>}^{1} "g44"+<-4ex,-5ex>;"g33"+<0ex,2.5ex>
& \bullet j \ar[l]_{g_{jk}}="g4" \ar@/^14ex/[llll]^{\partial(h_{k \ell m}){g_{\ell m} g_{k \ell}}g_{jk}}="g45" 
\ar@{=>}^{1} "g33"+<0ex,-2ex>;"g45"+<0ex,2.5ex> 
} \,,
\end{equation*}
which gives the interchanging $2$-arrow identity $3$-morphism:
\begin{equation}
    (1_G, h_{jk m})\#_1(g_{\ell m}g_{k \ell} g_{jk}, h_{k\ell m})=(g_{\ell m} g_{k\ell} g_{jk},h_{jkm}h_{k\ell m})\,.
\end{equation}

The two surfaces $\Sigma_1$ and $\Sigma_2$, depicted on the diagrams \eqref{diag4}  and \eqref{diag3:3}, have the same source and target, $\Sigma_1 : g_{\ell m} g_{k\ell}g_{jk} \to g_{jm}$ and $\Sigma_2 :g_{\ell m} g_{k\ell}g_{jk} \to g_{jm}$. Now, moving from surface shown on the diagram \eqref{diag4} to the surface shown on the diagram \eqref{diag3:3} is given by the volume morphism $\mathcal{V}: \Sigma_1 \to \Sigma_2$ determined by the group element $l_{jk \ell m}$, \ie\,,
\begin{equation}
    (g_{\ell m} g_{k\ell} g_{jk},h_{jkm}h_{k\ell m}) = \big(g_{\ell m} g_{k\ell}g_{jk}, \delta(l_{jk\ell m}) h_{j\ell m} (g_{\ell m}\rhd h_{jk\ell})\big)\,,
\end{equation}
that gives the relation,
\begin{equation}
    h_{jkm}h_{k\ell m}  = \delta(l_{jk\ell m}) h_{j\ell m} (g_{\ell m}\rhd h_{jk\ell})\,,
\end{equation}
giving rise to the $\delta$-function corresponding to the tetrahedron $(jk\ell m)$:
\begin{equation}\label{eq:deltaHhhhh}
    \delta_H(\delta(l_{jk\ell m}) h_{j\ell m} (g_{\ell m}\rhd h_{jk\ell})h_{k\ell m}^{-1}h_{jkm}^{-1})\,.
\end{equation}
Now, let us reverse the orientation of the curves, surfaces, and volume of a tetrahedron $(jk\ell m)$. The first diagram \eqref{diag4} now reduces to,
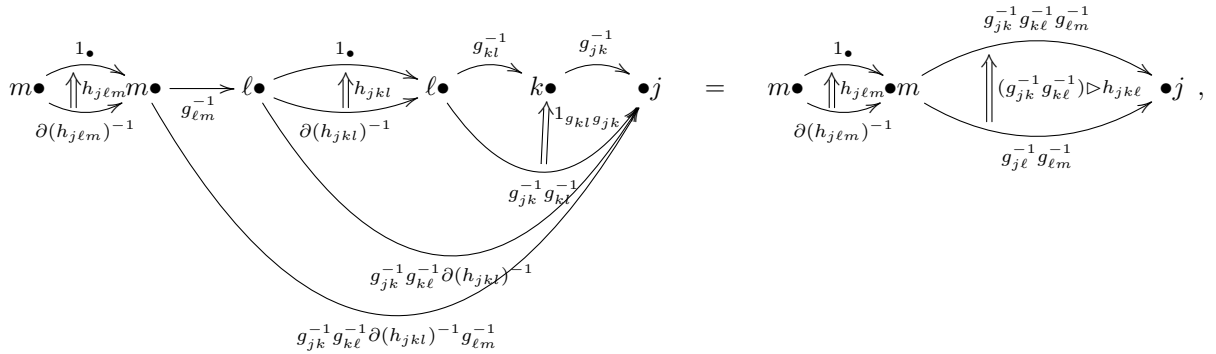
\begin{equation*}\label{diag111}
\xymatrix{\ar@/^2ex/[r]^{1_\bullet}="g9" \ar@/_2ex/[r]_{\partial(h_{j\ell m})^{-1}}="g10"
\ar@{<=}^{h_{j\ell m}} "g9"+<-1ex,-2ex>;"g10"+<-1ex,2ex>m \bullet &\ar[r]_{g_{\ell m}^{-1}}="g11" \ar@/_19ex/[rrrrr]_{g_{jk}^{-1}g_{k\ell}^{-1}\partial({h}_{jkl})^{-1}g_{\ell m}^{-1}}="g13"  m \bullet & \ar@/^2ex/[rr]^{1_\bullet}="g1" \ar@/_2ex/[rr]_{\partial(h_{jkl})^{-1}}="g2" \ar@/_14ex/[rrrr]_{g_{jk}^{-1}g_{k\ell}^{-1}\partial({h}_{jkl})^{-1}}="g12" \ell \bullet && \ell \bullet 
\ar@{<=}^{h_{jkl}} "g1"+<0ex,-2ex>;"g2"+<0ex,2ex>
\ar@/^2ex/[r]^{g_{kl}^{-1}}="g4" \ar@/_7ex/[rr]_{g_{jk}^{-1}g_{kl}^{-1}}="g3" & k\bullet
\ar@/^2ex/[r]^{g_{jk}^{-1}}="g5" & \bullet j 
\ar@{<=}^(0.25){1_{g_{kl}g_{jk}}} "g4"+<4.7ex,-4.8ex>;"g3"+<0ex,2.5ex>
}\quad=\quad \xymatrix{\ar@/^2ex/[r]^{1_\bullet}="g9" \ar@/_2ex/[r]_{\partial(h_{j\ell m})^{-1}}="g10"
\ar@{<=}^{h_{j\ell m}} "g9"+<-1ex,-2ex>;"g10"+<-1ex,2ex>m \bullet & \bullet m \ar@/^4ex/[rrr]^{g_{jk}^{-1}g_{k\ell}^{-1}g_{\ell m}^{-1}}="g10" \ar@/_4ex/[rrr]_{g_{j\ell}^{-1}g_{\ell m}^{-1}}="g20" \ar@{<=}^{(g_{jk}^{-1}g_{k\ell}^{-1})\rhd h_{j k \ell}} "g10"+<-4ex,-3ex>;"g20"+<-4ex,3ex> & & &\bullet j}\,,
\end{equation*}
giving rise to the interchanging $2$-arrow $3$-morphism with the source surface $\Sigma_1: (g_{jk}^{-1}g_{k\ell}^{-1})\rhd h_{j k \ell} (g_{j\ell}^{-1}g_{\ell m}^{-1})\rhd h_{j\ell m}$ and the target surface is given as $\Sigma_2: (g_{jk}^{-1}g_{k\ell}^{-1}g_{\ell m}^{-1})\rhd h_{j\ell m} (g_{jk}^{-1}g_{k\ell}^{-1})\rhd h_{j k \ell}  $,
\begin{equation}\label{eq:3morp1}
    \big(g_{jk}^{-1}g_{k\ell}^{-1}g_{\ell m}^{-1}, (g_{jk}^{-1}g_{k\ell}^{-1})\rhd h_{j k \ell} (g_{j\ell}^{-1}g_{\ell m}^{-1})\rhd h_{j\ell m}, \{(g_{jk}^{-1}g_{k\ell}^{-1})\rhd h_{j k \ell} ,(g_{j\ell}^{-1}g_{\ell m}^{-1})\rhd h_{j\ell m}\}_{\mathfrak{p}}^{-1} \big)\,.
\end{equation}
The next step is to calculate the $3$-morphism obtained by reversing  the orientation of the curves and surfaces of the diagram \eqref{diag3:3},
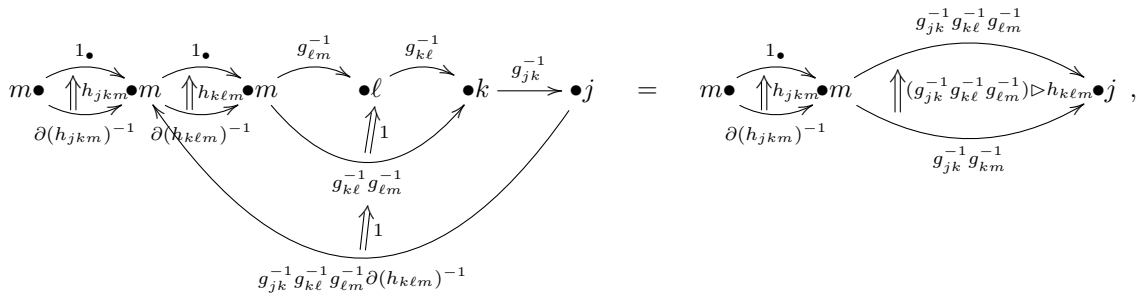
\begin{equation*}\label{diag311}
\xymatrix{ \ar@/^2ex/[r]^{1_\bullet}="g8" \ar@/_2ex/[r]_{\partial(h_{jk m})^{-1}}="g10" m \bullet & \bullet m  \ar@{<=}^{h_{j k m}} "g8"+<-1ex,-2ex>;"g10"+<-1ex,2ex>  \ar@/^2ex/[r]^{1_\bullet}="g5" \ar@/_2ex/[r]_{\partial(h_{k\ell m})^{-1}}="g6" & \bullet m  \ar@/_6ex/[rr]_{g_{k\ell}^{-1}g_{\ell m}^{-1}}="g33"
\ar@{<=}^{h_{k \ell m}} "g5"+<-1ex,-2ex>;"g6"+<-1ex,2ex>  \ar@/^2ex/[r]^{g_{\ell m}^{-1}}="g3" & \bullet \ell \ar@/^2ex/[r]^{g_{k\ell}^{-1}}="g44"
%\ar@/^15ex/[llll]^{g_{\ell m}\partial(h_{jk\ell})g_{k \ell} g_{j k}}="g5"&  \ar@/_2ex/[l]_{g_{\ell m}}="g2"  \bullet \ell 
& \bullet k
\ar@{<=}^{1} "g44"+<-4ex,-5ex>;"g33"+<0ex,2.5ex> \ar[r]^{g_{jk}^{-1}}="g4"
& \bullet j  \ar@/^14ex/[llll]^{g_{jk}^{-1}{ g_{k \ell}^{-1}}g_{\ell m}^{-1}\partial(h_{k \ell m})^{-1}}="g45" 
\ar@{<=}^{1} "g33"+<0ex,-2ex>;"g45"+<0ex,2.5ex> 
} \quad = \quad \xymatrix{\ar@/^2ex/[r]^{1_\bullet}="g9" \ar@/_2ex/[r]_{\partial(h_{j k m})^{-1}}="g10"
\ar@{<=}^{h_{j k m}} "g9"+<-1ex,-2ex>;"g10"+<-1ex,2ex>m \bullet & \bullet m \ar@/^4ex/[rrr]^{g_{jk}^{-1}g_{k\ell}^{-1}g_{\ell m}^{-1}}="g10" \ar@/_4ex/[rrr]_{g_{jk}^{-1}g_{km}^{-1}}="g20" \ar@{<=}^{(g_{jk}^{-1}g_{k\ell}^{-1}g_{\ell m}^{-1})\rhd h_{k \ell m}} "g10"+<-6ex,-4ex>;"g20"+<-6ex,4ex> & & &\bullet j}\,,
\end{equation*}
giving rise to the interchanging $2$-arrow $3$-morphism with the source surface $\Sigma_1: (g_{jk}^{-1}g_{k\ell}^{-1}g_{\ell m}^{-1})\rhd h_{k \ell m} (g_{jk}^{-1}g_{k m}^{-1})\rhd h_{jk m}$ and the target surface is given as $\Sigma_2: (g_{jk}^{-1}g_{k\ell}^{-1}g_{\ell m}^{-1})\rhd h_{j k m}(g_{jk}^{-1}g_{k\ell}^{-1}g_{\ell m}^{-1})\rhd h_{k \ell m}  $,
\begin{equation}\label{eq:3morp2}
    \big(g_{jk}^{-1}g_{k\ell}^{-1}g_{\ell m}^{-1},(g_{jk}^{-1}g_{k\ell}^{-1}g_{\ell m}^{-1})\rhd h_{k \ell m} (g_{jk}^{-1}g_{k m}^{-1})\rhd h_{jk m}, \{(g_{jk}^{-1}g_{k\ell}^{-1}g_{\ell m}^{-1})\rhd h_{k \ell m} , (g_{jk}^{-1}g_{k m}^{-1})\rhd h_{jk m}\}_{\mathfrak{p}}^{-1} \big)\,.
\end{equation}
The morphisms given by the equations \eqref{eq:3morp1} and\eqref{eq:3morp2} are connected by the $3$-morphism $(g_{jk}^{-1}g_{k\ell}^{-1}g_{\ell m}^{-1})\rhd l_{jk\ell m}^{-1}$,
which gives the relation 
\begin{equation}
    (g_{jk}^{-1}g_{k\ell}^{-1}g_{\ell m}^{-1})\rhd \delta(l_{jk\ell m})^{-1}  (g_{jk}^{-1}g_{k\ell}^{-1}g_{\ell m}^{-1})\rhd h_{j k m}(g_{jk}^{-1}g_{k\ell}^{-1}g_{\ell m}^{-1})\rhd h_{k \ell m} = (g_{jk}^{-1}g_{k\ell}^{-1}g_{\ell m}^{-1})\rhd h_{j\ell m} (g_{jk}^{-1}g_{k\ell}^{-1})\rhd h_{j k \ell}\,. 
\end{equation}
This gives the $\delta$-function,
\begin{equation}
\delta_H\big((g_{jk}^{-1}g_{k\ell}^{-1}g_{\ell m}^{-1})\rhd \delta(l_{jk\ell m})^{-1}  (g_{jk}^{-1}g_{k\ell}^{-1}g_{\ell m}^{-1})\rhd h_{j k m}(g_{jk}^{-1}g_{k\ell}^{-1}g_{\ell m}^{-1})\rhd h_{k \ell m}(g_{jk}^{-1}g_{k\ell}^{-1})\rhd h_{j k \ell}^{-1}(g_{jk}^{-1}g_{k\ell}^{-1}g_{\ell m}^{-1})\rhd h_{j\ell m}^{-1}\big)\,,
\end{equation}
which when multiplied with $g_{\ell m}g_{k\ell}g_{jk}$ reduces to
\begin{equation}
     \delta_H\big(\delta(l_{jk\ell m})^{-1}h_{j k m}h_{k \ell m} g_{\ell m}\rhd  h_{j k \ell}^{-1} h_{j\ell m}^{-1}\big)\,,
\end{equation}
which is the same as the $\delta$-function \eqref{eq:deltaHhhhh} for a tetrahedron before the orientation reversal.
\end{lemma}

\begin{lemma}[Orientation reversal for a $4$-simplex]\label{Th:03}
Given the third equation in (\ref{eq:DiracDeltaArguments}), we have
$$
\delta_L\bigl(l_{jk\ell mn}\bigr)=\delta_L\bigl(\overline{l_{jk\ell mn}}\bigr)\,,
$$
where $\overline{l_{jk\ell mn}}$ is the orientation reversal of $l_{jk\ell mn}$.

To prove this, let us consider a $4$-simplex, $(jk\ell m n)$. The edges $(jk)\,, j < k $, are labeled by group elements $g_{jk}\in G$, the triangles $(jk\ell)\,, j < k  < \ell$, by elements $h_{jk\ell}\in H$, and the tetrahedrons $(jk\ell m)\,, j<k <\ell <m$, by the group element $l_{jk\ell m}\in L$. The triangles $(j k \ell)$ are oriented so that the source curve is $g_{k \ell}g_{jk}$ and the target curve is $g_{j \ell}$, \ie\,, $g_{j \ell}=\partial(h_{j k \ell})g_{k\ell}g_{jk}$\,, and the tetrahedrons $(j k \ell m)$ are oriented so that the source surface is $h_{j\ell m} (g_{\ell m}\rhd h_{jk\ell})$ and the target surface is $ h_{jkm}h_{k\ell m} $, \ie\,, $ h_{jkm}h_{k\ell m}  = \delta(l_{jk\ell m}) h_{j\ell m} (g_{\ell m}\rhd h_{jk\ell})$. Now let us consider the diagrams from the lemma III.3 in \cite{Radenkovic2022_2} where all the morphisms, $2$-morphisms and $3$-morphisms have the inverted orientation.

Let us first cut the $4$-simplex volume along the surface $\Sigma_1: (g_{jk}^{-1}g_{k\ell}^{-1})\rhd h_{jk\ell} (g_{j\ell}^{-1}g_{\ell m}^{-1})\rhd h_{j\ell m} (g_{jm}^{-1}g_{mn}^{-1})\rhd h_{jmn}$. We want to construct the $3$-morphisms with the target surface $\Sigma_1$ and the source surface $\Sigma_2:(g_{jk}^{-1}g_{k\ell}^{-1}g_{\ell m}^{-1})\rhd h_{k\ell m}(g_{jk}^{-1}g_{km}^{-1})\rhd h_{jkm}(g_{jm}^{-1}g_{mn}^{-1})\rhd h_{jmn}$. Let us consider the diagram (\ref{diag51}). 
We first move the surface from $(g_{jk}^{-1}g_{k\ell}^{-1}g_{\ell m}^{-1})\rhd h_{k\ell m}(g_{jk}^{-1}g_{km}^{-1})\rhd h_{jkm}$ to surface $(g_{jk}^{-1}g_{k\ell}^{-1}g_{\ell m}^{-1})\rhd h_{jkm }(g_{jk}^{-1}g_{k\ell}^{-1}g_{\ell m}^{-1})\rhd h_{k\ell m} $ with the $3$-arrow $\{(g_{jk}^{-1}g_{k\ell}^{-1}g_{\ell m}^{-1})\rhd h_{k\ell m},(g_{jk}^{-1}g_{km}^{-1})\rhd h_{jkm}\}_{\mathrm{p}}^{-1}$. Then, one can move the surface to $(g_{jk}^{-1}g_{k\ell}^{-1}g_{\ell m}^{-1})\rhd h_{j\ell m}(g_{jk}^{-1}g_{k\ell}^{-1})\rhd h_{jk\ell} $ with the $3$-arrow $(g_{jk}^{-1}g_{k\ell}^{-1}g_{\ell m}^{-1})\rhd l_{jk\ell m}^{-1}$. From this surface, using the $3$-morphism $\{(g_{jk}^{-1}g_{k\ell}^{-1})\rhd h_{jk\ell},(g_{j\ell}^{-1}g_{\ell m}^{-1})\rhd h_{j\ell m}\}_\mathrm{p}$ one moves to the surface $(g_{jk}^{-1}g_{k\ell}^{-1})\rhd h_{jk\ell} (g_{j\ell}^{-1}g_{\ell m}^{-1})\rhd h_{j\ell m}$. To compose the resulting $3$-morphism with the surface $(g_{jm}^{-1}g_{mn}^{-1})\rhd h_{jmn}$ one must first whisker it from the left with $g_{m n}^{-1}$. The obtained $3$-morphism can be whiskered from below with this $2$-morphism, and the resulting $3$-morphism is $(g_{jk}^{-1}g_{k\ell}^{-1}g_{\ell m}^{-1}g_{mn}^{-1}, (g_{jk}^{-1}g_{k\ell}^{-1}g_{\ell m}^{-1})\rhd h_{k\ell m}(g_{jk}^{-1}g_{km}^{-1})\rhd h_{jkm}(g_{jm}^{-1}g_{mn}^{-1})\rhd h_{jmn}, \tilde{l}_{jk\ell m}^{-1})$, where the $3$-arrow is given as 
\begin{equation}
\resizebox{.99\hsize}{!}{$
\tilde{l}_{jk\ell m}^{-1}=\{(g_{jk}^{-1}g_{k\ell}^{-1})\rhd h_{jk\ell},(g_{j\ell}^{-1}g_{\ell m}^{-1})\rhd h_{j\ell m}\}_\mathrm{p}(g_{jk}^{-1}g_{k\ell}^{-1}g_{\ell m}^{-1})\rhd l_{jk\ell m}^{-1}\{(g_{jk}^{-1}g_{k\ell}^{-1}g_{\ell m}^{-1})\rhd h_{k\ell m},(g_{jk}^{-1}g_{km}^{-1})\rhd h_{jkm}\}_{\mathrm{p}}^{-1}\,,$}\label{eq:l_jkln}
\end{equation}
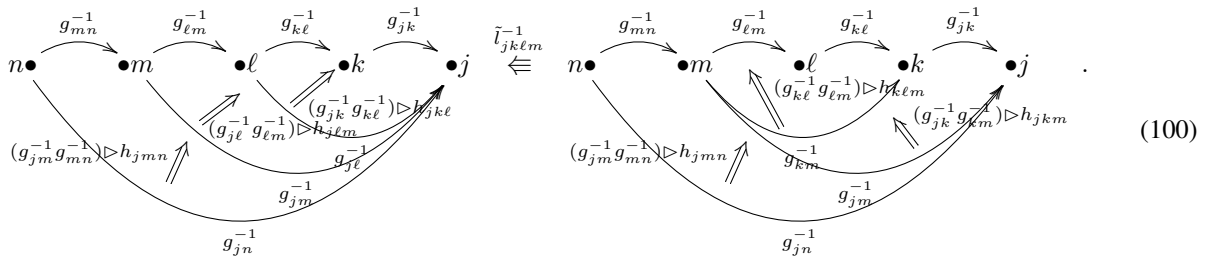
\begin{equation}\label{diag51}
\begin{aligned}
\xymatrix{ \ar@/^2ex/[r]^{g_{m n}^{-1}}="g6" \ar@/_13ex/[rrrr]_{g_{jn}^{-1}}="g7" n \bullet & \ar@/^2ex/[r]^{g_{\ell m}^{-1}}="g1" \ar@/_9ex/[rrr]_{g_{jm}^{-1}}="g5" \bullet m
& \bullet \ell 
\ar@/^2ex/[r]^{g_{k\ell }^{-1}}="g2" \ar@/_6ex/[rr]_{g_{j\ell }^{-1}}="g4"
& \bullet k
\ar@/^2ex/[r]^{g_{jk}^{-1}}="g3" 
& \bullet j
\ar@{<=}^{(g_{jk}^{-1}g_{k\ell}^{-1})\rhd h_{jk\ell }} "g2"+<3ex,-4ex>;"g4"+<-5ex,4.5ex>
\ar@{<=}^{(g_{j\ell}^{-1}g_{\ell m}^{-1})\rhd h_{j\ell m}} "g1"+<4.1ex,-6ex>;"g5"+<-8ex,6ex>
\ar@{<=}_{(g_{j m}^{-1}g_{mn}^{-1})\rhd h_{jm n}} "g6"+<9ex,-10ex>;"g7"+<-6ex,5ex>
} \; \stackrel{\tilde{l}_{jk\ell m}^{-1}}{\Lleftarrow} \; \xymatrix{\ar@/^2ex/[r]^{g_{m n}^{-1}}="g6" \ar@/_13ex/[rrrr]_{g_{jn}^{-1}}="g7" n \bullet 
 & \bullet m \ar@/^2ex/[r]^{g_{\ell m }^{-1}}="g4" \ar@/_6ex/[rr]_{g_{km}^{-1}}="g2" \ar@/_9ex/[rrr]_{g_{jm }^{-1}}="g5"
 & \ar@/^2ex/[r]^{g_{k\ell }^{-1}}="g3" \bullet \ell 
 & \bullet k
\ar@{<=}^{(g_{k \ell}^{-1}g_{\ell m}^{-1})\rhd h_{k\ell m }} "g4"+<0ex,-4ex>;"g2"+<-1.8ex,2.5ex> \ar@/^2ex/[r]^{g_{jk}^{-1}}="g1"
& \bullet j
\ar@{<=}^{(g_{j k}^{-1}g_{k m}^{-1})\rhd h_{jkm }} "g3"+<3ex,-8ex>;"g5"+<5ex,4ex>
\ar@{<=}_{(g_{j m}^{-1}g_{mn}^{-1})\rhd h_{jm n}} "g6"+<9ex,-10ex>;"g7"+<-6ex,5ex>
}\,.
\end{aligned}
\end{equation}

Let us now consider the diagram (\ref{diag61}). We want to construct the $3$-morphism with the source surface $\Sigma_3:(g_{jk}^{-1}g_{k\ell}^{-1}g_{\ell m}^{-1})\rhd h_{k\ell m}(g_{jk}^{-1}g_{km}^{-1}g_{mn}^{-1})\rhd h_{kmn} (g_{jk}^{-1}g_{kn}^{-1})\rhd h_{jkn}$ and the target surface $\Sigma_2:(g_{jk}^{-1}g_{k\ell}^{-1}g_{\ell m}^{-1})\rhd h_{k\ell m}(g_{jk}^{-1}g_{km}^{-1})\rhd h_{jkm}(g_{jm}^{-1}g_{mn}^{-1})\rhd h_{jmn}$. To do this, one again, similarly as it was done on the first diagram, passes from the surface $(g_{jk}^{-1}g_{km}^{-1}g_{mn}^{-1})\rhd h_{kmn} (g_{jk}^{-1}g_{kn}^{-1})\rhd h_{jkn}$ to the surface $(g_{jk}^{-1}g_{km}^{-1})\rhd h_{jkm}(g_{jm}^{-1}g_{mn}^{-1})\rhd h_{jmn}$ using the $3$-arrow $\tilde{l}_{jkmn}^{-1}$. This $3$-morphism can be whiskered from above with the $2$-morphism $(g_{jk}^{-1}g_{k\ell}^{-1}g_{\ell m }^{-1}g_{mn}^{-1},(g_{jk}^{-1}g_{k\ell}^{-1}g_{\ell m}^{-1})\rhd h_{k\ell m})$, and the obtained $3$-morphism is $(g_{jk}^{-1}g_{k\ell}^{-1}g_{\ell m }^{-1}g_{mn}^{-1},(g_{jk}^{-1}g_{k\ell}^{-1}g_{\ell m}^{-1})\rhd h_{k\ell m}(g_{jk}^{-1}g_{km}^{-1}g_{mn}^{-1})\rhd h_{kmn} (g_{jk}^{-1}g_{kn}^{-1})\rhd h_{jkn},((g_{jk}^{-1}g_{k\ell}^{-1}g_{\ell m}^{-1})\rhd h_{k\ell m})\rhd'\tilde{l}_{jkm n}^{-1})$, where the $3$-arrow is
$$
\tilde{l}_{jkm n}^{-1}= \{(g_{jk}^{-1}g_{km}^{-1})\rhd h_{jkm},(g_{jm}^{-1}g_{m n}^{-1})\rhd h_{jm n}\}_\mathrm{p}(g_{jk}^{-1}g_{km}^{-1}g_{mn}^{-1})\rhd l_{jkmn}^{-1}\{(g_{jk}^{-1}g_{km}^{-1}g_{mn}^{-1})\rhd h_{kmn},(g_{jk}^{-1}g_{kn}^{-1})\rhd h_{jkn}\}_{\mathrm{p}}^{-1},
$$
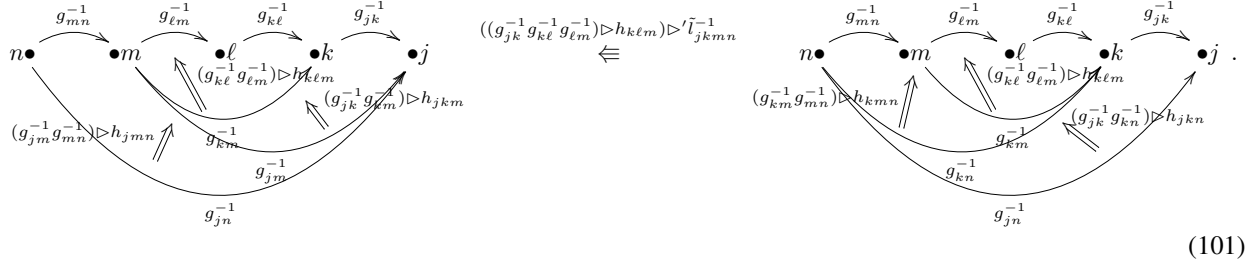
\begin{equation}\label{diag61}\resizebox{.99\hsize}{!}{$
    \begin{aligned}
  \xymatrix{\ar@/^2ex/[r]^{g_{m n}^{-1}}="g6" \ar@/_13ex/[rrrr]_{g_{jn}^{-1}}="g7" n \bullet 
 & \bullet m \ar@/^2ex/[r]^{g_{\ell m }^{-1}}="g4" \ar@/_6ex/[rr]_{g_{km}^{-1}}="g2" \ar@/_9ex/[rrr]_{g_{jm }^{-1}}="g5"
 & \ar@/^2ex/[r]^{g_{k\ell }^{-1}}="g3" \bullet \ell 
 & \bullet k
\ar@{<=}^{(g_{k \ell}^{-1}g_{\ell m}^{-1})\rhd h_{k\ell m }} "g4"+<0ex,-4ex>;"g2"+<-1.8ex,2.5ex> \ar@/^2ex/[r]^{g_{jk}^{-1}}="g1"
& \bullet j
\ar@{<=}^{(g_{j k}^{-1}g_{k m}^{-1})\rhd h_{jkm }} "g3"+<3ex,-8ex>;"g5"+<5ex,4ex>
\ar@{<=}_{(g_{j m}^{-1}g_{mn}^{-1})\rhd h_{jm n}} "g6"+<9ex,-10ex>;"g7"+<-6ex,5ex>
}\stackrel{((g_{jk}^{-1}g_{k\ell}^{-1}g_{\ell m}^{-1})\rhd h_{k\ell m})\rhd'\tilde{l}_{jkm n}^{-1}}{\Lleftarrow}  \xymatrix{\ar@/^2ex/[r]^{g_{m n}^{-1}}="g6" \ar@/_9ex/[rrr]_{g_{kn}^{-1}}="g5" \ar@/_13ex/[rrrr]_{g_{jn}^{-1}}="g7" n \bullet 
 &   \bullet m  \ar@/^2ex/[r]^{g_{\ell m }^{-1}}="g4" \ar@/_6ex/[rr]_{g_{km }^{-1}}="g2"
 & \bullet \ell \ar@/^2ex/[r]^{g_{k\ell }^{-1}}="g3"
 & \bullet k \ar@/^2ex/[r]^{g_{jk}^{-1}}="g1"
\ar@{<=}^{(g_{k \ell}^{-1}g_{\ell m}^{-1})\rhd h_{k\ell m }} "g4"+<0ex,-4ex>;"g2"+<-1.8ex,2.5ex>
& \bullet j
\ar@{<=}_{(g_{k m}^{-1}g_{mn}^{-1})\rhd h_{km n}} "g6"+<4.6ex,-5.6ex>;"g7"+<-10ex,8ex>
\ar@{<=}^{(g_{jk}^{-1}g_{kn}^{-1})\rhd  h_{jkn}} "g3"+<0ex,-10ex>;"g7"+<8ex,6ex>}\,.
\end{aligned}$}
\end{equation}

Next, we want to make the $3$-morphism with the source surface $\Sigma_4: (g_{jk}^{-1}g_{k\ell}^{-1}g_{\ell m}^{-1}g_{mn}^{-1})\rhd h_{\ell mn} (g_{jk}^{-1}g_{k\ell}^{-1}g_{\ell n}^{-1})\rhd h_{k\ell n}(g_{jk}^{-1}g_{kn}^{-1})\rhd h_{jkn}$ and target surface $\Sigma_3:(g_{jk}^{-1}g_{k\ell}^{-1}g_{\ell m}^{-1})\rhd h_{k\ell m}(g_{jk}^{-1}g_{km}^{-1}g_{mn}^{-1})\rhd h_{kmn} (g_{jk}^{-1}g_{kn}^{-1})\rhd h_{jkn}$, as shown on the diagram (\ref{diag71}). First, to move the surface from $(g_{k\ell}^{-1}g_{\ell m}^{-1}g_{mn}^{-1})\rhd h_{\ell mn} (g_{k\ell}^{-1}g_{\ell n}^{-1})\rhd h_{k\ell n}$ to the surface $(g_{k\ell}^{-1}g_{\ell m}^{-1})\rhd h_{k\ell m}(g_{km}^{-1}g_{mn}^{-1})\rhd h_{kmn}$ with the $3$-arrow $\tilde{l}_{k\ell mn}^{-1}$. To whisker this $3$-morphism with the surface $(g_{jk}^{-1}g_{kn}^{-1})\rhd h_{jkn}$ from below, one first needs to whisker it from the left with $g_{jk}^{-1}$. Finally, the obtained $3$-morphism is $(g_{jk}^{-1}g_{k\ell}^{-1}g_{\ell m }^{-1}g_{mn}^{-1},  (g_{jk}^{-1}g_{k\ell}^{-1}g_{\ell m}^{-1}g_{mn}^{-1})\rhd h_{\ell mn} (g_{jk}^{-1}g_{k\ell}^{-1}g_{\ell n}^{-1})\rhd h_{k\ell n}(g_{jk}^{-1}g_{kn}^{-1})\rhd h_{jkn}, g_{jk}^{-1}\rhd\tilde{l}_{k\ell mn}^{-1} $, where the $3$-arrow is given as
$$\tilde{l}_{k\ell mn}^{-1} = \{(g_{k\ell}^{-1}g_{\ell m}^{-1})\rhd h_{k\ell m},(g_{km}^{-1}g_{m n}^{-1})\rhd h_{km n}\}_\mathrm{p}(g_{k\ell}^{-1}g_{\ell m}^{-1}g_{mn}^{-1})\rhd l_{k\ell mn}^{-1}\{(g_{k\ell}^{-1}g_{\ell m}^{-1}g_{mn}^{-1})\rhd h_{\ell mn},(g_{k\ell}^{-1}g_{\ell n}^{-1})\rhd h_{k\ell n}\}_{\mathrm{p}}^{-1}\,,$$
\begin{equation}\label{diag71}
\begin{aligned}
 \xymatrix{\ar@/^2ex/[r]^{g_{m n}^{-1}}="g6" \ar@/_9ex/[rrr]_{g_{kn}^{-1}}="g5" \ar@/_13ex/[rrrr]_{g_{jn}^{-1}}="g7" n \bullet 
 &   \bullet m  \ar@/^2ex/[r]^{g_{\ell m }^{-1}}="g4" \ar@/_6ex/[rr]_{g_{km }^{-1}}="g2"
 & \bullet \ell \ar@/^2ex/[r]^{g_{k\ell }^{-1}}="g3"
 & \bullet k \ar@/^2ex/[r]^{g_{jk}^{-1}}="g1"
\ar@{<=}^{(g_{k \ell}^{-1}g_{\ell m}^{-1})\rhd h_{k\ell m }} "g4"+<0ex,-4ex>;"g2"+<-1.8ex,2.5ex>
& \bullet j
\ar@{<=}_{(g_{k m}^{-1}g_{mn}^{-1})\rhd h_{km n}} "g6"+<4.6ex,-5.6ex>;"g7"+<-10ex,8ex>
\ar@{<=}^{(g_{jk}^{-1}g_{kn}^{-1})\rhd  h_{jkn}} "g3"+<0ex,-10ex>;"g7"+<8ex,6ex>}\; \stackrel{g_{jk}^{-1}\rhd\tilde{l}_{k\ell mn}^{-1}}{\Lleftarrow} \; \xymatrix{ \ar@/^2ex/[r]^{g_{m n}^{-1}}="g6" \ar@/_6ex/[rr]_{g_{\ell n}^{-1}}="g2" \ar@/_9ex/[rrr]_{g_{kn}^{-1}}="g5" \ar@/_13ex/[rrrr]_{g_{jn}^{-1}}="g7" n \bullet 
 & \bullet m  \ar@/^2ex/[r]^{g_{\ell m }^{-1}}="g4"
 & \ar@/^2ex/[r]^{g_{k\ell }^{-1}}="g3" \bullet \ell & \bullet k \ar@/^2ex/[r]^{g_{jk}^{-1}}="g1"
\ar@{<=}^{(g_{k \ell}^{-1}g_{\ell n}^{-1})\rhd h_{k\ell n}} "g4"+<4ex,-5.6ex>;"g5"+<8ex,6ex>
& \bullet j
\ar@{<=}_{(g_{\ell m}^{-1}g_{m n}^{-1})\rhd  h_{\ell m n}} "g6"+<6ex,-4.6ex>;"g2"+<5ex,6ex>
\ar@{<=}^{(g_{jk}^{-1}g_{kn}^{-1})\rhd  h_{jkn}} "g3"+<0ex,-10ex>;"g7"+<8ex,6ex>}\,.
\end{aligned}
\end{equation}
The mapping of the surface $\Sigma_5:(g_{jk}^{-1}g_{k\ell}^{-1}g_{\ell m}^{-1}g_{mn}^{-1})\rhd h_{\ell m n}(g_{jk}^{-1}g_{k\ell}^{-1})\rhd h_{jk\ell} (g_{j\ell}^{-1}g_{\ell n}^{-1})\rhd h_{j\ell n}$ to the surface $\Sigma_4: (g_{jk}^{-1}g_{k\ell}^{-1}g_{\ell m}^{-1}g_{mn}^{-1})\rhd h_{\ell mn} (g_{jk}^{-1}g_{k\ell}^{-1}g_{\ell n}^{-1})\rhd h_{k\ell n}(g_{jk}^{-1}g_{kn}^{-1})\rhd h_{jkn}$ in shown on the diagram (\ref{diag81}). Moving of the surface from $(g_{jk}^{-1}g_{k\ell}^{-1})\rhd h_{jk\ell} (g_{j\ell}^{-1}g_{\ell n}^{-1})\rhd h_{j\ell n}$ to the surface $ (g_{jk}^{-1}g_{k\ell}^{-1}g_{\ell m}^{-1}g_{mn}^{-1})\rhd h_{\ell mn} (g_{jk}^{-1}g_{k\ell}^{-1}g_{\ell n}^{-1})\rhd h_{k\ell n}$ is done with the $3$-arrow $\tilde{l}_{jk\ell n}$. This $3$-morphism can now be whiskered from above with the surface $(g_{jk}^{-1}g_{k\ell}^{-1}g_{\ell m}^{-1}g_{mn}^{-1})\rhd h_{\ell m n}$ and the obtained $3$-morphism is $(g_{jk}^{-1}g_{k\ell}^{-1}g_{\ell m}^{-1}g_{mn}^{-1}, (g_{jk}^{-1}g_{k\ell}^{-1}g_{\ell m}^{-1}g_{mn}^{-1})\rhd h_{\ell m n}(g_{jk}^{-1}g_{k\ell}^{-1})\rhd h_{jk\ell} (g_{j\ell}^{-1}g_{\ell n}^{-1})\rhd h_{j\ell n},((g_{jk}^{-1}g_{k\ell}^{-1}g_{\ell m}^{-1}g_{mn}^{-1})\rhd h_{\ell m n})\rhd' \tilde{l}_{jk\ell n} )$, where
$$\tilde{l}_{jk\ell n}= \{(g_{jk}^{-1}g_{k\ell}^{-1}g_{\ell n}^{-1})\rhd h_{k\ell n},(g_{jk}^{-1}g_{kn}^{-1})\rhd h_{jkn}\}_{\mathrm{p}}(g_{jk}^{-1}g_{k\ell}^{-1}g_{\ell n}^{-1})\rhd l_{jk\ell n} \{(g_{jk}^{-1}g_{k\ell}^{-1})\rhd h_{jk\ell}, (g_{j\ell}^{-1}g_{\ell n}^{-1})\rhd h_{j\ell n}\}^{-1}_{\mathrm{p}}\,.$$
\begin{equation}\label{diag81}
\resizebox{.99\hsize}{!}{$
    \begin{aligned}
 \xymatrix{ \ar@/^2ex/[r]^{g_{m n}^{-1}}="g6" \ar@/_6ex/[rr]_{g_{\ell n}^{-1}}="g2" \ar@/_9ex/[rrr]_{g_{kn}^{-1}}="g5" \ar@/_13ex/[rrrr]_{g_{jn}^{-1}}="g7" n \bullet 
 & \bullet m  \ar@/^2ex/[r]^{g_{\ell m }^{-1}}="g4"
 & \ar@/^2ex/[r]^{g_{k\ell }^{-1}}="g3" \bullet \ell & \bullet k \ar@/^2ex/[r]^{g_{jk}^{-1}}="g1"
\ar@{<=}^{(g_{k \ell}^{-1}g_{\ell n}^{-1})\rhd h_{k\ell n}} "g4"+<4ex,-5.6ex>;"g5"+<8ex,6ex>
& \bullet j
\ar@{<=}_{(g_{\ell m}^{-1}g_{m n}^{-1})\rhd  h_{\ell m n}} "g6"+<6ex,-4.6ex>;"g2"+<5ex,6ex>
\ar@{<=}^{(g_{jk}^{-1}g_{kn}^{-1})\rhd  h_{jkn}} "g3"+<0ex,-10ex>;"g7"+<8ex,6ex>}\;\stackrel{((g_{jk}^{-1}g_{k\ell}^{-1}g_{\ell m}^{-1}g_{mn}^{-1})\rhd h_{\ell m n})\rhd' \tilde{l}_{jk\ell n}}{\Lleftarrow} \; \xymatrix{\ar@/^2ex/[r]^{g_{m n}^{-1}}="g6" \ar@/_6ex/[rr]_{g_{\ell n}^{-1}}="g2" \ar@/_13ex/[rrrr]_{g_{jn}^{-1}}="g7" n \bullet 
 & \bullet m  \ar@/^2ex/[r]^{g_{\ell m }^{-1}}="g4"
 & \bullet \ell \ar@/^2ex/[r]^{g_{k\ell}^{-1}}="g3" \ar@/_6ex/[rr]_{g_{j\ell}^{-1}}="g5"
& \bullet k
\ar@/^2ex/[r]^{g_{jk}^{-1}}="g1"
& \bullet j
\ar@{<=}^{(g_{jk}^{-1}g_{k\ell}^{-1})\rhd h_{jk\ell }} "g3"+<3ex,-4.6ex>;"g5"+<-5ex,6ex>
\ar@{<=}_{(g_{\ell m}^{-1}g_{m n}^{-1})\rhd  h_{\ell m n}} "g6"+<6ex,-4.6ex>;"g2"+<5ex,6ex>
    \ar@{<=}^{(g_{j \ell}^{-1}g_{\ell n}^{-1})\rhd  h_{j\ell n}} "g4"+<-0.5ex,-10ex>;"g7"+<-4.5ex,4ex>}\,.
\end{aligned}$}
\end{equation}
Next, we map the surface $\Sigma_6:(g_{jk}^{-1}g_{k\ell}^{-1})\rhd h_{jk\ell}(g_{j\ell}^{-1}g_{\ell m}^{-1}g_{mn}^{-1})\rhd h_{\ell m n} (g_{j\ell}^{-1}g_{\ell n}^{-1})\rhd h_{j\ell n}$ to the surface $\Sigma_5:(g_{jk}^{-1}g_{k\ell}^{-1}g_{\ell m}^{-1}g_{mn}^{-1})\rhd h_{\ell m n}(g_{jk}^{-1}g_{k\ell}^{-1})\rhd h_{jk\ell} (g_{j\ell}^{-1}g_{\ell n}^{-1})\rhd h_{j\ell n}$, see the diagram (\ref{diag91}). We use the interchanging $2$-arrow composition to map the surface $(g_{jk}^{-1}g_{k\ell}^{-1})\rhd h_{jk\ell}(g_{j\ell}^{-1}g_{\ell m}^{-1}g_{mn}^{-1})\rhd h_{\ell m n}$ to the surface $(g_{jk}^{-1}g_{k\ell}^{-1}g_{\ell m}^{-1}g_{mn}^{-1})\rhd h_{\ell m n}(g_{jk}^{-1}g_{k\ell}^{-1})\rhd h_{jk\ell}$, resulting in the $3$-arrow $\{(g_{jk}^{-1}g_{k\ell}^{-1})\rhd h_{jk\ell},(g_{j\ell}^{-1}g_{\ell m}^{-1}g_{mn}^{-1})\rhd h_{\ell mn}\}_{\mathrm{p}}^{-1}$. Next, we whisker the obtained $3$-morphism with the $2$-morphism $(g_{j\ell}^{-1}g_{\ell n}^{-1})\rhd h_{j\ell n}$ from below. The obtained $3$-morphism with the appropriate source and target surfaces is $(g_{jn}^{-1},(g_{jk}^{-1}g_{k\ell}^{-1})\rhd h_{jk\ell}(g_{j\ell}^{-1}g_{\ell m}^{-1}g_{mn}^{-1})\rhd h_{\ell m n} (g_{j\ell}^{-1}g_{\ell n}^{-1})\rhd h_{j\ell n}, \{(g_{jk}^{-1}g_{k\ell}^{-1})\rhd h_{jk\ell},(g_{j\ell}^{-1}g_{\ell m}^{-1}g_{mn}^{-1})\rhd h_{\ell mn}\}_{\mathrm{p}}^{-1})$,
\begin{equation}\resizebox{.99\hsize}{!}{$
    \begin{aligned}\label{diag91}
\xymatrix{\ar@/^2ex/[r]^{g_{m n}^{-1}}="g6" \ar@/_6ex/[rr]_{g_{\ell n}^{-1}}="g2" \ar@/_13ex/[rrrr]_{g_{jn}^{-1}}="g7" n \bullet 
 & \bullet m  \ar@/^2ex/[r]^{g_{\ell m }^{-1}}="g4"
 & \bullet \ell \ar@/^2ex/[r]^{g_{k\ell}^{-1}}="g3" \ar@/_6ex/[rr]_{g_{j\ell}^{-1}}="g5"
& \bullet k
\ar@/^2ex/[r]^{g_{jk}^{-1}}="g1"
& \bullet j
\ar@{<=}^{(g_{jk}^{-1}g_{k\ell}^{-1})\rhd h_{jk\ell }} "g3"+<3ex,-4.6ex>;"g5"+<-5ex,6ex>
\ar@{<=}_{(g_{\ell m}^{-1}g_{m n}^{-1})\rhd  h_{\ell m n}} "g6"+<6ex,-4.6ex>;"g2"+<5ex,6ex>
    \ar@{<=}^{(g_{j \ell}^{-1}g_{\ell n}^{-1})\rhd  h_{j\ell n}} "g4"+<-0.5ex,-10ex>;"g7"+<-4.5ex,4ex>}\;\stackrel{\{(g_{jk}^{-1}g_{k\ell}^{-1})\rhd h_{jk\ell},(g_{j\ell}^{-1}g_{\ell m}^{-1}g_{mn}^{-1})\rhd h_{\ell mn}\}_{\mathrm{p}}^{-1}}{\Lleftarrow} \; \xymatrix{\ar@/^2ex/[r]^{g_{m n}^{-1}}="g6" \ar@/_6ex/[rr]_{g_{\ell n}^{-1}}="g2" \ar@/_13ex/[rrrr]_{g_{jn}^{-1}}="g7" n \bullet 
 & \bullet m  \ar@/^2ex/[r]^{g_{\ell m }^{-1}}="g4"
 & \bullet \ell \ar@/^2ex/[r]^{g_{k\ell}^{-1}}="g3" \ar@/_6ex/[rr]_{g_{j\ell}^{-1}}="g5"
& \bullet k
\ar@/^2ex/[r]^{g_{jk}^{-1}}="g1"
& \bullet j
\ar@{<=}^{(g_{jk}^{-1}g_{k\ell}^{-1})\rhd h_{jk\ell }} "g3"+<3ex,-4.6ex>;"g5"+<-5ex,6ex>
\ar@{<=}_{(g_{\ell m}^{-1}g_{m n}^{-1})\rhd  h_{\ell m n}} "g6"+<6ex,-4.6ex>;"g2"+<5ex,6ex>
    \ar@{<=}^{(g_{j \ell}^{-1}g_{\ell n}^{-1})\rhd  h_{j\ell n}} "g4"+<-0.5ex,-10ex>;"g7"+<-4.5ex,4ex>}\,.
    \end{aligned}$}
\end{equation}
Finally, we construct the $3$-morphism that maps the starting surface $\Sigma_1: (g_{jk}^{-1}g_{k\ell}^{-1})\rhd h_{jk\ell} (g_{j\ell}^{-1}g_{\ell m}^{-1})\rhd h_{j\ell m} (g_{jm}^{-1}g_{mn}^{-1})\rhd h_{jmn}$ to the surface $\Sigma_6:(g_{jk}^{-1}g_{k\ell}^{-1})\rhd h_{jk\ell}(g_{j\ell}^{-1}g_{\ell m}^{-1}g_{mn}^{-1})\rhd h_{\ell m n} (g_{j\ell}^{-1}g_{\ell n}^{-1})\rhd h_{j\ell n}$. To obtain the $3$-morphism with the appropriate source and target surfaces we first move the surface $(g_{j\ell}^{-1}g_{\ell m}^{-1})\rhd h_{j\ell m} (g_{jm}^{-1}g_{mn}^{-1})\rhd h_{jmn}$ to the surface $(g_{j\ell}^{-1}g_{\ell m}^{-1}g_{mn}^{-1})\rhd h_{\ell m n} (g_{j\ell}^{-1}g_{\ell n}^{-1})\rhd h_{j\ell n}$ with the $3$-arrow $\tilde{l}_{j\ell m n}$. Next, we whisker the obtained $3$-morphism with the $2$-morphism $(g_{jk}^{-1}g_{k\ell}^{-1})\rhd h_{jk\ell}$ from above. The obtained $3$-morphism is $(g_{jn}^{-1},(g_{jk}^{-1}g_{k\ell}^{-1})\rhd h_{jk\ell} (g_{j\ell}^{-1}g_{\ell m}^{-1})\rhd h_{j\ell m} (g_{jm}^{-1}g_{mn}^{-1})\rhd h_{jmn},((g_{jk}^{-1}g_{k\ell}^{-1})\rhd h_{jk\ell})\rhd'\tilde{l}_{j\ell m n})$, as shown on the diagram (\ref{diag101}), where 
$$\tilde{l}_{j\ell m n}=\{ (g_{j\ell}^{-1}g_{\ell m}^{-1}g_{mn}^{-1})\rhd h_{\ell mn}, (g_{j\ell}^{-1}g_{\ell m}^{-1})\rhd h_{j\ell n}\}_{\mathrm{p}}(g_{j\ell}^{-1}g_{\ell m}^{-1}g_{mn}^{-1})\rhd l_{j\ell mn}\{(g_{j\ell}^{-1}g_{\ell m}^{-1})\rhd h_{j\ell m},(g_{jm}^{-1}g_{mn}^{-1 })\rhd h_{jmn}\}_{\mathrm{p}}^{-1}\,.$$
\begin{equation}
\resizebox{.99\hsize}{!}{$
\begin{aligned}\label{diag101}
\xymatrix{\ar@/^2ex/[r]^{g_{m n}^{-1}}="g6" \ar@/_6ex/[rr]_{g_{\ell n}^{-1}}="g2" \ar@/_13ex/[rrrr]_{g_{jn}^{-1}}="g7" n \bullet 
 & \bullet m  \ar@/^2ex/[r]^{g_{\ell m }^{-1}}="g4"
 & \bullet \ell \ar@/^2ex/[r]^{g_{k\ell}^{-1}}="g3" \ar@/_6ex/[rr]_{g_{j\ell}^{-1}}="g5"
& \bullet k
\ar@/^2ex/[r]^{g_{jk}^{-1}}="g1"
& \bullet j
\ar@{<=}^{(g_{jk}^{-1}g_{k\ell}^{-1})\rhd h_{jk\ell }} "g3"+<3ex,-4.6ex>;"g5"+<-5ex,6ex>
\ar@{<=}_{(g_{\ell m}^{-1}g_{m n}^{-1})\rhd  h_{\ell m n}} "g6"+<6ex,-4.6ex>;"g2"+<5ex,6ex>
    \ar@{<=}^{(g_{j \ell}^{-1}g_{\ell n}^{-1})\rhd  h_{j\ell n}} "g4"+<-0.5ex,-10ex>;"g7"+<-4.5ex,4ex>}\; \stackrel{((g_{jk}^{-1}g_{k\ell}^{-1})\rhd h_{jk\ell})\rhd'\tilde{l}_{j\ell m n}}{\Lleftarrow} \;\xymatrix{\ar@/^2ex/[r]^{g_{m n}^{-1}}="g6" \ar@/_13ex/[rrrr]_{g_{jn}^{-1}}="g7" n \bullet & \bullet m \ar@/^2ex/[r]^{g_{\ell m }^{-1}}="g1" \ar@/_9ex/[rrr]_{g_{jm }^{-1}}="g5"
& \bullet \ell \ar@/^2ex/[r]^{g_{k\ell }^{-1}}="g2" \ar@/_6ex/[rr]_{g_{j\ell }^{-1}}="g4"
& \bullet k
\ar@/^2ex/[r]^{g_{jk}^{-1}}="g3"
& \bullet j
\ar@{<=}^{(g_{jk}^{-1}g_{k\ell}^{-1})\rhd h_{jk\ell }} "g2"+<3ex,-4.6ex>;"g4"+<-5ex,6ex>
\ar@{<=}^{(g_{j\ell}^{-1}g_{\ell m}^{-1})\rhd h_{j\ell m }} "g1"+<5ex,-8ex>;"g5"+<-6ex,5.3ex>
\ar@{<=}_{(g_{j m}^{-1}g_{m n}^{-1})\rhd h_{jm n}} "g6"+<6ex,-10ex>;"g7"+<-10ex,7ex>
} \,.
\end{aligned}$}
\end{equation}
The obtained $3$-morphism is the identity morphism with source and target surface $\Sigma_1: (g_{jk}^{-1}g_{k\ell}^{-1})\rhd h_{jk\ell} (g_{j\ell}^{-1}g_{\ell m}^{-1})\rhd h_{j\ell m} (g_{jm}^{-1}g_{mn}^{-1})\rhd h_{jmn}$, and gives the $\delta$-function: 
\begin{equation}
    \begin{array}{lcl}\label{herm1}
    &&\delta_L(\tilde{l}_{jk\ell m}^{-1}((g_{jk}^{-1}g_{k\ell}^{-1}g_{\ell m}^{-1})\rhd h_{k\ell m})\rhd'\tilde{l}_{jkm n}^{-1}g_{jk}^{-1}\rhd\tilde{l}_{k\ell mn}^{-1} ((g_{jk}^{-1}g_{k\ell}^{-1}g_{\ell m}^{-1}g_{mn}^{-1})\rhd h_{\ell m n})\rhd' \tilde{l}_{jk\ell n}\vphantom{\ds\int}\\&& \{(g_{jk}^{-1}g_{k\ell}^{-1})\rhd h_{jk\ell},(g_{j\ell}^{-1}g_{\ell m}^{-1}g_{mn}^{-1})\rhd h_{\ell mn}\}_{\mathrm{p}}^{-1} ((g_{jk}^{-1}g_{k\ell}^{-1})\rhd h_{jk\ell})\rhd'\tilde{l}_{j\ell m n})\vphantom{\ds\int}\,.
    \end{array}
\end{equation}
One can explicitly check that this is equivalent to the $\delta$-function
 obtained for the $4$-simplex before the inversion of all morphisms, $2$-morphisms, and $3$-morphisms,
 \begin{equation}\label{herm2}
     \delta_L(l_{j\ell m n}^{-1}\,h_{j\ell n}\rhd'\{h_{\ell m n}, (g_{m n}g_{\ell m})\rhd h_{jk\ell }\}_{\mathrm{p}}\,l_{jk\ell n}^{-1}(h_{jkn}\rhd'l_{k\ell m n}) l_{jkm n} h_{jm n}\rhd'(g_{m n}\rhd l_{jk\ell m }))\,.
\end{equation}
The proof that the $\delta$-function given by the equation \eqref{herm1} is equal to the $\delta$-function given by the equation \eqref{herm2} is given in the Appendix \ref{app}.
\end{lemma}

\section{Proof of Lemma \ref{Th:03}}\label{app}

Let us prove that the $\delta(\tilde{l}_{jk\ell m n})$,
\begin{equation}
\resizebox{0.9\hsize}{!}{$
   \begin{array}{lcl}
\medmuskip=0mu
\thinmuskip=0mu
\thickmuskip=0mu
\delta_L\big(\tilde{l}_{jk\ell m n}\big)&=& \delta_L\big(\tilde{l}_{jk\ell m}^{-1}((g_{jk}^{-1}g_{k\ell}^{-1}g_{\ell m}^{-1})\rhd h_{k\ell m})\rhd'\tilde{l}_{jkm n}^{-1}g_{jk}^{-1}\rhd\tilde{l}_{k\ell mn}^{-1} ((g_{jk}^{-1}g_{k\ell}^{-1}g_{\ell m}^{-1}g_{mn}^{-1})\rhd h_{\ell m n})\rhd' \tilde{l}_{jk\ell n}\vphantom{\ds\int}\\&& \{(g_{jk}^{-1}g_{k\ell}^{-1})\rhd h_{jk\ell},(g_{j\ell}^{-1}g_{\ell m}^{-1}g_{mn}^{-1})\rhd h_{\ell mn}\}_{\mathrm{p}}^{-1} ((g_{jk}^{-1}g_{k\ell}^{-1})\rhd h_{jk\ell})\rhd'\tilde{l}_{j\ell m n}\big)\,,\vphantom{\ds\int}
    \end{array}$}
\end{equation}
obtained for a $4$-simplex after reversing the orientation of its edges, triangles, and tetrahedrons, is equal to the $\delta$-function $\delta({l}_{jk\ell m n})$ before the orientation reversal,
\begin{equation}
\resizebox{0.9\hsize}{!}{$
    \begin{array}{lcl}
\medmuskip=0mu
\thinmuskip=0mu
\thickmuskip=0mu
\delta_L\big({l}_{jk\ell m n}\big)&=&\delta_L\big(l_{jk\ell n}^{-1}\,(h_{jkn}\rhd' l_{k\ell mn})\,l_{jkmn}\,(h_{jmn}\rhd' (g_{mn}\rhd l_{jk\ell m}))\,h_{j\ell m n}^{-1} h_{j\ell n}\rhd' \{h_{\ell m n}, (g_{m n}g_{\ell m})\rhd h_{jk \ell}\}_{\mathrm{p}}\big)\,.
    \end{array}$}
\end{equation}
Substituting the definition of $\tilde{l}$ given in \eqref{eq:l_jkln} one obtains,
\begin{equation}
\resizebox{0.98\hsize}{!}{$
    \begin{array}{lcl}
    \delta_L\big(\tilde{l}_{jk\ell m n}\big)&=& \delta_L\big(
    \{(g_{jk}^{-1}g_{k\ell}^{-1})\rhd h_{jk\ell},(g_{j\ell}^{-1}g_{\ell m}^{-1})\rhd h_{j\ell m}\}_\mathrm{p}(g_{jk}^{-1}g_{k\ell}^{-1}g_{\ell m}^{-1})\rhd l_{jk\ell m}^{-1}\{(g_{jk}^{-1}g_{k\ell}^{-1}g_{\ell m}^{-1})\rhd h_{k\ell m},(g_{jk}^{-1}g_{km}^{-1})\rhd h_{jkm}\}_{\mathrm{p}}^{-1}\vphantom{\ds\int}\\&&
    ((g_{jk}^{-1}g_{k\ell}^{-1}g_{\ell m}^{-1})\rhd h_{k\ell m})\rhd'\big(  \{(g_{jk}^{-1}g_{km}^{-1})\rhd h_{jkm},(g_{jm}^{-1}g_{m n}^{-1})\rhd h_{jm n}\}_\mathrm{p}(g_{jk}^{-1}g_{km}^{-1}g_{mn}^{-1})\rhd l_{jkmn}^{-1}\vphantom{\ds\int}\\&& \{(g_{jk}^{-1}g_{km}^{-1}g_{mn}^{-1})\rhd h_{kmn},(g_{jk}^{-1}g_{kn}^{-1})\rhd h_{jkn}\}_{\mathrm{p}}^{-1} \big)\vphantom{\ds\int} \\ &&g^{-1}_{jk}\rhd \big(\{(g_{k\ell}^{-1}g_{\ell m}^{-1})\rhd h_{k\ell m},(g_{km}^{-1}g_{m n}^{-1})\rhd h_{km n}\}_\mathrm{p}(g_{k\ell}^{-1}g_{\ell m}^{-1}g_{mn}^{-1})\rhd l_{k\ell mn}^{-1}\{(g_{k\ell}^{-1}g_{\ell m}^{-1}g_{mn}^{-1})\rhd h_{\ell mn},(g_{k\ell}^{-1}g_{\ell n}^{-1})\rhd h_{k\ell n}\}_{\mathrm{p}}^{-1} \big)\vphantom{\ds\int}\\&&((g_{jk}^{-1}g_{k\ell}^{-1}g_{\ell m}^{-1}g_{mn}^{-1})\rhd h_{\ell m n})\rhd' \big(\{(g_{jk}^{-1}g_{k\ell}^{-1}g_{\ell n}^{-1})\rhd h_{k\ell n},(g_{jk}^{-1}g_{kn}^{-1})\rhd h_{jkn}\}_{\mathrm{p}}(g_{jk}^{-1}g_{k\ell}^{-1}g_{\ell n}^{-1})\rhd l_{jk\ell n}\vphantom{\ds\int} \\&& \{(g_{jk}^{-1}g_{k\ell}^{-1})\rhd h_{jk\ell}, (g_{j\ell}^{-1}g_{\ell n}^{-1})\rhd h_{j\ell n}\}^{-1}_{\mathrm{p}} \big)\{(g_{jk}^{-1}g_{k\ell}^{-1})\rhd h_{jk\ell},(g_{j\ell}^{-1}g_{\ell m}^{-1}g_{mn}^{-1})\rhd h_{\ell mn}\}_{\mathrm{p}}^{-1}\vphantom{\ds\int}\\ &&((g_{jk}^{-1}g_{k\ell}^{-1})\rhd h_{jk\ell})\rhd'\big(\{ (g_{j\ell}^{-1}g_{\ell m}^{-1}g_{mn}^{-1})\rhd h_{\ell mn}, (g_{j\ell}^{-1}g_{\ell m}^{-1})\rhd h_{j\ell n}\}_{\mathrm{p}}(g_{j\ell}^{-1}g_{\ell m}^{-1}g_{mn}^{-1})\rhd l_{j\ell mn}\vphantom{\ds\int}\\&& \{(g_{j\ell}^{-1}g_{\ell m}^{-1})\rhd h_{j\ell m},(g_{jm}^{-1}g_{mn}^{-1 })\rhd h_{jmn}\}_{\mathrm{p}}^{-1} \big)\big)\,.\vphantom{\ds\int}
    \end{array}$}
\end{equation}
Using the identity \eqref{eq:id02},
\begin{equation}
\resizebox{0.98\hsize}{!}{$
\begin{array}{lcl}
&& (g_{jk}^{-1}g_{k\ell}^{-1}g_{\ell m}^{-1}g_{mn}^{-1})\rhd h_{\ell m n}\rhd' \{(g_{jk}^{-1}g_{k\ell}^{-1})\rhd h_{jk\ell}, (g_{j\ell}^{-1}g_{\ell n}^{-1})\rhd h_{j\ell n}\}^{-1}_{\mathrm{p}}\{(g_{jk}^{-1}g_{k\ell}^{-1})\rhd h_{jk\ell},(g_{j\ell}^{-1}g_{\ell m}^{-1}g_{mn}^{-1})\rhd h_{\ell mn}\}_{\mathrm{p}}^{-1}\vphantom{\ds\int}\\&=& {\{(g_{jk}^{-1}g_{k\ell}^{-1})\rhd h_{jk\ell},(g_{j\ell}^{-1}g_{\ell m}^{-1}g_{mn}^{-1})\rhd h_{\ell mn} (g_{j\ell}^{-1}g_{\ell n}^{-1})\rhd h_{j\ell n}\}_{\mathrm{p}}^{-1}}\,,\vphantom{\ds\int}
\end{array}$}
\end{equation}
one obtains:
\begin{equation}
\resizebox{0.98\hsize}{!}{$
\begin{array}{lcl}
    \delta_L\big(\tilde{l}_{jk\ell m n}\big)&=& \delta_L\big(
    \{(g_{jk}^{-1}g_{k\ell}^{-1})\rhd h_{jk\ell},(g_{j\ell}^{-1}g_{\ell m}^{-1})\rhd h_{j\ell m}\}_\mathrm{p}(g_{jk}^{-1}g_{k\ell}^{-1}g_{\ell m}^{-1})\rhd l_{jk\ell m}^{-1}\{(g_{jk}^{-1}g_{k\ell}^{-1}g_{\ell m}^{-1})\rhd h_{k\ell m},(g_{jk}^{-1}g_{km}^{-1})\rhd h_{jkm}\}_{\mathrm{p}}^{-1}\vphantom{\ds\int}\\&&
    ((g_{jk}^{-1}g_{k\ell}^{-1}g_{\ell m}^{-1})\rhd h_{k\ell m})\rhd'  \{(g_{jk}^{-1}g_{km}^{-1})\rhd h_{jkm},(g_{jm}^{-1}g_{m n}^{-1})\rhd h_{jm n}\}_\mathrm{p}((g_{jk}^{-1}g_{k\ell}^{-1}g_{\ell m}^{-1})\rhd h_{k\ell m})\rhd'(g_{jk}^{-1}g_{km}^{-1}g_{mn}^{-1})\rhd l_{jkmn}^{-1}\vphantom{\ds\int}\\&&((g_{jk}^{-1}g_{k\ell}^{-1}g_{\ell m}^{-1})\rhd h_{k\ell m})\rhd' \{(g_{jk}^{-1}g_{km}^{-1}g_{mn}^{-1})\rhd h_{kmn},(g_{jk}^{-1}g_{kn}^{-1})\rhd h_{jkn}\}_{\mathrm{p}}^{-1}\vphantom{\ds\int}  \\ && \{(g^{-1}_{jk}g_{k\ell}^{-1}g_{\ell m}^{-1})\rhd h_{k\ell m},(g^{-1}_{jk}g_{km}^{-1}g_{m n}^{-1})\rhd h_{km n}\}_\mathrm{p}(g^{-1}_{jk}g_{k\ell}^{-1}g_{\ell m}^{-1}g_{mn}^{-1})\rhd l_{k\ell mn}^{-1}\{(g^{-1}_{jk}g_{k\ell}^{-1}g_{\ell m}^{-1}g_{mn}^{-1})\rhd h_{\ell mn},(g^{-1}_{jk}g_{k\ell}^{-1}g_{\ell n}^{-1})\rhd h_{k\ell n}\}_{\mathrm{p}}^{-1}\vphantom{\ds\int} \\&&((g_{jk}^{-1}g_{k\ell}^{-1}g_{\ell m}^{-1}g_{mn}^{-1})\rhd h_{\ell m n})\rhd' \{(g_{jk}^{-1}g_{k\ell}^{-1}g_{\ell n}^{-1})\rhd h_{k\ell n},(g_{jk}^{-1}g_{kn}^{-1})\rhd h_{jkn}\}_{\mathrm{p}}((g_{jk}^{-1}g_{k\ell}^{-1}g_{\ell m}^{-1}g_{mn}^{-1})\rhd h_{\ell m n})\rhd'(g_{jk}^{-1}g_{k\ell}^{-1}g_{\ell n}^{-1})\rhd l_{jk\ell n} \vphantom{\ds\int}\\ &&\{(g_{jk}^{-1}g_{k\ell}^{-1})\rhd h_{jk\ell},(g_{j\ell}^{-1}g_{\ell m}^{-1}g_{mn}^{-1})\rhd h_{\ell mn} (g_{j\ell}^{-1}g_{\ell n}^{-1})\rhd h_{j\ell n}\}_{\mathrm{p}}^{-1}\vphantom{\ds\int}\\ &&((g_{jk}^{-1}g_{k\ell}^{-1})\rhd h_{jk\ell})\rhd'\{ (g_{j\ell}^{-1}g_{\ell m}^{-1}g_{mn}^{-1})\rhd h_{\ell mn}, (g_{j\ell}^{-1}g_{\ell m}^{-1})\rhd h_{j\ell n}\}_{\mathrm{p}}((g_{jk}^{-1}g_{k\ell}^{-1})\rhd h_{jk\ell})\rhd'(g_{j\ell}^{-1}g_{\ell m}^{-1}g_{mn}^{-1})\rhd l_{j\ell mn}\vphantom{\ds\int}\\ &&((g_{jk}^{-1}g_{k\ell}^{-1})\rhd h_{jk\ell})\rhd'\{(g_{j\ell}^{-1}g_{\ell m}^{-1})\rhd h_{j\ell m},(g_{jm}^{-1}g_{mn}^{-1 })\rhd h_{jmn}\}_{\mathrm{p}}^{-1}\big)\,.\vphantom{\ds\int}
    \end{array}$}\vphantom{\ds\int}
\end{equation}
Then, using the identity \eqref{identitet} one writes,
\begin{equation}
\resizebox{0.98\hsize}{!}{$
\begin{array}{lcl}
&&((g_{jk}^{-1}g_{k\ell}^{-1})\rhd h_{jk\ell})\rhd'\{ (g_{j\ell}^{-1}g_{\ell m}^{-1}g_{mn}^{-1})\rhd h_{\ell mn}, (g_{j\ell}^{-1}g_{\ell m}^{-1})\rhd h_{j\ell n}\}_{\mathrm{p}}\vphantom{\ds\int}\\&=&(g_{jk}^{-1}g_{k\ell}^{-1}g_{j\ell})\rhd\{ (g_{j\ell}^{-1}g_{\ell m}^{-1}g_{mn}^{-1})\rhd h_{\ell mn}, (g_{j\ell}^{-1}g_{\ell m}^{-1})\rhd h_{j\ell n}\}_{\mathrm{p}} \{(g_{jk}^{-1}g_{k\ell}^{-1})\rhd h_{jk\ell}),\delta\{ (g_{j\ell}^{-1}g_{\ell m}^{-1}g_{mn}^{-1})\rhd h_{\ell mn}, (g_{j\ell}^{-1}g_{\ell m}^{-1})\rhd h_{j\ell n}\}_{\mathrm{p}}^{-1}\}_{\mathrm{p}}^{-1}\,,\vphantom{\ds\int}
\end{array}$}
\end{equation}
and by commuting the elements one obtains:
\begin{equation}\label{delta1}
\resizebox{0.98\hsize}{!}{$
    \begin{array}{lcl}
   \delta_L\big(\tilde{l}_{jk\ell m n}\big)&=&  \delta_L\big(\{(g_{jk}^{-1}g_{k\ell}^{-1})\rhd h_{jk\ell},(g_{j\ell}^{-1}g_{\ell m}^{-1})\rhd h_{j\ell m}\}_\mathrm{p}(g_{jk}^{-1}g_{k\ell}^{-1}g_{\ell m}^{-1})\rhd l_{jk\ell m}^{-1}\{(g_{jk}^{-1}g_{k\ell}^{-1}g_{\ell m}^{-1})\rhd h_{k\ell m},(g_{jk}^{-1}g_{km}^{-1})\rhd h_{jkm}\}_{\mathrm{p}}^{-1}\vphantom{\ds\int}\\&&
    ((g_{jk}^{-1}g_{k\ell}^{-1}g_{\ell m}^{-1})\rhd h_{k\ell m})\rhd'  \{(g_{jk}^{-1}g_{km}^{-1})\rhd h_{jkm},(g_{jm}^{-1}g_{m n}^{-1})\rhd h_{jm n}\}_\mathrm{p}((g_{jk}^{-1}g_{k\ell}^{-1}g_{\ell m}^{-1})\rhd h_{k\ell m})\rhd'(g_{jk}^{-1}g_{km}^{-1}g_{mn}^{-1})\rhd l_{jkmn}^{-1}\vphantom{\ds\int}\\&&((g_{jk}^{-1}g_{k\ell}^{-1}g_{\ell m}^{-1})\rhd h_{k\ell m})\rhd' \{(g_{jk}^{-1}g_{km}^{-1}g_{mn}^{-1})\rhd h_{kmn},(g_{jk}^{-1}g_{kn}^{-1})\rhd h_{jkn}\}_{\mathrm{p}}^{-1}\vphantom{\ds\int}  \\ && \{(g^{-1}_{jk}g_{k\ell}^{-1}g_{\ell m}^{-1})\rhd h_{k\ell m},(g^{-1}_{jk}g_{km}^{-1}g_{m n}^{-1})\rhd h_{km n}\}_\mathrm{p}(g^{-1}_{jk}g_{k\ell}^{-1}g_{\ell m}^{-1}g_{mn}^{-1})\rhd l_{k\ell mn}^{-1}\{(g^{-1}_{jk}g_{k\ell}^{-1}g_{\ell m}^{-1}g_{mn}^{-1})\rhd h_{\ell mn},(g^{-1}_{jk}g_{k\ell}^{-1}g_{\ell n}^{-1})\rhd h_{k\ell n}\}_{\mathrm{p}}^{-1} \vphantom{\ds\int}\\&&((g_{jk}^{-1}g_{k\ell}^{-1}g_{\ell m}^{-1}g_{mn}^{-1})\rhd h_{\ell m n})\rhd' \{(g_{jk}^{-1}g_{k\ell}^{-1}g_{\ell n}^{-1})\rhd h_{k\ell n},(g_{jk}^{-1}g_{kn}^{-1})\rhd h_{jkn}\}_{\mathrm{p}}((g_{jk}^{-1}g_{k\ell}^{-1}g_{\ell m}^{-1}g_{mn}^{-1})\rhd h_{\ell m n})\rhd'(g_{jk}^{-1}g_{k\ell}^{-1}g_{\ell n}^{-1})\rhd l_{jk\ell n}\vphantom{\ds\int}\\ &&(g_{jk}^{-1}g_{k\ell}^{-1}g_{j\ell})\rhd\{ (g_{j\ell}^{-1}g_{\ell m}^{-1}g_{mn}^{-1})\rhd h_{\ell mn}, (g_{j\ell}^{-1}g_{\ell m}^{-1})\rhd h_{j\ell n}\}_{\mathrm{p}}\vphantom{\ds\int} \\&&(g_{jk}^{-1}g_{k\ell}^{-1}g_{j\ell})\rhd \delta\left(\{ (g_{j\ell}^{-1}g_{\ell m}^{-1}g_{mn}^{-1})\rhd h_{\ell mn}, (g_{j\ell}^{-1}g_{\ell m}^{-1})\rhd h_{j\ell n}\}_{\mathrm{p}}^{-1}\right)\rhd \{(g_{jk}^{-1}g_{k\ell}^{-1})\rhd h_{jk\ell},(g_{j\ell}^{-1}g_{\ell m}^{-1}g_{mn}^{-1})\rhd h_{\ell mn} (g_{j\ell}^{-1}g_{\ell n}^{-1})\rhd h_{j\ell n}\}_{\mathrm{p}}^{-1}\vphantom{\ds\int}\\&& \{(g_{jk}^{-1}g_{k\ell}^{-1})\rhd h_{jk\ell}),\delta\{ (g_{j\ell}^{-1}g_{\ell m}^{-1}g_{mn}^{-1})\rhd h_{\ell mn}, (g_{j\ell}^{-1}g_{\ell m}^{-1})\rhd h_{j\ell n}\}_{\mathrm{p}}^{-1}\}_{\mathrm{p}}^{-1}\vphantom{\ds\int}\\ &&((g_{jk}^{-1}g_{k\ell}^{-1})\rhd h_{jk\ell})\rhd'(g_{j\ell}^{-1}g_{\ell m}^{-1}g_{mn}^{-1})\rhd l_{j\ell mn}((g_{jk}^{-1}g_{k\ell}^{-1})\rhd h_{jk\ell})\rhd'\{(g_{j\ell}^{-1}g_{\ell m}^{-1})\rhd h_{j\ell m},(g_{jm}^{-1}g_{mn}^{-1 })\rhd h_{jmn}\}_{\mathrm{p}}^{-1}\big)\,.\vphantom{\ds\int}
    \end{array}$}
\end{equation}
Now, using the identity \eqref{eq:id02}, and then the identity \eqref{prop1}, one can write
\begin{equation}
\resizebox{0.98\hsize}{!}{$
\begin{array}{lcl}
&&(g_{jk}^{-1}g_{k\ell}^{-1}g_{j\ell})\rhd \delta ( \{ (g_{j\ell}^{-1}g_{\ell m}^{-1}g_{mn}^{-1})\rhd h_{\ell mn}, (g_{j\ell}^{-1}g_{\ell m}^{-1})\rhd h_{j\ell n}\}_{\mathrm{p}}^{-1})\rhd \{(g_{jk}^{-1}g_{k\ell}^{-1})\rhd h_{jk\ell},(g_{j\ell}^{-1}g_{\ell m}^{-1}g_{mn}^{-1})\rhd h_{\ell mn} (g_{j\ell}^{-1}g_{\ell n}^{-1})\rhd h_{j\ell n}\}_{\mathrm{p}}^{-1}\vphantom{\ds\int}\\&& \{(g_{jk}^{-1}g_{k\ell}^{-1})\rhd h_{jk\ell}),\delta\{ (g_{j\ell}^{-1}g_{\ell m}^{-1}g_{mn}^{-1})\rhd h_{\ell mn}, (g_{j\ell}^{-1}g_{\ell m}^{-1})\rhd h_{j\ell n}\}_{\mathrm{p}}^{-1}\}_{\mathrm{p}}^{-1}\vphantom{\ds\int}\\&=&\{(g_{jk}^{-1}g_{k\ell}^{-1})\rhd h_{jk\ell},\delta\{ (g_{j\ell}^{-1}g_{\ell m}^{-1}g_{mn}^{-1})\rhd h_{\ell mn}, (g_{j\ell}^{-1}g_{\ell m}^{-1})\rhd h_{j\ell n}\}_{\mathrm{p}}^{-1}(g_{j\ell}^{-1}g_{\ell m}^{-1}g_{mn}^{-1})\rhd h_{\ell mn} (g_{j\ell}^{-1}g_{\ell n}^{-1})\rhd h_{j\ell n}\}_{\mathrm{p}}^{-1}\vphantom{\ds\int}\\&=&\{(g_{jk}^{-1}g_{k\ell}^{-1})\rhd h_{jk\ell},(g_{j\ell}^{-1}g_{\ell m}^{-1}g_{ mn}^{-1})\rhd h_{j\ell n}(g_{j\ell}^{-1}g_{\ell m}^{-1}g_{mn}^{-1})\rhd h_{\ell mn}\}_{\mathrm{p}}^{-1}\,.\vphantom{\ds\int}
\end{array}$}
\end{equation}
Substituting this expression in \eqref{delta1} one obtains,
\begin{equation}\label{delta2}
\resizebox{0.98\hsize}{!}{$
    \begin{array}{lcl}
    \delta_L\big(\tilde{l}_{jk\ell m n}\big)&=& \delta_L\big(\{(g_{jk}^{-1}g_{k\ell}^{-1})\rhd h_{jk\ell},(g_{j\ell}^{-1}g_{\ell m}^{-1})\rhd h_{j\ell m}\}_\mathrm{p}(g_{jk}^{-1}g_{k\ell}^{-1}g_{\ell m}^{-1})\rhd l_{jk\ell m}^{-1}\{(g_{jk}^{-1}g_{k\ell}^{-1}g_{\ell m}^{-1})\rhd h_{k\ell m},(g_{jk}^{-1}g_{km}^{-1})\rhd h_{jkm}\}_{\mathrm{p}}^{-1}\vphantom{\ds\int}\\&&
    ((g_{jk}^{-1}g_{k\ell}^{-1}g_{\ell m}^{-1})\rhd h_{k\ell m})\rhd'  \{(g_{jk}^{-1}g_{km}^{-1})\rhd h_{jkm},(g_{jm}^{-1}g_{m n}^{-1})\rhd h_{jm n}\}_\mathrm{p}((g_{jk}^{-1}g_{k\ell}^{-1}g_{\ell m}^{-1})\rhd h_{k\ell m})\rhd'(g_{jk}^{-1}g_{km}^{-1}g_{mn}^{-1})\rhd l_{jkmn}^{-1}\vphantom{\ds\int}\\&&((g_{jk}^{-1}g_{k\ell}^{-1}g_{\ell m}^{-1})\rhd h_{k\ell m})\rhd' \{(g_{jk}^{-1}g_{km}^{-1}g_{mn}^{-1})\rhd h_{kmn},(g_{jk}^{-1}g_{kn}^{-1})\rhd h_{jkn}\}_{\mathrm{p}}^{-1} \vphantom{\ds\int} \\ && \{(g^{-1}_{jk}g_{k\ell}^{-1}g_{\ell m}^{-1})\rhd h_{k\ell m},(g^{-1}_{jk}g_{km}^{-1}g_{m n}^{-1})\rhd h_{km n}\}_\mathrm{p}(g^{-1}_{jk}g_{k\ell}^{-1}g_{\ell m}^{-1}g_{mn}^{-1})\rhd l_{k\ell mn}^{-1}\{(g^{-1}_{jk}g_{k\ell}^{-1}g_{\ell m}^{-1}g_{mn}^{-1})\rhd h_{\ell mn},(g^{-1}_{jk}g_{k\ell}^{-1}g_{\ell n}^{-1})\rhd h_{k\ell n}\}_{\mathrm{p}}^{-1} \vphantom{\ds\int}\\&&((g_{jk}^{-1}g_{k\ell}^{-1}g_{\ell m}^{-1}g_{mn}^{-1})\rhd h_{\ell m n})\rhd' \{(g_{jk}^{-1}g_{k\ell}^{-1}g_{\ell n}^{-1})\rhd h_{k\ell n},(g_{jk}^{-1}g_{kn}^{-1})\rhd h_{jkn}\}_{\mathrm{p}}((g_{jk}^{-1}g_{k\ell}^{-1}g_{\ell m}^{-1}g_{mn}^{-1})\rhd h_{\ell m n})\rhd'(g_{jk}^{-1}g_{k\ell}^{-1}g_{\ell n}^{-1})\rhd l_{jk\ell n}\vphantom{\ds\int}\\ &&(g_{jk}^{-1}g_{k\ell}^{-1}g_{j\ell})\rhd\{ (g_{j\ell}^{-1}g_{\ell m}^{-1}g_{mn}^{-1})\rhd h_{\ell mn}, (g_{j\ell}^{-1}g_{\ell m}^{-1})\rhd h_{j\ell n}\}_{\mathrm{p}}\vphantom{\ds\int} \\ &&\{(g_{jk}^{-1}g_{k\ell}^{-1})\rhd h_{jk\ell},(g_{j\ell}^{-1}g_{\ell m}^{-1}g_{ mn}^{-1})\rhd h_{j\ell n}(g_{j\ell}^{-1}g_{\ell m}^{-1}g_{mn}^{-1})\rhd h_{\ell mn}\}_{\mathrm{p}}^{-1}\vphantom{\ds\int}\\&& ((g_{jk}^{-1}g_{k\ell}^{-1})\rhd h_{jk\ell})\rhd'(g_{j\ell}^{-1}g_{\ell m}^{-1}g_{mn}^{-1})\rhd l_{j\ell mn}((g_{jk}^{-1}g_{k\ell}^{-1})\rhd h_{jk\ell})\rhd'\{(g_{j\ell}^{-1}g_{\ell m}^{-1})\rhd h_{j\ell m},(g_{jm}^{-1}g_{mn}^{-1 })\rhd h_{jmn}\}_{\mathrm{p}}^{-1}\big)\,.\vphantom{\ds\int}
    \end{array}$}
\end{equation}
Now, using the identity \eqref{identitet}, one writes:
\begin{equation}
\resizebox{0.98\hsize}{!}{$
 ((g_{jk}^{-1}g_{k\ell}^{-1})\rhd h_{jk\ell})\rhd'(g_{j\ell}^{-1}g_{\ell m}^{-1}g_{mn}^{-1})\rhd l_{j\ell mn}=(g_{jk}^{-1}g_{k\ell}^{-1}g_{\ell m}^{-1}g_{mn}^{-1})\rhd l_{j\ell mn} \{(g_{jk}^{-1}g_{k\ell}^{-1})\rhd h_{jk\ell}),(g_{j\ell}^{-1}g_{\ell m}^{-1}g_{mn}^{-1})\rhd \delta(l_{j\ell mn})^{-1}\}_{\mathrm{p}}^{-1}\,.$}
\end{equation}
Substituting this expression in \eqref{delta2} one obtains,
\begin{equation}\label{eq:delta3}
\resizebox{0.98\hsize}{!}{$
\begin{array}{lcl}
   \delta_L\big(\tilde{l}_{jk\ell m n}\big)&=& \delta_L\big( \{(g_{jk}^{-1}g_{k\ell}^{-1})\rhd h_{jk\ell},(g_{j\ell}^{-1}g_{\ell m}^{-1})\rhd h_{j\ell m}\}_\mathrm{p}(g_{jk}^{-1}g_{k\ell}^{-1}g_{\ell m}^{-1})\rhd l_{jk\ell m}^{-1}\{(g_{jk}^{-1}g_{k\ell}^{-1}g_{\ell m}^{-1})\rhd h_{k\ell m},(g_{jk}^{-1}g_{km}^{-1})\rhd h_{jkm}\}_{\mathrm{p}}^{-1}\vphantom{\ds\int}\\&&
    ((g_{jk}^{-1}g_{k\ell}^{-1}g_{\ell m}^{-1})\rhd h_{k\ell m})\rhd'  \{(g_{jk}^{-1}g_{km}^{-1})\rhd h_{jkm},(g_{jm}^{-1}g_{m n}^{-1})\rhd h_{jm n}\}_\mathrm{p}((g_{jk}^{-1}g_{k\ell}^{-1}g_{\ell m}^{-1})\rhd h_{k\ell m})\rhd'(g_{jk}^{-1}g_{km}^{-1}g_{mn}^{-1})\rhd l_{jkmn}^{-1}\vphantom{\ds\int}\\&&((g_{jk}^{-1}g_{k\ell}^{-1}g_{\ell m}^{-1})\rhd h_{k\ell m})\rhd' \{(g_{jk}^{-1}g_{km}^{-1}g_{mn}^{-1})\rhd h_{kmn},(g_{jk}^{-1}g_{kn}^{-1})\rhd h_{jkn}\}_{\mathrm{p}}^{-1}\vphantom{\ds\int}  \\ && \{(g^{-1}_{jk}g_{k\ell}^{-1}g_{\ell m}^{-1})\rhd h_{k\ell m},(g^{-1}_{jk}g_{km}^{-1}g_{m n}^{-1})\rhd h_{km n}\}_\mathrm{p}(g^{-1}_{jk}g_{k\ell}^{-1}g_{\ell m}^{-1}g_{mn}^{-1})\rhd l_{k\ell mn}^{-1}\{(g^{-1}_{jk}g_{k\ell}^{-1}g_{\ell m}^{-1}g_{mn}^{-1})\rhd h_{\ell mn},(g^{-1}_{jk}g_{k\ell}^{-1}g_{\ell n}^{-1})\rhd h_{k\ell n}\}_{\mathrm{p}}^{-1}\vphantom{\ds\int} \\&&((g_{jk}^{-1}g_{k\ell}^{-1}g_{\ell m}^{-1}g_{mn}^{-1})\rhd h_{\ell m n})\rhd' \{(g_{jk}^{-1}g_{k\ell}^{-1}g_{\ell n}^{-1})\rhd h_{k\ell n},(g_{jk}^{-1}g_{kn}^{-1})\rhd h_{jkn}\}_{\mathrm{p}}((g_{jk}^{-1}g_{k\ell}^{-1}g_{\ell m}^{-1}g_{mn}^{-1})\rhd h_{\ell m n})\rhd'(g_{jk}^{-1}g_{k\ell}^{-1}g_{\ell n}^{-1})\rhd l_{jk\ell n}\vphantom{\ds\int}\\ &&(g_{jk}^{-1}g_{k\ell}^{-1}g_{j\ell})\rhd\{ (g_{j\ell}^{-1}g_{\ell m}^{-1}g_{mn}^{-1})\rhd h_{\ell mn}, (g_{j\ell}^{-1}g_{\ell m}^{-1})\rhd h_{j\ell n}\}_{\mathrm{p}}\vphantom{\ds\int}\\ &&\{(g_{jk}^{-1}g_{k\ell}^{-1})\rhd h_{jk\ell},(g_{j\ell}^{-1}g_{\ell m}^{-1}g_{ mn}^{-1})\rhd h_{j\ell n}(g_{j\ell}^{-1}g_{\ell m}^{-1}g_{mn}^{-1})\rhd h_{\ell mn}\}_{\mathrm{p}}^{-1}\vphantom{\ds\int}\\
\end{array}$}
\end{equation}
Commuting the elements and then using the identities \eqref{eq:id02} and \eqref{prop1} one writes:
\begin{equation}
\resizebox{0.98\hsize}{!}{$
\begin{array}{lcl}
&&\{(g_{jk}^{-1}g_{k\ell}^{-1})\rhd h_{jk\ell},(g_{j\ell}^{-1}g_{\ell m}^{-1}g_{ mn}^{-1})\rhd h_{j\ell n}(g_{j\ell}^{-1}g_{\ell m}^{-1}g_{mn}^{-1})\rhd h_{\ell mn}\}_{\mathrm{p}}^{-1}{(g_{jk}^{-1}g_{k\ell}^{-1}g_{\ell m}^{-1}g_{mn}^{-1})\rhd l_{j\ell mn} \{(g_{jk}^{-1}g_{k\ell}^{-1})\rhd h_{jk\ell}),(g_{j\ell}^{-1}g_{\ell m}^{-1}g_{mn}^{-1})\rhd \delta(l_{j\ell mn})^{-1}\}_{\mathrm{p}}^{-1}}\vphantom{\ds\int}\\&=&
{(g_{jk}^{-1}g_{k\ell}^{-1}g_{\ell m}^{-1}g_{mn}^{-1})\rhd l_{j\ell mn} (g_{jk}^{-1}g_{k\ell}^{-1}g_{\ell m}^{-1}g_{mn}^{-1})\rhd \delta (l_{j\ell mn})^{-1} \rhd'} {\{(g_{jk}^{-1}g_{k\ell}^{-1})\rhd h_{jk\ell},(g_{j\ell}^{-1}g_{\ell m}^{-1}g_{ mn}^{-1})\rhd h_{j\ell n}(g_{j\ell}^{-1}g_{\ell m}^{-1}g_{mn}^{-1})\rhd h_{\ell mn}\}_{\mathrm{p}}^{-1}}\vphantom{\ds\int}\\ &&{\{(g_{jk}^{-1}g_{k\ell}^{-1})\rhd h_{jk\ell}),(g_{j\ell}^{-1}g_{\ell m}^{-1}g_{mn}^{-1})\rhd \delta(l_{j\ell mn})^{-1}\}_{\mathrm{p}}^{-1}}\vphantom{\ds\int}\\&=& (g_{jk}^{-1}g_{k\ell}^{-1}g_{\ell m}^{-1}g_{mn}^{-1})\rhd l_{j\ell mn} \{(g_{jk}^{-1}g_{k\ell}^{-1})\rhd h_{jk\ell}),(g_{j\ell}^{-1}g_{\ell m}^{-1}g_{mn}^{-1})\rhd \delta(l_{j\ell mn})^{-1}(g_{j\ell}^{-1}g_{\ell m}^{-1}g_{ mn}^{-1})\rhd h_{j\ell n}(g_{j\ell}^{-1}g_{\ell m}^{-1}g_{mn}^{-1})\rhd h_{\ell mn}\}_{\mathrm{p}}^{-1}\vphantom{\ds\int}\\&=& (g_{jk}^{-1}g_{k\ell}^{-1}g_{\ell m}^{-1}g_{mn}^{-1})\rhd l_{j\ell mn} \{(g_{jk}^{-1}g_{k\ell}^{-1})\rhd h_{jk\ell},(g_{j\ell}^{-1}g_{\ell m}^{-1}g_{mn}^{-1})\rhd\left(h_{jmn}(g_{mn}\rhd h_{j\ell m})\right)\}_{\mathrm{p}}^{-1}\,.\vphantom{\ds\int}
\end{array}$}
\end{equation}
Then, substituting this expression in the equation \eqref{eq:delta3} one obtains:
\begin{equation}
\resizebox{0.98\hsize}{!}{$
    \begin{array}{lcl}
   \delta_L\big(\tilde{l}_{jk\ell m n}\big)&=& \delta_L\big(\{(g_{jk}^{-1}g_{k\ell}^{-1})\rhd h_{jk\ell},(g_{j\ell}^{-1}g_{\ell m}^{-1})\rhd h_{j\ell m}\}_\mathrm{p}(g_{jk}^{-1}g_{k\ell}^{-1}g_{\ell m}^{-1})\rhd l_{jk\ell m}^{-1}\{(g_{jk}^{-1}g_{k\ell}^{-1}g_{\ell m}^{-1})\rhd h_{k\ell m},(g_{jk}^{-1}g_{km}^{-1})\rhd h_{jkm}\}_{\mathrm{p}}^{-1}\vphantom{\ds\int}\\&&
    ((g_{jk}^{-1}g_{k\ell}^{-1}g_{\ell m}^{-1})\rhd h_{k\ell m})\rhd'  \{(g_{jk}^{-1}g_{km}^{-1})\rhd h_{jkm},(g_{jm}^{-1}g_{m n}^{-1})\rhd h_{jm n}\}_\mathrm{p}((g_{jk}^{-1}g_{k\ell}^{-1}g_{\ell m}^{-1})\rhd h_{k\ell m})\rhd'(g_{jk}^{-1}g_{km}^{-1}g_{mn}^{-1})\rhd l_{jkmn}^{-1}\vphantom{\ds\int}\\&&((g_{jk}^{-1}g_{k\ell}^{-1}g_{\ell m}^{-1})\rhd h_{k\ell m})\rhd' \{(g_{jk}^{-1}g_{km}^{-1}g_{mn}^{-1})\rhd h_{kmn},(g_{jk}^{-1}g_{kn}^{-1})\rhd h_{jkn}\}_{\mathrm{p}}^{-1} \vphantom{\ds\int} \\ && \{(g^{-1}_{jk}g_{k\ell}^{-1}g_{\ell m}^{-1})\rhd h_{k\ell m},(g^{-1}_{jk}g_{km}^{-1}g_{m n}^{-1})\rhd h_{km n}\}_\mathrm{p}(g^{-1}_{jk}g_{k\ell}^{-1}g_{\ell m}^{-1}g_{mn}^{-1})\rhd l_{k\ell mn}^{-1}\{(g^{-1}_{jk}g_{k\ell}^{-1}g_{\ell m}^{-1}g_{mn}^{-1})\rhd h_{\ell mn},(g^{-1}_{jk}g_{k\ell}^{-1}g_{\ell n}^{-1})\rhd h_{k\ell n}\}_{\mathrm{p}}^{-1}\vphantom{\ds\int} \\&&((g_{jk}^{-1}g_{k\ell}^{-1}g_{\ell m}^{-1}g_{mn}^{-1})\rhd h_{\ell m n})\rhd' \{(g_{jk}^{-1}g_{k\ell}^{-1}g_{\ell n}^{-1})\rhd h_{k\ell n},(g_{jk}^{-1}g_{kn}^{-1})\rhd h_{jkn}\}_{\mathrm{p}}((g_{jk}^{-1}g_{k\ell}^{-1}g_{\ell m}^{-1}g_{mn}^{-1})\rhd h_{\ell m n})\rhd'(g_{jk}^{-1}g_{k\ell}^{-1}g_{\ell n}^{-1})\rhd l_{jk\ell n}\vphantom{\ds\int}\\ &&{(g_{jk}^{-1}g_{k\ell}^{-1}g_{j\ell})\rhd\{ (g_{j\ell}^{-1}g_{\ell m}^{-1}g_{mn}^{-1})\rhd h_{\ell mn}, (g_{j\ell}^{-1}g_{\ell m}^{-1})\rhd h_{j\ell n}\}_{\mathrm{p}}} \vphantom{\ds\int}\\
\end{array}$}
\end{equation}
Using the identity \eqref{identitet} again has
\begin{equation}
\resizebox{0.98\hsize}{!}{$
\begin{array}{lcl}
&&((g_{jk}^{-1}g_{k\ell}^{-1})\rhd h_{jk\ell})\rhd'\{(g_{j\ell}^{-1}g_{\ell m}^{-1})\rhd h_{j\ell m},(g_{jm}^{-1}g_{mn}^{-1 })\rhd h_{jmn}\}_{\mathrm{p}}^{-1}\vphantom{\ds\int}\\&=&\{(g_{jk}^{-1}g_{k\ell}^{-1})\rhd h_{jk\ell},\delta(\{(g_{j\ell}^{-1}g_{\ell m}^{-1})\rhd h_{j\ell m},(g_{jm}^{-1}g_{mn}^{-1 })\rhd h_{jmn}\}_{\mathrm{p}}^{-1})\}_{\mathrm{p}}(g_{jk}^{-1}g_{k\ell}^{-1}g_{j\ell})\rhd\{(g_{j\ell}^{-1}g_{\ell m}^{-1})\rhd h_{j\ell m},(g_{jm}^{-1}g_{mn}^{-1 })\rhd h_{jmn}\}_{\mathrm{p}}^{-1}\,,\vphantom{\ds\int}
\end{array}$}
\end{equation}
so that the $\delta$-function $\delta_L\big(\tilde{l}_{jk\ell m n}\big)$ becomes:
\begin{equation}
\resizebox{0.98\hsize}{!}{$
    \begin{array}{lcl}
    \delta_L\big(\tilde{l}_{jk\ell m n}\big)&=& \delta_L\big(\{(g_{jk}^{-1}g_{k\ell}^{-1})\rhd h_{jk\ell},(g_{j\ell}^{-1}g_{\ell m}^{-1})\rhd h_{j\ell m}\}_\mathrm{p}(g_{jk}^{-1}g_{k\ell}^{-1}g_{\ell m}^{-1})\rhd l_{jk\ell m}^{-1}\{(g_{jk}^{-1}g_{k\ell}^{-1}g_{\ell m}^{-1})\rhd h_{k\ell m},(g_{jk}^{-1}g_{km}^{-1})\rhd h_{jkm}\}_{\mathrm{p}}^{-1}\vphantom{\ds\int}\\&&
    ((g_{jk}^{-1}g_{k\ell}^{-1}g_{\ell m}^{-1})\rhd h_{k\ell m})\rhd'  \{(g_{jk}^{-1}g_{km}^{-1})\rhd h_{jkm},(g_{jm}^{-1}g_{m n}^{-1})\rhd h_{jm n}\}_\mathrm{p}((g_{jk}^{-1}g_{k\ell}^{-1}g_{\ell m}^{-1})\rhd h_{k\ell m})\rhd'(g_{jk}^{-1}g_{km}^{-1}g_{mn}^{-1})\rhd l_{jkmn}^{-1}\vphantom{\ds\int}\\&&((g_{jk}^{-1}g_{k\ell}^{-1}g_{\ell m}^{-1})\rhd h_{k\ell m})\rhd' \{(g_{jk}^{-1}g_{km}^{-1}g_{mn}^{-1})\rhd h_{kmn},(g_{jk}^{-1}g_{kn}^{-1})\rhd h_{jkn}\}_{\mathrm{p}}^{-1} \vphantom{\ds\int} \\ && \{(g^{-1}_{jk}g_{k\ell}^{-1}g_{\ell m}^{-1})\rhd h_{k\ell m},(g^{-1}_{jk}g_{km}^{-1}g_{m n}^{-1})\rhd h_{km n}\}_\mathrm{p}(g^{-1}_{jk}g_{k\ell}^{-1}g_{\ell m}^{-1}g_{mn}^{-1})\rhd l_{k\ell mn}^{-1}\{(g^{-1}_{jk}g_{k\ell}^{-1}g_{\ell m}^{-1}g_{mn}^{-1})\rhd h_{\ell mn},(g^{-1}_{jk}g_{k\ell}^{-1}g_{\ell n}^{-1})\rhd h_{k\ell n}\}_{\mathrm{p}}^{-1} \vphantom{\ds\int}\\&&((g_{jk}^{-1}g_{k\ell}^{-1}g_{\ell m}^{-1}g_{mn}^{-1})\rhd h_{\ell m n})\rhd' \{(g_{jk}^{-1}g_{k\ell}^{-1}g_{\ell n}^{-1})\rhd h_{k\ell n},(g_{jk}^{-1}g_{kn}^{-1})\rhd h_{jkn}\}_{\mathrm{p}}((g_{jk}^{-1}g_{k\ell}^{-1}g_{\ell m}^{-1}g_{mn}^{-1})\rhd h_{\ell m n})\rhd'(g_{jk}^{-1}g_{k\ell}^{-1}g_{\ell n}^{-1})\rhd l_{jk\ell n}\vphantom{\ds\int}\\ &&{(g_{jk}^{-1}g_{k\ell}^{-1}g_{j\ell})\rhd\{ (g_{j\ell}^{-1}g_{\ell m}^{-1}g_{mn}^{-1})\rhd h_{\ell mn}, (g_{j\ell}^{-1}g_{\ell m}^{-1})\rhd h_{j\ell n}\}_{\mathrm{p}}} \vphantom{\ds\int}\\
\end{array}$}
\end{equation}
Now, again using the identities \eqref{eq:id06} and \eqref{eq:id02} one has:
\begin{equation}
\resizebox{0.98\hsize}{!}{$
\begin{array}{lcl}
&&{\{(g_{jk}^{-1}g_{k\ell}^{-1})\rhd h_{jk\ell},(g_{j\ell}^{-1}g_{\ell m}^{-1}g_{mn}^{-1})\rhd\left(h_{jmn}(g_{mn}\rhd h_{j\ell m})\right)\}_{\mathrm{p}}^{-1}} { \{(g_{jk}^{-1}g_{k\ell}^{-1})\rhd h_{jk\ell},\delta(\{(g_{j\ell}^{-1}g_{\ell m}^{-1})\rhd h_{j\ell m},(g_{jm}^{-1}g_{mn}^{-1 })\rhd h_{jmn}\}_{\mathrm{p}}^{-1})\}_{\mathrm{p}}}\vphantom{\ds\int}\\&=&\big((g_{jk}^{-1}g_{k\ell}^{-1})\rhd \partial h_{jk\ell}\rhd \delta(\{(g_{j\ell}^{-1}g_{\ell m}^{-1})\rhd h_{j\ell m},(g_{jm}^{-1}g_{mn}^{-1 })\rhd h_{jmn}\}_{\mathrm{p}}^{-1})\big)\rhd' \{(g_{jk}^{-1}g_{k\ell}^{-1})\rhd h_{jk\ell},(g_{j\ell}^{-1}g_{\ell m}^{-1})\rhd h_{j\ell m}(g_{jm}^{-1}g_{mn}^{-1 })\rhd h_{jmn}\}_{\mathrm{p}}^{-1}\,.\vphantom{\ds\int}
\end{array}$}
\end{equation}
Substituting this expression and commuting the elements, one obtains:
\begin{equation}
\resizebox{0.98\hsize}{!}{$
    \begin{array}{lcl}
    \delta_L\big(\tilde{l}_{jk\ell m n}\big)&=& \delta_L\big(\{(g_{jk}^{-1}g_{k\ell}^{-1})\rhd h_{jk\ell},(g_{j\ell}^{-1}g_{\ell m}^{-1})\rhd h_{j\ell m}\}_\mathrm{p}(g_{jk}^{-1}g_{k\ell}^{-1}g_{\ell m}^{-1})\rhd l_{jk\ell m}^{-1}\{(g_{jk}^{-1}g_{k\ell}^{-1}g_{\ell m}^{-1})\rhd h_{k\ell m},(g_{jk}^{-1}g_{km}^{-1})\rhd h_{jkm}\}_{\mathrm{p}}^{-1}\vphantom{\ds\int}\\&&
    ((g_{jk}^{-1}g_{k\ell}^{-1}g_{\ell m}^{-1})\rhd h_{k\ell m})\rhd'  \{(g_{jk}^{-1}g_{km}^{-1})\rhd h_{jkm},(g_{jm}^{-1}g_{m n}^{-1})\rhd h_{jm n}\}_\mathrm{p}((g_{jk}^{-1}g_{k\ell}^{-1}g_{\ell m}^{-1})\rhd h_{k\ell m})\rhd'(g_{jk}^{-1}g_{km}^{-1}g_{mn}^{-1})\rhd l_{jkmn}^{-1}\vphantom{\ds\int}\\&&((g_{jk}^{-1}g_{k\ell}^{-1}g_{\ell m}^{-1})\rhd h_{k\ell m})\rhd' \{(g_{jk}^{-1}g_{km}^{-1}g_{mn}^{-1})\rhd h_{kmn},(g_{jk}^{-1}g_{kn}^{-1})\rhd h_{jkn}\}_{\mathrm{p}}^{-1} \vphantom{\ds\int} \\ && \{(g^{-1}_{jk}g_{k\ell}^{-1}g_{\ell m}^{-1})\rhd h_{k\ell m},(g^{-1}_{jk}g_{km}^{-1}g_{m n}^{-1})\rhd h_{km n}\}_\mathrm{p}(g^{-1}_{jk}g_{k\ell}^{-1}g_{\ell m}^{-1}g_{mn}^{-1})\rhd l_{k\ell mn}^{-1}\{(g^{-1}_{jk}g_{k\ell}^{-1}g_{\ell m}^{-1}g_{mn}^{-1})\rhd h_{\ell mn},(g^{-1}_{jk}g_{k\ell}^{-1}g_{\ell n}^{-1})\rhd h_{k\ell n}\}_{\mathrm{p}}^{-1} \vphantom{\ds\int}\\&&((g_{jk}^{-1}g_{k\ell}^{-1}g_{\ell m}^{-1}g_{mn}^{-1})\rhd h_{\ell m n})\rhd' \{(g_{jk}^{-1}g_{k\ell}^{-1}g_{\ell n}^{-1})\rhd h_{k\ell n},(g_{jk}^{-1}g_{kn}^{-1})\rhd h_{jkn}\}_{\mathrm{p}}((g_{jk}^{-1}g_{k\ell}^{-1}g_{\ell m}^{-1}g_{mn}^{-1})\rhd h_{\ell m n})\rhd'(g_{jk}^{-1}g_{k\ell}^{-1}g_{\ell n}^{-1})\rhd l_{jk\ell n}\vphantom{\ds\int}\\&& {(g_{jk}^{-1}g_{k\ell}^{-1}g_{j\ell})\rhd\{ (g_{j\ell}^{-1}g_{\ell m}^{-1}g_{mn}^{-1})\rhd h_{\ell mn}, (g_{j\ell}^{-1}g_{\ell m}^{-1})\rhd h_{j\ell n}\}_{\mathrm{p}}} \vphantom{\ds\int}\\
\end{array}$}
\end{equation}
Then, using the identity \eqref{eq:id02} again, one has:
\begin{equation}
\begin{array}{lcl}
&& \{{(g_{jk}^{-1}g_{k\ell}^{-1})\rhd h_{jk\ell},(g_{j\ell}^{-1}g_{\ell m}^{-1})\rhd h_{j\ell m}(g_{jm}^{-1}g_{mn}^{-1 })\rhd h_{jmn}\}_{\mathrm{p}}^{-1}}\{(g_{jk}^{-1}g_{k\ell}^{-1})\rhd h_{jk\ell},(g_{j\ell}^{-1}g_{\ell m}^{-1})\rhd h_{j\ell m}\}_\mathrm{p}\vphantom{\ds\int}\\&=&(g_{jk}^{-1}g_{k\ell}^{-1}g_{\ell m}^{-1})\rhd h_{j\ell m} \rhd'\{ {(g_{jk}^{-1}g_{k\ell}^{-1})\rhd h_{jk\ell},(g_{jm}^{-1}g_{mn}^{-1 })\rhd h_{jmn}\}_{\mathrm{p}}^{-1}}\,.\vphantom{\ds\int}
\end{array}
\end{equation}
Substituting this expression and commuting the elements, one obtains:
\begin{equation}
\resizebox{0.98\hsize}{!}{$
    \begin{array}{lcl}
\delta_L\big(\tilde{l}_{jk\ell m n}\big)&=& \delta_L\big((g_{jk}^{-1}g_{k\ell}^{-1}g_{\ell m}^{-1})\rhd l_{jk\ell m}^{-1}\{(g_{jk}^{-1}g_{k\ell}^{-1}g_{\ell m}^{-1})\rhd h_{k\ell m},(g_{jk}^{-1}g_{km}^{-1})\rhd h_{jkm}\}_{\mathrm{p}}^{-1}\vphantom{\ds\int}\\&&
    ((g_{jk}^{-1}g_{k\ell}^{-1}g_{\ell m}^{-1})\rhd h_{k\ell m})\rhd'  \{(g_{jk}^{-1}g_{km}^{-1})\rhd h_{jkm},(g_{jm}^{-1}g_{m n}^{-1})\rhd h_{jm n}\}_\mathrm{p}((g_{jk}^{-1}g_{k\ell}^{-1}g_{\ell m}^{-1})\rhd h_{k\ell m})\rhd'(g_{jk}^{-1}g_{km}^{-1}g_{mn}^{-1})\rhd l_{jkmn}^{-1}\vphantom{\ds\int}\\&&((g_{jk}^{-1}g_{k\ell}^{-1}g_{\ell m}^{-1})\rhd h_{k\ell m})\rhd' \{(g_{jk}^{-1}g_{km}^{-1}g_{mn}^{-1})\rhd h_{kmn},(g_{jk}^{-1}g_{kn}^{-1})\rhd h_{jkn}\}_{\mathrm{p}}^{-1}\vphantom{\ds\int}  \\&&  \{(g^{-1}_{jk}g_{k\ell}^{-1}g_{\ell m}^{-1})\rhd h_{k\ell m},(g^{-1}_{jk}g_{km}^{-1}g_{m n}^{-1})\rhd h_{km n}\}_\mathrm{p}(g^{-1}_{jk}g_{k\ell}^{-1}g_{\ell m}^{-1}g_{mn}^{-1})\rhd l_{k\ell mn}^{-1}\{(g^{-1}_{jk}g_{k\ell}^{-1}g_{\ell m}^{-1}g_{mn}^{-1})\rhd h_{\ell mn},(g^{-1}_{jk}g_{k\ell}^{-1}g_{\ell n}^{-1})\rhd h_{k\ell n}\}_{\mathrm{p}}^{-1} \vphantom{\ds\int}\\&&((g_{jk}^{-1}g_{k\ell}^{-1}g_{\ell m}^{-1}g_{mn}^{-1})\rhd h_{\ell m n})\rhd' \{(g_{jk}^{-1}g_{k\ell}^{-1}g_{\ell n}^{-1})\rhd h_{k\ell n},(g_{jk}^{-1}g_{kn}^{-1})\rhd h_{jkn}\}_{\mathrm{p}}((g_{jk}^{-1}g_{k\ell}^{-1}g_{\ell m}^{-1}g_{mn}^{-1})\rhd h_{\ell m n})\rhd'(g_{jk}^{-1}g_{k\ell}^{-1}g_{\ell n}^{-1})\rhd l_{jk\ell n}\vphantom{\ds\int}\\ &&{(g_{jk}^{-1}g_{k\ell}^{-1}g_{j\ell})\rhd\{ (g_{j\ell}^{-1}g_{\ell m}^{-1}g_{mn}^{-1})\rhd h_{\ell mn}, (g_{j\ell}^{-1}g_{\ell m}^{-1})\rhd h_{j\ell n}\}_{\mathrm{p}}} \vphantom{\ds\int}\\ &&{(g_{jk}^{-1}g_{k\ell}^{-1}g_{\ell m}^{-1}g_{mn}^{-1})\rhd l_{j\ell mn}}{(g_{jk}^{-1}g_{k\ell}^{-1}g_{j\ell})\rhd\{(g_{j\ell}^{-1}g_{\ell m}^{-1})\rhd h_{j\ell m},(g_{jm}^{-1}g_{mn}^{-1 })\rhd h_{jmn}\}_{\mathrm{p}}^{-1}}\vphantom{\ds\int}\\
\end{array}$}
\end{equation}
Then, using the identity \eqref{eq:id01},
\begin{equation}
\resizebox{0.98\hsize}{!}{$
\begin{array}{lcl}
&&{(g_{jk}^{-1}g_{k\ell}^{-1}g_{j\ell})\rhd\{(g_{j\ell}^{-1}g_{\ell m}^{-1})\rhd h_{j\ell m},(g_{jm}^{-1}g_{mn}^{-1 })\rhd h_{jmn}\}_{\mathrm{p}}^{-1}}{(g_{jk}^{-1}g_{k\ell}^{-1}g_{\ell m}^{-1})\rhd h_{j\ell m} \rhd' \{(g_{jk}^{-1}g_{k\ell}^{-1})\rhd h_{jk\ell},(g_{jm}^{-1}g_{mn}^{-1 })\rhd h_{jmn}\}_{\mathrm{p}}^{-1}}\vphantom{\ds\int}\\&=&\{(g_{jk}^{-1}g_{k\ell}^{-1}g_{\ell m}^{-1})\rhd h_{j\ell m}(g_{jk}^{-1}g_{k\ell}^{-1})\rhd h_{jk\ell},(g_{jm}^{-1}g_{mn}^{-1 })\rhd h_{jmn}\}_{\mathrm{p}}^{-1}\,,\vphantom{\ds\int}
\end{array}$}
\end{equation}
and substituting this in the above expression one obtains:
\begin{equation}\label{eq:delta4}
\resizebox{0.98\hsize}{!}{$
    \begin{array}{lcl}
\delta_L\big(\tilde{l}_{jk\ell m n}\big)&=& \delta_L\big((g_{jk}^{-1}g_{k\ell}^{-1}g_{\ell m}^{-1})\rhd l_{jk\ell m}^{-1}\{(g_{jk}^{-1}g_{k\ell}^{-1}g_{\ell m}^{-1})\rhd h_{k\ell m},(g_{jk}^{-1}g_{km}^{-1})\rhd h_{jkm}\}_{\mathrm{p}}^{-1}\vphantom{\ds\int}\\&&
    ((g_{jk}^{-1}g_{k\ell}^{-1}g_{\ell m}^{-1})\rhd h_{k\ell m})\rhd'  \{(g_{jk}^{-1}g_{km}^{-1})\rhd h_{jkm},(g_{jm}^{-1}g_{m n}^{-1})\rhd h_{jm n}\}_\mathrm{p}((g_{jk}^{-1}g_{k\ell}^{-1}g_{\ell m}^{-1})\rhd h_{k\ell m})\rhd'(g_{jk}^{-1}g_{km}^{-1}g_{mn}^{-1})\rhd l_{jkmn}^{-1}\vphantom{\ds\int}\\&&((g_{jk}^{-1}g_{k\ell}^{-1}g_{\ell m}^{-1})\rhd h_{k\ell m})\rhd' \{(g_{jk}^{-1}g_{km}^{-1}g_{mn}^{-1})\rhd h_{kmn},(g_{jk}^{-1}g_{kn}^{-1})\rhd h_{jkn}\}_{\mathrm{p}}^{-1} \vphantom{\ds\int} \\&&  \{(g^{-1}_{jk}g_{k\ell}^{-1}g_{\ell m}^{-1})\rhd h_{k\ell m},(g^{-1}_{jk}g_{km}^{-1}g_{m n}^{-1})\rhd h_{km n}\}_\mathrm{p}(g^{-1}_{jk}g_{k\ell}^{-1}g_{\ell m}^{-1}g_{mn}^{-1})\rhd l_{k\ell mn}^{-1}\{(g^{-1}_{jk}g_{k\ell}^{-1}g_{\ell m}^{-1}g_{mn}^{-1})\rhd h_{\ell mn},(g^{-1}_{jk}g_{k\ell}^{-1}g_{\ell n}^{-1})\rhd h_{k\ell n}\}_{\mathrm{p}}^{-1} \vphantom{\ds\int}\\&&((g_{jk}^{-1}g_{k\ell}^{-1}g_{\ell m}^{-1}g_{mn}^{-1})\rhd h_{\ell m n})\rhd' \{(g_{jk}^{-1}g_{k\ell}^{-1}g_{\ell n}^{-1})\rhd h_{k\ell n},(g_{jk}^{-1}g_{kn}^{-1})\rhd h_{jkn}\}_{\mathrm{p}}{((g_{jk}^{-1}g_{k\ell}^{-1}g_{\ell m}^{-1}g_{mn}^{-1})\rhd h_{\ell m n})\rhd'(g_{jk}^{-1}g_{k\ell}^{-1}g_{\ell n}^{-1})\rhd l_{jk\ell n}}\vphantom{\ds\int}\\&& {(g_{jk}^{-1}g_{k\ell}^{-1}g_{j\ell})\rhd\{ (g_{j\ell}^{-1}g_{\ell m}^{-1}g_{mn}^{-1})\rhd h_{\ell mn}, (g_{j\ell}^{-1}g_{\ell m}^{-1})\rhd h_{j\ell n}\}_{\mathrm{p}}} \vphantom{\ds\int}\\ &&{(g_{jk}^{-1}g_{k\ell}^{-1}g_{\ell m}^{-1}g_{mn}^{-1})\rhd l_{j\ell mn}}{\{(g_{jk}^{-1}g_{k\ell}^{-1}g_{\ell m}^{-1})\rhd h_{j\ell m} (g_{jk}^{-1}g_{k\ell}^{-1})\rhd h_{jk\ell},(g_{jm}^{-1}g_{mn}^{-1 })\rhd h_{jmn}\}_{\mathrm{p}}^{-1}}\big)\,.\vphantom{\ds\int}
    \end{array}$}
\end{equation}
Next, one can write:
\begin{equation}
\resizebox{0.98\hsize}{!}{$
\begin{array}{lcl}
&&((g_{jk}^{-1}g_{k\ell}^{-1}g_{\ell m}^{-1}g_{mn}^{-1})\rhd h_{\ell m n})\rhd'(g_{jk}^{-1}g_{k\ell}^{-1}g_{\ell n}^{-1})\rhd l_{jk\ell n}\vphantom{\ds\int}\\&=&(g_{jk}^{-1}g_{k\ell}^{-1}g_{\ell m}^{-1}g_{mn}^{-1})\rhd  l_{jk\ell n} \{(g_{jk}^{-1}g_{k\ell}^{-1}g_{\ell m}^{-1}g_{mn}^{-1})\rhd h_{\ell m n},(g_{jk}^{-1}g_{k\ell}^{-1}g_{\ell n}^{-1})\rhd \delta( l_{jk\ell n})^{-1}\}_{\mathrm{p}}^{-1}\vphantom{\ds\int}\\&=&(g_{jk}^{-1}g_{k\ell}^{-1}g_{\ell m}^{-1}g_{mn}^{-1})\rhd  l_{jk\ell n} \{(g_{jk}^{-1}g_{k\ell}^{-1}g_{\ell m}^{-1}g_{mn}^{-1})\rhd h_{\ell m n},(g_{jk}^{-1}g_{k\ell}^{-1}g_{\ell n}^{-1})\rhd ( h_{j\ell n} g_{\ell n}\rhd h_{jk\ell}h_{k\ell n}^{-1}h_{jkn}^{-1})\}_{\mathrm{p}}^{-1}\vphantom{\ds\int}\\&=&(g_{jk}^{-1}g_{k\ell}^{-1}g_{\ell m}^{-1}g_{mn}^{-1})\rhd  l_{jk\ell n} \left((g_{jk}^{-1}g_{k\ell}^{-1}g_{\ell m}^{-1}g_{mn}^{-1})\rhd h_{j\ell n}\right)\rhd'\{(g_{jk}^{-1}g_{k\ell}^{-1}g_{\ell m}^{-1}g_{mn}^{-1})\rhd h_{\ell m n},(g_{jk}^{-1}g_{k\ell}^{-1}g_{\ell n}^{-1})\rhd ( g_{\ell n}\rhd h_{jk\ell}h_{k\ell n}^{-1}h_{jkn}^{-1})\}_{\mathrm{p}}^{-1}\vphantom{\ds\int}\\&&\{(g_{jk}^{-1}g_{k\ell}^{-1}g_{\ell m}^{-1}g_{mn}^{-1})\rhd h_{\ell m n},(g_{jk}^{-1}g_{k\ell}^{-1}g_{\ell n}^{-1})\rhd h_{j\ell n}\}_{\mathrm{p}}^{-1}\,.\vphantom{\ds\int}
\end{array}$}
\end{equation}
Using the previous expression in equation \eqref{eq:delta4} one can write:
\begin{equation}\label{eq:delta5}
\resizebox{0.98\hsize}{!}{$
    \begin{array}{lcl}
\delta_L\big(\tilde{l}_{jk\ell m n}\big)&=& \delta_L\big((g_{jk}^{-1}g_{k\ell}^{-1}g_{\ell m}^{-1})\rhd l_{jk\ell m}^{-1}\{(g_{jk}^{-1}g_{k\ell}^{-1}g_{\ell m}^{-1})\rhd h_{k\ell m},(g_{jk}^{-1}g_{km}^{-1})\rhd h_{jkm}\}_{\mathrm{p}}^{-1}\vphantom{\ds\int}\\&&
    ((g_{jk}^{-1}g_{k\ell}^{-1}g_{\ell m}^{-1})\rhd h_{k\ell m})\rhd'  \{(g_{jk}^{-1}g_{km}^{-1})\rhd h_{jkm},(g_{jm}^{-1}g_{m n}^{-1})\rhd h_{jm n}\}_\mathrm{p}((g_{jk}^{-1}g_{k\ell}^{-1}g_{\ell m}^{-1})\rhd h_{k\ell m})\rhd'(g_{jk}^{-1}g_{km}^{-1}g_{mn}^{-1})\rhd l_{jkmn}^{-1}\vphantom{\ds\int}\\&&((g_{jk}^{-1}g_{k\ell}^{-1}g_{\ell m}^{-1})\rhd h_{k\ell m})\rhd' \{(g_{jk}^{-1}g_{km}^{-1}g_{mn}^{-1})\rhd h_{kmn},(g_{jk}^{-1}g_{kn}^{-1})\rhd h_{jkn}\}_{\mathrm{p}}^{-1} \vphantom{\ds\int} \\ && \{(g^{-1}_{jk}g_{k\ell}^{-1}g_{\ell m}^{-1})\rhd h_{k\ell m},(g^{-1}_{jk}g_{km}^{-1}g_{m n}^{-1})\rhd h_{km n}\}_\mathrm{p}(g^{-1}_{jk}g_{k\ell}^{-1}g_{\ell m}^{-1}g_{mn}^{-1})\rhd l_{k\ell mn}^{-1}\{(g^{-1}_{jk}g_{k\ell}^{-1}g_{\ell m}^{-1}g_{mn}^{-1})\rhd h_{\ell mn},(g^{-1}_{jk}g_{k\ell}^{-1}g_{\ell n}^{-1})\rhd h_{k\ell n}\}_{\mathrm{p}}^{-1}\vphantom{\ds\int} \\&&((g_{jk}^{-1}g_{k\ell}^{-1}g_{\ell m}^{-1}g_{mn}^{-1})\rhd h_{\ell m n})\rhd' \{(g_{jk}^{-1}g_{k\ell}^{-1}g_{\ell n}^{-1})\rhd h_{k\ell n},(g_{jk}^{-1}g_{kn}^{-1})\rhd h_{jkn}\}_{\mathrm{p}}\vphantom{\ds\int}\\&&{(g_{jk}^{-1}g_{k\ell}^{-1}g_{\ell m}^{-1}g_{mn}^{-1})\rhd  l_{jk\ell n} \left((g_{jk}^{-1}g_{k\ell}^{-1}g_{\ell m}^{-1}g_{mn}^{-1})\rhd h_{j\ell n}\right)\rhd'\{(g_{jk}^{-1}g_{k\ell}^{-1}g_{\ell m}^{-1}g_{mn}^{-1})\rhd h_{\ell m n},(g_{jk}^{-1}g_{k\ell}^{-1}g_{\ell n}^{-1})\rhd ( g_{\ell n}\rhd h_{jk\ell}h_{k\ell n}^{-1}h_{jkn}^{-1})\}_{\mathrm{p}}^{-1}}\vphantom{\ds\int}\\
\end{array}$}
\end{equation}
Using the identity \eqref{identitet} one has:
\begin{equation}
\resizebox{0.98\hsize}{!}{$
    \begin{array}{lcl}
&&((g_{jk}^{-1}g_{k\ell}^{-1}g_{\ell m}^{-1}g_{mn}^{-1})\rhd h_{\ell m n})\rhd' \{(g_{jk}^{-1}g_{k\ell}^{-1}g_{\ell n}^{-1})\rhd h_{k\ell n},(g_{jk}^{-1}g_{kn}^{-1})\rhd h_{jkn}\}_{\mathrm{p}}\vphantom{\ds\int}\\&=& \{((g_{jk}^{-1}g_{k\ell}^{-1}g_{\ell m}^{-1}g_{mn}^{-1})\rhd h_{k\ell n},(g_{jk}^{-1}g_{k\ell}^{-1}g_{\ell m}^{-1}g_{mn}^{-1}g_{\ell n} g_{k\ell} g_{kn}^{-1})\rhd h_{jkn}\}_{\mathrm{p}} \{(g_{jk}^{-1}g_{k\ell}^{-1}g_{\ell m}^{-1}g_{mn}^{-1})\rhd h_{\ell mn}, \delta (\{(g_{jk}^{-1}g_{k\ell}^{-1}g_{\ell n}^{-1})\rhd h_{k\ell n},(g_{jk}^{-1}g_{kn}^{-1})\rhd h_{jkn}\}_{\mathrm{p}}^{-1})\}_{\mathrm{p}}^{-1}\,.\vphantom{\ds\int}
    \end{array}$}
    \end{equation}
Then, substituting this expression in \eqref{eq:delta5} one obtains:
\begin{equation}
\resizebox{0.98\hsize}{!}{$
    \begin{array}{lcl}
\delta_L\big(\tilde{l}_{jk\ell m n}\big)&=& \delta_L\big((g_{jk}^{-1}g_{k\ell}^{-1}g_{\ell m}^{-1})\rhd l_{jk\ell m}^{-1}\{(g_{jk}^{-1}g_{k\ell}^{-1}g_{\ell m}^{-1})\rhd h_{k\ell m},(g_{jk}^{-1}g_{km}^{-1})\rhd h_{jkm}\}_{\mathrm{p}}^{-1}\vphantom{\ds\int}\\&&
    ((g_{jk}^{-1}g_{k\ell}^{-1}g_{\ell m}^{-1})\rhd h_{k\ell m})\rhd'  \{(g_{jk}^{-1}g_{km}^{-1})\rhd h_{jkm},(g_{jm}^{-1}g_{m n}^{-1})\rhd h_{jm n}\}_\mathrm{p}((g_{jk}^{-1}g_{k\ell}^{-1}g_{\ell m}^{-1})\rhd h_{k\ell m})\rhd'(g_{jk}^{-1}g_{km}^{-1}g_{mn}^{-1})\rhd l_{jkmn}^{-1}\vphantom{\ds\int}\\&&((g_{jk}^{-1}g_{k\ell}^{-1}g_{\ell m}^{-1})\rhd h_{k\ell m})\rhd' \{(g_{jk}^{-1}g_{km}^{-1}g_{mn}^{-1})\rhd h_{kmn},(g_{jk}^{-1}g_{kn}^{-1})\rhd h_{jkn}\}_{\mathrm{p}}^{-1} \vphantom{\ds\int} \\&&  \{(g^{-1}_{jk}g_{k\ell}^{-1}g_{\ell m}^{-1})\rhd h_{k\ell m},(g^{-1}_{jk}g_{km}^{-1}g_{m n}^{-1})\rhd h_{km n}\}_\mathrm{p}(g^{-1}_{jk}g_{k\ell}^{-1}g_{\ell m}^{-1}g_{mn}^{-1})\rhd l_{k\ell mn}^{-1}\{(g^{-1}_{jk}g_{k\ell}^{-1}g_{\ell m}^{-1}g_{mn}^{-1})\rhd h_{\ell mn},(g^{-1}_{jk}g_{k\ell}^{-1}g_{\ell n}^{-1})\rhd h_{k\ell n}\}_{\mathrm{p}}^{-1}\vphantom{\ds\int} \\&&\{((g_{jk}^{-1}g_{k\ell}^{-1}g_{\ell m}^{-1}g_{mn}^{-1})\rhd h_{k\ell n},(g_{jk}^{-1}g_{k\ell}^{-1}g_{\ell m}^{-1}g_{mn}^{-1}g_{\ell n} g_{k\ell} g_{kn}^{-1})\rhd h_{jkn}\}_{\mathrm{p}} \{(g_{jk}^{-1}g_{k\ell}^{-1}g_{\ell m}^{-1}g_{mn}^{-1})\rhd h_{\ell mn}, \delta (\{(g_{jk}^{-1}g_{k\ell}^{-1}g_{\ell n}^{-1})\rhd h_{k\ell n},(g_{jk}^{-1}g_{kn}^{-1})\rhd h_{jkn}\}_{\mathrm{p}}^{-1})\}_{\mathrm{p}}^{-1}\vphantom{\ds\int}\\&&{(g_{jk}^{-1}g_{k\ell}^{-1}g_{\ell m}^{-1}g_{mn}^{-1})\rhd  l_{jk\ell n} \left((g_{jk}^{-1}g_{k\ell}^{-1}g_{\ell m}^{-1}g_{mn}^{-1})\rhd h_{j\ell n}\right)\rhd'\{(g_{jk}^{-1}g_{k\ell}^{-1}g_{\ell m}^{-1}g_{mn}^{-1})\rhd h_{\ell m n},(g_{jk}^{-1}g_{k\ell}^{-1}g_{\ell n}^{-1})\rhd ( g_{\ell n}\rhd h_{jk\ell}h_{k\ell n}^{-1}h_{jkn}^{-1})\}_{\mathrm{p}}^{-1}}\vphantom{\ds\int}\\
\end{array}$}
\end{equation}
After commuting the elements the $\delta$-function $\delta_L(\tilde{l}_{jk\ell m n})$ becomes:
\begin{equation}\label{eq:delta6}
\resizebox{0.98\hsize}{!}{$
    \begin{array}{lcl}
\delta_L\big(\tilde{l}_{jk\ell m n}\big)&=& \delta_L\big((g_{jk}^{-1}g_{k\ell}^{-1}g_{\ell m}^{-1})\rhd l_{jk\ell m}^{-1}\{(g_{jk}^{-1}g_{k\ell}^{-1}g_{\ell m}^{-1})\rhd h_{k\ell m},(g_{jk}^{-1}g_{km}^{-1})\rhd h_{jkm}\}_{\mathrm{p}}^{-1}\vphantom{\ds\int}\\&&
    ((g_{jk}^{-1}g_{k\ell}^{-1}g_{\ell m}^{-1})\rhd h_{k\ell m})\rhd'  \{(g_{jk}^{-1}g_{km}^{-1})\rhd h_{jkm},(g_{jm}^{-1}g_{m n}^{-1})\rhd h_{jm n}\}_\mathrm{p}((g_{jk}^{-1}g_{k\ell}^{-1}g_{\ell m}^{-1})\rhd h_{k\ell m})\rhd'(g_{jk}^{-1}g_{km}^{-1}g_{mn}^{-1})\rhd l_{jkmn}^{-1}\vphantom{\ds\int}\\&&((g_{jk}^{-1}g_{k\ell}^{-1}g_{\ell m}^{-1})\rhd h_{k\ell m})\rhd' \{(g_{jk}^{-1}g_{km}^{-1}g_{mn}^{-1})\rhd h_{kmn},(g_{jk}^{-1}g_{kn}^{-1})\rhd h_{jkn}\}_{\mathrm{p}}^{-1} \vphantom{\ds\int} \\&&  \{(g^{-1}_{jk}g_{k\ell}^{-1}g_{\ell m}^{-1})\rhd h_{k\ell m},(g^{-1}_{jk}g_{km}^{-1}g_{m n}^{-1})\rhd h_{km n}\}_\mathrm{p}(g^{-1}_{jk}g_{k\ell}^{-1}g_{\ell m}^{-1}g_{mn}^{-1})\rhd l_{k\ell mn}^{-1}\{(g^{-1}_{jk}g_{k\ell}^{-1}g_{\ell m}^{-1}g_{mn}^{-1})\rhd h_{\ell mn},(g^{-1}_{jk}g_{k\ell}^{-1}g_{\ell n}^{-1})\rhd h_{k\ell n}\}_{\mathrm{p}}^{-1}\vphantom{\ds\int} \\&&\{((g_{jk}^{-1}g_{k\ell}^{-1}g_{\ell m}^{-1}g_{mn}^{-1})\rhd h_{k\ell n},(g_{jk}^{-1}g_{k\ell}^{-1}g_{\ell m}^{-1}g_{mn}^{-1}g_{\ell n} g_{k\ell} g_{kn}^{-1})\rhd h_{jkn}\}_{\mathrm{p}}{(g_{jk}^{-1}g_{k\ell}^{-1}g_{\ell m}^{-1}g_{mn}^{-1})\rhd  l_{jk\ell n} }\vphantom{\ds\int}\\&&\left((g_{jk}^{-1}g_{k\ell}^{-1}g_{\ell m}^{-1}g_{mn}^{-1})\rhd (h_{j\ell n}g_{\ell n}\rhd h_{j k \ell }h_{k \ell n}^{-1}h_{jkn}^{-1})\right)\rhd \{(g_{jk}^{-1}g_{k\ell}^{-1}g_{\ell m}^{-1}g_{mn}^{-1})\rhd h_{\ell mn}, \delta (\{(g_{jk}^{-1}g_{k\ell}^{-1}g_{\ell n}^{-1})\rhd h_{k\ell n},(g_{jk}^{-1}g_{kn}^{-1})\rhd h_{jkn}\}_{\mathrm{p}}^{-1})\}_{\mathrm{p}}^{-1}\vphantom{\ds\int}\\
\end{array}$}
\end{equation}
Using the identity \eqref{eq:id02} one writes:
\begin{equation}
\resizebox{0.98\hsize}{!}{$
    \begin{array}{lcl}
&&\left((g_{jk}^{-1}g_{k\ell}^{-1}g_{\ell m}^{-1}g_{mn}^{-1})\rhd (h_{j\ell n}g_{\ell n}\rhd h_{j k \ell }h_{k \ell n}^{-1}h_{jkn}^{-1})\right)\rhd \{(g_{jk}^{-1}g_{k\ell}^{-1}g_{\ell m}^{-1}g_{mn}^{-1})\rhd h_{\ell mn}, \delta (\{(g_{jk}^{-1}g_{k\ell}^{-1}g_{\ell n}^{-1})\rhd h_{k\ell n},(g_{jk}^{-1}g_{kn}^{-1})\rhd h_{jkn}\}_{\mathrm{p}}^{-1})\}_{\mathrm{p}}^{-1}\vphantom{\ds\int}\\&&{\left((g_{jk}^{-1}g_{k\ell}^{-1}g_{\ell m}^{-1}g_{mn}^{-1})\rhd h_{j\ell n}\right)\rhd'\{(g_{jk}^{-1}g_{k\ell}^{-1}g_{\ell m}^{-1}g_{mn}^{-1})\rhd h_{\ell m n},(g_{jk}^{-1}g_{k\ell}^{-1}g_{\ell n}^{-1})\rhd ( g_{\ell n}\rhd h_{jk\ell}h_{k\ell n}^{-1}h_{jkn}^{-1})\}_{\mathrm{p}}^{-1}}\vphantom{\ds\int}\\&=&\left((g_{jk}^{-1}g_{k\ell}^{-1}g_{\ell m}^{-1}g_{mn}^{-1})\rhd h_{j\ell n}\right)\rhd'\{(g_{jk}^{-1}g_{k\ell}^{-1}g_{\ell m}^{-1}g_{mn}^{-1})\rhd h_{\ell m n}, (g_{jk}^{-1}g_{k\ell}^{-1})\rhd h_{jk\ell} (g_{jk}^{-1}g_{kn}^{-1})\rhd h_{jkn}^{-1} (g_{jk}^{-1}g_{k\ell}^{-1}g_{\ell n}^{-1})\rhd h_{k\ell n}^{-1} \}_{\mathrm{p}}^{-1}\,.\vphantom{\ds\int}
    \end{array}$}
    \end{equation}
Then, substituting the previous expression in the equation \eqref{eq:delta6} and commuting the elements one obtains:
\begin{equation}
\resizebox{0.98\hsize}{!}{$
    \begin{array}{lcl}
\delta_L\big(\tilde{l}_{jk\ell m n}\big)&=& \delta_L\big((g_{jk}^{-1}g_{k\ell}^{-1}g_{\ell m}^{-1})\rhd l_{jk\ell m}^{-1}\{(g_{jk}^{-1}g_{k\ell}^{-1}g_{\ell m}^{-1})\rhd h_{k\ell m},(g_{jk}^{-1}g_{km}^{-1})\rhd h_{jkm}\}_{\mathrm{p}}^{-1}\vphantom{\ds\int}\\&&
    ((g_{jk}^{-1}g_{k\ell}^{-1}g_{\ell m}^{-1})\rhd h_{k\ell m})\rhd'  \{(g_{jk}^{-1}g_{km}^{-1})\rhd h_{jkm},(g_{jm}^{-1}g_{m n}^{-1})\rhd h_{jm n}\}_\mathrm{p}((g_{jk}^{-1}g_{k\ell}^{-1}g_{\ell m}^{-1})\rhd h_{k\ell m})\rhd'(g_{jk}^{-1}g_{km}^{-1}g_{mn}^{-1})\rhd l_{jkmn}^{-1}\vphantom{\ds\int}\\&&((g_{jk}^{-1}g_{k\ell}^{-1}g_{\ell m}^{-1})\rhd h_{k\ell m})\rhd' \{(g_{jk}^{-1}g_{km}^{-1}g_{mn}^{-1})\rhd h_{kmn},(g_{jk}^{-1}g_{kn}^{-1})\rhd h_{jkn}\}_{\mathrm{p}}^{-1} \vphantom{\ds\int} \\ && \{(g^{-1}_{jk}g_{k\ell}^{-1}g_{\ell m}^{-1})\rhd h_{k\ell m},(g^{-1}_{jk}g_{km}^{-1}g_{m n}^{-1})\rhd h_{km n}\}_\mathrm{p}(g^{-1}_{jk}g_{k\ell}^{-1}g_{\ell m}^{-1}g_{mn}^{-1})\rhd l_{k\ell mn}^{-1}\{(g^{-1}_{jk}g_{k\ell}^{-1}g_{\ell m}^{-1}g_{mn}^{-1})\rhd h_{\ell mn},(g^{-1}_{jk}g_{k\ell}^{-1}g_{\ell n}^{-1})\rhd h_{k\ell n}\}_{\mathrm{p}}^{-1}\vphantom{\ds\int} \\&&\{((g_{jk}^{-1}g_{k\ell}^{-1}g_{\ell m}^{-1}g_{mn}^{-1})\rhd h_{k\ell n},(g_{jk}^{-1}g_{k\ell}^{-1}g_{\ell m}^{-1}g_{mn}^{-1}g_{\ell n} g_{k\ell} g_{kn}^{-1})\rhd h_{jkn}\}_{\mathrm{p}}{(g_{jk}^{-1}g_{k\ell}^{-1}g_{\ell m}^{-1}g_{mn}^{-1})\rhd  l_{jk\ell n} }\vphantom{\ds\int}\\&& {\left((g_{jk}^{-1}g_{k\ell}^{-1}g_{\ell m}^{-1}g_{mn}^{-1})\rhd h_{j\ell n}\right)\rhd'\{(g_{jk}^{-1}g_{k\ell}^{-1}g_{\ell m}^{-1}g_{mn}^{-1})\rhd h_{\ell m n}, (g_{jk}^{-1}g_{k\ell}^{-1})\rhd h_{jk\ell} (g_{jk}^{-1}g_{kn}^{-1})\rhd h_{jkn}^{-1} (g_{jk}^{-1}g_{k\ell}^{-1}g_{\ell n}^{-1})\rhd h_{k\ell n}^{-1} \}_{\mathrm{p}}^{-1}}\vphantom{\ds\int}\\
\end{array}$}
\end{equation}
After commuting the elements one obtains: 
\begin{equation}
\resizebox{0.98\hsize}{!}{$
    \begin{array}{lcl}
\delta_L\big(\tilde{l}_{jk\ell m n}\big)&=& \delta_L\big((g_{jk}^{-1}g_{k\ell}^{-1}g_{\ell m}^{-1})\rhd l_{jk\ell m}^{-1}\{(g_{jk}^{-1}g_{k\ell}^{-1}g_{\ell m}^{-1})\rhd h_{k\ell m},(g_{jk}^{-1}g_{km}^{-1})\rhd h_{jkm}\}_{\mathrm{p}}^{-1}\vphantom{\ds\int}\\&&
    ((g_{jk}^{-1}g_{k\ell}^{-1}g_{\ell m}^{-1})\rhd h_{k\ell m})\rhd'  \{(g_{jk}^{-1}g_{km}^{-1})\rhd h_{jkm},(g_{jm}^{-1}g_{m n}^{-1})\rhd h_{jm n}\}_\mathrm{p}((g_{jk}^{-1}g_{k\ell}^{-1}g_{\ell m}^{-1})\rhd h_{k\ell m})\rhd'(g_{jk}^{-1}g_{km}^{-1}g_{mn}^{-1})\rhd l_{jkmn}^{-1}\vphantom{\ds\int}\\&&((g_{jk}^{-1}g_{k\ell}^{-1}g_{\ell m}^{-1})\rhd h_{k\ell m})\rhd' \{(g_{jk}^{-1}g_{km}^{-1}g_{mn}^{-1})\rhd h_{kmn},(g_{jk}^{-1}g_{kn}^{-1})\rhd h_{jkn}\}_{\mathrm{p}}^{-1} \vphantom{\ds\int} \\ && \{(g^{-1}_{jk}g_{k\ell}^{-1}g_{\ell m}^{-1})\rhd h_{k\ell m},(g^{-1}_{jk}g_{km}^{-1}g_{m n}^{-1})\rhd h_{km n}\}_\mathrm{p}(g^{-1}_{jk}g_{k\ell}^{-1}g_{\ell m}^{-1}g_{mn}^{-1})\rhd l_{k\ell mn}^{-1}\vphantom{\ds\int}\\&&\{((g_{jk}^{-1}g_{k\ell}^{-1}g_{\ell m}^{-1}g_{mn}^{-1})\rhd h_{k\ell n},(g_{jk}^{-1}g_{k\ell}^{-1}g_{\ell m}^{-1}g_{mn}^{-1}g_{\ell n} g_{k\ell} g_{kn}^{-1})\rhd h_{jkn}\}_{\mathrm{p}}{(g_{jk}^{-1}g_{k\ell}^{-1}g_{\ell m}^{-1}g_{mn}^{-1})\rhd  l_{jk\ell n} }\vphantom{\ds\int}\\&&{\left((g_{jk}^{-1}g_{k\ell}^{-1}g_{\ell m}^{-1}g_{mn}^{-1})\rhd (h_{j\ell n}g_{\ell n}\rhd h_{jk\ell} (g_{\ell n}g_{k\ell}g_{kn}^{-1})\rhd h_{jkn}^{-1}  h_{k\ell n}^{-1})\right)\rhd'}\{(g^{-1}_{jk}g_{k\ell}^{-1}g_{\ell m}^{-1}g_{mn}^{-1})\rhd h_{\ell mn},(g^{-1}_{jk}g_{k\ell}^{-1}g_{\ell n}^{-1})\rhd h_{k\ell n}\}_{\mathrm{p}}^{-1}\vphantom{\ds\int} \\  &&{ \big((g_{jk}^{-1}g_{k\ell}^{-1}g_{\ell m}^{-1}g_{mn}^{-1})\rhd h_{j\ell n}\big)\rhd' \{(g_{jk}^{-1}g_{k\ell}^{-1}g_{\ell m}^{-1}g_{mn}^{-1})\rhd h_{\ell m n}, (g_{jk}^{-1}g_{k\ell}^{-1})\rhd h_{jk\ell} (g_{jk}^{-1}g_{kn}^{-1})\rhd h_{jkn}^{-1} (g_{jk}^{-1}g_{k\ell}^{-1}g_{\ell n}^{-1})\rhd h_{k\ell n}^{-1} \}_{\mathrm{p}}^{-1}}\vphantom{\ds\int}\\
\end{array}$}
\end{equation}
Then, using the identity \eqref{eq:id02} one obtains:
\begin{equation}\label{eq:delta7}
\resizebox{0.98\hsize}{!}{$
    \begin{array}{lcl}
\delta_L\big(\tilde{l}_{jk\ell m n}\big)&=& \delta_L\big((g_{jk}^{-1}g_{k\ell}^{-1}g_{\ell m}^{-1})\rhd l_{jk\ell m}^{-1}\{(g_{jk}^{-1}g_{k\ell}^{-1}g_{\ell m}^{-1})\rhd h_{k\ell m},(g_{jk}^{-1}g_{km}^{-1})\rhd h_{jkm}\}_{\mathrm{p}}^{-1}\vphantom{\ds\int}\\&&
    ((g_{jk}^{-1}g_{k\ell}^{-1}g_{\ell m}^{-1})\rhd h_{k\ell m})\rhd'  \{(g_{jk}^{-1}g_{km}^{-1})\rhd h_{jkm},(g_{jm}^{-1}g_{m n}^{-1})\rhd h_{jm n}\}_\mathrm{p}((g_{jk}^{-1}g_{k\ell}^{-1}g_{\ell m}^{-1})\rhd h_{k\ell m})\rhd'(g_{jk}^{-1}g_{km}^{-1}g_{mn}^{-1})\rhd l_{jkmn}^{-1}\vphantom{\ds\int}\\&&((g_{jk}^{-1}g_{k\ell}^{-1}g_{\ell m}^{-1})\rhd h_{k\ell m})\rhd' \{(g_{jk}^{-1}g_{km}^{-1}g_{mn}^{-1})\rhd h_{kmn},(g_{jk}^{-1}g_{kn}^{-1})\rhd h_{jkn}\}_{\mathrm{p}}^{-1}\vphantom{\ds\int}  \\ && \{(g^{-1}_{jk}g_{k\ell}^{-1}g_{\ell m}^{-1})\rhd h_{k\ell m},(g^{-1}_{jk}g_{km}^{-1}g_{m n}^{-1})\rhd h_{km n}\}_\mathrm{p}(g^{-1}_{jk}g_{k\ell}^{-1}g_{\ell m}^{-1}g_{mn}^{-1})\rhd l_{k\ell mn}^{-1}\vphantom{\ds\int}\\&&\{((g_{jk}^{-1}g_{k\ell}^{-1}g_{\ell m}^{-1}g_{mn}^{-1})\rhd h_{k\ell n},(g_{jk}^{-1}g_{k\ell}^{-1}g_{\ell m}^{-1}g_{mn}^{-1}g_{\ell n} g_{k\ell} g_{kn}^{-1})\rhd h_{jkn}\}_{\mathrm{p}}{(g_{jk}^{-1}g_{k\ell}^{-1}g_{\ell m}^{-1}g_{mn}^{-1})\rhd  l_{jk\ell n} }\vphantom{\ds\int}\\&&  {\big((g_{jk}^{-1}g_{k\ell}^{-1}g_{\ell m}^{-1}g_{mn}^{-1})\rhd h_{j\ell n}\big)\rhd' \{(g_{jk}^{-1}g_{k\ell}^{-1}g_{\ell m}^{-1}g_{mn}^{-1})\rhd h_{\ell m n}, (g_{jk}^{-1}g_{k\ell}^{-1})\rhd h_{jk\ell} (g_{jk}^{-1}g_{kn}^{-1})\rhd h_{jkn}^{-1} \}_{\mathrm{p}}^{-1}}\vphantom{\ds\int}\\ &&{(g_{jk}^{-1}g_{k\ell}^{-1}g_{\ell m}^{-1}g_{mn}^{-1})\rhd l_{j\ell mn}}{\{(g_{jk}^{-1}g_{k\ell}^{-1}g_{\ell m}^{-1})\rhd h_{j\ell m}(g_{jk}^{-1}g_{k\ell}^{-1})\rhd h_{jk\ell},(g_{jm}^{-1}g_{mn}^{-1 })\rhd h_{jmn}\}_{\mathrm{p}}^{-1}}\big)\,.\vphantom{\ds\int}
    \end{array}$}
\end{equation}
One can use the identity,
\begin{equation}
\{h_1,h_2 h_3 \}^{-1}_{\mathrm{p}}=(\partial h_1 \rhd h_2) \rhd' \{h_1, h_3\}^{-1}_{\mathrm{p}} \{h_1, h_2\}^{-1}_{\mathrm{p}},
\end{equation}
to obtain
\begin{equation}
\resizebox{0.98\hsize}{!}{$
    \begin{array}{lcl}
 &&\{(g_{jk}^{-1}g_{k\ell}^{-1}g_{\ell m}^{-1}g_{mn}^{-1})\rhd h_{\ell m n}, (g_{jk}^{-1}g_{k\ell}^{-1})\rhd h_{jk\ell} (g_{jk}^{-1}g_{kn}^{-1})\rhd h_{jkn}^{-1} \}_{\mathrm{p}}^{-1}\vphantom{\ds\int}\\&=&(g_{jk}^{-1}g_{k\ell}^{-1}g_{\ell m}^{-1}g_{mn}^{-1}g_{ln})\rhd h_{jk\ell}\rhd'\{(g_{jk}^{-1}g_{k\ell}^{-1}g_{\ell m}^{-1}g_{mn}^{-1})\rhd h_{\ell m n}, (g_{jk}^{-1}g_{kn}^{-1})\rhd h_{jkn}^{-1} \}_{\mathrm{p}}^{-1}\{(g_{jk}^{-1}g_{k\ell}^{-1}g_{\ell m}^{-1}g_{mn}^{-1})\rhd h_{\ell m n},  (g_{jk}^{-1}g_{k\ell}^{-1})\rhd h_{jk\ell} \}_{\mathrm{p}}^{-1}\,.\vphantom{\ds\int}
\end{array}$}
\end{equation}
Substituting in \eqref{eq:delta7} one obtains:
\begin{equation}
\resizebox{0.98\hsize}{!}{$
    \begin{array}{lcl}
\delta_L\big(\tilde{l}_{jk\ell m n}\big)&=& \delta_L\big((g_{jk}^{-1}g_{k\ell}^{-1}g_{\ell m}^{-1})\rhd l_{jk\ell m}^{-1}\{(g_{jk}^{-1}g_{k\ell}^{-1}g_{\ell m}^{-1})\rhd h_{k\ell m},(g_{jk}^{-1}g_{km}^{-1})\rhd h_{jkm}\}_{\mathrm{p}}^{-1}\vphantom{\ds\int}\\&&
    ((g_{jk}^{-1}g_{k\ell}^{-1}g_{\ell m}^{-1})\rhd h_{k\ell m})\rhd'  \{(g_{jk}^{-1}g_{km}^{-1})\rhd h_{jkm},(g_{jm}^{-1}g_{m n}^{-1})\rhd h_{jm n}\}_\mathrm{p}((g_{jk}^{-1}g_{k\ell}^{-1}g_{\ell m}^{-1})\rhd h_{k\ell m})\rhd'(g_{jk}^{-1}g_{km}^{-1}g_{mn}^{-1})\rhd l_{jkmn}^{-1}\vphantom{\ds\int}\\&&((g_{jk}^{-1}g_{k\ell}^{-1}g_{\ell m}^{-1})\rhd h_{k\ell m})\rhd' \{(g_{jk}^{-1}g_{km}^{-1}g_{mn}^{-1})\rhd h_{kmn},(g_{jk}^{-1}g_{kn}^{-1})\rhd h_{jkn}\}_{\mathrm{p}}^{-1} \vphantom{\ds\int} \\ && \{(g^{-1}_{jk}g_{k\ell}^{-1}g_{\ell m}^{-1})\rhd h_{k\ell m},(g^{-1}_{jk}g_{km}^{-1}g_{m n}^{-1})\rhd h_{km n}\}_\mathrm{p}(g^{-1}_{jk}g_{k\ell}^{-1}g_{\ell m}^{-1}g_{mn}^{-1})\rhd l_{k\ell mn}^{-1}\vphantom{\ds\int}\\&&\{((g_{jk}^{-1}g_{k\ell}^{-1}g_{\ell m}^{-1}g_{mn}^{-1})\rhd h_{k\ell n},(g_{jk}^{-1}g_{k\ell}^{-1}g_{\ell m}^{-1}g_{mn}^{-1}g_{\ell n} g_{k\ell} g_{kn}^{-1})\rhd h_{jkn}\}_{\mathrm{p}}{(g_{jk}^{-1}g_{k\ell}^{-1}g_{\ell m}^{-1}g_{mn}^{-1})\rhd  l_{jk\ell n} }\vphantom{\ds\int}\\&&  {\big((g_{jk}^{-1}g_{k\ell}^{-1}g_{\ell m}^{-1}g_{mn}^{-1})\rhd h_{j\ell n}\big)\rhd' (g_{jk}^{-1}g_{k\ell}^{-1}g_{\ell m}^{-1}g_{mn}^{-1}g_{ln})\rhd h_{jk\ell}\rhd'\{(g_{jk}^{-1}g_{k\ell}^{-1}g_{\ell m}^{-1}g_{mn}^{-1})\rhd h_{\ell m n}, (g_{jk}^{-1}g_{kn}^{-1})\rhd h_{jkn}^{-1} \}_{\mathrm{p}}^{-1}}\vphantom{\ds\int}\\&&{\big((g_{jk}^{-1}g_{k\ell}^{-1}g_{\ell m}^{-1}g_{mn}^{-1})\rhd h_{j\ell n}\big) \rhd' \{(g_{jk}^{-1}g_{k\ell}^{-1}g_{\ell m}^{-1}g_{mn}^{-1})\rhd h_{\ell m n},  (g_{jk}^{-1}g_{k\ell}^{-1})\rhd h_{jk\ell} \}_{\mathrm{p}}^{-1}}\vphantom{\ds\int}\\
\end{array}$}
\end{equation}
Then, commuting the elements one obtains:
\begin{equation}
\resizebox{0.98\hsize}{!}{$
    \begin{array}{lcl}
\delta_L\big(\tilde{l}_{jk\ell m n}\big)&=& \delta_L\big((g_{jk}^{-1}g_{k\ell}^{-1}g_{\ell m}^{-1})\rhd l_{jk\ell m}^{-1}\{(g_{jk}^{-1}g_{k\ell}^{-1}g_{\ell m}^{-1})\rhd h_{k\ell m},(g_{jk}^{-1}g_{km}^{-1})\rhd h_{jkm}\}_{\mathrm{p}}^{-1}\vphantom{\ds\int}\\&&
    ((g_{jk}^{-1}g_{k\ell}^{-1}g_{\ell m}^{-1})\rhd h_{k\ell m})\rhd'  \{(g_{jk}^{-1}g_{km}^{-1})\rhd h_{jkm},(g_{jm}^{-1}g_{m n}^{-1})\rhd h_{jm n}\}_\mathrm{p}((g_{jk}^{-1}g_{k\ell}^{-1}g_{\ell m}^{-1})\rhd h_{k\ell m})\rhd'(g_{jk}^{-1}g_{km}^{-1}g_{mn}^{-1})\rhd l_{jkmn}^{-1}\vphantom{\ds\int}\\&&((g_{jk}^{-1}g_{k\ell}^{-1}g_{\ell m}^{-1})\rhd h_{k\ell m})\rhd' \{(g_{jk}^{-1}g_{km}^{-1}g_{mn}^{-1})\rhd h_{kmn},(g_{jk}^{-1}g_{kn}^{-1})\rhd h_{jkn}\}_{\mathrm{p}}^{-1} \vphantom{\ds\int} \\ && \{(g^{-1}_{jk}g_{k\ell}^{-1}g_{\ell m}^{-1})\rhd h_{k\ell m},(g^{-1}_{jk}g_{km}^{-1}g_{m n}^{-1})\rhd h_{km n}\}_\mathrm{p}(g^{-1}_{jk}g_{k\ell}^{-1}g_{\ell m}^{-1}g_{mn}^{-1})\rhd l_{k\ell mn}^{-1}\vphantom{\ds\int}\\&&\{((g_{jk}^{-1}g_{k\ell}^{-1}g_{\ell m}^{-1}g_{mn}^{-1})\rhd h_{k\ell n},(g_{jk}^{-1}g_{k\ell}^{-1}g_{\ell m}^{-1}g_{mn}^{-1}g_{\ell n} g_{k\ell} g_{kn}^{-1})\rhd h_{jkn}\}_{\mathrm{p}}\vphantom{\ds\int}\\  &&{\big((g_{jk}^{-1}g_{k\ell}^{-1}g_{\ell m}^{-1}g_{mn}^{-1})\rhd (h_{jkn}h_{k\ell n})\big)\rhd'\{(g_{jk}^{-1}g_{k\ell}^{-1}g_{\ell m}^{-1}g_{mn}^{-1})\rhd h_{\ell m n}, (g_{jk}^{-1}g_{kn}^{-1})\rhd h_{jkn}^{-1} \}_{\mathrm{p}}^{-1}}\vphantom{\ds\int}\\&&{(g_{jk}^{-1}g_{k\ell}^{-1}g_{\ell m}^{-1}g_{mn}^{-1})\rhd  l_{jk\ell n} }{\big((g_{jk}^{-1}g_{k\ell}^{-1}g_{\ell m}^{-1}g_{mn}^{-1})\rhd h_{j\ell n}\big) \rhd' \{(g_{jk}^{-1}g_{k\ell}^{-1}g_{\ell m}^{-1}g_{mn}^{-1})\rhd h_{\ell m n},  (g_{jk}^{-1}g_{k\ell}^{-1})\rhd h_{jk\ell} \}_{\mathrm{p}}^{-1}}\vphantom{\ds\int}\\
\end{array}$}
\end{equation}
Using the identity \eqref{eq:id06} one can rewrite the term as
\begin{equation}
\resizebox{0.98\hsize}{!}{$
    \begin{aligned}
    \{(g_{jk}^{-1}g_{k\ell}^{-1}g_{\ell m}^{-1}g_{mn}^{-1})\rhd h_{\ell m n}, (g_{jk}^{-1}g_{kn}^{-1})\rhd h_{jkn}^{-1} \}_{\mathrm{p}}^{-1}=(g_{jk}^{-1}g_{k\ell}^{-1}g_{\ell m}^{-1}g_{mn}^{-1}g_{\ell n} g_{k\ell} g_{kn}^{-1})\rhd h_{jkn}^{-1}\rhd'\{(g_{jk}^{-1}g_{k\ell}^{-1}g_{\ell m}^{-1}g_{mn}^{-1})\rhd h_{\ell m n}, (g_{jk}^{-1}g_{kn}^{-1})\rhd h_{jkn} \}_{\mathrm{p}}\,,
    \end{aligned}$}
    \end{equation}
so that after commuting the elements one obtains:
\begin{equation}\label{eq:delta8}
\resizebox{0.98\hsize}{!}{$
    \begin{array}{lcl}
\delta_L\big(\tilde{l}_{jk\ell m n}\big)&=& \delta_L\big((g_{jk}^{-1}g_{k\ell}^{-1}g_{\ell m}^{-1})\rhd l_{jk\ell m}^{-1}\{(g_{jk}^{-1}g_{k\ell}^{-1}g_{\ell m}^{-1})\rhd h_{k\ell m},(g_{jk}^{-1}g_{km}^{-1})\rhd h_{jkm}\}_{\mathrm{p}}^{-1}\vphantom{\ds\int}\\&&
    ((g_{jk}^{-1}g_{k\ell}^{-1}g_{\ell m}^{-1})\rhd h_{k\ell m})\rhd'  \{(g_{jk}^{-1}g_{km}^{-1})\rhd h_{jkm},(g_{jm}^{-1}g_{m n}^{-1})\rhd h_{jm n}\}_\mathrm{p}((g_{jk}^{-1}g_{k\ell}^{-1}g_{\ell m}^{-1})\rhd h_{k\ell m})\rhd'(g_{jk}^{-1}g_{km}^{-1}g_{mn}^{-1})\rhd l_{jkmn}^{-1}\vphantom{\ds\int}\\&&((g_{jk}^{-1}g_{k\ell}^{-1}g_{\ell m}^{-1})\rhd h_{k\ell m})\rhd' \{(g_{jk}^{-1}g_{km}^{-1}g_{mn}^{-1})\rhd h_{kmn},(g_{jk}^{-1}g_{kn}^{-1})\rhd h_{jkn}\}_{\mathrm{p}}^{-1} \vphantom{\ds\int} \\  &&\{(g^{-1}_{jk}g_{k\ell}^{-1}g_{\ell m}^{-1})\rhd h_{k\ell m},(g^{-1}_{jk}g_{km}^{-1}g_{m n}^{-1})\rhd h_{km n}\}_\mathrm{p}(g^{-1}_{jk}g_{k\ell}^{-1}g_{\ell m}^{-1}g_{mn}^{-1})\rhd l_{k\ell mn}^{-1}\vphantom{\ds\int}\\&&  {\big((g_{jk}^{-1}g_{k\ell}^{-1}g_{\ell m}^{-1}g_{mn}^{-1})\rhd h_{k\ell n}\big)\rhd'\{(g_{jk}^{-1}g_{k\ell}^{-1}g_{\ell m}^{-1}g_{mn}^{-1})\rhd h_{\ell m n}, (g_{jk}^{-1}g_{kn}^{-1})\rhd h_{jkn} \}_{\mathrm{p}}}\vphantom{\ds\int}\\&&\{((g_{jk}^{-1}g_{k\ell}^{-1}g_{\ell m}^{-1}g_{mn}^{-1})\rhd h_{k\ell n},(g_{jk}^{-1}g_{k\ell}^{-1}g_{\ell m}^{-1}g_{mn}^{-1}g_{\ell n} g_{k\ell} g_{kn}^{-1})\rhd h_{jkn}\}_{\mathrm{p}}\vphantom{\ds\int}\\&&{(g_{jk}^{-1}g_{k\ell}^{-1}g_{\ell m}^{-1}g_{mn}^{-1})\rhd  l_{jk\ell n} }{\big((g_{jk}^{-1}g_{k\ell}^{-1}g_{\ell m}^{-1}g_{mn}^{-1})\rhd h_{j\ell n}\big) \rhd' \{(g_{jk}^{-1}g_{k\ell}^{-1}g_{\ell m}^{-1}g_{mn}^{-1})\rhd h_{\ell m n},  (g_{jk}^{-1}g_{k\ell}^{-1})\rhd h_{jk\ell} \}_{\mathrm{p}}^{-1}}\vphantom{\ds\int}\\&& {(g_{jk}^{-1}g_{k\ell}^{-1}g_{\ell m}^{-1}g_{mn}^{-1})\rhd l_{j\ell mn}}{\{(g_{jk}^{-1}g_{k\ell}^{-1}g_{\ell m}^{-1})\rhd h_{j\ell m}(g_{jk}^{-1}g_{k\ell}^{-1})\rhd h_{jk\ell},(g_{jm}^{-1}g_{mn}^{-1 })\rhd h_{jmn}\}_{\mathrm{p}}^{-1}}\big)\,.\vphantom{\ds\int}
    \end{array}$}
\end{equation}
Using the identity \eqref{eq:id01} one has the following equation:
\begin{equation}
\resizebox{0.98\hsize}{!}{$
    \begin{array}{lcl}
&& {\big((g_{jk}^{-1}g_{k\ell}^{-1}g_{\ell m}^{-1}g_{mn}^{-1})\rhd h_{k\ell n}\big)\rhd'\{(g_{jk}^{-1}g_{k\ell}^{-1}g_{\ell m}^{-1}g_{mn}^{-1})\rhd h_{\ell m n}, (g_{jk}^{-1}g_{kn}^{-1})\rhd h_{jkn} \}_{\mathrm{p}}}\{((g_{jk}^{-1}g_{k\ell}^{-1}g_{\ell m}^{-1}g_{mn}^{-1})\rhd h_{k\ell n},(g_{jk}^{-1}g_{k\ell}^{-1}g_{\ell m}^{-1}g_{mn}^{-1}g_{\ell n} g_{k\ell} g_{kn}^{-1})\rhd h_{jkn}\}_{\mathrm{p}}\vphantom{\ds\int}\\&=& \{(g_{jk}^{-1}g_{k\ell}^{-1}g_{\ell m}^{-1}g_{mn}^{-1})\rhd h_{k\ell n} (g_{jk}^{-1}g_{k\ell}^{-1}g_{\ell m}^{-1}g_{mn}^{-1})\rhd h_{\ell m n}, (g_{jk}^{-1}g_{kn}^{-1})\rhd h_{jkn} \}_{\mathrm{p}}\,.\vphantom{\ds\int}
    \end{array}$}
    \end{equation}
Substituting this expression in \eqref{eq:delta8} one obtains:
\begin{equation}
\resizebox{0.98\hsize}{!}{$
    \begin{array}{lcl}
\delta_L\big(\tilde{l}_{jk\ell m n}\big)&=& \delta_L\big((g_{jk}^{-1}g_{k\ell}^{-1}g_{\ell m}^{-1})\rhd l_{jk\ell m}^{-1}\{(g_{jk}^{-1}g_{k\ell}^{-1}g_{\ell m}^{-1})\rhd h_{k\ell m},(g_{jk}^{-1}g_{km}^{-1})\rhd h_{jkm}\}_{\mathrm{p}}^{-1}\vphantom{\ds\int}\\&&
    ((g_{jk}^{-1}g_{k\ell}^{-1}g_{\ell m}^{-1})\rhd h_{k\ell m})\rhd'  \{(g_{jk}^{-1}g_{km}^{-1})\rhd h_{jkm},(g_{jm}^{-1}g_{m n}^{-1})\rhd h_{jm n}\}_\mathrm{p}((g_{jk}^{-1}g_{k\ell}^{-1}g_{\ell m}^{-1})\rhd h_{k\ell m})\rhd'(g_{jk}^{-1}g_{km}^{-1}g_{mn}^{-1})\rhd l_{jkmn}^{-1}\vphantom{\ds\int}\\&&((g_{jk}^{-1}g_{k\ell}^{-1}g_{\ell m}^{-1})\rhd h_{k\ell m})\rhd' \{(g_{jk}^{-1}g_{km}^{-1}g_{mn}^{-1})\rhd h_{kmn},(g_{jk}^{-1}g_{kn}^{-1})\rhd h_{jkn}\}_{\mathrm{p}}^{-1}\vphantom{\ds\int}  \\ && \{(g^{-1}_{jk}g_{k\ell}^{-1}g_{\ell m}^{-1})\rhd h_{k\ell m},(g^{-1}_{jk}g_{km}^{-1}g_{m n}^{-1})\rhd h_{km n}\}_\mathrm{p}(g^{-1}_{jk}g_{k\ell}^{-1}g_{\ell m}^{-1}g_{mn}^{-1})\rhd l_{k\ell mn}^{-1}\vphantom{\ds\int}\\&& {\{(g_{jk}^{-1}g_{k\ell}^{-1}g_{\ell m}^{-1}g_{mn}^{-1})\rhd h_{k\ell n} (g_{jk}^{-1}g_{k\ell}^{-1}g_{\ell m}^{-1}g_{mn}^{-1})\rhd h_{\ell m n}, (g_{jk}^{-1}g_{kn}^{-1})\rhd h_{jkn} \}_{\mathrm{p}}}\vphantom{\ds\int}\\&&{(g_{jk}^{-1}g_{k\ell}^{-1}g_{\ell m}^{-1}g_{mn}^{-1})\rhd  l_{jk\ell n} }{\big((g_{jk}^{-1}g_{k\ell}^{-1}g_{\ell m}^{-1}g_{mn}^{-1})\rhd h_{j\ell n}\big) \rhd' \{(g_{jk}^{-1}g_{k\ell}^{-1}g_{\ell m}^{-1}g_{mn}^{-1})\rhd h_{\ell m n},  (g_{jk}^{-1}g_{k\ell}^{-1})\rhd h_{jk\ell} \}_{\mathrm{p}}^{-1}}\vphantom{\ds\int}\\ &&{(g_{jk}^{-1}g_{k\ell}^{-1}g_{\ell m}^{-1}g_{mn}^{-1})\rhd l_{j\ell mn}}{\{(g_{jk}^{-1}g_{k\ell}^{-1}g_{\ell m}^{-1})\rhd h_{j\ell m}(g_{jk}^{-1}g_{k\ell}^{-1})\rhd h_{jk\ell},(g_{jm}^{-1}g_{mn}^{-1 })\rhd h_{jmn}\}_{\mathrm{p}}^{-1}}\big)\,.\vphantom{\ds\int}
    \end{array}$}
\end{equation}
One has, using the identity \eqref{identitet} the equations
\begin{equation}\label{ident1}
\resizebox{0.98\hsize}{!}{$
((g_{jk}^{-1}g_{k\ell}^{-1}g_{\ell m}^{-1})\rhd h_{k\ell m})\rhd'(g_{jk}^{-1}g_{km}^{-1}g_{mn}^{-1})\rhd l_{jkmn}^{-1}=\{(g_{jk}^{-1}g_{k\ell}^{-1}g_{\ell m}^{-1})\rhd h_{k\ell m},(g_{jk}^{-1}g_{km}^{-1}g_{mn}^{-1})\rhd \delta(l_{jkmn})^{-1}\}_{\mathrm{p}} (g_{jk}^{-1}g_{k\ell}^{-1}g_{\ell m}^{-1}g_{mn}^{-1})\rhd l_{jkmn}^{-1}\,,$}
\end{equation}
and
\begin{equation}\label{ident2}
\resizebox{0.98\hsize}{!}{$
\begin{array}{lcl}
&&((g_{jk}^{-1}g_{k\ell}^{-1}g_{\ell m}^{-1})\rhd h_{k\ell m})\rhd'  \{(g_{jk}^{-1}g_{km}^{-1})\rhd h_{jkm},(g_{jm}^{-1}g_{m n}^{-1})\rhd h_{jm n}\}_\mathrm{p}\vphantom{\ds\int}\\&=&\{(g_{jk}^{-1}g_{k\ell}^{-1}g_{\ell m}^{-1})\rhd h_{jkm},(g_{jk}^{-1}g_{k\ell}^{-1}g_{\ell m}^{-1}g_{km}g_{jk}g_{jm}^{-1}g_{m n}^{-1})\rhd h_{jm n}\}_\mathrm{p}\{(g_{jk}^{-1}g_{k\ell}^{-1}g_{\ell m}^{-1})\rhd h_{k\ell m}, \delta( \{(g_{jk}^{-1}g_{km}^{-1})\rhd h_{jkm},(g_{jm}^{-1}g_{m n}^{-1})\rhd h_{jm n}\}_\mathrm{p}^{-1}) \}_\mathrm{p}^{-1}\,.\vphantom{\ds\int}
\end{array}
$}
\end{equation}
Using the identity,
\begin{equation}
\{h_1,h_2h_3\}_\mathrm{p}=(\partial h_1 \rhd h_2) \rhd' \big( \{h_1,h_2^{-1}\}_\mathrm{p}^{-1} \{h_1,h_3\}_\mathrm{p} \big)\,,
\end{equation}
and the equations \eqref{ident1} and \eqref{ident2} one can write:
\begin{equation}
\resizebox{0.98\hsize}{!}{$
    \begin{array}{lcl}
\delta_L\big(\tilde{l}_{jk\ell m n}\big)&=& \delta_L\big((g_{jk}^{-1}g_{k\ell}^{-1}g_{\ell m}^{-1})\rhd l_{jk\ell m}^{-1}\{(g_{jk}^{-1}g_{k\ell}^{-1}g_{\ell m}^{-1})\rhd h_{k\ell m},(g_{jk}^{-1}g_{km}^{-1})\rhd h_{jkm}\}_{\mathrm{p}}^{-1}\vphantom{\ds\int}\\ &&{\{(g_{jk}^{-1}g_{k\ell}^{-1}g_{\ell m}^{-1})\rhd h_{jkm},(g_{jk}^{-1}g_{k\ell}^{-1}g_{\ell m}^{-1}g_{km}g_{jk}g_{jm}^{-1}g_{m n}^{-1})\rhd h_{jm n}\}_\mathrm{p}}\vphantom{\ds\int}\\&& \big((g_{jk}^{-1}g_{k\ell}^{-1}g_{\ell m}^{-1}g_{km}g_{jk})\rhd \delta \{(g_{jk}^{-1}g_{km}^{-1})\rhd h_{jkm},(g_{jm}^{-1}g_{m n}^{-1})\rhd h_{jm n}\}_\mathrm{p}^{-1} \big)\rhd'\vphantom{\ds\int}\\&&\{(g_{jk}^{-1}g_{k\ell}^{-1}g_{\ell m}^{-1})\rhd h_{k\ell m},\delta\{(g_{jk}^{-1}g_{km}^{-1})\rhd h_{jkm},(g_{jm}^{-1}g_{m n}^{-1})\rhd h_{jm n}\}_\mathrm{p} { (g_{jk}^{-1}g_{km}^{-1}g_{mn}^{-1})\rhd \delta(l_{jkmn})^{-1}\}_{\mathrm{p}}}(g_{jk}^{-1}g_{k\ell}^{-1}g_{\ell m}^{-1}g_{mn}^{-1})\rhd l_{jkmn}^{-1}\vphantom{\ds\int}\\&&((g_{jk}^{-1}g_{k\ell}^{-1}g_{\ell m}^{-1})\rhd h_{k\ell m})\rhd' \{(g_{jk}^{-1}g_{km}^{-1}g_{mn}^{-1})\rhd h_{kmn},(g_{jk}^{-1}g_{kn}^{-1})\rhd h_{jkn}\}_{\mathrm{p}}^{-1} \vphantom{\ds\int} \\&&  \{(g^{-1}_{jk}g_{k\ell}^{-1}g_{\ell m}^{-1})\rhd h_{k\ell m},(g^{-1}_{jk}g_{km}^{-1}g_{m n}^{-1})\rhd h_{km n}\}_\mathrm{p}(g^{-1}_{jk}g_{k\ell}^{-1}g_{\ell m}^{-1}g_{mn}^{-1})\rhd l_{k\ell mn}^{-1}\vphantom{\ds\int}\\ &&\{(g_{jk}^{-1}g_{k\ell}^{-1}g_{\ell m}^{-1}g_{mn}^{-1})\rhd h_{k\ell n} (g_{jk}^{-1}g_{k\ell}^{-1}g_{\ell m}^{-1}g_{mn}^{-1})\rhd h_{\ell m n}, (g_{jk}^{-1}g_{kn}^{-1})\rhd h_{jkn} \}_{\mathrm{p}}\vphantom{\ds\int}\\&&(g_{jk}^{-1}g_{k\ell}^{-1}g_{\ell m}^{-1}g_{mn}^{-1})\rhd  l_{jk\ell n} \big((g_{jk}^{-1}g_{k\ell}^{-1}g_{\ell m}^{-1}g_{mn}^{-1})\rhd h_{j\ell n}\big) \rhd' \{(g_{jk}^{-1}g_{k\ell}^{-1}g_{\ell m}^{-1}g_{mn}^{-1})\rhd h_{\ell m n},  (g_{jk}^{-1}g_{k\ell}^{-1})\rhd h_{jk\ell} \}_{\mathrm{p}}^{-1}\vphantom{\ds\int}\\
\end{array}$}
\end{equation}
Then, using the identity \eqref{prop1}:
\begin{equation}
\resizebox{0.98\hsize}{!}{$
    \begin{array}{lcl}
\delta_L\big(\tilde{l}_{jk\ell m n}\big)&=& \delta_L\big((g_{jk}^{-1}g_{k\ell}^{-1}g_{\ell m}^{-1})\rhd l_{jk\ell m}^{-1}\{(g_{jk}^{-1}g_{k\ell}^{-1}g_{\ell m}^{-1})\rhd h_{k\ell m},(g_{jk}^{-1}g_{km}^{-1})\rhd h_{jkm}\}_{\mathrm{p}}^{-1}\vphantom{\ds\int}\\&& \{(g_{jk}^{-1}g_{k\ell}^{-1}g_{\ell m}^{-1})\rhd h_{jkm},(g_{jk}^{-1}g_{k\ell}^{-1}g_{\ell m}^{-1}g_{km}g_{jk}g_{jm}^{-1}g_{m n}^{-1})\rhd h_{jm n}\}_\mathrm{p}\vphantom{\ds\int}\\ &&\big((g_{jk}^{-1}g_{k\ell}^{-1}g_{\ell m}^{-1}g_{km}g_{jk})\rhd \delta \{(g_{jk}^{-1}g_{km}^{-1})\rhd h_{jkm},(g_{jm}^{-1}g_{m n}^{-1})\rhd h_{jm n}\}_\mathrm{p}^{-1} \big)\rhd'\vphantom{\ds\int}\\&&\{(g_{jk}^{-1}g_{k\ell}^{-1}g_{\ell m}^{-1})\rhd h_{k\ell m},(g_{jk}^{-1}g_{km}^{-1})\rhd h_{jkm}(g_{jm}^{-1}g_{m n}^{-1})\rhd h_{jm n} (g_{jk}^{-1}g_{km}^{-1}g_{mn}^{-1}\rhd(h_{kmn}^{-1}h_{jkn}^{-1}))\}_{\mathrm{p}} { (g_{jk}^{-1}g_{k\ell}^{-1}g_{\ell m}^{-1}g_{mn}^{-1})\rhd l_{jkmn}^{-1}}\vphantom{\ds\int}\\&&((g_{jk}^{-1}g_{k\ell}^{-1}g_{\ell m}^{-1})\rhd h_{k\ell m})\rhd' \{(g_{jk}^{-1}g_{km}^{-1}g_{mn}^{-1})\rhd h_{kmn},(g_{jk}^{-1}g_{kn}^{-1})\rhd h_{jkn}\}_{\mathrm{p}}^{-1}  \vphantom{\ds\int}\\&&  \{(g^{-1}_{jk}g_{k\ell}^{-1}g_{\ell m}^{-1})\rhd h_{k\ell m},(g^{-1}_{jk}g_{km}^{-1}g_{m n}^{-1})\rhd h_{km n}\}_\mathrm{p}(g^{-1}_{jk}g_{k\ell}^{-1}g_{\ell m}^{-1}g_{mn}^{-1})\rhd l_{k\ell mn}^{-1}\vphantom{\ds\int}\\ &&{\{(g_{jk}^{-1}g_{k\ell}^{-1}g_{\ell m}^{-1}g_{mn}^{-1})\rhd h_{k\ell n} (g_{jk}^{-1}g_{k\ell}^{-1}g_{\ell m}^{-1}g_{mn}^{-1})\rhd h_{\ell m n}, (g_{jk}^{-1}g_{kn}^{-1})\rhd h_{jkn} \}_{\mathrm{p}}}\vphantom{\ds\int}\\&&{(g_{jk}^{-1}g_{k\ell}^{-1}g_{\ell m}^{-1}g_{mn}^{-1})\rhd  l_{jk\ell n} }{\big((g_{jk}^{-1}g_{k\ell}^{-1}g_{\ell m}^{-1}g_{mn}^{-1})\rhd h_{j\ell n}\big) \rhd' \{(g_{jk}^{-1}g_{k\ell}^{-1}g_{\ell m}^{-1}g_{mn}^{-1})\rhd h_{\ell m n},  (g_{jk}^{-1}g_{k\ell}^{-1})\rhd h_{jk\ell} \}_{\mathrm{p}}^{-1}}\vphantom{\ds\int}\\
\end{array}$}
\end{equation}
Commuting the elements one obtains:
\begin{equation}\label{eq:delta77}
\resizebox{0.98\hsize}{!}{$
    \begin{array}{lcl}
\delta_L\big(\tilde{l}_{jk\ell m n}\big)&=& \delta_L\big((g_{jk}^{-1}g_{k\ell}^{-1}g_{\ell m}^{-1})\rhd l_{jk\ell m}^{-1}\{(g_{jk}^{-1}g_{k\ell}^{-1}g_{\ell m}^{-1})\rhd h_{k\ell m},(g_{jk}^{-1}g_{km}^{-1})\rhd h_{jkm}\}_{\mathrm{p}}^{-1}\vphantom{\ds\int}\\ &&{\{(g_{jk}^{-1}g_{k\ell}^{-1}g_{\ell m}^{-1})\rhd h_{k\ell m},(g_{jk}^{-1}g_{km}^{-1})\rhd h_{jkm}(g_{jm}^{-1}g_{m n}^{-1})\rhd h_{jm n} (g_{jk}^{-1}g_{km}^{-1}g_{mn}^{-1}\rhd(h_{kmn}^{-1}h_{jkn}^{-1}))\}_\mathrm{p}}\vphantom{\ds\int} \\&&{ \{(g_{jk}^{-1}g_{k\ell}^{-1}g_{\ell m}^{-1})\rhd h_{jkm},(g_{jk}^{-1}g_{k\ell}^{-1}g_{\ell m}^{-1}g_{km}g_{jk}g_{jm}^{-1}g_{m n}^{-1})\rhd h_{jm n}\}_\mathrm{p}(g_{jk}^{-1}g_{k\ell}^{-1}g_{\ell m}^{-1}g_{mn}^{-1})\rhd l_{jkmn}^{-1}}\vphantom{\ds\int}\\&&((g_{jk}^{-1}g_{k\ell}^{-1}g_{\ell m}^{-1})\rhd h_{k\ell m})\rhd' \{(g_{jk}^{-1}g_{km}^{-1}g_{mn}^{-1})\rhd h_{kmn},(g_{jk}^{-1}g_{kn}^{-1})\rhd h_{jkn}\}_{\mathrm{p}}^{-1}  \vphantom{\ds\int}\\ && \{(g^{-1}_{jk}g_{k\ell}^{-1}g_{\ell m}^{-1})\rhd h_{k\ell m},(g^{-1}_{jk}g_{km}^{-1}g_{m n}^{-1})\rhd h_{km n}\}_\mathrm{p}(g^{-1}_{jk}g_{k\ell}^{-1}g_{\ell m}^{-1}g_{mn}^{-1})\rhd l_{k\ell mn}^{-1}\vphantom{\ds\int}\\ &&{\{(g_{jk}^{-1}g_{k\ell}^{-1}g_{\ell m}^{-1}g_{mn}^{-1})\rhd h_{k\ell n} (g_{jk}^{-1}g_{k\ell}^{-1}g_{\ell m}^{-1}g_{mn}^{-1})\rhd h_{\ell m n}, (g_{jk}^{-1}g_{kn}^{-1})\rhd h_{jkn} \}_{\mathrm{p}}}\vphantom{\ds\int}\\&&{(g_{jk}^{-1}g_{k\ell}^{-1}g_{\ell m}^{-1}g_{mn}^{-1})\rhd  l_{jk\ell n} }{\big((g_{jk}^{-1}g_{k\ell}^{-1}g_{\ell m}^{-1}g_{mn}^{-1})\rhd h_{j\ell n}\big) \rhd' \{(g_{jk}^{-1}g_{k\ell}^{-1}g_{\ell m}^{-1}g_{mn}^{-1})\rhd h_{\ell m n},  (g_{jk}^{-1}g_{k\ell}^{-1})\rhd h_{jk\ell} \}_{\mathrm{p}}^{-1}}\vphantom{\ds\int}\\
\end{array}$}
\end{equation}
Then, by using the identities \eqref{identitet} and \eqref{eq:id02} one has,
\begin{equation}
\begin{array}{lcl}
&&{\{(g_{jk}^{-1}g_{k\ell}^{-1}g_{\ell m}^{-1})\rhd h_{k\ell m},(g_{jk}^{-1}g_{km}^{-1})\rhd h_{jkm}(g_{jm}^{-1}g_{m n}^{-1})\rhd h_{jm n} (g_{jk}^{-1}g_{km}^{-1}g_{mn}^{-1}\rhd(h_{kmn}^{-1}h_{jkn}^{-1}))\}_\mathrm{p}}\vphantom{\ds\int} \\&&{ \{(g_{jk}^{-1}g_{k\ell}^{-1}g_{\ell m}^{-1})\rhd h_{jkm},(g_{jk}^{-1}g_{k\ell}^{-1}g_{\ell m}^{-1}g_{km}g_{jk}g_{jm}^{-1}g_{m n}^{-1})\rhd h_{jm n}\}_\mathrm{p}(g_{jk}^{-1}g_{k\ell}^{-1}g_{\ell m}^{-1}g_{mn}^{-1})\rhd l_{jkmn}^{-1}}\vphantom{\ds\int}\\&&((g_{jk}^{-1}g_{k\ell}^{-1}g_{\ell m}^{-1})\rhd h_{k\ell m})\rhd' \{(g_{jk}^{-1}g_{km}^{-1}g_{mn}^{-1})\rhd h_{kmn},(g_{jk}^{-1}g_{kn}^{-1})\rhd h_{jkn}\}_{\mathrm{p}}^{-1}\vphantom{\ds\int} \\&=&{\{(g_{jk}^{-1}g_{k\ell}^{-1}g_{\ell m}^{-1})\rhd h_{k\ell m},(g_{jk}^{-1}g_{km}^{-1})\rhd h_{jkm}(g_{jm}^{-1}g_{m n}^{-1})\rhd h_{jm n} (g_{jk}^{-1}g_{kn}^{-1})\rhd h_{jkn}^{-1} 
(g_{jk}^{-1}g_{km}^{-1}g_{mn}^{-1}) \rhd h_{kmn}^{-1})\}_\mathrm{p}}\vphantom{\ds\int}\\&&{ \{(g_{jk}^{-1}g_{k\ell}^{-1}g_{\ell m}^{-1})\rhd h_{jkm},(g_{jk}^{-1}g_{k\ell}^{-1}g_{\ell m}^{-1}g_{km}g_{jk}g_{jm}^{-1}g_{m n}^{-1})\rhd h_{jm n}\}_\mathrm{p}(g_{jk}^{-1}g_{k\ell}^{-1}g_{\ell m}^{-1}g_{mn}^{-1})\rhd l_{jkmn}^{-1}}\vphantom{\ds\int}\\&&{ \{ (g_{jk}^{-1}g_{k\ell}^{-1}g_{\ell m}^{-1}g_{mn}^{-1}) \rhd h_{kmn}, (g_{jk}^{-1}g_{k\ell}^{-1}g_{\ell m}^{-1}g_{km}g_{kn}^{-1}) \rhd h_{jkn}  \}^{-1}}\,,\vphantom{\ds\int}
\end{array}
\end{equation}
so that substituting this in the equation \eqref{eq:delta77} one obtains:
\begin{equation}
\resizebox{0.98\hsize}{!}{$
    \begin{array}{lcl}
   \delta_L\big(\tilde{l}_{jk\ell m n}\big)&=&  \delta_L\big((g_{jk}^{-1}g_{k\ell}^{-1}g_{\ell m}^{-1})\rhd l_{jk\ell m}^{-1}\{(g_{jk}^{-1}g_{k\ell}^{-1}g_{\ell m}^{-1})\rhd h_{k\ell m},(g_{jk}^{-1}g_{km}^{-1})\rhd h_{jkm}\}_{\mathrm{p}}^{-1}\vphantom{\ds\int}\\&& {\{(g_{jk}^{-1}g_{k\ell}^{-1}g_{\ell m}^{-1})\rhd h_{k\ell m},(g_{jk}^{-1}g_{km}^{-1})\rhd h_{jkm}(g_{jm}^{-1}g_{m n}^{-1})\rhd h_{jm n} (g_{jk}^{-1}g_{kn}^{-1})\rhd h_{jkn}^{-1} 
(g_{jk}^{-1}g_{km}^{-1}g_{mn}^{-1}) \rhd h_{kmn}^{-1})\}_\mathrm{p}}\vphantom{\ds\int}\\&&{ \{(g_{jk}^{-1}g_{k\ell}^{-1}g_{\ell m}^{-1})\rhd h_{jkm},(g_{jk}^{-1}g_{k\ell}^{-1}g_{\ell m}^{-1}g_{km}g_{jk}g_{jm}^{-1}g_{m n}^{-1})\rhd h_{jm n}\}_\mathrm{p}(g_{jk}^{-1}g_{k\ell}^{-1}g_{\ell m}^{-1}g_{mn}^{-1})\rhd l_{jkmn}^{-1}}\vphantom{\ds\int}\\&&{ \{ (g_{jk}^{-1}g_{k\ell}^{-1}g_{\ell m}^{-1}g_{mn}^{-1}) \rhd h_{kmn}, (g_{jk}^{-1}g_{k\ell}^{-1}g_{\ell m}^{-1}g_{km}g_{kn}^{-1}) \rhd h_{jkn}  \}^{-1}}\vphantom{\ds\int} \\  &&\{(g^{-1}_{jk}g_{k\ell}^{-1}g_{\ell m}^{-1})\rhd h_{k\ell m},(g^{-1}_{jk}g_{km}^{-1}g_{m n}^{-1})\rhd h_{km n}\}_\mathrm{p}(g^{-1}_{jk}g_{k\ell}^{-1}g_{\ell m}^{-1}g_{mn}^{-1})\rhd l_{k\ell mn}^{-1}\vphantom{\ds\int}\\ &&{\{(g_{jk}^{-1}g_{k\ell}^{-1}g_{\ell m}^{-1}g_{mn}^{-1})\rhd h_{k\ell n} (g_{jk}^{-1}g_{k\ell}^{-1}g_{\ell m}^{-1}g_{mn}^{-1})\rhd h_{\ell m n}, (g_{jk}^{-1}g_{kn}^{-1})\rhd h_{jkn} \}_{\mathrm{p}}}\vphantom{\ds\int}\\&&{(g_{jk}^{-1}g_{k\ell}^{-1}g_{\ell m}^{-1}g_{mn}^{-1})\rhd  l_{jk\ell n} }{\big((g_{jk}^{-1}g_{k\ell}^{-1}g_{\ell m}^{-1}g_{mn}^{-1})\rhd h_{j\ell n}\big) \rhd' \{(g_{jk}^{-1}g_{k\ell}^{-1}g_{\ell m}^{-1}g_{mn}^{-1})\rhd h_{\ell m n},  (g_{jk}^{-1}g_{k\ell}^{-1})\rhd h_{jk\ell} \}_{\mathrm{p}}^{-1}}\vphantom{\ds\int}\\
\end{array}$}
\end{equation}
Using the identity \eqref{eq:id02}, and after the cancellation of terms, one obtains:
\begin{equation}
\resizebox{0.98\hsize}{!}{$
    \begin{array}{lcl}
\delta_L\big(\tilde{l}_{jk\ell m n}\big)&=& \delta_L\big({\big((g_{jk}^{-1}g_{k\ell}^{-1}g_{\ell m}^{-1} )\rhd h_{jkm}\big)\rhd' \{(g_{jk}^{-1}g_{k\ell}^{-1}g_{\ell m}^{-1})\rhd h_{k\ell m},(g_{jm}^{-1}g_{m n}^{-1})\rhd h_{jm n} (g_{jk}^{-1}g_{kn}^{-1})\rhd h_{jkn}^{-1} 
(g_{jk}^{-1}g_{km}^{-1}g_{mn}^{-1}) \rhd h_{kmn}^{-1})\}_\mathrm{p}}\vphantom{\ds\int}\\&&{ \{(g_{jk}^{-1}g_{k\ell}^{-1}g_{\ell m}^{-1})\rhd h_{jkm},(g_{jk}^{-1}g_{k\ell}^{-1}g_{\ell m}^{-1}g_{km}g_{jk}g_{jm}^{-1}g_{m n}^{-1})\rhd h_{jm n}\}_\mathrm{p}(g_{jk}^{-1}g_{k\ell}^{-1}g_{\ell m}^{-1}g_{mn}^{-1})\rhd l_{jkmn}^{-1}}\vphantom{\ds\int}\\&&{ \{ (g_{jk}^{-1}g_{k\ell}^{-1}g_{\ell m}^{-1}g_{mn}^{-1}) \rhd h_{kmn}, (g_{jk}^{-1}g_{k\ell}^{-1}g_{\ell m}^{-1}g_{km}g_{kn}^{-1}) \rhd h_{jkn}  \}^{-1}} \vphantom{\ds\int}\\  &&\{(g^{-1}_{jk}g_{k\ell}^{-1}g_{\ell m}^{-1})\rhd h_{k\ell m},(g^{-1}_{jk}g_{km}^{-1}g_{m n}^{-1})\rhd h_{km n}\}_\mathrm{p}(g^{-1}_{jk}g_{k\ell}^{-1}g_{\ell m}^{-1}g_{mn}^{-1})\rhd l_{k\ell mn}^{-1}\vphantom{\ds\int}\\ &&{\{(g_{jk}^{-1}g_{k\ell}^{-1}g_{\ell m}^{-1}g_{mn}^{-1})\rhd h_{k\ell n} (g_{jk}^{-1}g_{k\ell}^{-1}g_{\ell m}^{-1}g_{mn}^{-1})\rhd h_{\ell m n}, (g_{jk}^{-1}g_{kn}^{-1})\rhd h_{jkn} \}_{\mathrm{p}}}\vphantom{\ds\int}\\&&{(g_{jk}^{-1}g_{k\ell}^{-1}g_{\ell m}^{-1}g_{mn}^{-1})\rhd  l_{jk\ell n} }{\big((g_{jk}^{-1}g_{k\ell}^{-1}g_{\ell m}^{-1}g_{mn}^{-1})\rhd h_{j\ell n}\big) \rhd' \{(g_{jk}^{-1}g_{k\ell}^{-1}g_{\ell m}^{-1}g_{mn}^{-1})\rhd h_{\ell m n},  (g_{jk}^{-1}g_{k\ell}^{-1})\rhd h_{jk\ell} \}_{\mathrm{p}}^{-1}}\vphantom{\ds\int}\\&& {(g_{jk}^{-1}g_{k\ell}^{-1}g_{\ell m}^{-1}g_{mn}^{-1})\rhd l_{j\ell mn}}{\{(g_{jk}^{-1}g_{k\ell}^{-1}g_{\ell m}^{-1})\rhd h_{j\ell m}(g_{jk}^{-1}g_{k\ell}^{-1})\rhd h_{jk\ell},(g_{jm}^{-1}g_{mn}^{-1 })\rhd h_{jmn}\}_{\mathrm{p}}^{-1}}(g_{jk}^{-1}g_{k\ell}^{-1}g_{\ell m}^{-1})\rhd l_{jk\ell m}^{-1} \big)\,.\vphantom{\ds\int}
    \end{array}$}
\end{equation}
Commuting the elements, one obtains:
\begin{equation}
\resizebox{0.98\hsize}{!}{$
    \begin{array}{lcl}
\delta_L\big(\tilde{l}_{jk\ell m n}\big)&=& \delta_L\big((g_{jk}^{-1}g_{k\ell}^{-1}g_{\ell m}^{-1})\rhd l_{jk\ell m}^{-1} {\big((g_{jk}^{-1}g_{k\ell}^{-1}g_{\ell m}^{-1} )\rhd h_{jkm}\big)\rhd' \{(g_{jk}^{-1}g_{k\ell}^{-1}g_{\ell m}^{-1})\rhd h_{k\ell m},(g_{jm}^{-1}g_{m n}^{-1})\rhd h_{jm n} (g_{jk}^{-1}g_{kn}^{-1})\rhd h_{jkn}^{-1} 
(g_{jk}^{-1}g_{km}^{-1}g_{mn}^{-1}) \rhd h_{kmn}^{-1})\}_\mathrm{p}}\vphantom{\ds\int}\\ &&\left((g_{jk}^{-1}g_{k\ell}^{-1}g_{\ell m}^{-1})\rhd h_{jkm}\,(g_{jk}^{-1}g_{k\ell}^{-1}g_{\ell m}^{-1}g_{km}g_{jk}g_{jm}^{-1}g_{m n}^{-1})\rhd h_{jm n}(g_{jk}^{-1}g_{k\ell}^{-1}g_{\ell m}^{-1}g_{km}g_{kn}^{-1}) \rhd h_{jkn}  \}^{-1} (g_{jk}^{-1}g_{k\ell}^{-1}g_{\ell m}^{-1}g_{mn}^{-1}) \rhd h_{kmn}\right) \rhd'\vphantom{\ds\int}\\&& \{(g^{-1}_{jk}g_{k\ell}^{-1}g_{\ell m}^{-1})\rhd h_{k\ell m},(g^{-1}_{jk}g_{km}^{-1}g_{m n}^{-1})\rhd h_{km n}\}_\mathrm{p}{ \{(g_{jk}^{-1}g_{k\ell}^{-1}g_{\ell m}^{-1})\rhd h_{jkm},(g_{jk}^{-1}g_{k\ell}^{-1}g_{\ell m}^{-1}g_{km}g_{jk}g_{jm}^{-1}g_{m n}^{-1})\rhd h_{jm n}\}_\mathrm{p}(g_{jk}^{-1}g_{k\ell}^{-1}g_{\ell m}^{-1}g_{mn}^{-1})\rhd l_{jkmn}^{-1}}\vphantom{\ds\int}\\&&{ \{ (g_{jk}^{-1}g_{k\ell}^{-1}g_{\ell m}^{-1}g_{mn}^{-1}) \rhd h_{kmn}, (g_{jk}^{-1}g_{k\ell}^{-1}g_{\ell m}^{-1}g_{km}g_{kn}^{-1}) \rhd h_{jkn}  \}^{-1}} (g^{-1}_{jk}g_{k\ell}^{-1}g_{\ell m}^{-1}g_{mn}^{-1})\rhd l_{k\ell mn}^{-1}\vphantom{\ds\int}\\ &&{\{(g_{jk}^{-1}g_{k\ell}^{-1}g_{\ell m}^{-1}g_{mn}^{-1})\rhd h_{k\ell n} (g_{jk}^{-1}g_{k\ell}^{-1}g_{\ell m}^{-1}g_{mn}^{-1})\rhd h_{\ell m n}, (g_{jk}^{-1}g_{kn}^{-1})\rhd h_{jkn} \}_{\mathrm{p}}}\vphantom{\ds\int}\\&&{(g_{jk}^{-1}g_{k\ell}^{-1}g_{\ell m}^{-1}g_{mn}^{-1})\rhd  l_{jk\ell n} }{\big((g_{jk}^{-1}g_{k\ell}^{-1}g_{\ell m}^{-1}g_{mn}^{-1})\rhd h_{j\ell n}\big) \rhd' \{(g_{jk}^{-1}g_{k\ell}^{-1}g_{\ell m}^{-1}g_{mn}^{-1})\rhd h_{\ell m n},  (g_{jk}^{-1}g_{k\ell}^{-1})\rhd h_{jk\ell} \}_{\mathrm{p}}^{-1}}\vphantom{\ds\int}\\
\end{array}$}
\end{equation}
Using the identities \eqref{eq:id02} and \eqref{eq:id06} one can write,
\begin{equation}
\resizebox{0.98\hsize}{!}{$
\begin{array}{lcl}
&&\big((g_{jk}^{-1}g_{k\ell}^{-1}g_{\ell m}^{-1} )\rhd h_{jkm}\big)\rhd' \{(g_{jk}^{-1}g_{k\ell}^{-1}g_{\ell m}^{-1})\rhd h_{k\ell m},(g_{jm}^{-1}g_{m n}^{-1})\rhd h_{jm n} (g_{jk}^{-1}g_{kn}^{-1})\rhd h_{jkn}^{-1} 
(g_{jk}^{-1}g_{km}^{-1}g_{mn}^{-1}) \rhd h_{kmn}^{-1})\}_\mathrm{p}\vphantom{\ds\int}\\ &&\left((g_{jk}^{-1}g_{k\ell}^{-1}g_{\ell m}^{-1})\rhd h_{jkm}\,(g_{jk}^{-1}g_{k\ell}^{-1}g_{\ell m}^{-1}g_{km}g_{jk}g_{jm}^{-1}g_{m n}^{-1})\rhd h_{jm n}(g_{jk}^{-1}g_{k\ell}^{-1}g_{\ell m}^{-1}g_{km}g_{kn}^{-1}) \rhd h_{jkn}  \}^{-1} (g_{jk}^{-1}g_{k\ell}^{-1}g_{\ell m}^{-1}g_{mn}^{-1}) \rhd h_{kmn}\right) \rhd'\vphantom{\ds\int}\\&& \{(g^{-1}_{jk}g_{k\ell}^{-1}g_{\ell m}^{-1})\rhd h_{k\ell m},(g^{-1}_{jk}g_{km}^{-1}g_{m n}^{-1})\rhd h_{km n}\}_\mathrm{p} 
\vphantom{\ds\int}\\&=&  {\big((g_{jk}^{-1}g_{k\ell}^{-1}g_{\ell m}^{-1} )\rhd h_{jkm}\big)\rhd' \{(g_{jk}^{-1}g_{k\ell}^{-1}g_{\ell m}^{-1})\rhd h_{k\ell m},(g_{jm}^{-1}g_{m n}^{-1})\rhd h_{jm n} (g_{jk}^{-1}g_{kn}^{-1})\rhd h_{jkn}^{-1} 
)\}_\mathrm{p}}\vphantom{\ds\int}\,,
\end{array}$}
\end{equation}
which leads to the result:
\begin{equation}
\resizebox{0.98\hsize}{!}{$
    %
$}
\end{equation}
Then, using again the identity \eqref{eq:id02},
\begin{equation}
\resizebox{0.98\hsize}{!}{$
\begin{array}{lcl}
&&\{(g_{jk}^{-1}g_{k\ell}^{-1}g_{\ell m}^{-1})\rhd h_{k\ell m},(g_{jm}^{-1}g_{m n}^{-1})\rhd h_{jm n} (g_{jk}^{-1}g_{kn}^{-1})\rhd h_{jkn}^{-1} 
)\}_\mathrm{p}\vphantom{\ds\int}\\&=&\{(g_{jk}^{-1}g_{k\ell}^{-1}g_{\ell m}^{-1})\rhd h_{k\ell m},(g_{jm}^{-1}g_{m n}^{-1})\rhd h_{jm n} \}_\mathrm{p} \big((g_{jk}^{-1}g_{k\ell}^{-1}g_{\ell m}^{-1}g_{km}g_{jk}g_{jm}^{-1}g_{mn}^{-1})\rhd h_{jmn} \big)\rhd' \{(g_{jk}^{-1}g_{k\ell}^{-1}g_{\ell m}^{-1})\rhd h_{k\ell m},(g_{jk}^{-1}g_{kn}^{-1})\rhd h_{jkn}^{-1} 
)\}_\mathrm{p}\vphantom{\ds\int}
\end{array}%
$}
\end{equation}
Commuting the elements and using the identity \eqref{prop1} one obtains:
\begin{equation}
\resizebox{0.98\hsize}{!}{$
    %
$}
\end{equation}
Using the identity \eqref{eq:id01} one writes,
\begin{equation}
\resizebox{0.98\hsize}{!}{$
    %
$}
\end{equation}    
Commuting the elements one obtains:
\begin{equation}
\resizebox{0.98\hsize}{!}{$
    %
$}
\end{equation}    
Using the identity \eqref{eq:id06} one obtains:
\begin{equation}
\resizebox{0.98\hsize}{!}{$
    %
$}
\end{equation}    
Then, by calculating the inverse of the Peiffer lifting by using the identity \eqref{eq:id03},
\begin{equation}
\resizebox{0.98\hsize}{!}{$
\begin{array}{lcl}
&&\{ (g_{jk}^{-1}g_{k\ell}^{-1}g_{\ell m}^{-1}g_{mn}^{-1}) \rhd h_{kmn} (g_{jk}^{-1}g_{k\ell}^{-1}g_{\ell m}^{-1})\rhd h_{k\ell m}, (g_{jk}^{-1}g_{kn}^{-1}) \rhd h_{jkn}  \}\}_\mathrm{p}^{-1}\vphantom{\ds\int}\\&=&\big((g_{jk}^{-1}g_{k\ell}^{-1}g_{\ell m}^{-1}g_{mn}^{-1}) \rhd h_{kmn} (g_{jk}^{-1}g_{k\ell}^{-1}g_{\ell m}^{-1})\rhd h_{k\ell m}\big)\rhd \{  (g_{jk}^{-1}g_{k\ell}^{-1}g_{\ell m}^{-1})\rhd h_{k\ell m}^{-1}(g_{jk}^{-1}g_{k\ell}^{-1}g_{\ell m}^{-1}g_{mn}^{-1}) \rhd h_{kmn}^{-1}, (g_{jk}^{-1}g_{k\ell}^{-1}g_{\ell m}^{-1}g_{mn}^{-1}) \rhd h_{jkn}  \}\}_\mathrm{p},\vphantom{\ds\int}
\end{array}$}
\end{equation}
and commuting the elements one has:
\begin{equation}
\resizebox{0.98\hsize}{!}{$
    %
$}
\end{equation}    
Then, using the identity \eqref{eq:id01} one obtains:
\begin{equation}
\resizebox{0.98\hsize}{!}{$
    %
$}
\end{equation}  
Then, using the identity \eqref{eq:id01} one can write,
\begin{equation}
\resizebox{0.98\hsize}{!}{$
%
$}
\end{equation}
which leads to the equation:
\begin{equation}\label{eq:delta9}
\resizebox{0.98\hsize}{!}{$
    %
$}
\end{equation}  
Then, using the identity \eqref{eq:id03} one writes
\begin{equation}
\resizebox{0.98\hsize}{!}{$
%
$}
\end{equation}
and substituting this expression in the equation \eqref{eq:delta9} obtains:
\begin{equation}
\resizebox{0.98\hsize}{!}{$
    %
$}
\end{equation}  
By commuting the elements, one obtains:
\begin{equation}
\resizebox{0.98\hsize}{!}{$
    %
$}
\end{equation}  
Then, using the identity \eqref{eq:id01} one obtains,
\begin{equation}
\resizebox{0.98\hsize}{!}{$
%
$}
\end{equation} 
Finally, using the identity  \eqref{identitet2},
\begin{equation}
\resizebox{0.98\hsize}{!}{$
\big((g_{jk}^{-1}g_{k\ell}^{-1}g_{\ell m}^{-1}g_{mn}^{-1})\rhd h_{jmn}\big)\rhd'(g_{jk}^{-1}g_{k\ell}^{-1}g_{\ell m}^{-1})\rhd l_{jk\ell m}^{-1}=(g_{jk}^{-1}g_{k\ell}^{-1}g_{\ell m}^{-1})\rhd l_{jk\ell m}^{-1}\{(g_{jk}^{-1}g_{k\ell}^{-1}g_{\ell m}^{-1})\rhd \delta(l_{jk\ell m}),(g_{jk}^{-1}g_{k\ell}^{-1}g_{\ell m}^{-1}g_{mn}^{-1 })\rhd h_{jmn}\}_{\mathrm{p}}$}\,,
\end{equation}
one obtains:
\begin{equation}
\resizebox{0.98\hsize}{!}{$
    \begin{array}{lcl}
 \delta_L\big(\tilde{l}_{jk\ell m n}\big)&=& \delta_L\big(\big((g_{jk}^{-1}g_{k\ell}^{-1}g_{\ell m}^{-1}g_{mn}^{-1})\rhd h_{jmn}\big)\rhd'(g_{jk}^{-1}g_{k\ell}^{-1}g_{\ell m}^{-1})\rhd l_{jk\ell m}^{-1}  {(g_{jk}^{-1}g_{k\ell}^{-1}g_{\ell m}^{-1}g_{mn}^{-1})\rhd l_{jkmn}^{-1}} (g_{jk}^{-1}g_{k\ell}^{-1}g_{\ell m}^{-1}g_{mn}^{-1}) \rhd h_{jkn}  \rhd' (g^{-1}_{jk}g_{k\ell}^{-1}g_{\ell m}^{-1}g_{mn}^{-1})\rhd l_{k\ell mn}^{-1} \vphantom{\ds\int}\\&&{(g_{jk}^{-1}g_{k\ell}^{-1}g_{\ell m}^{-1}g_{mn}^{-1})\rhd  l_{jk\ell n} }{\big((g_{jk}^{-1}g_{k\ell}^{-1}g_{\ell m}^{-1}g_{mn}^{-1})\rhd h_{j\ell n}\big) \rhd' \{(g_{jk}^{-1}g_{k\ell}^{-1}g_{\ell m}^{-1}g_{mn}^{-1})\rhd h_{\ell m n},  (g_{jk}^{-1}g_{k\ell}^{-1})\rhd h_{jk\ell} \}_{\mathrm{p}}^{-1}} {(g_{jk}^{-1}g_{k\ell}^{-1}g_{\ell m}^{-1}g_{mn}^{-1})\rhd l_{j\ell mn}}\big)\,.\vphantom{\ds\int}
    \end{array}$}
\end{equation}
Acting on the expression with $g_{mn}g_{\ell m}g_{k\ell}g_{jk}$ gives the $\delta$-function:
\begin{equation}
    \begin{aligned}
 \delta_L\big(\tilde{l}_{jk\ell m n}\big)={ l_{j\ell mn}^{-1}}{ h_{j\ell n}\rhd' \{ h_{\ell m n},  (g_{mn}g_{\ell m})\rhd h_{jk\ell} \}_{\mathrm{p}}}{  l_{jk\ell n}^{-1}}h_{jkn}  \rhd' l_{k\ell mn} { l_{jkmn}} h_{jmn}\rhd'(g_{mn}\rhd l_{jk\ell m})\,.
    \end{aligned}
\end{equation}
This expression is exactly equal to $\delta_L(l_{jk\ell mn})$. This concludes the proof.

\end{document}